\documentclass[11pt, double, phd]{osudiss-2} 

\usepackage{amsmath, amssymb, amsthm} 
\usepackage{graphics, graphicx} 
\usepackage{epsfig} 
\usepackage{lipsum} 
\usepackage{rotating} 
\usepackage{overpic} 
\usepackage{tikz} 
\usepackage{multirow} 
\usepackage{subfig} 
\usepackage{bm} 
\usepackage{booktabs} 
\usepackage{CJKutf8} 
\usepackage{hyperref} 
\usepackage{color}
\usepackage{slashed}
\usepackage{pifont}

\usepackage[notlof,notlot,nottoc]{tocbibind} 

\input glyphtounicode
\definecolor{excelgreen}{RGB}{0,176,80}
\definecolor{excelpurple}{RGB}{112,48,160}
\definecolor{excelblue}{RGB}{31,73,125}
\definecolor{excelred}{RGB}{192,80,77}
\definecolor{originpurple}{RGB}{132,0,255}

\usepackage{bookmark} 
\hypersetup{colorlinks=true,linkcolor=blue, breaklinks} 
\RequirePackage[hyperpageref]{backref} 
   \renewcommand*{\backref}[1]{}  
   \renewcommand*{\backrefalt}[4]{
      \ifcase #1
         Not cited.
      \or
         Cited on page #2.
      \else
         Cited on pages #2.
      \fi}
\usepackage[all]{hypcap}

\usepackage{cite} 

\newcommand{\as}{\alpha_s}
\newcommand{\ul}[1]{\underline{#1}}
\newcommand{\ord}[1]{\mathcal{O}\left(#1\right)}
\newcommand{\Tr}{\mathrm{Tr}}
\newcommand{\ubar}[1]{\overline{U}_{#1}}
\newcommand{\vbar}[1]{\overline{V}_{#1}}
\newcommand{\psibar}{\overline{\psi}}

\newcommand{\bra}[1]{\left\langle #1 \right|}
\newcommand{\ket}[1]{\left| #1 \right\rangle}

\def\eq#1{{Eq.~(\ref{#1})}}
\def\fig#1{{Fig.~\ref{#1}}}
\def\peq#1{{(\ref{#1})}}

\title{Transverse Spin and Classical Gluon Fields: Combining Two Perspectives on Hadronic Structure} 
\author{Matthew D. Sievert}
\advisorname{Yuri Kovchegov}
\degree{Doctor of Philosophy} 
\member{Eric Braaten}
\member{Michael Lisa}
\member{Samir Mathur}
\authordegrees{M.S. , B.S. , B.A.}
\graduationyear{2014}
\unit{Graduate Program in Physics} 

\begin{document}

\frontmatter

\begin{abstract}

In recent decades, the spin and transverse momentum of quarks and gluons were found to play integral roles in the structure of the nucleon.  Simultaneously, the onset of gluon saturation in hadrons and nuclei at high energies was predicted to result in a new state of matter dominated by classical gluon fields.  Understanding both of these contributions to hadronic structure is essential for current and future collider phenomenology.  In this Dissertation, we study the combined effects of transverse spin and gluon saturation using the Glauber-Gribov-Mueller / McLerran-Venugopalan model of a heavy nucleus in the quasi-classical approximation.  We investigate the use of a transversely-polarized projectile as a probe of the saturated gluon fields in the nucleus, finding that the transverse spin asymmetry of produced particles couples to the component of the gluon fields which is antisymmetric under both time reversal and charge conjugation.  We also analyze the effects of saturation on the transverse spin asymmetry (Sivers function) of quarks within the wave function of the nucleus, finding that gluon saturation preferentially generates the asymmetry through the orbital angular momentum of the nucleons, together with nuclear shadowing.

\end{abstract}

\dedication
 {
\textit{To all the teachers, professors, and mentors who have invested so much of themselves and their passion in me and my future.  I can see further than I ever thought possible because I stand on the shoulders of giants.}} 
\begin{acknowledgments}

First and foremost, I owe great thanks to my advisor Yuri Kovchegov.  Your amazing ability to explain both the technical details and the underlying physics principles makes you a phenomenal physicist, teacher, and advisor.  You have been equally willing to help me wrestle with the big questions and to get your hands dirty helping me search for my minus signs and factors of 2, and you have somehow always found time for me in your busy schedule.  Learning a new field of study together with you has taught me by example what it means to be a physicist.

To James and Veronica and Logan, to Jer and Kristie and Chloe, to Aidan, and especially to my love Jesse -- I cannot thank you enough for keeping me sane, for giving me perspective, and for believing in me when I have had nothing left but doubts.  Your strength and love make me a whole person and fill my life with meaning and purpose.  

To Adam and David and Alexis, and to all my fellow karateka -- thank you beyond measure for pushing me, for challenging me, for accepting and respecting me, for inspiring and competing with me, for not hesitating to throw the punch when I leave my guard down.  You help me to hold myself to a higher standard and to continually seek perfection of character.

And to Mom and Dad -- thank you for everything.  For fostering my curiosity and my love of learning.  For teaching me division with sunflower seeds.  For pushing me to work harder in school when I just wanted to play video games.  For teaching me tolerance and responsibility.  For celebrating my successes and forgiving my failures.  For loving and accepting me and respecting me as an adult, and for always believing in me.  I love you from the bottom of my heart.

\end{acknowledgments}

\begin{vita}
\dateitem{August 5, 1984}{Born---Lexington, VA}
\dateitem{May, 2002}{Chesterfield County Mathematics and Science High School at Clover Hill, Midlothian, VA}
\dateitem{May, 2006}{B.S. Physics, B.A. Spanish, Virginia Commonwealth University, Richmond, VA}
\dateitem{August, 2007}{M.S. Physics, Virginia Commonwealth University, Richmond, VA}
\dateitem{2007 - 2008}{William A. Fowler Graduate Fellow in Physics, The Ohio State University, Columbus, OH}
\dateitem{2007 - 2008 ; 2010 - present}{Susan L. Huntington Distinguished University Fellow, The Ohio State University, Columbus, OH}
\end{vita}


\begin{publist}
\pubitem{Y. V. Kovchegov and M. D. Sievert, ``A New Mechanism for Generating a Single Transverse Spin Asymmetry'', Phys. Rev. {\bf D86}, 034028 (2012)}
\pubitem{Y. V. Kovchegov and M. D. Sievert, ``Single Spin Asymmetry in High Energy QCD'', Int. J. Mod. Phys. Conf. Ser. {\bf 20}, 177 (2012)}
\pubitem{M. D. Sievert, ``A New Mechanism for Generating a Single Transverse Spin Asymmetry'', Nucl. Phys. {\bf A 904-905}, 833c (2013)}
\pubitem{S. J. Brodsky, D. S. Hwang, Y. V. Kovchegov, I. Schmidt, and M. D. Sievert, ``Single-Spin Asymmetries in Semi-Inclusive Deep Inelastic Scattering and Drell-Yan Processes'', Phys. Rev. {\bf D88} 014032 (2013)}
\pubitem{M. D. Sievert, ``Single-Spin Asymmetries in Semi-Inclusive Deep Inelastic Scattering and Drell-Yan Processes'', Int. J. Mod. Phys. Conf. Ser. {\bf 25} 1460015 (2014)}
\pubitem{Y. V. Kovchegov and M. D. Sievert, ``Sivers Function in the Quasi-Classical Approximation'', Phys. Rev. {\bf D89}, 054035 (2014)
}
\end{publist}

\begin{fieldsstudy}
\majorfield{Physics}
 \begin{studieslist}
  \studyitem{Gluon saturation in high-energy QCD}{Yuri V. Kovchegov}
 \end{studieslist}
\end{fieldsstudy}

\vspace{1cm}
This research is sponsored in part by the U.S. Department of Energy under Grant No. DE-SC0004286.

\tableofcontents

\clearpage 
\listoffigures 

\clearpage 
\listoftables 


\mainmatter 

\chapter{Overview}
\label{chap-Overview}


Since the proton was first discovered a century ago, the quest to understand its quantum structure has revealed layer upon layer of new mysteries.  Each successive breakthrough we make in resolving its subcomponents and their properties opens the door to new puzzles which fundamentally challenge our understanding of hadronic physics.  The quark model \cite{GellMann:1964nj, Zweig:1981pd, Zweig:1964jf} organized the baffling zoo of hadronic particles in the proton's family tree into a logical ``periodic table,'' but the absence of free quarks in nature thwarted early attempts to formulate them into a quantum theory.  The advent of the theory of quantum chromodynamics \cite{Gross:1973id, Politzer:1973fx} with the remarkable property of asymptotic freedom bridged this gap, reconciling long-standing challenges to the validity of quantum field theory itself, but it also implied a fundamental disconnect between the elementary quarks and gluons of the theory and the hadronic degrees of freedom seen in nature.  Each step in the development of our understanding has revealed another hidden layer of structure and complexity, driving still further advances in both theory and experiment.  

In this Dissertation, we present an analysis of two such layers of structure which have been developed largely in parallel over the last 30 years, together with our recent work to understand their interconnection.  The first of these layers involves the complex spin-orbit and spin-spin coupling mechanisms that translate the spin of hadrons like the nucleon (proton or neutron) into the spins and orbital motion of its quark and gluon constituents.  This paradigm has been driven largely by past and current experiments which continue to demonstrate the importance of spin and transverse momentum to our understanding of hadronic structure.  The second involves the coherent nonlinear interactions that occur in high-energy scattering when the density of quarks and gluons inside the hadron is large.  This saturation paradigm has been driven largely by theoretical considerations which demand that new physical processes take over at high energies and densities in order to preserve the internal consistency of the theory.  Our efforts to combine these two paradigms have resulted in new insights, both into how spin physics can be used as a probe of saturation, as well as how saturation mechanisms can mediate the exchange of spin and transverse momentum.  These insights open the door to future work extending the interplay of spin and saturation, and the analysis presented here represents only a small portion of the active frontiers of research into hadronic structure.  

\section{The Evolving Picture of the Nucleon}
\label{sec-EvolveNucleon}


\subsection{Quark and Gluon Degrees of Freedom}
\label{subsec-QuarkGluon}

\begin{figure}
 \centering
 \includegraphics[width=0.15\textwidth]{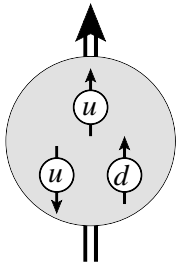}
 \caption{Bookkeeping of the proton's quantum numbers within the quark model.  Two ``up'' quarks carry an electric charge of $+\tfrac{2}{3} e$ each and one ``down'' quark carries a charge of $-\tfrac{1}{3} e$, yielding a total electric charge of $+e$.  One can similarly account for the spin $= \tfrac{1}{2}$ of the proton (in units of $\hbar$) if two of the constituent quarks have their spins aligned with the proton spin, carrying projections $+\tfrac{1}{2}$ each, and one has its spin anti-aligned, carrying projection $-\tfrac{1}{2}$. }
 \label{fig-Const_Quark}
\end{figure}

The \textit{quark model} proposed by Gell-Mann and Zweig in 1964 \cite{GellMann:1964nj, Zweig:1981pd, Zweig:1964jf} explained the relationship between the proton, neutron, and other hadrons based on the charges of their constituent quarks.  In this picture, the proton is composed of two ``up'' quarks with electric charge $+\tfrac{2}{3} e$ and one ``down'' quark with electric charge $-\tfrac{1}{3} e$, where $e$ is the magnitude of the electron charge.  Similarly, the neutron is composed of two ``down'' quarks and one ``up'' quark.  Each constituent quark would have a mass of around $\tfrac{1}{3}$ of the proton mass, on the order of $\sim 300~\mathrm{MeV}$.  Since the postulated quarks are spin-$\tfrac{1}{2}$ fermions, they can also simply account for the total spin $\tfrac{1}{2}$ of the nucleon if two constituent quarks have their spins aligned parallel to the nucleon and one antiparallel, as illustrated in Fig.~\ref{fig-Const_Quark}.  While successful at capturing the relationships between the masses and charges of the hadrons, the quark model was not a quantum theory and did not describe the interactions of the quarks which bind them into hadrons.  


The key to determining the properties of the quark interactions came from the SLAC-MIT experiment \cite{Bloom:1969kc, Breidenbach:1969kd} in 1968, which studied the \textit{deep inelastic scattering} of high-energy electrons on hadronic targets.  The hard electromagnetic scattering was consistent with the nucleon being composed of a number of pointlike constituents collectively called \textit{partons} \cite{Feynman:1969ej}.  The scaling properties observed in the SLAC-MIT experiment \cite{Bjorken:1968dy, Callan:1969uq} confirmed that the charged partons are spin-$\tfrac{1}{2}$ fermions and suggested that their interactions over short times and distances are weak, but strong enough over long times and distances to bind them together into hadrons.  These properties are in stark contrast to those of quantum electrodynamics (QED), whose quantum self-interactions become \textit{strongest} at short distances.  

The insight that the interaction of quarks must possess \textit{asymptotic freedom} at short distances led to the development in 1973 of a full-fledged quantum field theory known as \textit{quantum chromodynamics (QCD)} \cite{Gross:1973id, Politzer:1973fx}.  Like its close cousin quantum electrodynamics, QCD is a quantization of a classical theory of charges and fields, but whereas QED quantizes the linear Maxwell equations, QCD quantizes the highly-nonlinear Yang-Mills equations \cite{Yang:1954ek}.  The Yang-Mills equations are structurally similar to the Maxwell equations, with one essential difference: the field itself is charged and can act as a source for further radiation.  In the quantum analog, this means that the gluon fields of QCD interact with themselves and each other, unlike the photons of QED.  This additional self-interaction is essential to generating asymptotic freedom, and the rigorous development of its quantum origins from the QCD Lagrangian helped put quantum field theory itself on a firm footing.  The intrinsic breakdown of QED and similar theories at short distances (or high energies) had cast fundamental doubt on the validity of quantum field theory itself as a framework for quantum mechanics \cite{Landau:1955aaa, Landau:1956zr}, but the asymptotic freedom embodied by QCD resolved this crisis by providing a theory which is ``UV complete'' -- self-consistent up to arbitrarily high energies.  

The price of asymptotic freedom is that the interactions between quarks and gluons become strongest at \textit{low} energies, such as those relevant for the calculation of the nucleon wave function.  In principle this information is encoded in the QCD Lagrangian, but because of the strong coupling it cannot be calculated perturbatively from the fundamental theory.  This reflects the physics of \textit{quark confinement}: while quarks and gluons are the relevant degrees of freedom at short distances and high energies, the emergent degrees of freedom at long distances and low energies are their bound states: nucleons, pions, and the whole zoo of hadronic particles.  When the wave function of the nucleon is probed by a high-energy projectile as in deep inelastic scattering, the short-distance interactions with the quarks and gluons can be calculated perturbatively, but the observables are always contaminated by nonperturbative, incalculable low-energy quantities.  The bridge between the perturbative and nonperturbative elements of QCD is provided in the form of \textit{factorization theorems} (see, e.g. \cite{Libby:1978qf, Libby:1978bx, Sterman:1994ce, Collins:2011zzd}) which show that the distribution of quarks and gluons within a hadron can be measured with one process and used predictively in another.


\begin{figure}
 \centering
 \includegraphics[width=0.6\textwidth]{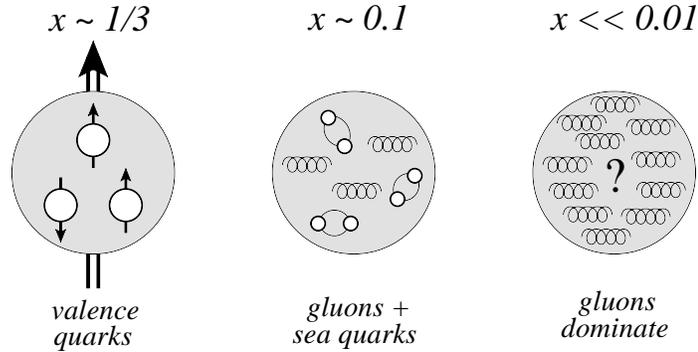}
 \caption{Picture of the nucleon as a collinear distribution of quarks and gluons which changes depending on the kinematics used to probe it.  Pictured here is the variation as the momentum fraction $x$ of the partons changes.  Left panel: At $x \sim \tfrac{1}{3}$, the nucleon is predominantly composed of the three valence quarks which can account for the spin of the nucleon.  Center panel: As $x$ is decreased, the momentum of the nucleon is shared among more of the gluons and ``sea quarks'' produced by radiation and pair production.  Right panel: When $x$ becomes very small, the number of soft gluons sharing a small fraction of the nucleon momentum becomes huge (see also Fig.~\ref{fig-Small_x}).}
 \label{fig-Coll_Distr}
\end{figure}


The picture of the nucleon changes depending on the kinematics of the scattering process used to probe its wave function; in particular, the scaling variable denoted $x$ describes the fraction of the nucleon's longitudinal momentum carried by the partons (see Chapter~\ref{chap-TMD}).  This picture is visualized in Fig.~\ref{fig-Coll_Distr}.  At large $x \sim \tfrac{1}{3}$, the nucleon appears to be composed of three valence quarks sharing the nucleon's momentum, as in the quark model.  As $x$ decreases, the scattering probe becomes less sensitive to the valence quarks and more sensitive to the partons produced by radiation, which tend to share smaller fractions of the nucleon momentum among a larger number of particles.  At very small $x$, the nucleon wave function is dominated by a large number of gluons, each sharing a very small fraction of the total momentum.  This picture of the nucleon as resolved into a collinear beam of quarks and gluons \cite{Bjorken:1968dy, Feynman:1969ej} was immensely successful in explaining the structure observed in deep inelastic scattering experiments over a wide range of kinematics \cite{Aaron:2009aa, Benvenuti:1989rh, Adams:1996gu, Arneodo:1996qe, Whitlow:1991uw}.


\subsection{Spin and Transverse Momentum}
\label{subsec-SpinTransv}

This simple one-dimensional picture of nucleon structure was shattered by groundbreaking experiments that revealed a far more complex role played by spin and partonic transverse momentum.  The European Muon Collaboration performed measurements in 1988 on longitudinally-polarized protons that measured the spin contribution carried by the quarks; in striking contradiction to the naive expectation shown in Figs.~\ref{fig-Const_Quark} and \ref{fig-Coll_Distr}, they found that only ``$14 \pm 9 \pm 21 \%$ of the proton spin is carried by the spin of the quarks'' \cite{Ashman:1987hv}.  This shocking result came to be known as the \textit{proton spin crisis}, and their conclusion that ``the remaining spin must be carried by gluons or orbital angular momentum'' inaugurated a worldwide effort to find the missing angular momentum which continues to this day.  

In general, the spin $\tfrac{1}{2}$ of the nucleon can be decomposed into a part coming from the net polarization $\Delta\Sigma$ of the spin-$\tfrac{1}{2}$ quarks, a part from the net polarization $\Delta G$ of the spin-1 gluons, and the orbital angular momentum $L_q$ and $L_g$ of quarks and gluons, respectively \cite{Jaffe:1989jz}:
\begin{align}
 \label{eq-JaffeManohar}
 \frac{1}{2} = \frac{1}{2} \Delta\Sigma + \Delta G + L_q + L_g .
\end{align}
Modern determinations \cite{deFlorian:2009vb} of the quark spin $\Delta\Sigma$ estimate the contribution at about $\Delta\Sigma \approx 0.25$, corresponding to $\sim 25\%$ of the proton spin, and very recent measurements \cite{deFlorian:2014yva, Adare:2014hsq} of the gluon polarization $\Delta G$ find a contribution of about $\Delta G \approx 0.20$, corresponding to $\sim 40 \%$ of the proton spin.  Although these values are not very precisely determined, they still leave significant room for the contribution of the angular momentum of quarks $L_q$ and gluons $L_g$ coming from their transverse motion.

\begin{figure}[ht]
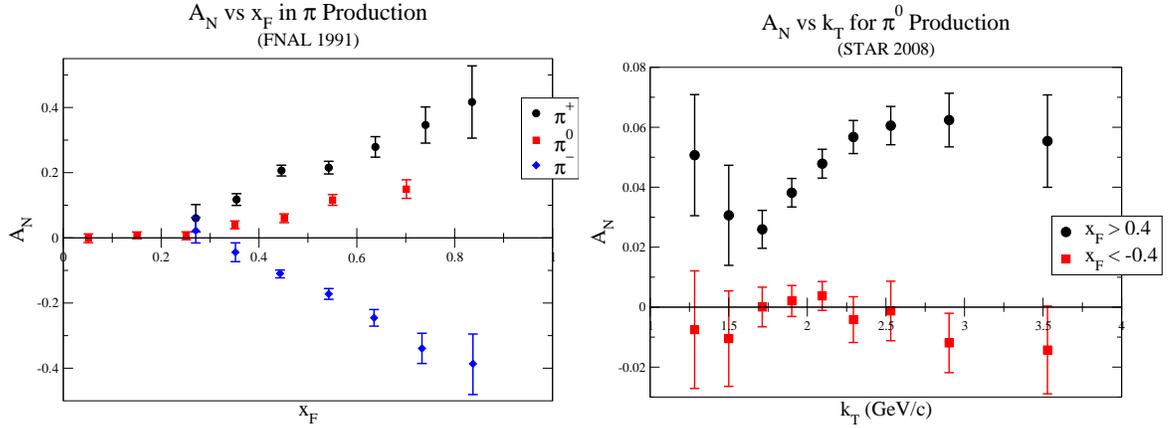

\centering
 \begin{tabular}{c c}
 \includegraphics[width=0.5\textwidth]{FNAL_91_xF_Improved2.eps}
 \includegraphics[width=0.5\textwidth]{STAR_08_pT.eps}
 \end{tabular}
  \caption{Experimental data on the pion single transverse spin asymmetry 
    $A_N$ as a function of $x$ reported by E581 and E704
    collaborations (graphically reconstructed from
    \cite{Adams:1991cs}, shown in the left panel) for $0.7 \le k_T \le
    2.0$~GeV/c, and as a function of the pion transverse momentum
    $k_T$ collected by the STAR collaboration \cite{Abelev:2008qb}
    (right panel).}
 \label{fig-Experimental Data}  
\end{figure}

A similarly dramatic revelation occurred with regard to the role of transverse polarization.  An expectation dating as far back as Feynman \cite{Feynman:1973xc} predicted that transverse polarization effects are universally suppressed at high energies (see also \cite{Kane:1978nd} and the discussion in \cite{Collins:2011zzd}).  But in 1991 when the E581/E704 Collaborations at Fermi National Accelerator Laboratory (FNAL) measured the single transverse spin asymmetry (STSA) produced in collisions of transversely-polarized protons, they found strikingly large asymmetries of up to 30-40\% \cite{Adams:1991rw,Adams:1991cs,Adams:1991ru,Adams:1995gg,Bravar:1996ki,Adams:1994yu}.  More recently, the  PHENIX, STAR, and BRAHMS collaborations at the Relativistic Heavy Ion Collider (RHIC) have studied transverse spin asymmetries at a higher energy and over a wide kinematic range \cite{Abelev:2007ii,Nogach:2006gm,Adler:2005in,Lee:2007zzh}.  The data they have presented \cite{Abelev:2007ii,Nogach:2006gm,Adler:2005in} confirmed and extended the Fermilab results, and also indicated a non-monotonic dependence of STSA on the transverse momentum of the produced hadron \cite{Abelev:2008qb,Wei:2011nt}.  Some of these plots are reproduced in Fig.~\ref{fig-Experimental Data}; for a useful review of STSA physics see \cite{D'Alesio:2007jt}.

\begin{figure}
 \centering
 \includegraphics[width=0.20\textwidth]{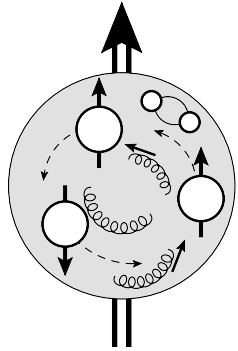}
 \caption{Modern picture of the nucleon's ``spin budget'' allocated among the spins and orbital angular momentum of quarks and gluons.}
 \label{fig-TMD_Distr}
\end{figure}

The implications of these experiments have led to a considerable broadening of our picture of nucleon structure.  The nucleon's ``spin budget'' (Fig.~\ref{fig-TMD_Distr}) is distributed among the polarizations and transverse orbital motion of quarks and gluons, which can manifest themselves in a number of spin and momentum correlations that are observable in hadronic collisions.  The generalization to a three-dimensional picture of nucleon structure, including transverse momentum and its correlations with the nucleon and parton spins is made possible by more inclusive versions of factorization theorems (see, e.g. \cite{Ralston:1979ys, Collins:1981uk, Collins:1985ue, Collins:1984kg, Collins:2011zzd}).  While this formalism brings the intricate spin-orbit and spin-spin correlations in the nucleon within reach of theory and experiment, it also opens the door to still further challenges and opportunities.  The universality of the parton distributions, for example, is lost by the inclusion of transverse momentum \cite{Collins:2002kn}, and in some processes the properties of the target and projectile seem to be entangled so that they cannot even be separately defined \cite{Gamberg:2010tj, Rogers:2010dm}.  As with the other historic advances in our understanding of nucleon structure, the inclusion of spin and transverse momentum answers our questions with still more questions.  We discuss the spin and transverse momentum paradigm of hadronic structure in detail in Chapter~\ref{chap-TMD}.


\subsection{Gluon Saturation}
\label{subsec-Saturation}

While the distributions of quarks and gluons within the nucleon are fundamentally low-energy quantities that are not perturbatively calculable in QCD, the manner in which these distributions \textit{evolve} with external parameters are.  These \textit{quantum evolution equations} describe how QCD radiative processes modify the distribution of partons, such as by the collinear emission of gluons or pair-production of quarks  \cite{Gribov:1972ri, Altarelli:1977zs, Dokshitzer:1977sg}.  The momentum fraction $x$ of partons resolved in a hadronic collision is kinematically related to the center-of-mass energy at which the collision occurs.  As the energy is increased, $x$ is decreased, and the nucleon structure becomes dominated more and more by gluons as shown in the right panel of Fig.~\ref{fig-Coll_Distr}.  


\begin{figure}
 \centering
 \includegraphics[width=0.6\textwidth]{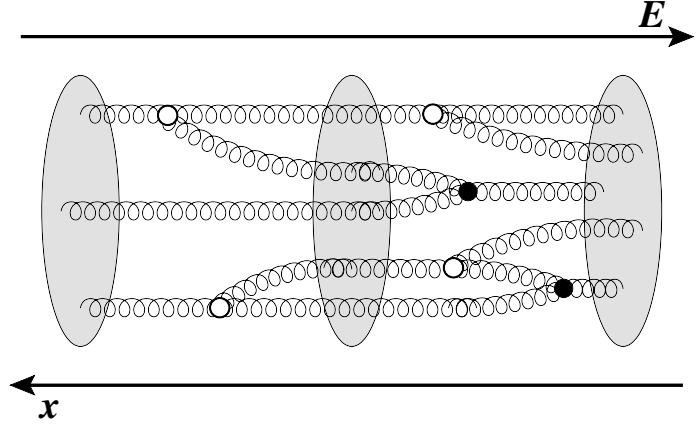}
 \caption{Evolution of the gluon distribution as the energy $E$ is increased or the momentum fraction $x$ is decreased.  At lower energies, the effect of increasing energy is to induce gluon bremsstrahlung (open vertices) which rapidly increases the gluon density.  At higher energies when the gluon density is large, gluon fusion (solid vertices) begins to compete with bremsstrahlung.  The result is a \textit{saturation} of the gluon density at high energies.  }
 \label{fig-Small_x}
\end{figure}


The origin of these additional gluons is through the quantum evolution \cite{Kuraev:1977fs, Balitsky:1978ic} of the pre-existing partons, which radiate gluons through bremsstrahlung when the energy is increased (Fig.~\ref{fig-Small_x}, open circles).  The effect of this bremsstrahlung is to increase the density of gluons in the nucleon as a function of the collision energy.  But as the energy is increased further, these new gluons themselves undergo bremsstrahlung, increasing the gluon density even faster.  The result of this rapid proliferation of gluons is to make the nucleon more and more opaque to a high-energy projectile as the energy continues to increase.  If the successive gluon radiation were to continue unabated, it would lead to scattering probabilities greater than $100\%$ at high energies, which would violate the fundamental principle of unitarity in quantum mechanics \cite{Froissart:1961ux, Martin:1969a, Lukaszuk:1967zz}.

Unitarity is a non-negotiable element of quantum field theory, and its preservation demands that the structure of the nucleon at high energies must be very different from the low-energy structure that gives rise to this exploding gluon density.  At sufficiently high density the spatial distribution of gluons begins to overlap, leaving no available space for additional independent bremsstrahlung.  When this happens, the nonlinear interactions between the radiated gluons become important, including their recombination through gluon fusion (Fig.~\ref{fig-Small_x}, solid circles).  Gluon fusion tends to decrease the gluon density and so competes with the increase due to bremsstrahlung; when these effects become comparable, the gluon density \textit{saturates} and cuts off the growth with energy \cite{Gribov:1984tu, Mueller:1986wy}.  The result is that the structure of the nucleon at very high energies is driven overwhelmingly by gluon fields with high densities and occupation numbers; in this saturation regime the dominant degrees of freedom are the \textit{classical} gluon fields obtained from the Yang-Mills equations \cite{Kovchegov:1997pc}.

The onset of saturation at high densities, although not yet observed unambiguously in experiment, is necessary for the consistency of nucleon structure with the unitarity of quantum field theory.  When these high-density effects are included into the small-$x$ quantum evolution equations, the resulting nonlinear evolution explicitly preserves unitarity \cite{Balitsky:1996ub, Kovchegov:1999yj, Jalilian-Marian:1997jx, Jalilian-Marian:1997gr, Jalilian-Marian:1997dw, Iancu:2001ad, Iancu:2000hn}.  The same limit of high gluon densities can also be obtained in a different physical system: a heavy nucleus with a large number of nucleons \cite{McLerran:1994vd , McLerran:1993ni , McLerran:1993ka, Glauber:1955qq, Glauber:1970jm, Franco:1965wi, Gribov:1968jf, Gribov:1968gs, Mueller:1989st}.  This approach provides a much more direct route to obtaining the physical properties of the saturation regime, without the need to solve the complicated small-$x$ quantum evolution equations.  We discuss the onset of saturation in a heavy nucleus and the emergence of classical gluon fields in detail in Chapter~\ref{chap-CGC}.


\section{Organization of this Document}
\label{sec-Organization}

This document is structured as follows.  In Chapter~\ref{chap-TMD} we introduce the formalism for describing the transverse-momentum-dependent distributions of quarks and gluons in a hadron, in the context of deep inelastic scattering.  After reviewing the foundational knowledge in the field, we present original work analyzing one of these transverse-momentum-dependent parton distribution functions in detail in Sec.~\ref{sec-Sivers}.  In Chapter~\ref{chap-CGC} we lay out the saturation formalism relevant for the resummation of high-density effects at high energies, emphasizing the role of multiple scattering and the emergence of classical gluon fields as the relevant degrees of freedom.  Then we present original work analyzing the interplay between these paradigms in two ways, demonstrating how each provides the tools to acquire new understanding of the other.  In Chapter~\ref{chap-odderon} we show how transverse spin can be used as a tool to study novel aspects of the saturation regime by accessing a different component of the dense gluon fields.  Then in Chapter~\ref{chap-MVspin} we use the saturation formalism to elucidate a new relationship between the transverse-momentum-dependent quark distributions and their orbital angular momentum.  We conclude with a brief outlook in Chapter~\ref{chap-Outlook} which summarizes the main results presented here and proposes the next logical steps which can be taken to extend them.


\subsection{Notation and Conventions}
\label{subsec-Notation}

Wherever possible, we choose our conventions to correspond with those of \cite{Kovchegov:2012mbw}.  As is standard, we work in natural units in which $\hbar = c = 1$.  

We work with the QCD Lagrangian in the form
\begin{align}
 \label{LQCD}
 \mathcal{L}_{QCD} &= \sum_f \psibar_i^{(f)} (i \slashed{D} - m_f)_{i j} \psi_j^{(f)} - \frac{1}{4}
 F_{\mu \nu}^a F^{\mu \nu a} 
 \\ \nonumber
 D_\mu &\equiv \partial_\mu - i g T^a A_\mu^a
 \\ \nonumber
 F^{\mu \nu a} &\equiv \partial^\mu A^{\nu a} - \partial^\nu A^{\mu a} + g f^{a b c} A^{\mu b} A^{\nu c}
\end{align}
where $f$ denotes the quark flavor, $(i, j)$ are color indices in the fundamental representation of the gauge group, $(a, b, c)$ are color indices in the adjoint representation of the gauge group, and $f^{a b c}$ are the structure constants.  The gauge group of QCD is $SU(3)$ with generators $T^a$ in the fundamental representation related to the Gell-Mann matrices $\lambda^a$ by $T^a = \tfrac{1}{2} \lambda^a$; however, it is convenient to work in the more general gauge group $SU(N_c)$ with $N_c$ the number of colors.  This makes the group-theoretical structure of the formulas more explicit and allows us to take advantage of 't Hooft's large-$N_c$ limit \cite{'tHooft:1973jz} in Chapter~\ref{chap-odderon}.  The sum of the squares of the generators of the group is known as the quadratic Casimir invariant \cite{Casimir:1931a}.  In the fundamental representation this is given by
\begin{align}
 \label{Casimir}
 (T^a)_{i j} \, (T^a)_{j k} \equiv C_F \, \delta_{i k} = \frac{N_c^2 - 1}{2 N_c} \, \delta_{i k}
\end{align}
which gives $C_F = \tfrac{4}{3}$ for $SU(3)$.  The quadratic Casimir in the adjoint representation is just equal to the number of colors, $C_A = N_C$.


\begin{figure}
 \centering
 \includegraphics[width=0.3\textwidth]{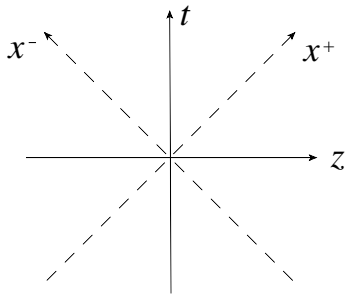}
 \caption{Illustration of the light-cone coordinates $x^+$ and $x^-$ defined in \eqref{lcmetric}.}
 \label{fig-LC_coords}
\end{figure}


In high-energy interactions with particles traveling very close to the speed of light, it is convenient to work in \textit{light-cone coordinates} as in Fig.~\ref{fig-LC_coords} which are linear combinations of $t$ and $z$.  We choose the normalization of the coordinates and the corresponding metric tensor to be
\begin{align}
\label{lcmetric}
 x^\pm & \equiv x^0 \pm x^3 = t \pm z
 \\ \nonumber
 p^\mu q_\mu &\equiv p^\mu g_{\mu\nu} q^\nu = \frac{1}{2} p^+ q^- + \frac{1}{2} p^- q^+ - \ul{p} \cdot \ul{q}
\end{align}
where vectors which are underlined represent transverse vectors in the $x y-$plane.  Thus we label the components of a 4-vector as
\begin{align}
 \label{vecdefn}
 p^\mu &= \left( p^+ \, , \, p^- \, , \, \ul{p} \right) \\ \nonumber
 \ul{p} &\equiv \left( p_\bot^1 \, , \, p_\bot^2 \right)
\end{align}
and use the subscript $\bot$ to denote the component of a transverse vector.  For the magnitude of a transverse vector, we use the subscript $T$, as in
\begin{align}
 \label{tvecdefn}
 x_T^2 &\equiv (x_\bot^1)^2 + (x_\bot^2)^2 \\ \nonumber
 |x-y|_T^2 &= (x_\bot^1 - y_\bot^1)^2 + (x_\bot^2 - y_\bot^2)^2 .
\end{align}
The antisymmetric (``cross'') product of two transverse vectors is given in terms of the two-dimensional antisymmetric Levi-Civita tensor as
\begin{align}
 \label{tcross}
 \ul{x} \times \ul{y} \equiv x_\bot^i \epsilon^{i j} y_\bot^j \equiv x_\bot^1 y_\bot^2 - 
 x_\bot^2 y_\bot^1 .
\end{align}
The normalization of the light-cone coordinates and the corresponding metric in \eqref{lcmetric} is not universal; often one uses coordinates normalized by $\tfrac{1}{\sqrt{2}}$ so that there are no factors of $1/2$ appearing in the metric.  At times the factors of 2 in our convention will be a convenience and at other times an inconvenience.

For brevity's sake, in many of the formulas we will abbreviate the notation for multi-dimensional integration.  For integration over a transverse variable we use the notation $d^2 x \equiv d x_\bot^1 \, d x_\bot^2$ which is commonplace, and sometimes we wish to integrate over transverse variables and one light-cone coordinate, for which we will use the notation
\begin{align}
 \label{mattintegral}
 d^{2+} p &\equiv d^2 p \, d p^+ \\ \nonumber
 d^{2-} x &\equiv d^2 x \, d x^-
\end{align}
and similarly for the arguments of Dirac delta functions like $\delta^{2+}(p-q)$ or $\delta^{2-}(x-y)$.

A common variable in high-energy collisions is the \textit{rapidity} $y$ of an on-shell particle with momentum $k^\mu$, denoted
\begin{align}
 \label{rapidity}
 y \equiv \frac{1}{2} \ln \frac{k^+}{k^-} = \ln \frac{k^+}{\sqrt{k_T^2 + m^2}} = \ln \frac{\sqrt{k_T^2 + m^2}}{k^-},
\end{align}
where we have made use of the on-shell condition $k^+ k^- - k_T^2 = m^2$.  In terms of rapidity, the invariant differential cross-section is written as
\begin{align}
 \label{rapcross}
 E_k \frac{d\sigma}{d^3 k} = \frac{d\sigma}{d^2 k \, d y} = k^+ \frac{d\sigma}{d^2 k \, d k^+} = 
 k^- \frac{d\sigma}{d^2 k \, d k^-} .
\end{align}

When identifying the power-counting in large kinematic quantities such as the center-of-mass energy squared $s$, we will often compare them to smaller quantities such as masses or transverse momenta, which we generically assume to be of the order of the masses unless otherwise specified.  We denote the order of such quantities generically as $\bot$, as in ``$s \gg \bot^2 \gg \bot^4 / s$.''

Finally, in some cases we will perform calculations using ordinary Feynman perturbation theory through the use of Feynman diagrams.  In others, particularly in the high-energy limit, it is more convenient to calculate observables through the use of \textit{light-cone perturbation theory (LCPT)}.  LCPT corresponds to time-ordered perturbation theory in the ordinary sense, but with the light-cone coordinate $x^+$ playing the role of time (for a particle moving along the $x^+$ axis with high energy).  The natural ordering of high-energy processes in $x^+$ makes this a useful tool, and we use the conventions of \cite{Kovchegov:2012mbw} unless otherwise specified.

\chapter{The Spin and Transverse Momentum Paradigm of Hadronic Structure}
\label{chap-TMD}


The most natural way to measure the structure of the nucleon is through its interaction with an electromagnetic probe.  If an incident electron scatters off the nucleon by exchanging a spacelike virtual photon, the injected momentum $q^\mu$ may cause the nucleon to break up inelastically into a multi-hadron final state: $\ell + N \rightarrow \ell' + X$.  The uncertainty principle suggests that the larger the momentum transfer, the smaller the distance scales on which the charge distribution is measured.  Thus, for the case of \textit{deep inelastic scattering (DIS)} when the photon virtuality $Q^2 = - q_\mu q^\mu$ is large, the electron interacts with the charged sub-components of the nucleon.  In this way, deep inelastic scattering gives a direct window into the substructure of the nucleon, and it played a key role in the historical establishment of QCD as the fundamental theory of the strong nuclear force (see, e.g. \cite{Gross:1973id, Politzer:1973fx, Poucher:1973rg} and \cite{Friedman:1991ip} for a review).

The simplest application of this idea is in the form of the \textit{parton model} \cite{Feynman:1973xc, Bjorken:1969ja}, in which the virtual photon is assumed to interact with a single charged sub-component of the nucleon.  These pointlike constituents are collectively referred to as ``partons,'' and it is possible to determine whether they are bosons or fermions from the form of the resulting cross-section \cite{Callan:1969uq}.  In this way, ``partons'' were identified as charged fermions (quarks) and their associated QCD gauge field (gluons).  

The parton model reduces the process of deep inelastic scattering to a fixed, short-distance electromagnetic vertex \cite{Mueller:1970fa, Mueller:1981sg} that effectively ``measures'' the distribution of quarks within the nucleon wave function.  This allows the nucleon structure to be parameterized in terms of \textit{parton distribution functions} (PDF's) which resolve the nucleon into a collinear beam of quarks and gluons.  Although significantly modified by QCD corrections, this essential concept survives in the form of \textit{collinear factorization} (see, e.g. \cite{Libby:1978qf, Libby:1978bx, Bassetto:1982ma} and the textbooks \cite{Sterman:1994ce, Collins:2011zzd}).  Once suitably generalized, these parton distribution functions can be shown to be intrinsic, universal properties of the nucleon which can be measured in one experiment and then used predictively in another.  These theoretical cornerstones form the basis of the \textit{collinear paradigm} of hadronic structure, which has been immensely successful in describing experimental data over many orders of magnitude in $Q^2$ \cite{Aaron:2009aa, Benvenuti:1989rh, Adams:1996gu, Arneodo:1996qe, Whitlow:1991uw}.

A series of revolutionary experiments in the early 90's (see \cite{Ashman:1987hv, Adams:1991rw, Adams:1991cs, Adams:1991ru, Adams:1995gg}, among others) revealed the surprising importance of spin and transverse-momentum dynamics, which are not captured in the collinear paradigm.  Differential observables that describe the azimuthal distribution of produced hadrons provide another external ``lever'' to parameterize the nucleon's substructure.  One such differential observable is \textit{semi-inclusive deep inelastic scattering (SIDIS)}, in which both the scattered lepton and one final-state hadron are tagged: $\ell + N \rightarrow \ell' + h + X$.  Like the fully-inclusive case, SIDIS couples to parton distribution functions, but now with the transverse momentum of the active parton accessible through the momentum of the tagged hadron $h$.  These \textit{transverse-momentum-dependent parton distribution functions (TMD's)} are capable of resolving both the transverse and longitudinal structure of the nucleon, adding another dimension to the parameter space of parton distributions.  As with the collinear case, the naive parton model is heavily modified by QCD corrections into the modern form of \textit{TMD factorization} (see, e.g. \cite{Ralston:1979ys, Collins:1981uk, Collins:1985ue, Collins:1984kg, Collins:2011zzd}) .

The inclusion of dependence on the nucleon spin, parton spin, and parton transverse momentum permits a wealth of  new spin-momentum correlations in the SIDIS cross-section and in the TMD's.  The potential for such nontrivial spin-orbit and spin-spin coupling in the nucleon mirrors the role of the fine and hyperfine structure in atomic physics.  Among these spin correlations, transverse spin plays a distinct role from that of longitudinal spin (helicity) because it introduces a preferred direction in the transverse plane.  For a singly-polarized process such as SIDIS on a polarized nucleon, rotational invariance uniquely couples the transverse spin direction of the nucleon to the transverse momentum direction of the produced hadron, resulting in a \textit{single transverse spin asymmetry (STSA)} of the detected hadrons.  Furthermore, the discrete symmetries of QCD: charge-conjugation, parity, and time-reversal ($C$, $P$, and $T$), strongly constrain the form of spin correlations such as STSA - and the partonic mechanisms that can generate them.  The combination $PT$ of parity and time reversal (sometimes called ``naive time reversal'') plays a particularly important role in the origin of STSA because this symmetry operation flips the direction of the transverse spin, while leaving the momenta of the colliding particles unchanged; thus STSA is odd under $PT$.

The partonic analog of STSA is a TMD called the \textit{Sivers function} \cite{Sivers:1989cc}, a correlation between the transverse spin of the nucleon and the transverse orbital momentum of its partons.  Like STSA, the Sivers function is by definition odd under ``naive time reversal''; but unlike STSA, the Sivers function is interpreted as an intrinsic property of the nucleon.  Since the nucleon wave function is an eigenstate of the QCD Hamiltonian, it must be $PT$-even, so it is natural to expect that the Sivers function is identically zero.  However, this expectation is wrong because of an essential difference between the collinear and transverse-momentum paradigms; while the collinear PDF's can be simply written as densities of partons, in TMD's the parton densities are fundamentally entangled with initial- and final-state interactions.  The nontrivial role played by initial- and final-state interactions permits a nonzero Sivers function, and time reversal further implies that the Sivers function measured in processes with final-state interactions is equal in magnitude and opposite in sign from processes with initial-state interactions \cite{Collins:2002kn}.  When TMD factorization holds, this gives rise to the prediction of an exact sign reversal between the Sivers functions in SIDIS and its mirror image: the Drell-Yan process.  The diagrammatic mechanism of this sign reversal can be examined within the context of a simple model for the nucleon \cite{Brodsky:2013oya}.  Although the model verifies the predicted sign flip at leading order in the hard scale $Q^2$, it suggests that violations may occur at subleading orders.  This model calculation also motivates a physical interpretation of the sign-flip relation in terms of the ``QCD lensing'' of quarks due to initial- or final-state interactions \cite{Brodsky:2002cx, Brodsky:2002rv}.


 \section{Spin- and Momentum-Dependent Observables}

 \subsection{Inclusive and Semi-Inclusive Deep Inelastic Scattering}
 \label{subsec-SIDIS}

\begin{figure} [tb]
 \centering
 \includegraphics[width=0.7\textwidth]{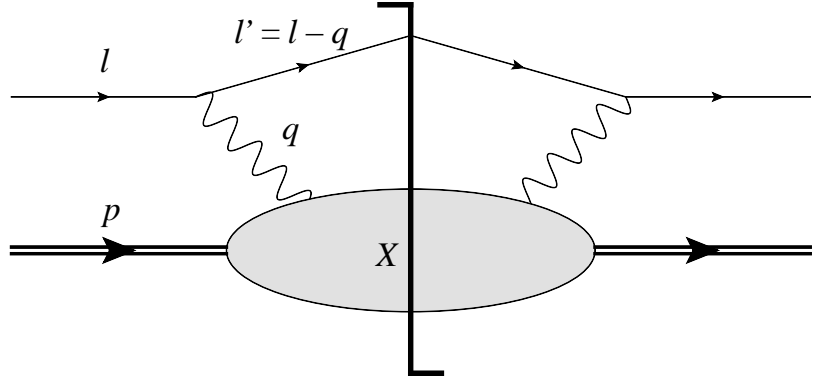}
 \caption{The kinematics of deep inelastic scattering.  An incident electron with momentum $\ell$ scatters electromagnetically off a nucleon with momentum $p$ by the exchange of a virtual photon with momentum $q$.  If the injected momentum $q$ is large enough, it can shatter the proton into a multi-hadron final state, generically denoted $\ket{X}$.}
 \label{figDIS1}
\end{figure}

Because of asymptotic freedom and quark confinement \cite{Gross:1973id, Politzer:1973fx}, one of the most direct ways to experimentally probe the quark and gluon degrees of freedom is through deep inelastic scattering (DIS).  In this process, a high energy lepton, which we take here to be an electron, scatters off a nucleon by exchanging a spacelike virtual photon $q$.  The deep inelastic regime occurs when the magnitude of the momentum transfer $Q^2 \equiv -q^2$ is large; then the injected hard momentum scatters perturbatively off the wave function of the nucleon.  Experimentally, measuring the recoil of the electron fixes the kinematics of the scattering, and one may consider the \textit{inclusive DIS} process $e^- + N \rightarrow e^- + X$ containing any final hadronic state $X$ or the \textit{semi-inclusive DIS} process in which one final-state hadronic particle $h$ is tagged $e^- + N \rightarrow e^- + h + X$.  The kinematics of DIS are illustrated in Fig.~\ref{figDIS1}.
Two conventional choices of Lorentz invariants used to characterize the kinematics of DIS are the photon's virtuality $Q^2$ and the Bjorken variable $x_{B}$:
\begin{eqnarray}
Q^2 &\equiv& - q^2 \label{eq-virtuality}
\\ \label{eq-Bjorken}
x_{B} &\equiv& \frac{Q^2}{2 p \cdot q} .
\end{eqnarray}
We will work in the \textit{Bjorken limit} of the kinematics, in which the virtuality is large compared to the typical scale of the nucleon $Q^2 \gg m_N^2$ and $x_B$ is held fixed and $\ord{1}$.  The total invariant mass $m_X$ of the hadronic final state $X$ is equal to the photon/nucleon center-of-mass energy $\sqrt{s}$; since this quantity is positive-definite, we have
\begin{equation}
\label{eq-missing mass}
s \equiv (p+q)^2 = m_N^2 - Q^2 + 2 p \cdot q \approx Q^2 \left( \frac{1}{x_{Bj}} - 1 \right) \geq 0
\end{equation}
which gives the kinematic range of $x_B$ as $0 \leq x_{B} \leq 1$ in the Bjorken limit.

The emission and propagation of the virtual photon can be expressed using perturbative QED, and, without knowing anything about the structure of the nucleon, we can express its interaction with the virtual photon as a transition matrix element of the electromagnetic current:
\begin{align}
 \label{DIS-ampl}
 i \mathcal{M} = \frac{i e^2}{q^2} \ubar{}(\ell') \gamma_\mu U(\ell) \,
 \bra{X} J^\mu (0) \ket{p S}.
\end{align}
Here $\ell' \equiv \ell - q$ is the momentum of the outgoing electron, and the electromagnetic current of a system of quarks with various flavors $f$ is
\begin{align}
 \label{current1}
 J^\mu(x) = Z_f \psibar(x) \gamma^\mu \psi(x)
\end{align}
where $Z_f$ is the electric charge of quark flavor $f$ in units of the electron charge $e$, and a sum over such flavors is implied.

By squaring the amplitude \eqref{DIS-ampl} and including the associated flux factors and phase-space integrals, we can compute the invariant cross-section in the ordinary way \cite{Peskin:1995ev, Kovchegov:2012mbw}, obtaining the standard result
\begin{align}
 \label{SIDIS1}
 E_{\ell'} \frac{d\sigma}{d^3 \ell'} = \frac{d\sigma}{d^2 \ell' d y_\ell '} = \frac{\alpha_{EM}^2}{E_\ell Q^4} L_{\mu \nu} W^{\mu \nu},
\end{align}
where the leptonic tensor for an unpolarized lepton is
\begin{align}
 \label{SIDIS2}
 L_{\mu\nu} = 2 \ell_\mu \ell'_\nu + 2 \ell'_\mu \ell_\nu - Q^2 g_{\mu \nu}
\end{align}
and the hadronic tensor $W^{\mu\nu}$ expresses the interaction with the nucleon in terms of a current-current correlation function:
\begin{align}
 \label{SIDIS3}
 W^{\mu\nu} &\equiv \frac{1}{4\pi m_N} \int d^4r \, e^{i q \cdot r} \bra{p S} J^\mu (r) J^\nu (0) \ket{p S} \\ \nonumber
 &= \frac{1}{4\pi m_N} \sum_X (2\pi)^4 \delta^4 (p_X - p - q) \bra{p S} J^\mu(0) \ket{X} \bra{X} J^\nu(0) \ket{p S} .
\end{align}

\begin{figure}
 \centering
 \includegraphics[width=0.7\textwidth]{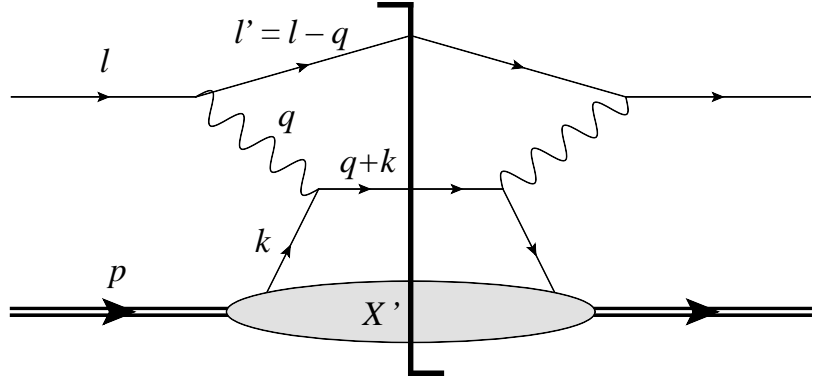}
 \caption{A parton model picture of DIS; the virtual photon resolves an individual quark in the nucleon wave function.  The outgoing quark fragments into a collimated jet of hadrons which can be observed and tagged in the final state.}
 \label{figDIS2}
\end{figure}

When the photon virtuality $Q^2$ is large, the photon resolves an individual parton in the nucleon wave function as shown in Fig.~\ref{figDIS2}.  If we write the state $\ket{X}$ as the product of an active quark with momentum $(q+k)^\mu$ and an arbitrary state $\ket{X'}$ containing the other nucleon remnants, the parton model corresponds to the interaction of the virtual photon with just the $(q+k)$ quark line.

To proceed from here, it is useful to specify a frame for the process in which to work out the kinematics.  Let us for the moment work in the photon-nucleon center-of-mass frame, in which 
\begin{align}
 \label{SIDIS8}
 q^\mu &= \left( -\frac{Q^2}{q^-} , q^- , \ul{0} \right) \\ \nonumber
 p^\mu &= \left( p^+ , \frac{m_N^2}{p^+} , \ul{0} \right).
\end{align}
Momentum conservation, together with the on-shell conditions, fix the values of $k^+ , k^-$ in terms of $k_T$ and various constants:
\begin{align}
 \label{SIDIS9}
 k^+ &= (q+k)^+ - q^+ = \frac{k_T^2 + m_q^2}{q^- + k^-} + \frac{Q^2}{q^-} \sim \ord{Q} \\ \nonumber
 k^- &= p^- - p_X'^- \sim \ord{\frac{\bot^2}{Q}}.
\end{align}
The large (negative) light-cone plus momentum $q^+$ flowing through the virtual photon ensures that $k^+$ is large, and to avoid sending a large invariant mass $\sim k^+ k^-$ through the $t$-channel, $k^-$ must be small.  The dominant kinematic regime thus has the large light-cone momenta $(p^+ , |q^+| ,  q^- , k^+ , p_X'^+ ) \sim\ord{Q}$ and the corresponding small light-cone momenta $ (p^- , k^-, p_X'^- ) \sim\ord{\frac{\bot^2}{Q}}$; the transverse momentum $\ul{k}$ is an intermediate scale, which we take comparable to the masses like $m_N$ and denote as $\bot$ in the power-counting.  The expansion of the kinematics in powers of $\bot^2 / Q^2$ is referred to as the \textit{twist expansion}, with the contributions that are ``leading twist'' unsuppressed at large $Q^2$.  The twist expansion can be given a precise operator meaning through the use of the operator product expansion (\cite{Balitsky:1988fi, Balitsky:1987bk}, see also \cite{Peskin:1995ev}).

The on-shell condition for the final-state quark $(k+q)$ allows us to relate the longitudinal momentum $k^+$ of the active quark to the observable parameter $x_B$ to leading-twist accuracy:
\begin{align}
 \label{SIDISkin1}
 0 &= (k+q)^2 - m_q^2  \\ \nonumber
 &\approx 2 \left(\frac{1}{2} k^+ q^- \right) - Q^2 \\ \nonumber
 &= 2 \left(\frac{k^+}{p^+}\right) p \cdot q - Q^2 \\ \nonumber 0
 &= 2 (p \cdot q) \left( \frac{k^+}{p^+} - x_B \right).
\end{align}
Thus in the partonic picture, the longitudinal momentum fraction of the active quark (sometimes known as ``Feynman x'' $x_F$) is equal to the Bjorken variable:
\begin{align}
 \label{SIDISkin2}
 x_F \equiv \frac{k^+}{p^+} = x_B \equiv \frac{Q^2}{2 p \cdot q} \equiv x .
\end{align}
Combining \eqref{SIDIS8}, \eqref{SIDIS9} , \eqref{SIDISkin1} , and \eqref{SIDISkin2} , we can summarize the kinematics to leading twist as
\begin{align}
 \label{SIDISkin3}
 q^\mu &= \left( - x p^+ , q^- , \ul{0} \right) \\ \nonumber
 p^\mu &= \left( p^+ , 0 , \ul{0} \right) \\ \nonumber
 k^\mu &= \left( x p^+ , 0 , \ul{k} \right) \\ \nonumber
 (p_X')^\mu = (p-k)^\mu &= \left( (1-x) p^+ , 0 , -\ul{k} \right) .
\end{align}


To evaluate the parton model contribution to \eqref{SIDIS3}, we need to simplify the matrix elements of the electromagnetic current by contracting the current operator with the active quark state $\bra{(q+k) \sigma}$ having momentum $(q+k)^\mu$ and spin $\sigma$:
\begin{align}
 \label{SIDIS4}
 \bra{X} J^\nu(0) \ket{pS} &= \bigg( \bra{(q+k) \sigma} \otimes \bra{X'} \bigg) \: \psibar(0) \: (Z_f \gamma^\nu) \: \psi(0) \: \ket{pS} 
 \\ \nonumber &=
 \bra{X'} b_{(q+k) , \sigma} \:\: \psibar(0) \: (Z_f \gamma^\nu) \: \psi(0) \: \ket{pS} 
 \\ \nonumber &=
 \bra{X'} \: \bigg\{ b_{(q+k) , \sigma} \: , \: \psibar(0) \bigg\} \: (Z_f \gamma^\nu) \: \psi(0) \: \ket{pS} 
 \\ \nonumber &=
 Z_f \ubar{\sigma}(q+k) \, \gamma^\nu \: \bra{X'} \psi(0) \ket{p S} ,
\end{align}
where we have used the definition of the quark field $\bar\psi$ and the anticommutation relations
\begin{align}
 \label{anticomm}
 \psibar(y) &= \int \frac{d^{2+}\ell}{2(2\pi)^3 \ell^+} \sum_\tau \left( b_{\ell \tau}^\dagger
 \ubar{\tau}(\ell) \, e^{i \ell \cdot y} + d_{\ell \tau} \vbar{\tau}(\ell) \, e^{-i \ell \cdot y} 
 \right)
 \\ \nonumber
 \bigg\{ b_{p , \sigma} \: , \: b^\dagger_{\ell , \tau} \bigg\} &= 2(2\pi)^3 p^+ \delta_{\sigma \tau}
 \delta^{2+} (\ell-p)
 \\ \nonumber
 \bigg\{ b_{p , \sigma} \: , \: d_{\ell , \tau} \bigg\} &= 0 ,
\end{align}
with $b (b^\dagger)$ and $d (d^\dagger)$ the annihilation (creation) operators for quarks and antiquarks, respectively.  

Separating out the phase space of the full hadronic final state $\ket{X}$ into the phase space of the active quark $(q+k)$ and the other remnants $\ket{X'}$ gives
\begin{align}
 \label{SIDIS5}
 p_X &\rightarrow p_X' + q + k \\ \nonumber
 \sum_X &\rightarrow \int \frac{d^{2-}(q+k)}{2(2\pi)^3 (q+k)^-} \,\sum_\sigma \, \sum_{X'},
\end{align}
which, together with \eqref{anticomm}, can be used to rewrite \eqref{SIDIS3} as
\begin{align}
 \label{SIDIS6}
  W^{\mu\nu} &= \frac{Z_f^2}{4\pi m_N} \sum_{X'} \int\frac{d^{2-}(q+k)}{2(2\pi)^3 (q+k)^-} (2\pi)^4 \delta^4 (p_X' + (q+k) - p - q)
	\\ \nonumber &\times
	\bra{p S} \psibar(0) \ket{X'} \gamma^\mu \: \bigg( \sum_\sigma U_{\sigma}(q+k) \ubar{\sigma}(q+k) \bigg) \: \gamma^\nu \bra{X'} \psi(0) \ket{p S}.
\end{align}
Performing the sum over the spins of the active quark yields
\begin{align}
 \label{SIDIS7}
 \Gamma^{\mu\nu} &\equiv \sum_{\sigma} \gamma^\mu U_{\sigma}(q+k) \ubar{\sigma}(q+k) \gamma^\nu =
 \gamma^\mu (\slashed{q}+\slashed{k}+m_q) \gamma^\nu.
 \\ \nonumber &\approx
 \frac{1}{2} q^- \left( \gamma^\mu \gamma^+ \gamma^\nu \right)
\end{align}
where we have simplified the expression using the leading-twist kinematics \eqref{SIDISkin3} .
Specifying the kinematics is also important in order to discern the effect of the momentum-conserving delta function in \eqref{SIDIS6}.  Three of the momentum components - say, the transverse and plus components - are conserved in the ordinary fashion.  The fourth component - say, the minus component - is fixed by the on-shell conditions in terms of the other three.  Thus we write the delta function as
\begin{align}
 \label{SIDIS10}
 \delta^4(p_X' + (q+k) - p - q) &= 2 \delta^{2+}(p_X' - p + k) \; \delta( p_X'^- + (q+k)^- - p^- - q^-) \\ \nonumber
 &\approx 2 \delta^{2+}(p_X' - p + k) \; \delta((q+k)^- - q^-),
\end{align}
where we have simplified the minus-component delta function using the power-counting of \eqref{SIDIS9} and the factor of 2 comes from the choice of metric.

Plugging this back into \eqref{SIDIS6} gives
\begin{align}
 \label{SIDIS11}
   W^{\mu\nu} = \frac{Z_f^2}{4\pi m_N} \sum_{X'} \int&\frac{d^{2-}(q+k)}{2(2\pi)^3 (q+k)^-}
	 \left[ 2 (2\pi)^4 \delta^{2+} (p_X' - p + k) \; \delta((q+k)^- - q^-) \right]	
	 \\ \nonumber &\times
	\bra{p S} \psibar(0) \ket{X'} \Gamma^{\mu\nu} \bra{X'} \psi(0) \ket{p S}.
\end{align}
Integrating $d(q+k)^-$ picks up the delta function and sets $k^- \approx 0$, while the other delta function can be rewritten as a Fourier integral:
\begin{align}
 \label{SIDIS12}
  W^{\mu\nu} &= \frac{Z_f^2}{4\pi m_N} \sum_{X'} \int\frac{d^2k}{2(2\pi)^3 q^-} 
	\left[(2\pi) \int d^{2-}r \, e^{i (k-p+p_X') \cdot r} \right]
	\\ \nonumber &\times
	\bra{p S} \psibar(0) \ket{X'} \Gamma^{\mu\nu} \bra{X'} \psi(0) \ket{p S},
\end{align}
where the dot product in the Fourier exponent represents $v \cdot r = \frac{1}{2} v^+ r^- - \ul{v}\cdot\ul{r}$ since there is no $r^+$ component in the integration.  We can further absorb the Fourier factor into the translation of one of the quark operators:
\begin{align}
 \label{SIDIS13}
 e^{-i (p-p_X')\cdot r} \bra{X'} \psi(0) \ket{pS} &= \bra{X'} \: e^{+ i \hat{P} \cdot r} \: \psi(0) \: 
 e^{- i \hat{P} \cdot r} \: \ket{pS}
 \\ \nonumber &=
 \bra{X'} \psi(r) \ket{pS}_{r^+ = 0} ,
\end{align}
where $\hat{P}$ denotes the momentum operator, and the plus coordinate of the operator has not been shifted away from zero because there is no $r^+$ term in the Fourier factor.  This allows us to write
\begin{align}
 \label{SIDIS14}
 W^{\mu\nu} &= \frac{Z_f^2}{2 m_N} \frac{1}{2(2\pi)^3 q^-} \int d^2k \, d^{2-}r \, e^{i k \cdot r}
 \left[ \sum_{X'} \bra{p S} \psibar_j(0) \ket{X'} \bra{X'} \psi_i(r) \ket{p S}_{r^+ = 0} \right] \Gamma^{\mu\nu}_{j i} 
 \\ \nonumber &=
 \frac{Z_f^2}{2 m_N} \frac{1}{q^-} \int d^2k \, \left[ \frac{1}{2(2\pi)^3} \int d^{2-}r \, e^{i k \cdot r} \bra{p S} \psibar_j(0) \, \psi_i(r) \ket{p S} \right]_{r^+ = 0} \Gamma^{\mu\nu}_{j i} 
 \\ \nonumber &\equiv
 \frac{Z_f^2}{2 m_N} \frac{1}{q^-} \int d^2k \, \Tr\left[\Phi(x,\ul{k}) \Gamma^{\mu\nu} \right],
\end{align}
where we have used completeness to sum over the unrestricted states $\ket{X'}$.  The quantity 
\begin{align}
 \label{phiprelim}  
 \Phi_{i j} (x,\ul{k}) \equiv \frac{1}{2(2\pi)^3} \int d^{2-}r \, e^{i k \cdot r} \bra{p S} \psibar_j(0) \, \psi_i(r) \ket{p S}_{r^+ = 0}
\end{align}
is a quark-quark correlation function in the nucleon state, reflecting the transverse-momentum dependence of the ``quark content'' of the nucleon.

\begin{figure}
 \centering
 \includegraphics[width=0.7\textwidth]{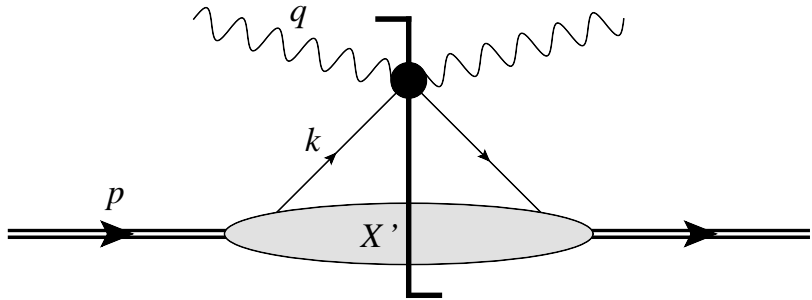}
 \caption{Illustration of the hadronic tensor \eqref{SIDIS16}.  The interaction of the virtual photon with the nucleon has been reduced to a short-distance vertex $\Gamma^{\mu\nu}$ which probes the quark degrees of freedom $\Phi$ in the nucleon.}
 \label{figDIS3}
\end{figure}

Using \eqref{SIDIS7} for the vertex $\Gamma^{\mu\nu}$ gives the final expression for the hadronic tensor as
\begin{align}
 \label{SIDIS16}
 W^{\mu\nu} = \frac{Z_f^2}{4 m_N} \int d^2 k \, \Tr\left[\Phi(x,\ul{k}) \gamma^\mu \gamma^+ \gamma^\nu \right],
\end{align}
which shows that the interaction of the virtual photon with the nucleon has been reduced to an effective short-distance vertex as illustrated in Fig.~\ref{figDIS3}.  This concept of an effective operator description at short distances can be formalized into the \textit{operator product expansion}, as has been done for inclusive DIS \cite{Balitsky:1988fi, Balitsky:1987bk}.
Back-substituting \eqref{SIDIS16} into \eqref{SIDIS1} lets us write the cross-section as
\begin{align}
 \label{SIDIS17}
 \frac{d\sigma}{d^2\ell' dy_\ell'} = \frac{\alpha_{EM}^2}{E_\ell Q^4} L_{\mu\nu} \left[\frac{Z_f^2}{4 m_N} \int d^2 k \, \Tr\left[\Phi(x,\ul{k}) \gamma^\mu \gamma^+ \gamma^\nu \right] \right].
\end{align}
The fully-inclusive DIS cross-section, summed over all hadronic final states $\ket{X}$, is thus proportional to an integral over the transverse momentum of the active quark.  By simply moving the differential $d^2k$ to the left-hand side, we obtain an expression for the SIDIS cross section for the production of a quark:
\begin{align}
 \label{SIDIS18}
 \frac{d\sigma}{d^2\ell' dy_\ell' d^2k} = \frac{\alpha_{EM}^2}{E_\ell Q^4} L_{\mu\nu} \left(\frac{Z_f^2}{4 m_N}\right)
 \Tr\left[\Phi(x,\ul{k}) \gamma^\mu \gamma^+ \gamma^\nu \right].
\end{align}
Because of confinement, the final-state quark is not directly observable; instead, it undergoes the (nonperturbative) process of \textit{fragmentation} into a jet of collimated hadrons.  The SIDIS cross-section \eqref{SIDIS18} can therefore be interpreted as the semi-inclusive distribution of jets produced from the deep inelastic scattering.  If one wanted to write the semi-inclusive distribution of a particular hadron (say, a pion), then the inclusion of a fragmentation function would be necessary to account for the probability of the outgoing quark fragmenting into the desired hadron (in this case, a pion) \cite{Collins:1981uw}.

\subsection{Longitudinal and Transverse Spin}

The hadronic tensor given in \eqref{SIDIS16} describes the response of the nucleon to a highly-virtual photon as appropriate for SIDIS.  But a photon is not the only exchanged particle that can probe the structure of the nucleon.  In neutrino deep inelastic scattering ($\nu$DIS) for example, an incident neutrino scatters off the nucleon by the exchange of an electroweak $W$ or $Z$ boson at high $Q^2$.  The analysis of this process follows along the same lines as for conventional DIS, with one modification to the hadronic tensor $W^{\mu\nu}$: the electroweak bosons couple to the left-handed chiral current 
\begin{align}
 \label{diss-spin1}
 J^\mu_L (x) \sim \psibar(x) \gamma^\mu P_L \psi(x) = \psibar(x) \left[\frac{1}{2}\left( 1 - \gamma^5 \right)\right] \psi(x)
\end{align}
rather than the electromagnetic current $J^\mu (x) \sim \psibar(x) \gamma^\mu \psi(x)$.  When this change is propagated forward into the hadronic tensor, one again obtains (c.f. \eqref{SIDIS16})
\begin{align}
 \label{diss-spin2}
 W^{\mu\nu} \sim \int d^2k \, \Tr\left[\Phi(x,\ul{k}) \Gamma^{\mu\nu}_L \right]
\end{align}
but with a new effective vertex
\begin{align}
 \label{diss-spin3}
 \Gamma^{\mu\nu}_L = \gamma^\mu P_L \gamma^+ \gamma^\nu = \frac{1}{2} \bigg[ \big( \gamma^\mu \gamma^+ \gamma^\nu \big) + \big( \gamma^\mu \gamma^+ \gamma^5 \gamma^\nu \big) \bigg]
\end{align}
that couples to the quark-quark correlation function $\Phi(x,\ul{k})$.  The $\nu$DIS vertex contains one term which corresponds to the same $\gamma^+$ operator present in ordinary DIS, but it also contains a new chiral operator $\gamma^+ \gamma^5$.  This operator has different quantum numbers than the usual DIS vertex and instead couples to the parity-odd part of the correlator $\Phi$; as we will see in Sec.~\ref{subsec-partdens}, this is closely related to the distribution of \textit{longitudinally-polarized} quarks in the nucleon.  

This illustrates a general principle: when working in the high-$Q^2$ Bjorken kinematics, the scattering of some incident particle probes the correlation function $\Phi$ with an effective vertex $\Gamma$.  Depending on the probe, this vertex projects out the part of the correlator with the appropriate quantum numbers and symmetries.  Furthermore, we need not restrict ourselves to considering only the physical particles known to exist in nature; by imagining the scattering of fictitious particles on the nucleon, we could construct a vertex $\Gamma$ for all possible projections of $\Phi$.  A complete set of such operators $\Gamma$ makes it possible to formulate parton distributions corresponding to quarks with net longitudinal or transverse polarization as well as their various correlations with the quark transverse momentum $\ul{k}$.  Additionally, the nucleon itself may possess an explicit polarization $S$ which can couple to both the momentum and the spin of the active quark.  

As a preliminary step to formulating the parton distribution functions for these spin-momentum correlations, let us explicitly establish a spinor basis for longitudinal and transverse polarizations and formulate their properties.  We will work with the spinors defined in Ref. \cite{Lepage:1980fj}, which are given in the ``standard'' (Dirac) representation of the Clifford algebra as
\begin{eqnarray}
\label{spinors1}
U_{+z} (p) &= \frac{1}{\sqrt{2p^+}} \left[ 
 \begin{array}{c} p^+ + m \\ p_\bot^1 + i p_\bot^2 \\ p^+ - m \\ p_\bot^1 + i p_\bot^2 \end{array}
 \right] \hspace{1cm}
U_{-z}(p) &= \frac{1}{\sqrt{2p^+}} \left[ 
 \begin{array}{c} -p_\bot^1 + i p_\bot^2 \\ p^+ + m \\ p_\bot^1 - i p_\bot^2 \\ -p^+ + m \end{array}
 \right]
\\ \nonumber
V_{+z} (p) &= \frac{1}{\sqrt{2p^+}} \left[ 
 \begin{array}{c} -p_\bot^1 + i p_\bot^2 \\ p^+ - m \\ p_\bot^1 - i p_\bot^2 \\ -p^+ - m \end{array}
 \right] \hspace{1cm}
V_{-z}(p) &= \frac{1}{\sqrt{2p^+}} \left[ 
 \begin{array}{c} p^+ - m \\ p_\bot^1 + i p_\bot^2 \\ p^+ + m \\ p_\bot^1 + i p_\bot^2 \end{array}
 \right] .
\end{eqnarray}
For a particle moving along the $z$-axis with $\ul{p} = \ul{0}$, these spinors have definite spin projections along the $z$-axis.  One natural Lorentz-covariant generalization of spin is the Pauli-Lubanski vector
\begin{align}
 \label{diss-spin4}
 W_\mu \equiv -\frac{1}{2} \epsilon_{\mu\nu\rho\sigma} S^{\nu\rho} p^\sigma
\end{align}
where $\epsilon_{\mu\nu\rho\sigma}$ is the 4-dimensional antisymmetric Levi-Civita tensor with the convention $\epsilon_{0123}=+1$ and $S^{\nu\rho} \equiv \frac{i}{4} [\gamma^\nu , \gamma^\rho]$ is the generator of Lorentz transformations for spinors.  
When $\ul{p}=\ul{0}$, the z-component of the Pauli-Lubanski vector is $W_3 = \frac{i}{2} E \gamma^1 \gamma^2 = \frac{1}{2} E \Sigma^3$, where $\Sigma^i = diag(\sigma^i, \sigma^i)$ is just the block-diagonal implementation of the Pauli matrices for 4-component spinors.  As can be explicitly verified from \eqref{spinors1}, these spinors are eigenstates of $W_3$ for $\ul{p}=\ul{0}$,
\begin{align}
 \label{diss-spin5}
 W_3 U_{\pm z} &= \left(\pm \frac{E}{2} \right) U_{\pm z} \\ \nonumber
 W_3 V_{\pm z} &= \left(\mp \frac{E}{2} \right) V_{\pm z}
\end{align}
and correspond to longitudinal (or helicity) spin states.  Note that the eigenvalue of the spinors $V$ is opposite to the spin of the physical antiparticle.

Like any spinor basis for solutions of the Dirac equation, the
spinors \eqref{spinors1} satisfy identities that embody the discrete $\mathrm{C}$,
$\mathrm{P}$, and $\mathrm{T}$ symmetries of the theory.  As can be
explicitly verified from \eqref{spinors1}, these spinors obey the
identities
\begin{eqnarray}
\label{CPT1}
\mathrm{C:} \hspace{1cm} & -i \gamma^2 V_{\pm z}^* (p) = U_{\pm z} (p) \\ \nonumber
\mathrm{PT:} \hspace{1cm} & \gamma^1 \gamma^3 \gamma^0 U_{\pm z}^* (p) = \mp U_{\mp z} (p) \\ \nonumber
                          & \gamma^1 \gamma^3 \gamma^0 V_{\pm z}^* (p) = \pm V_{\mp z} (p) \\ \nonumber
\mathrm{CPT:} \hspace{1cm} & U_{\pm z} (p) = \pm \gamma^5 V_{\mp z} (p) ,
\end{eqnarray}
where the final $\mathrm{CPT}$ identity combines the other two in a
compact form.

We are also interested in transverse spin states, which are a superposition of longitudinal spin states.  For particles with $\ul{p}=\ul{0}$, we can construct such transverse spinors in analogy to \eqref{spinors1} by diagonalizing one of the transverse components of $W_\mu$, say, $W_1$ (cf. e.g. \cite{Cortes:1991ja}).  Doing so gives spinors corresponding to polarization along the $x$-axis:
\begin{eqnarray}
\label{spinors2}
U_\chi &= \frac{1}{\sqrt 2} (U_{+z} + \chi U_{-z}) \\ \nonumber
V_\chi &= \frac{1}{\sqrt 2} (V_{+z} - \chi V_{-z}) ,
\end{eqnarray}
where $\chi = \pm 1$ is the spin eigenvalue along the $x$-axis: 
\begin{align}
 \label{diss-spin6}
  W_1 U_{\chi} &= \left(\chi \frac{m}{2} \right) U_{\chi} \\ \nonumber
  W_1 V_{\chi} &= \left(-\chi \frac{m}{2} \right) V_{\chi}
\end{align}
Combining \eqref{CPT1} and \eqref{spinors2} gives the somewhat different
$\mathrm{C/P/T}$ identities satisfied by the transverse spinors:
\begin{eqnarray}
\label{CPT2}
\mathrm{C}: \hspace{1cm} &-i \gamma^2 V_\chi^* (p) = U_{-\chi} (p) \\ \nonumber
\mathrm{CPT}: \hspace{1cm} & U_\chi (p) = - \chi \gamma^5 V_\chi (p) .
\end{eqnarray}

The $\mathrm{C/P/T}$ properties of these spinors translate into corresponding properties of spin-dependent observables.  This is particularly true for transverse spin states; as we will now show, these $\mathrm{C/P/T}$ properties strongly constrain the processes that can give rise to transverse-spin dependence.  Employing and generalizing  \eqref{CPT2} allows us to write a
complete set of identities for any transverse spinor matrix element:
\begin{eqnarray}
\label{CPT3}
\mathrm{C:} \hspace{1cm}
\vbar{\chi'}(k) \gamma^{\mu_1} \cdots \gamma^{\mu_n} V_{\chi} (p)  
 &=& \left[ \vbar{\chi}(p) \gamma^{\mu_n} \cdots \gamma^{\mu_1} V_{\chi'} (k) \right]^* \\ \nonumber 
 &=& (-1)^{n-1} \ubar{-\chi}(p) \gamma^{\mu_n} \cdots \gamma^{\mu_1} U_{-\chi'}(k) \\ \nonumber
 &=& (-1)^{n-1} \left[ \ubar{-\chi'}(k) \gamma^{\mu_1} \cdots \gamma^{\mu_n} U_{-\chi}(p) \right]^*
\\ \nonumber \\ \label{CPT4}
\mathrm{C:} \hspace{1cm}
\ubar{\chi'}(k) \gamma^{\mu_1} \cdots \gamma^{\mu_n} V_{\chi}(p) 
 &=& \left[ \vbar{\chi}(p) \gamma^{\mu_n} \cdots \gamma^{\mu_1} U_{\chi'} (k) \right]^* \\ \nonumber
 &=& (-1)^{n-1} \ubar{-\chi}(p) \gamma^{\mu_n} \cdots \gamma^{\mu_1} V_{-\chi'}(k) \\ \nonumber
 &=& (-1)^{n-1} \left[ \vbar{-\chi'}(k) \gamma^{\mu_1} \cdots \gamma^{\mu_n} U_{-\chi}(p) \right]^*
\\ \nonumber \\ \label{CPT5}
\mathrm{CPT:} \hspace{1cm}
\vbar{\chi'}(k) \gamma^{\mu_1} \cdots \gamma^{\mu_n} U_\chi (p) 
 &=& \chi \chi' \left[ \vbar{-\chi'}(k) \gamma^{\mu_1} \cdots \gamma^{\mu_n} U_{-\chi}(p) \right]^* \\ \nonumber
\ubar{\chi'}(k) \gamma^{\mu_1} \cdots \gamma^{\mu_n} U_\chi (p)
 &=& \chi \chi' \left[ \ubar{-\chi'}(k) \gamma^{\mu_1} \cdots \gamma^{\mu_n} U_{-\chi}(p) \right]^* .
\end{eqnarray}

These identities allow us to explicitly determine the rigid
constraints on the form of transverse spinor products.  In
particular, consider the parameterizations of both classes of spinor
products:
\begin{eqnarray}
\label{spinors3}
\vbar{\chi'}(k) \gamma^{\mu_1} \cdots \gamma^{\mu_n} U_\chi(p) \, &\equiv& \,
 \delta_{\chi \chi'} [a(k,p) + \chi a'(k,p)] + \delta_{\chi, -\chi'} [b(k,p) + \chi b'(k,p)] \\ \nonumber
\ubar{\chi'}(k) \gamma^{\mu_1} \cdots \gamma^{\mu_n} U_\chi (p) \, &\equiv& \,
 \delta_{\chi \chi'} [c(k,p) + \chi c'(k,p)] + \delta_{\chi, -\chi'} [d(k,p) + \chi d'(k,p)] ;
\end{eqnarray}
applying \eqref{CPT5}, one readily concludes that $\mathrm{C/P/T}$
constraints imply that:
\begin{itemize}
 \item $a$, $b'$, $c$, and $d'$ are real-valued.
 \item $a'$, $b$, $c'$, and $d$ are pure imaginary.
\end{itemize}
Furthermore, this implies that if we multiply any two of these spinor
matrix elements and sum over one of the spins ($\chi'$), e.g.,
\begin{eqnarray}
\label{spinors4}
& \sum_{\chi'} &  \,  [\vbar{\chi'}(k) \gamma^{\mu_1} \cdots \gamma^{\mu_n} U_\chi(p)] \;
  [\ubar{\chi'}(k) \gamma^{\mu_1} \cdots \gamma^{\mu_n} U_\chi (p)]^* = \\ \nonumber
  &=& \underbrace{[a c^* + a' (c')^* + b d^* + b' (d')^*]}_{\mathrm{real}} \, + \,
      \chi \, \underbrace{[ a (c')^* + a' c^* +  b (d')^* + b' d^* ]}_{\mathrm{imaginary}},
\end{eqnarray}
the result naturally
partitions into an unpolarized, real contribution, and a polarized,
imaginary contribution.  Thus in particular, the \textit{spin-dependent part} of any product of two
transverse matrix elements (say $S_1(\chi)$ and $S_2^*(\chi)$), summed over final-state polarizations, is
always \textit{pure imaginary}:
\begin{equation}
\label{spinors5}
S_1 (\chi) S_2^* (\chi) - S_1 (-\chi) S_2^* (-\chi) = - [ S_1^* (\chi) S_2 (\chi) - S_1^* (-\chi) S_2 (-\chi) ].
\end{equation}


\subsection{The Single Transverse Spin Asymmetry}
\label{subsubsec-STSA}

The conclusion \eqref{spinors5} has direct application to cross-sections with transversely-polarized targets.  In SIDIS with an unpolarized lepton beam on a polarized target, for example, one can study the effect of the polarization by measuring the difference in particle production when the target is polarized ``up'' versus ``down.''  While this can be done for either longitudinally- or transversely-polarized targets, the latter case introduces a preferred azimuthal direction that transforms under rotations about the beam axis (Fig.~\ref{STSArot}).  Because of rotational invariance, this implies that producing a particle moving to the left when the transverse spin is pointing ``up'' (top panel) is identical to producing a particle moving to the right when the transverse spin is pointing ``down'' (bottom panel).  This transverse-spin dependence can therefore be expressed either as the difference between spin-up and spin-down cross-sections for producing a particle at fixed transverse momentum, or as the left-right asymmetry in the particle production cross-section with fixed transverse spin.  The ratio of this spin-dependent cross-section to the unpolarized cross-section $d\sigma_{unp}$ is known as the \textit{single transverse spin asymmetry (STSA)} $A_N$: 
\begin{figure}
 \centering
 \includegraphics[width=0.7\textwidth]{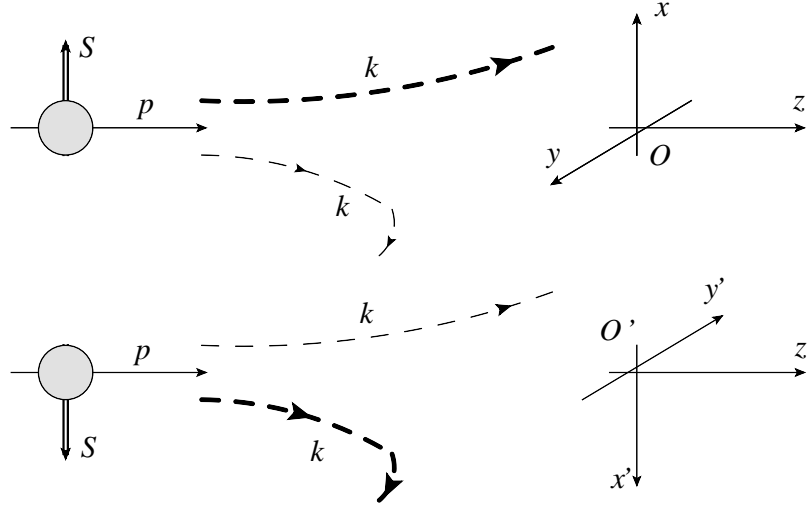}
 \caption{Illustration of rotational invariance in the single transverse spin asymmetry.  The STSA, as seen from the coordinate system $O$, measures the tendency of a nucleon polarized along the $(+x)$ axis to produce more particles moving in the $(-y)$ direction than the $(+y)$ direction.  This is equivalent to a process viewed from the coordinate system $O'$ in which a nucleon polarized in the $(-x) = (+x')$ direction produces more particles moving in the $(+y) = (-y')$ direction than in the $(-y) = (+y')$ direction.}
 \label{STSArot}
\end{figure}
\begin{equation}
\label{AN1}
A_N \equiv \frac{d\sigma^\uparrow (\ul{k}) - d\sigma^\downarrow (\ul{k})} {2 \, d\sigma_{unp}} =
 \frac{d\sigma^\uparrow (\ul{k}) - d\sigma^\uparrow(-\ul{k})}{2 \, d\sigma_{unp}}
\end{equation}
where $d\sigma(\ul{k})$ stands for the invariant production cross section,
e.g. $\frac{d\sigma}{d^2 k \, dy}$, for a particle with transverse momentum $\ul{k}$ coming from
scattering on a target with transverse spin $\uparrow , \downarrow$. 

The STSA $A_N$ in \eqref{AN1} expresses a correlation between the transverse spin of the polarized target, the transverse momentum of the produced particle, and the longitudinal direction defined by the beam axis.  This correlation changes sign when either the spin $\ul{S}$ or the momentum $\ul{k}$ are reversed, and it can be expressed as a vector triple product of the 3-vectors $\vec{S}$, $\vec{k}$, and, say, the momentum $\vec{p}$ of the polarized target seen in the center-of-mass frame:
\begin{align}
 \label{diss-spin7}
 A_N \propto (\vec{S} \times \vec{k}) \cdot \vec{p} \sim S_x k_y p_z .
\end{align}
Unlike the familiar unpolarized cross-section, the STSA possesses an unusual property for an observable: it is odd under ``naive time-reversal'' (time reversal $T$ followed by parity inversion $P$).  Under $PT$, spin vectors are reflected, but momentum components are left unchanged.  Since the asymmetry has different quantum numbers than the $PT$-even unpolarized cross-section, it must couple to different underlying production mechanisms; thus transverse spin observables like the STSA can give access to properties of the nucleon inaccessible to unpolarized processes.

From \eqref{AN1}, we see that the asymmetry $A_N$ is proportional to the
difference between the amplitude-squared $|\mathcal{A}|^2$ for $\chi =
+1$ and $\chi = -1$.  We denote this spin-difference amplitude squared
as $\Delta |\mathcal{A}|^2$:
\begin{equation}
\label{AN2}
A_N \propto |\mathcal{A}|^2 (\chi = +1) - |\mathcal{A}|^2 (\chi = -1) \equiv \Delta |\mathcal{A}|^2.
\end{equation}
Now let us identify the types of Feynman diagrams from which $A_N$ can arise.
Suppose there is a contribution from the square of the Born-level amplitude
$\mathcal{A}_{(0)} (\chi) = F_0 \, S_0 (\chi)$ which is factorized into a spinor product
$S_0 (\chi)$ which depends on the transverse spin eigenvalue $\chi$ and a
factor $F_0$ coming from the rest of the diagram.  Then the contribution
of the square of $\mathcal{A}_{(0)}$ to the asymmetry would be
\begin{eqnarray}
\Delta |\mathcal{A}|^2 &=& |\mathcal{A}_{(0)}|^2 (+1) - |\mathcal{A}_{(0)}|^2 (-1) \\ \nonumber
 &=& |F_0^2| \, \left[ |S_0(+1)|^2 - |S_0(-1)|^2 \right] .
\end{eqnarray}
But, substituting $S_1 = S_2 = S_0$ into \eqref{spinors5}, we see that the $\mathrm{C/P/T}$ constraints
 imply that
\begin{eqnarray}
 \label{spinors-squared}
 |S_0 (+1)|^2 - |S_0 (-1)|^2 = 0.
\end{eqnarray}
This is easy to understand mathematically: the spin-dependent part of a given term
must be pure imaginary, but any amplitude squared is explicitly real.
Hence, the square of any factorized amplitude (such as the Born-level amplitude)
 is independent of $\chi$ and cannot generate the asymmetry; $A_N$ can only be 
generated by the \textit{quantum interference} between two different diagrams.

This also implies that a relative $\ord{\alpha_s}$ correction to the Born amplitude coming from
the real emission of another particle cannot generate the asymmetry either.  Such a contribution would
again be factorized, and its square must be purely real and spin-independent for the same reasons as
\eqref{spinors-squared}.  Thus at lowest order in perturbation theory, the asymmetry can be
generated by the interference between the Born-level amplitude and a relative $\mathcal{O}(\alpha_s)$
virtual correction.  So let us consider a similar exercise to determine the contribution to $\Delta
|\mathcal{A}|^2$ from this $\mathcal{O}(\alpha_s)$ correction.  Writing the tree-level amplitude 
$\mathcal{A}_{(0)}$ and the one-loop amplitude $\mathcal{A}_{(1)}$ as
\begin{eqnarray}
\label{ImPart1}
\mathcal{A}_{(1)}(\chi) &\equiv& F_1 \int d^4 k \frac{S_1(k,\chi)}{D_1(k)} \\ \nonumber
\mathcal{A}_{(0)}(\chi) &\equiv& F_0 \, S_0(\chi)
\end{eqnarray}
where the factor $S_1$ includes all momentum and spin-dependent
numerators, and the factor $D_1$ contains all the propagator
denominators, the spin-dependent contribution is
\begin{eqnarray}
\label{ImPart2}
\Delta|\mathcal{A}|^2 
 &=& \mathcal{A}_{(1)}(+1) \mathcal{A}_{(0)}^*(+1) + 
  \mathcal{A}_{(1)}^*(+1) \mathcal{A}_{(0)}(+1) \, - \, (\chi \rightarrow - \chi) \\ \nonumber
 &=& F_1 F_0^* \int d^4 k \frac{S_1(k,+1) S_0^*(+1)}{D_1(k)} + 
  F_1^* F_0 \int d^4 k \frac{S_1^*(k,+1) S_0(+1)}{D_1^*(k)} - (\chi \rightarrow -\chi) \\ \nonumber
 &=& F_1 F_0^* \int d^4 k \frac{S_1(k,+1) S_0^*(+1) - S_1(k,-1) S_0^*(-1)}{D_1(k)} 
  + \mathrm{c.c.} \, .
\end{eqnarray}
But from the $\mathrm{C/P/T}$ constraints \eqref{spinors5}, we see
that the numerator of \eqref{ImPart2} is pure imaginary, giving
\begin{eqnarray}
\label{ImPart3}
\Delta|\mathcal{A}|^2 &=& \int d^4 k \left[ \frac{F_1 F_0^*}{D_1(k)} - \mathrm{c.c.} \right] \,
 \left[S_1(k,+1) S_0^*(+1) - S_1(k,-1) S_0^*(-1)\right] \\ \nonumber
&=& 2 i \int d^4 k \, \mathrm{Im} \left[ \frac{F_1 F_0^*}{D_1(k)} \right] \,
 \left[S_1(k,+1) S_0^*(+1) - S_1(k,-1) S_0^*(-1)\right] .
\end{eqnarray}

Thus we conclude that the spin-dependent part which contributes to the
asymmetry requires an \textit{imaginary part} from the remainder of the 
expression (aside from the spinor matrix elements themselves).  This 
is also easy to understand mathematically: if the spin-dependent part of
the spinor matrix elements is pure imaginary, then it must multiply
another imaginary factor to generate a real contribution to the asymmetry.  
This imaginary part picks out terms with a relative complex phase, and a 
complex phase is automatically $PT$-odd because of the antilinearity of time-reversal.  

In this way, \eqref{ImPart3} codifies the statements made previously: 
since $A_N$ is a (naive) T-odd observable, it must couple to scattering processes 
in which a T-odd complex phase is present.  This complex phase is not 
simply the imaginary part of any one diagram, but rather a
relative phase between the tree-level and one-loop amplitudes.  If
there is no relative phase present in the pre-factors,
e.g. $\mathrm{Im}(F_1 F_0^*)=0$, then the imaginary part comes from
the denominator of the loop integral $D_1(k)$.  In that case, taking
the imaginary part corresponds to putting an intermediate virtual
state on shell \cite{Cutkosky:1960sp}.  
The imaginary part generated this way was discussed in
\cite{Brodsky:2002cx} and \cite{Brodsky:2002rv} as a possible source of the
STSA.


\section{Transverse-Momentum-Dependent Parton Distributions}

\subsection{Relation to Parton Densities}
\label{subsec-partdens}

The quark-quark correlation function which couples to SIDIS, defined in \eqref{phiprelim} as
\begin{align}
 \label{DISS-TMD1}
 \Phi_{ij}(x,\ul{k}) \equiv \frac{1}{2(2\pi)^3} \int d^{2-}r \, e^{i k\cdot r} \bra{p S} 
 \psibar_j(0) \psi_i (r) \ket{pS}_{r^+ = 0},
\end{align}
is a measure of the ``quark content'' of the nucleon state with quantum numbers $p, S$.  To quantify this statement more precisely, we need to rewrite the quark fields in terms of creation and annihilation operators:
\begin{align}
 \label{DISS-TMD2}
 \psi(x) &= \int \frac{d^{2+}q}{2(2\pi)^3 q^+} \sum_\sigma \left( b_{q\sigma} U_\sigma (q) \, 
 e^{-i q\cdot x} + d_{q \sigma}^\dagger V_\sigma (q) \, e^{i q \cdot x} \right) \\ \nonumber
 \psibar(y) &= \int \frac{d^{2+}\ell}{2(2\pi)^3 \ell^+} \sum_\tau \left( b_{\ell \tau}^\dagger
 \ubar{\tau}(\ell) \, e^{i \ell \cdot y} + d_{\ell \tau} \vbar{\tau}(\ell) \, e^{-i \ell \cdot y} 
 \right).
\end{align}
It is convenient to rewrite \eqref{DISS-TMD1} by multiplying and dividing by a volume factor $\mathcal{V}^-~\equiv~\frac{1}{2}\int d^{2-}y$.  
\begin{align}
 \label{DISS-TMD3}
 \Phi_{ij}(x,\ul{k}) = \frac{1}{4(2\pi)^3 \mathcal{V}^-} \int d^{2-}r \, d^{2-}y \, e^{i k\cdot r} 
 \bra{p S} \psibar_j(0) \psi_i (r) \ket{pS}.
\end{align}
Since the nucleon is in a plane-wave state, the expectation value possesses translational invariance, allowing us to shift the operators by a displacement $y$:
\begin{align}
 \label{DISS-TMD4}
 \bra{p S} \psibar_j(0) \psi_i (r) \ket{pS} &= \bra{p S} \: e^{i p \cdot y} \: \psibar_j(0) \psi_i (r) \: e^{-i p \cdot y} \: \ket{pS}
 \\ \nonumber &=
 \bra{p S} \: e^{i \hat{P} \cdot y} \: \psibar_j(0) \: e^{-i \hat{P} \cdot y} \: 
 e^{i \hat{P} \cdot y} \: \psi_i (r) \: e^{-i \hat{P} \cdot y} \: \ket{pS}
 \\ \nonumber &=
 \bra{p S} \psibar_j(y) \psi_i (y+r) \ket{pS}.
\end{align}
Then we can change variables from $r$ to $x \equiv y+r$ and apply \eqref{DISS-TMD2}
\begin{align}
 \label{DISS-TMD5}
 \Phi_{ij}(x,\ul{k}) &= \frac{1}{4(2\pi)^3 \mathcal{V}^-} \int d^{2-}x \, d^{2-}y \, e^{i k\cdot (x-y)} 
 \bra{p S} \psibar_j(y) \psi_i (x) \ket{pS} \\ \nonumber
 &= \frac{1}{4(2\pi)^3 \mathcal{V}^-} \int d^{2-}x \, d^{2-}y \left[\frac{d^{2+}q}{2(2\pi)^3 q^+} \right]
 \left[\frac{d^{2+}\ell}{2(2\pi)^3 \ell^+}\right] e^{i k\cdot(x-y)} \\ \nonumber
 &\times \sum_{\sigma\tau} \bra{p S} \left( b_{\ell \tau}^\dagger
 \ubar{\tau}^j (\ell) \, e^{i \ell \cdot y} + d_{\ell \tau} \vbar{\tau}^j (\ell) \, e^{-i \ell \cdot y} 
 \right)  \left( b_{q\sigma} U_\sigma^i (q) \, 
 e^{-i q\cdot x} + d_{q \sigma}^\dagger V_\sigma^i (q) \, e^{i q \cdot x} \right) \ket{p S}.
\end{align}
Only combinations of operators that do not change the net particle content of the state $\ket{p S}$ can contribute; that is, only $b^\dagger b$ and $d d^\dagger$:
\begin{align}
 \label{DISS-TMD6}
\Phi_{ij}(x,\ul{k}) &= \frac{1}{4(2\pi)^3 \mathcal{V}^-} \int d^{2-}x \, d^{2-}y \left[\frac{d^{2+}q}
 {2(2\pi)^3 q^+} \right] \left[\frac{d^{2+}\ell}{2(2\pi)^3 \ell^+}\right]  
 \\ \nonumber &\times 
 \sum_{\sigma\tau} \bra{p S} \bigg( 
 b_{\ell \tau}^\dagger b_{q\sigma} \left[\ubar{\tau}^j (\ell) U_\sigma^i (q) \right] \, 
 e^{i (k-q) \cdot x} \, e^{- i (k-\ell) \cdot y} 
 \\ \nonumber &+ 
 d_{\ell \tau} d_{q \sigma}^\dagger \left[ \vbar{\tau}^j(\ell) V_\sigma^i (q) \right] \, 
 e^{-i (k + \ell) \cdot y} \, e^{i (k+q) \cdot x}
 \bigg) \ket{p S}.
\end{align}
Now we can integrate over the coordinates $d^{2-} x$, $d^{2-} y$, generating delta functions, e.g. $2(2\pi)^3 \delta^{2+} (k \pm q)$, from the Fourier factors.  In the $b^\dagger b$ term, this sets $k = q = \ell$, and we can use the resulting delta functions to integrate out $d^{2+}\ell$ and $d^{2+} q$.  In the $d d^\dagger$ term, this similarly sets $k = - q = - \ell$, but in this case it is impossible to pick up the singularity of the delta function because $k^+ = x p^+$ is constrained to be positive due to the kinematic condition \eqref{eq-missing mass} and $q^+$ and $\ell^+$ are also positive because they correspond to on-shell quark fields \eqref{DISS-TMD2}.
\footnote{
This suggests a useful generalization of \eqref{DISS-TMD1} in which we extend the range of $x$ to be $-1 \leq x \leq 1$.  Then the negative $x$ range picks out the antiquark terms in \eqref{DISS-TMD6} so that the quark and antiquark distributions can be combined into a single correlator.}

After performing all of the integrations, the only contribution that remains is
\begin{align}
 \label{DISS-TMD7}
 \Phi_{ij}(x,\ul{k}) &= \frac{1}{4 (2\pi)^3 \mathcal{V}^-} \frac{1}{(k^+)^2} \sum_{\sigma \tau}
 \bra{p S} b_{k \tau}^\dagger b_{k \sigma} \ket{p S} \left[ \ubar{\tau}^j (k) U_\sigma^i (k) \right],
\end{align}
which shows that the correlator $\Phi$ is proportional to the expectation value of the quark number density operator $n_k = b_k^\dagger b_k$.  This correlator is a matrix in Dirac space, with the spinors $\ubar , U$ free to be contracted with another matrix $\Gamma$ as in Sec.~\ref{subsec-SIDIS}.  The choice of the vertex $\Gamma$ which couples to the correlator $\Phi$ will pick out a particular superposition of quark spins $\tau, \sigma$.  

In this way, scattering processes like SIDIS that interact with the nucleon under high-$Q^2$ kinematics are directly probing the parton densities in the wave function of the nucleon.  For SIDIS, the hadronic tensor in \eqref{SIDIS16} couples to the correlator $\Phi$ by $W^{\mu\nu} \sim \Tr[\Phi \gamma^\mu \gamma^+ \gamma^\nu].$  If the lepton beam in SIDIS is unpolarized, then the leptonic tensor \eqref{SIDIS2} is symmetric and couples only to the symmetric part of $W^{\mu\nu}$, with the dominant components being $\mu , \nu = \bot$ (see, e.g. \cite{Kovchegov:2012mbw}).  Then the vertex that couples to the correlator $\Phi$ is
\begin{align}
 \label{DISS-TMD8}
 \Gamma^{ij} &= \frac{1}{2}\left[\gamma_\bot^i \gamma^+ \gamma_\bot^j + \gamma_\bot^i \gamma^+ \gamma_\bot^j \right] \\ \nonumber
 & = - \frac{1}{2} \left\{ \gamma_\bot^i , \gamma_\bot^j \right\} \gamma^+ \\ \nonumber
 \Gamma^{ij} & = (-g^{ij}) \gamma^+.
\end{align}
Thus the SIDIS vertex to probe the parton density of the nucleon is essentially $\gamma^+$.  Substituting this into \eqref{DISS-TMD7} gives
\begin{align}
 \label{DISS-TMD9}
 \Tr[\Phi(x,\ul{k}) \gamma^+] &= \frac{1}{4 (2\pi)^3 \mathcal{V}^-} \frac{1}{(k^+)^2} \sum_{\sigma \tau}
 \bra{p S} b_{k \tau}^\dagger b_{k \sigma} \ket{p S} \left[ \ubar{\tau} (k) \gamma^+ U_\sigma (k) \right] \\ \nonumber 
 &= \frac{1}{4 (2\pi)^3 \mathcal{V}^-} \frac{1}{(k^+)^2} \sum_{\sigma \tau}
 \bra{p S} b_{k \tau}^\dagger b_{k \sigma} \ket{p S} \left[ 2 k^+ \delta_{\sigma \tau} \right] \\ \nonumber
 &= \frac{1}{2 (2\pi)^3} \frac{1}{p^+ \mathcal{V}^-} \frac{1}{x} \sum_{\sigma}
 \bra{p S} b_{k \sigma}^\dagger b_{k \sigma} \ket{p S},
\end{align}
where the spinor product was evaluated using the spinors \eqref{spinors1} and we have used $k^+ = x p^+$.  

To interpret this expression, consider evaluating it in the state $\ket{p S} = b_{p S}^\dagger \ket{0}$ consisting of a single quark.  Then we have
\begin{align}
 \label{newTMD1}
 \Tr[\Phi(x,\ul{k}) \gamma^+] &= \frac{1}{2 (2\pi)^3} \frac{1}{p^+ \mathcal{V}^-} \frac{1}{x} \sum_{\sigma} \bra{0} \: b_{p S} \: b_{k \sigma}^\dagger b_{k \sigma} \: b_{p S}^\dagger \: \ket{0}
 \\ \nonumber &=
 \frac{1}{2 (2\pi)^3} \frac{1}{p^+ \mathcal{V}^-} \frac{1}{x} \sum_{\sigma} \bra{0} \: 
 \bigg\{ b_{p S} \: , \: b_{k \sigma}^\dagger \bigg\} \: \bigg\{ b_{k \sigma} \: , \: b_{p S}^\dagger \bigg\} \: \ket{0}
 \\ \nonumber &=
 \frac{1}{2 (2\pi)^3} \frac{1}{p^+ \mathcal{V}^-} \frac{1}{x} \sum_{\sigma}
 \left[2(2\pi)^3 p^+ \delta_{\sigma S} \delta^{2+}(k-p) \right] 
 \left[2(2\pi)^3 p^+ \delta_{\sigma S} \delta^{2+}(k-p) \right]
 \\ \nonumber &=
 \frac{2}{x} \sum_\sigma \left(\frac{1}{p^+ \mathcal{V}^-} \: \delta_{\sigma S} \: \delta(1-x) \: 
 \delta^2(\ul{k}) \bigg[(2\pi)^3 p^+ \delta^{2+}(p-k) \bigg] \right)
\end{align}
where we have used the anticommutation relations \eqref{anticomm} and rewritten one of the delta functions as $p^+ \delta(p^+ - k^+) = \delta(1-x)$.  By rewriting the factor in brackets as a Fourier transform and then imposing the conditions $p^+ = k^+ , \ul{p} = \ul{k}$ from the other delta function, we can identify it as simply the volume factor $p^+ \mathcal{V}^-$:
\begin{align}
 \label{newTMD2}
 \bigg[(2\pi)^3 p^+ \delta^{2+}(p-k) \bigg] &= \frac{1}{2} p^+ \: \int d^{2-}x \, e^{i (p-k)\cdot x}
 \\ \nonumber & \rightarrow
 \frac{1}{2} p^+ \: \int d^{2-}x = p^+ \mathcal{V}^- .
\end{align}
This cancels the factor of volume in the denominator, yielding
\begin{align}
 \label{newTMD3}
 \frac{1}{2} \Tr[\Phi(x,\ul{k}) \gamma^+] = \delta(1-x) \delta^2(\ul{k}) =
 \frac{d N}{d^2 k \, d x}
\end{align}
which is just the number of quarks in the state $\ket{p S} = b_{p S}^\dagger \ket{0}$ per unit $x$, per unit transverse momentum.  This allows us to interpret the expectation value in \eqref{DISS-TMD9} as
\begin{align}
 \label{newTMD4}
 \frac{d N_\sigma}{d^2 k \, d x} = \frac{1}{2 p^+ \mathcal{V}^-} \, \frac{1}{x} \,
 \frac{1}{2 (2\pi)^3} \, \bra{p S} b_{k \sigma}^\dagger b_{k \sigma} \ket{p S} ,
\end{align}
which is just the number of quarks of a given polarization $\sigma$ per unit $x$, per unit transverse momentum.  The factor of $1/2 p^+ \mathcal{V}^-$ reflects the density of one quark in an infinite volume and normalizes the integral of \eqref{newTMD4} to unity.  Therefore we see that $\tfrac{1}{2} \Tr [\Phi \gamma^+]$ is just the \textit{transverse-momentum-dependent distribution of unpolarized quarks} in the state $\ket{p S}$.

Similarly, for $\nu$DIS, there is an additional term discussed in \eqref{diss-spin3} that couples the correlator $\Phi$ to a chiral vertex $\gamma^+ \gamma^5$.  Using this vertex in \eqref{DISS-TMD7} gives the part of the quark distribution accessed by this vertex:
\begin{align}
 \label{DISS-TMD10}
 \frac{1}{2} \Tr[\Phi(x,\ul{k}) \gamma^+ \gamma^5] &= \frac{1}{8 (2\pi)^3 \mathcal{V}^-} \frac{1}{(k^+)^2} \sum_{\sigma \tau} \bra{p S} b_{k \tau}^\dagger b_{k \sigma} \ket{p S} \left[ \ubar{\tau} (k) \gamma^+ \gamma^5 U_\sigma (k) \right] 
 \\ \nonumber &= 
 \frac{1}{8 (2\pi)^3 \mathcal{V}^-} \frac{1}{(k^+)^2} \sum_{\sigma \tau} \bra{p S} b_{k \tau}^\dagger b_{k \sigma} \ket{p S} \left[ 2 k^+ \sigma \delta_{\sigma \tau} \right] 
 \\ \nonumber &= 
 \frac{1}{4 (2\pi)^3} \frac{1}{p^+ \mathcal{V}^-} \frac{1}{x}
 \bra{p S} \left( b_{k, +z}^\dagger b_{k, +z} - b_{k, -z}^\dagger b_{k, -z} \right) \ket{p S}
 \\ \nonumber &= 
 \frac{d N_{+ z}}{d^2 k d x} - \frac{d N_{- z}}{d^2 k d x} ,
\end{align}
which is the \textit{transverse-momentum-dependent distribution of longitudinally-polarized quarks} in the state $\ket{p S}$.  Thus the chiral interaction due to electroweak boson exchange in $\nu$DIS measures the longitudinal polarization of quarks in the nucleon.

By extending this procedure to other effective vertices $\Gamma$, we can project out the distributions of unpolarized, longitudinally-polarized, and transversely-polarized quarks.  And by further parameterizing the correlations of these spins with the quark transverse momentum, we can generate a complete decomposition of nucleon structure into TMD parton distribution functions.


\subsection{Gauge Invariance: The Importance of the Glue}

There is one important flaw in the interpretation of the correlator $\Phi$ as a density of partons in the nucleon: the definition \eqref{DISS-TMD1} is not gauge-invariant.  Under an $SU(N_c)$ color rotation, the quark fields transform as
\begin{align}
 \label{DISS-GAUGE1}
 \psi(r) &\rightarrow S(r) \psi(r) \equiv e^{i \theta^a (r) T^a} \psi(r) \\ \nonumber
 \psibar(0) &\rightarrow \psibar(0) S^{-1}(0) \equiv \psibar(0) e^{-i \theta^a (0) T^a},
\end{align}
where $T^a$ are the $SU(N_c)$ generators in the fundamental representation and $\theta^a(x)$ are the parameters of the local gauge transformation.  Because the quark fields in \eqref{DISS-TMD1} are evaluated at different spacetime points, their gauge transformations do not cancel, leaving a nontrivial gauge-dependence to $\Phi$.  

To remedy this and define gauge-invariant parton distributions, we will need to modify the definition \eqref{DISS-TMD1} of $\Phi$ to compensate for the gauge transformations \eqref{DISS-GAUGE1}.  The quantity with the appropriate gauge transformation rule is the \textit{gauge-link operator} $U_C [0,r]$
\begin{align}
 \label{DISS-GAUGE2}
 U_C [y,x] &\equiv \mathcal{P} \exp \left[ i g {\int\limits_{C ; \, x}^{y}} dz_\mu \, A^{\mu a}(z) \, T^a \right] \\ \nonumber
 U_C [0,r] &\rightarrow S(0) \, U_C [0,r] \, S^{-1}(r) = e^{i \theta^a (0) T^a} \, U_C [0,r] \, e^{i \theta^a (r) T^a}
\end{align}
where $\mathcal{P}$ stands for a path ordering of the non-abelian factors in the expansion of the exponential and the integration runs from point $x^\mu$ to point $y^\mu$ along a contour $C$.  Then, modifying \eqref{DISS-TMD1} to include the gauge link \eqref{DISS-GAUGE2}, we write the new gauge-invariant definition as
\begin{align}
 \label{DISS-GAUGE3}
 \Phi_{ij}^C (x,\ul{k}) \equiv \frac{1}{2(2\pi)^3} \int d^{2-} r \, e^{i k \cdot r}
 \bra{p S} \psibar_j (0) \, U_C [0,r] \, \psi_i (r) \ket{p S}_{r^+ = 0}.
\end{align}

The presence of the gauge link introduces the gluon field $A^{\mu a}$ into the quark correlator, reflecting the fact that one cannot measure a state's quark distribution in isolation.  Rather, the quark distribution is always ``dressed'' by an accompanying gluon field that modifies the behavior of the quarks before or after the hard SIDIS interaction.  At zeroth order in the QCD coupling $g$ (or $\alpha_s$), the gauge link reduces to unity, recovering the parton model correlator \eqref{DISS-TMD1}.  But at higher orders, the quark-quark operator mixes with the gluon fields:
\begin{align}
 \label{DISS-GAUGE4}
 \bra{p S} \psibar(0) \,& U_C[0,r] \, \psi (r) \ket{p S} = \bra{p S} \psibar(0) \psi (r) \ket{p S} + 
 \\ \nonumber &+
 i g {\int\limits_{C ; \, r}^{0}} dz_\mu \, T^a \, \bra{p S} \psibar(0) A^{\mu a}(z) \psi (r) \ket{p S} + 
 \ord{g^2}.
\end{align}
This mixing indicates that the straightforward interpretation of $\Phi^C$ as a quark density as in \eqref{DISS-TMD7} breaks down beyond the leading order, when contributions arise such as the quark-gluon-quark correlation function in \eqref{DISS-GAUGE4} and other multi-parton correlators at higher orders.  

\begin{figure}
 \centering
 \includegraphics[width=0.8\textwidth]{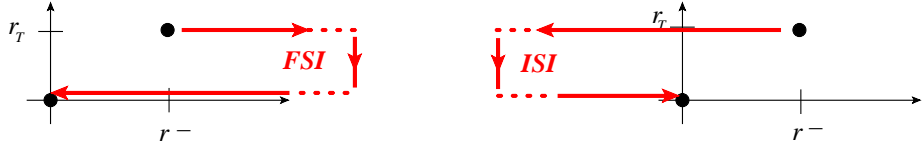}
 \caption{The gauge links $U_C[0,r]$ for processes with final-state interactions such as SIDIS (left-hand side, \eqref{DISS-GAUGE5}) and initial-state interactions such as DY (right-hand side, \eqref{DISS-GAUGE6}).  The direction of the gauge link follows the natural color flow for the process under consideration.} 
 \label{fig-GaugeLinks}
\end{figure}

The gauge link $U_C [0,r]$ in \eqref{DISS-GAUGE3} flows from the point $r^\mu = (0^+, r^-, \ul{r})$ to the origin along a contour $C$, in accordance with the color flow for the process under consideration \cite{Collins:2002kn}.  In the SIDIS scattering amplitude for example, an active quark is knocked out of the nucleon by the virtual photon, possibly rescattering by gluon exchange with the other ``spectator'' remnants of the nucleon.  SIDIS is therefore characterized by \textit{final-state interactions (FSI)}, with its color flow extending from the quark field $\psi(r)$ to \textit{future infinity}, as in the left-hand side of Fig.~\ref{fig-GaugeLinks}.  As we see from the high-$Q^2$ kinematics \eqref{SIDIS8} , \eqref{SIDIS9}, the active quark travels with a large momentum along the minus light-cone direction $(q+k)^- \sim Q$; these kinematics fix the precise direction of the contour $C$.  Indeed, when a formal factorization theorem is derived, the momentum of the outgoing quark is deformed to follow a light-like trajectory along the minus direction from $r^\mu = (0^+, r^-, \ul{r})$ to a point at future light-cone infinity $(0^+, \infty, \ul{r})$ \cite{Collins:2011zzd}.  The same discussion applies to the complex-conjugated SIDIS amplitude, with the color flow (after complex conjugation) going from a point at future light-cone infinity $(0^+, \infty, \ul{0})$ to the origin.  These two light-like gauge links are connected by a transverse gauge link at future infinity flowing from $(0^+, \infty, \ul{r})$ to $(0^+, \infty, \ul{0})$ that completes the contour $C$ describing the color flow in SIDIS.  Thus we can write the gauge link with final-state interactions appropriate for SIDIS as
\begin{align}
 \label{DISS-GAUGE5}
 U_{FSI} [0,r] &\equiv U_{LC} [(0^+, 0^-, \ul{0}), (0^+, \infty, \ul{0})] \times
                       U_\bot [(0^+, \infty, \ul{0}), (0^+, \infty, \ul{r})] 
											 \\ \nonumber &\times
                       U_{LC} [(0^+, \infty, \ul{r}), (0^+, r^-, \ul{r})] \\ \nonumber
 & \!\!\! = \left[ \mathcal{P} \exp \left( \frac{i g}{2} \int\limits_{\infty}^{0^-} dz^- A^{+a} (0^+, z^-, \ul{0}) T^a \right)    
    \right] 
    \left[ \mathcal{P} \exp \left(- i g \int\limits_{\ul{r}}^{\ul{0}} d \ul{z} \cdot \ul{A}^{a} (0^+, \infty, \ul{z}) T^a \right) \right] 
		\\ \nonumber &\times
		\left[ \mathcal{P} \exp \left( \frac{i g}{2} \int\limits_{r^-}^{\infty} dz^- A^{+a} (0^+, z^-, \ul{r}) T^a \right) \right]
\end{align}

This future-pointing ``light-cone staple'' contour is not unique, however.  It applies for processes like SIDIS that only have final-state QCD interactions (FSI).  The opposite applies to processes for which there are only initial-state QCD interactions (ISI), such as the Drell-Yan process.  In the Drell-Yan process (DY), an incident antiquark may scatter on the gluon field of the nucleon before annihilating with a quark from its parton distribution function to produce a dilepton pair.  Analogous to the case of FSI in SIDIS, the color flow in the DY scattering amplitude travels from the point $r^\mu = (0^+, r^-, \ul{r})$ along a light-like trajectory in the minus direction to a point at \textit{past} light-cone infinity $(0^+, -\infty, \ul{r})$, as in the right-hand side of Fig.~\ref{fig-GaugeLinks} \cite{Collins:2002kn}.  Combined with the complex-conjugated amplitude and a transverse gauge link to complete the contour, this gives a past-pointing ``light-cone staple'' contour characterizing ISI:
\begin{align}
 \label{DISS-GAUGE6}
 U_{ISI} [0,r] &\equiv U_{LC} [(0^+, 0^-, \ul{0}), (0^+, -\infty, \ul{0})] \times
                       U_\bot [(0^+, -\infty, \ul{0}), (0^+, -\infty, \ul{r})] 
											 \\ \nonumber &\times
                       U_{LC} [(0^+, -\infty, \ul{r}), (0^+, r^-, \ul{r})] \\ \nonumber
 & \!\!\!\!\!\!\!\! = \left[ \mathcal{P} \exp \left( \frac{i g}{2} \int\limits_{-\infty}^{0^-} dz^- A^{+a} (0^+, z^-, \ul{0}) T^a \right)    
    \right] 
    \left[ \mathcal{P} \exp \left(- i g \int\limits_{\ul{r}}^{\ul{0}} d \ul{z} \cdot \ul{A}^{a} (0^+, -\infty, \ul{z}) T^a \right) \right] 
		\\ \nonumber &\times
		\left[ \mathcal{P} \exp \left( \frac{i g}{2} \int\limits_{r^-}^{-\infty} dz^- A^{+a} (0^+, z^-, \ul{r}) T^a \right) \right].
\end{align}

\begin{figure}
 \centering
 \includegraphics[width=0.3\textwidth]{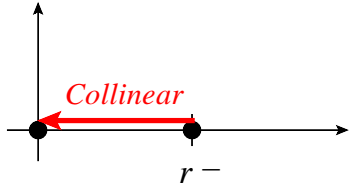}
 \caption{The collinear gauge link $U_{coll}[0,r]$ \eqref{DISS-GAUGE7}.  Both the ISI and FSI gauge links of Fig.~\ref{fig-GaugeLinks} reduce to this collinear gauge link after integration over the transverse momentum.  This gauge link is universal, and it can be eliminated entirely by the choice of light-cone gauge $A^+ = 0$.}
 \label{fig-collGaugeLinks}
\end{figure}

The fact that the color flow in SIDIS and DY generates different gauge links $U_C [0,r]$ is a sign of the process-dependence of the associated transverse-momentum-dependent parton distributions contained within the correlator $\Phi^C$.  This process dependence is a unique feature of the transverse-momentum paradigm \cite{Collins:2011zzd}.  If we integrate \eqref{DISS-GAUGE3} over the transverse momentum $d^2k$ to recover the collinear limit, we obtain a delta function $\delta^2 (\ul{r})$ that eliminates the transverse separation between the quark fields $\psibar(0) , \psi(r)$.  In this limit, the width of the ``light-cone staple'' gauge links shrinks to zero, and the parts of the gauge links extending out to light cone $\pm \infty$ cancel exactly.  What remains is just a gauge link along the light cone that connects the point $(0^+, r^-, \ul{0})$ to the origin, as shown in Fig.~\ref{fig-collGaugeLinks}:
\begin{align}
 \label{DISS-GAUGE7}
 U_{coll} = U_{ISI} = U_{FSI} = \mathcal{P} \exp \left( i g \int\limits_{r^-}^{0^-} dz^- A^{+a} (0^+, z^-, \ul{0}) \right).
\end{align}
Thus the gauge link that enters the collinear parton distribution functions is universal from one process to another, allowing one to measure the PDF in a DIS experiment and use it predictively in a DY experiment \cite{Sterman:1994ce}.  Furthermore, the collinear gauge link can be gauged away entirely by working in the $A^+ = 0$ light-cone gauge; this gives the collinear PDF's a simple interpretation as pure parton densities, without mixing due to initial- or final-state interactions.  The fact that in the transverse-momentum paradigm the gauge links and parton distributions are non-universal is a substantial threat to the predictive power of the theory.  It is only because of time-reversal symmetry, which relates the future-pointing FSI gauge link to the past-pointing ISI gauge link that one recovers the ability to apply this ``controlled process-dependence'' predictively.

These two cases - only ISI or only FSI - are the only two which are presently under solid theoretical control.  The general case in which both ISI and FSI contribute, as would be appropriate for hadron production from nucleon-nucleon collisions, is a frontier of active research at this time (see \cite{Gamberg:2010tj, Rogers:2010dm} and many others).  At $\ord{g^2}$ in the expansion of the gauge link \eqref{DISS-GAUGE2} for such hadronic collisions, the color flow becomes entangled between the correlators $\Phi$ in the projectile and in the target.  This color entanglement violates factorization in a fundamental way, making it impossible to separately define the properties of the projectile and target for this process \cite{Rogers:2010dm}.  Thus TMD factorization is known to hold for SIDIS and DY, but known to fail for nucleon-nucleon collisions.  Within the scope of this document, we will restrict ourselves to studying the effects of FSI, as exemplified by SIDIS, and ISI, as exemplified by DY.

\subsection{TMD Decomposition of Hadronic Structure}

The correlation function $\Phi_{ij}^C (x, \ul{k}; p, S)$ defined in \eqref{DISS-GAUGE3} can be decomposed in terms of its Dirac structure and its transverse-momentum dependence.  Using the conditions of hermiticity and $C/P/T$ symmetry, we can write a complete decomposition of $\Phi^C$ at leading twist as \cite{Meissner:2007rx}
\begin{align}
 \label{DISS-DECOMP1}
 \Phi^C (x, \ul{k} &; p, S) \equiv \left[f_1^q (x, k_T) - \frac{(\ul{k} \times \ul{S})}{m_N} f_{1T}^{\bot q} 
 (x, k_T) \right] \left[\frac{1}{4} \gamma^- \right] 
 \\ \nonumber &+
 \left[S_L g_1^q (x, k_T) + \frac{(\ul{k} \cdot \ul{S})}
 {m_N} g_{1T}^q (x, k_T) \right] \left[\frac{1}{4} \gamma^5 \gamma^- \right] 
 \\ \nonumber &+
 \left[S_\bot^i h_{1T}^q (x, k_T) + \left(\frac{k_\bot^i}{m_N}\right) S_L h_{1L}^{\bot q} (x, k_T) +
 \left(\frac{k_\bot^i}{m_N} \right) \left(\frac{\ul{k}\cdot\ul{S}}{m_N}\right) h_{1T}^{\bot q} \right]
 \left[\frac{1}{4} \gamma^5 \gamma_{\bot i} \gamma^- \right] 
 \\ \nonumber &+
 \left[\left(\frac{k_\bot^i}{m_N}\right) h_1^{\bot q} (x, k_T) \right] \left[\frac{1}{4} i \gamma_{\bot i}
 \gamma^- \right].
\end{align}
The 8 quantities $\{f_1^q , f_{1T}^{\bot q} , g_1^q , g_{1T}^q , h_{1T}^q , h_{1L}^q , h_{1T}^{\bot q} ,
h_1^{\bot q} \}$ are the independent leading-twist TMD parton distribution functions that parameterize the structure of the nucleon.  The TMD's are defined such that the azimuthal correlations with the direction of the transverse momentum $\ul{k}$ are explicitly contained in the pre-factors; thus the TMD's themselves are functions only of the magnitude $k_T$ and measure the strength of these azimuthal correlations.  The nomenclature of the TMD's is chosen so that the distributions labeled by $f, g$ and $h$ correspond to unpolarized, longitudinally-polarized, and transversely-polarized quarks, respectively.  This can be seen by projecting \eqref{DISS-DECOMP1} onto various Dirac structures \cite{Meissner:2007rx}
\begin{align}
 \label{DISS-DECOMP2}
 \frac{1}{2} \Tr \left[ \Phi^C \gamma^+ \right] &= f_1^q - \frac{(\ul{k} \times \ul{S})}{m_N} f_{1T}^{\bot q} 
 \\ \nonumber
 \frac{1}{2} \Tr \left[ \Phi^C \gamma^+ \gamma^5 \right] &= S_L g_1^q + \frac{(\ul{k} \cdot \ul{S})}
 {m_N} g_{1T}^q 
 \\ \nonumber
 \frac{1}{2} \Tr \left[ \Phi^C \gamma^+ \gamma_\bot^i \gamma^5 \right] &= S_\bot^i h_{1T}^q + 
 \left(\frac{k_\bot^i}{m_N}\right) S_L h_{1L}^{\bot q} + \left(\frac{k_\bot^i}{m_N} \right) 
 \left(\frac{\ul{k}\cdot\ul{S}}{m_N}\right) h_{1T}^{\bot q} + \epsilon_T^{ij} \left(\frac{k_\bot^j}{m_N}
 \right) h_1^{\bot q}
\end{align}
and comparing with the partonic interpretations derived at lowest order in \eqref{newTMD3} and \eqref{DISS-TMD10}.  

Of the 8 leading-twist TMD's, only 3 contributions remain in the collinear limit after integration over $d^2k$:
\begin{align}
 \label{DISS-DECOMP3}
 \Phi_{coll}(x; p, S) &\equiv \int d^2k \, \Phi^C (x,\ul{k}; p, S) 
 \\ \nonumber &= 
 \left[ \int d^2k \, f_1^q (x, k_T) \right] \left[\frac{1}{4} \gamma^- \right] + 
 \left[ S_L \int d^2k \, g_1^q (x, k_T) \right] \left[\frac{1}{4} \gamma^5 \gamma^- \right] 
 \\ \nonumber &+
 \left[S_\bot^i \int d^2k \left( h_{1T}(x,k_T) + \frac{1}{2} \frac{k_T^2}{m_N^2} h_{1T}^\bot (x,k_T) \right) \right] \left[ \frac{1}{4} \gamma^5 \gamma_{\bot i} \gamma^- \right] 
 \\ \nonumber &\equiv
 \bigg[f_{1,coll}^q (x) \bigg] \left[\frac{1}{4} \gamma^- \right] + \bigg[S_L \, g_{1,coll}^q(x) \bigg]\left[\frac{1}{4} \gamma^5 \gamma^- \right] + \bigg[S_\bot^i \, h_{1,coll}^q (x) \bigg] \left[ \frac{1}{4} \gamma^5 \gamma_{\bot i} \gamma^- \right].
\end{align}
These three quantities describe spin-spin correlations and correspond to the collinear distribution $f_{1, coll}^q$ of unpolarized quarks in an unpolarized nucleon, the collinear distribution $g_{1, coll}^q$ of longitudinally-polarized quarks in a longitudinally-polarized nucleon (\textit{helicity} distribution), and the collinear distribution $h_{1, coll}^q$ of transversely-polarized quarks in a transversely-polarized nucleon (\textit{transversity} distribution).  Thus $f_1^q$, $g_1^q$, and the linear combination $h_{1T} + \frac{1}{2} \frac{k_T^2}{m_N^2} h_{1T}^\bot$ correspond to the TMD distributions of the same quantities.

The other 5 TMD's describe new spin-orbit correlations between the azimuthal direction of the quark momentum $\ul{k}$ and the spin of either the quark or the nucleon.  These spin-orbit TMD's are \cite{Mulders:1995dh, Boer:1997nt}:
\begin{itemize}
 \item The \textit{Sivers function} $f_{1T}^{\bot q}$ - the azimuthal distribution of unpolarized quarks in a transversely-polarized nucleon.
 \item The \textit{``Worm-Gear'' g-function} $g_{1T}^q$ - the azimuthal distribution of longitudinally-polarized quarks in a transversely-polarized nucleon.  The name ``worm-gear'' refers to an axial gear shaped like a screw, which in combination with a conventional gear converts between longitudinal and transverse rotation.
 \item The \textit{``Worm-Gear'' h-function} $h_{1L}^{\bot q}$ - the azimuthal distribution of transversely-polarized quarks in a longitudinally-polarized nucleon.
 \item The \textit{``Pretzelosity'' function} $h_{1T}^{\bot q}$ - the azimuthal distribution of transversely-polarized quarks with spins aligned perpendicular to the transversely-polarized nucleon.  Strictly speaking, the pretzelosity contribution is the part of the $h_{1T}^{\bot q}$ term which does not contribute to the transversity distribution $h_1^{q}$ in \eqref{DISS-DECOMP3}.
 \item The \textit{Boer-Mulders function} $h_1^{\bot q}$ - the azimuthal distribution of transversely-polarized quarks in an unpolarized nucleon.
\end{itemize}
The 8 leading-order TMD's can be conveniently organized into the form of Table~\ref{table-qTMD}.

\begin{table}
\centering
\begin{tabular}{|c|c|c|c|}
 \hline  & $U$ & $L$ & $T$ \\ \hline
 $U$ & $f_1^q$ &  & $h_1^{\bot q} $ \\ \hline
 $L$ &  & $g_1^q$ & $h_{1L}^{\bot q}$ \\ \hline
 $T$ & $f_{1T}^{\bot q}$ & $g_{1T}^q$ & $h_1^q , h_{1T}^{\bot q}$ \\ \hline
\end{tabular}
\caption{Table classifying the 8 leading-order quark TMD's \eqref{DISS-DECOMP1}.  The columns labeled $U, L, T$ refer to unpolarized, longitudinally-polarized, and transversely-polarized quarks, respectively; the rows labeled $U, L, T$ refer to these quark distributions within an unpolarized, longitudinally-polarized, and transversely-polarized nucleon, respectively.}
\label{table-qTMD} 
\end{table}

For completeness, we should also briefly discuss the TMD parton distributions for gluons.  Consider a correlator of the field-strength tensors analogous to \eqref{DISS-GAUGE3}:
\begin{align}
 \label{gluonTMD1}
 \tilde{\Phi}^{\mu \nu}_C (x, \ul{k}) &\equiv \frac{1}{x p^+} \frac{1}{2(2\pi)^3} \int d^{2-} r \,
 e^{i k \cdot r} \bra{p S} F^{+ \nu a}(0) \, \mathcal{U}^{a b}_C[0,r] \, F^{+ \mu b}(r) \ket{p S}
 \\ \nonumber &=
 \frac{1}{x p^+} \frac{1}{4(2\pi)^3 \mathcal{V}^-} \int d^{2-} x \, d^{2-} y \,
 e^{i k \cdot (x-y)} \bra{p S} F^{+ \nu a}(y) \, \mathcal{U}^{a b}_C[y,x] \, F^{+ \mu b}(x) \ket{p S},
\end{align}
where $\mathcal{U}_C$ is a gauge link in the adjoint representation of $SU(N_c)$ and we have again multiplied and divided by $\mathcal{V}^- \equiv \frac{1}{2}\int d^{2-} y$.  While this correlator is gauge-invariant, its particle interpretation in terms of the number of gluons is not.  To obtain a gluon density interpretation, it is necessary to work in the light-cone gauge $A^+ =0$, for which the field-strength tensors appearing in \eqref{gluonTMD1} reduce to
\begin{align}
 \label{gluonTMD2}
 F^{+ \nu a}(y) \equiv &\partial^+ A^{\nu a} - \partial^\nu A^{+ a} + g f^{a b c} A^{+ b} A^{\nu c} \\ \nonumber
 \stackrel{(A^+ = 0)}{=} &\partial^+ A^{\nu a} = 2 \frac{\partial A^{\nu a}(y)}{\partial y^-} .
\end{align}
After rewriting the field-strength tensors in \eqref{gluonTMD1} as derivatives with respect to $x^- , y^-$, we can integrate by parts so that the derivatives act on the Fourier exponential, yielding a factor of $\frac{1}{4}(k^+)^2 = \frac{1}{4} x^2 (p^+)^2$.  This gives an expression for $\tilde{\Phi}_C^{\mu\nu}$ in terms of the gauge fields $A^{\mu a}$:
\begin{align}
 \label{gluonTMD3}
 \tilde{\Phi}^{\mu \nu}_C (x, \ul{k}) &=
 (x p^+) \frac{1}{4(2\pi)^3 \mathcal{V}^-} \int d^{2-} x \, d^{2-} y \,
 e^{i k \cdot (x-y)} \bra{p S} A^{\nu a}(y) \, \mathcal{U}^{a b}_C[y,x] \, A^{\mu b}(x) \ket{p S}.
\end{align}

To obtain a partonic density interpretation, we need to expand the correlator $\tilde{\Phi}_C^{\mu\nu}$ to lowest order in $\alpha_s$, which replaces the gauge link $\mathcal{U}_C^{ab}$ with unity.  Then we can use the relation between the gauge fields and gluon creation/annihilation operators
\begin{align}
 \label{gluonTMD4}
 A^{\nu a}(y) = \int\frac{d^{2+}\ell}{2(2\pi)^3 \ell^+} \sum_{\lambda}^{phys}
 \left( a_{\ell \lambda}^a \, \epsilon_\lambda^\nu (\ell) \, e^{-i \ell \cdot y} + a_{\ell \lambda}^{\dagger a} \, \epsilon_\lambda^{\nu *} (\ell) \, e^{i \ell \cdot y} \right),
\end{align}
where $\epsilon_\lambda^\mu$ are the physical gluon polarization vectors, to rewrite the correlator as
\begin{align}
 \label{gluonTMD5}
 \tilde{\Phi}^{\mu\nu}&(x,\ul{k}) = (x p^+) \frac{1}{4(2\pi)^3 \mathcal{V}^-} \int d^{2-} x \, d^{2-} y \,
 \left[\frac{d^{2+}\ell}{2(2\pi)^3 \ell^+}\right] \left[\frac{d^{2+}q}{2(2\pi)^3 q^+}\right]
 e^{i k \cdot (x-y)} 
 \\ \nonumber &\times
 \sum_{\lambda \eta}^{phys} \bra{p S} \left( a_{\ell \lambda}^a \, \epsilon_\lambda^\nu (\ell) \, e^{-i \ell \cdot y} + a_{\ell \lambda}^{\dagger a} \, \epsilon_\lambda^{\nu *} (\ell) \, e^{i \ell \cdot y} \right) 
 \left( a_{q \eta}^a \, \epsilon_\eta^\mu (q) \, e^{-i q \cdot x} + a_{q \eta}^{\dagger a} \, \epsilon_\eta^{\mu *} (q) \, e^{i q \cdot x} \right) \ket{p S}.
\end{align}
Dropping the terms $a a$ and $a^\dagger a^\dagger$ which change particle number, we obtain
\begin{align}
 \label{gluonTMD6}
 \tilde{\Phi}^{\mu\nu}&(x,\ul{k}) = (x p^+) \frac{1}{4(2\pi)^3 \mathcal{V}^-} \int d^{2-} x \, d^{2-} y \,
 \left[\frac{d^{2+}\ell}{2(2\pi)^3 \ell^+}\right] \left[\frac{d^{2+}q}{2(2\pi)^3 q^+}\right] 
 \\ \nonumber &\times
 \sum_{\lambda \eta}^{phys} \bra{p S} \bigg(
 a_{\ell \lambda}^a \,  a_{q \eta}^{\dagger a}
 \left[\epsilon_\lambda^\nu (\ell)  \epsilon_\eta^{\mu *} (q) \right]
 e^{-i (k+\ell) \cdot y} \, e^{i (k+q) \cdot x} +
 \\ \nonumber &+
 a_{\ell \lambda}^{\dagger a} \, a_{q \eta}^a
 \left[ \epsilon_\lambda^{\nu *} (\ell) \, \epsilon_\eta^\mu (q) \right]
  e^{-i (k-\ell) \cdot y} \, e^{i (k-q) \cdot x}
 \bigg) \ket{p S},
\end{align}
and, as in \eqref{DISS-TMD6}, we can perform the integrals over $d^{2-}x , d^{2-} y$ to obtain delta functions.  In the $a^\dagger a$ term, the delta functions set $k = q = \ell$, allowing us to perform the $d^{2+} \ell , d^{2+} q$ integrals.  In the $a a^\dagger$ term, the delta functions would set $k = -q = -\ell$, which is prohibited since kinematically all of $k^+, q^+, \ell^+$ are constrained to be positive.  Thus the only contribution that survives is the $a^\dagger a$ term:
\begin{align}
 \label{gluonTMD7}
 \tilde{\Phi}^{\mu\nu}&(x,\ul{k}) = \frac{1}{4(2\pi)^3} \frac{1}{p^+ \mathcal{V}^-} \frac{1}{x}
 \sum_{\lambda \eta}^{phys} \bra{p S} a_{k \lambda}^{\dagger a} \, a_{k \eta}^a \ket{p S}
 \left[ \epsilon_\lambda^{\nu *} (k) \, \epsilon_\eta^\mu (k) \right],
\end{align}
which has the interpretation of gluon number density in the nucleon state $\ket{p S}$.  As with the quark distribution, higher-order contributions from the gauge link mix this gluon density with multi-gluon correlation functions.

The gluon correlator $\tilde{\Phi}^{\mu\nu}_C$ is a Lorentz tensor, and it can be expanded both in terms of its Lorentz structure and dependence on the transverse momentum.  A decomposition of \eqref{gluonTMD7} is most easily expressed by promoting transverse vectors to 4-vectors, e.g. $k_\bot^\mu \equiv (0^+, 0^-, \ul{k})$ and by using invariant tensors $g_{T}^{\mu\nu} , \epsilon_T^{\mu\nu}$ appropriate for the transverse sector,
\begin{align}
 \label{transvtensors}
 g_T^{\mu\nu} &\equiv g^{\mu\nu} - \hat{t}^\mu \hat{t}^\nu + \hat{z}^\mu \hat{z}^\nu \\ \nonumber
 \epsilon_T^{\mu\nu} &\equiv \frac{1}{2} \epsilon^{+ - \mu \nu},
\end{align}
where $\hat{t}^\mu$ and $\hat{z}^\mu$ are unit vectors along the time and $z$ axes, respectively.  
A complete decomposition of \eqref{gluonTMD7} using hermiticity and $C/P/T$ symmetry at leading twist can be parameterized as \cite{Meissner:2007rx, Buffing:2013eka}:
\begin{align}
 \label{gluonTMD8}
 \tilde{\Phi}^{\mu\nu}_C (&x, \ul{k}; p, S) \equiv 
 \left[f_1^g (x,k_T) - \frac{(\ul{k} \times \ul{S})}{m_N} f_{1T}^{\bot g}(x,k_T)\right] \left[ -\frac{1}{2} g_T^{\mu\nu} \right]
 \\ \nonumber &+
 \left[S_L g_1^g (x,k_T) + \frac{(\ul{k}\cdot\ul{S})}{m_N} g_{1T}^g (x,k_T) \right] \left[-\frac{i}{2} \epsilon_T^{\mu\nu} \right] +
 \frac{1}{2 m_N^2} \left[ k_\bot^\mu k_\bot^\nu + \frac{1}{2} g_T^{\mu\nu} k_T^2 \right] h_1^{\bot g} (x,k_T) 
 \\ \nonumber &+
 \frac{1}{4 m_N^2} \bigg[k_\bot^\mu \epsilon_T^{\nu\alpha} k_{\bot\alpha} + k_\bot^\nu \epsilon_T^{\mu\alpha} k_{\bot\alpha} \bigg] \left[S_L h_{1L}^{\bot g} (x,k_T) + \frac{(\ul{k}\cdot\ul{S})}{m_N} h_{1T}^{\bot g} (x,k_T) \right]
 \\ \nonumber &+
 \frac{1}{8 m_N} \bigg[k_\bot^\mu \epsilon_T^{\nu\alpha} S_{\bot \alpha} + k_\bot^\nu \epsilon_T^{\mu\alpha} S_{\bot \alpha} + S_\bot^\mu \epsilon_T^{\nu\alpha} k_{\bot\alpha} + S_\bot^\nu \epsilon_T^{\mu\alpha} k_{\bot\alpha} \bigg] h_{1T}^g (x,k_T).
\end{align}
This defines the 8 leading-order gluon TMD's, which are analogous to their quark counterparts \eqref{DISS-DECOMP1}, with circularly-polarized gluons playing the role of longitudinally-polarized quarks and linearly-polarized gluons playing the role of transversely-polarized quarks.  As with the quark sector, these 8 leading-order gluon TMD's can be summarized in the form of Table~\ref{table-gTMD}.

\begin{table}
\centering
\begin{tabular}{|c|c|c|c|}
 \hline  & $U$ & $Circ$ & $Lin$ \\ \hline
 $U$ & $f_1^g$ &  & $h_1^{\bot g}$ \\ \hline
 $L$ &  & $g_1^g$ & $h_{1L}^{\bot g}$ \\ \hline
 $T$ & $f_{1T}^{\bot g}$ & $g_{1T}^g$ & $h_1^g , h_{1T}^{\bot g}$ \\ \hline
\end{tabular}
\caption{Table classifying the 8 leading-order gluon TMD's \eqref{gluonTMD8}.  The columns labeled $U, Circ, Lin$ refer to unpolarized, circularly-polarized, and linearly-polarized gluons, respectively; the rows labeled $U, L, T$ refer to these gluon distributions within an unpolarized, longitudinally-polarized, and transversely-polarized nucleon, respectively.}
\label{table-gTMD} 
\end{table}

To get a feel for the structure of the quark TMD's, it is useful to consider a toy model in which the distributions are calculable in perturbation theory.  One such toy model is the scalar diquark model, in which the nucleon is regarded as a fundamental pointlike field $\psi_N$ that couples to the quark field $\psi_q$ and a pointlike scalar ``diquark'' field $\phi$ through a Yukawa vertex.  We take the Lagrangian for the diquark model to be \cite{Meissner:2007rx, Brodsky:2000ii, Brodsky:2002cx}
\begin{align}
 \label{DISS-DECOMP4}
 \mathcal{L}_{diquark} &= \psibar_N (i \slashed{\partial} - m_N) \psi_N + \psibar_q (i \slashed{D}) \psi_q
 + \phi^* (\overleftarrow{D}_\mu \overrightarrow{D}^\mu - \lambda^2) \phi - \frac{1}{4} F^{\mu \nu a} F_{\mu \nu}^a \\ \nonumber &+ G (\psibar_N \psi_q + \psibar_q \psi_N ) \phi
\end{align}
where the covariant derivative is $i D_\mu = i \partial_\mu + g \mathcal{Q} A_\mu^a T^a$.  Here $g$ represents the QCD coupling strength of the quark and diquark fields, $G$ represents the coupling strength of the nucleon to a quark + diquark, and we have included the nucleon mass $m_N$ and diquark mass $\lambda$ but considered the quarks to be massless.  The charges $\mathcal{Q}$ (in units of $g$) of the quark and diquark are $+1$ and $-1$, respectively, which follows from the requirement that the nucleon be color-neutral.  In order to prevent the spontaneous decay of the nucleon, one can impose the mass ordering $\lambda > m_N$.

\begin{figure}
 \centering
 \includegraphics[width=0.6\textwidth]{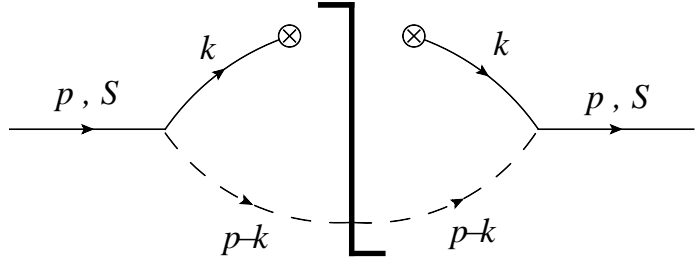}
 \caption{The quark-quark correlation function $\Phi$ evaluated in the scalar diquark model \eqref{DISS-DECOMP4} to lowest order $G^2$ in the coupling, \eqref{DISS-DECOMP5}.  This consists of the square of the wave function for a nucleon with momentum $p$ and spin $S$ splitting into a quark with momentum $k$ and diquark with momentum $p-k$.}
 \label{fig-diquark}
\end{figure}

In this implementation of the diquark model, it is straightforward to evaluate the distribution $\Phi_{ij}$ to lowest order (without rescattering) using \eqref{DISS-TMD1}, as shown in Fig.~\ref{fig-diquark}:
\begin{align}
 \label{DISS-DECOMP5}
 \Phi_{ij}(x,\ul{k}) = \frac{1}{2(2\pi)^3} \frac{1}{(1-x)p^+} \frac{G^2}{(k^2)^2} 
 \left[ \ubar{S} (p) \slashed{k} \right]_j \left[\slashed{k} U_S (p) \right]_i .
\end{align}
Using this, we can evaluate any of the TMD's defined in \eqref{DISS-DECOMP1}; the results to lowest order for the unpolarized distribution, the helicity distribution, and the Sivers function are shown below \cite{Meissner:2007rx}.
\begin{align}
 \label{DISS-DECOMP6}
 f_1^q (x, k_T) &= \frac{G^2}{2(2\pi)^3} \, (1-x) \, \frac{k_T^2 + x^2 m_N^2}{[k_T^2 + x \lambda^2 - x(1-x) m_N^2]^2} \\ \nonumber
 g_1^q (x, k_T) &= - \frac{G^2}{2(2\pi)^3} \, (1-x) \, \frac{k_T^2 - x^2 m_N^2}{[k_T^2 + x \lambda^2 - x(1-x) m_N^2]^2} \\ \nonumber
 f_{1T}^{\bot q} (x,k_T) &= 0 \;\; + \; \ord{G^2 \alpha_s}
\end{align}
At large transverse momentum $(k_T \gg m_N , \lambda)$, both $f_1^q$ and $g_1^q$ fall off as $1/k_T^2$, with $g_1^q = - f_1^q$.  From the partonic interpretations \eqref{newTMD3} and \eqref{DISS-TMD10}, we see that this implies that $\bra{p, (+z)} b^\dagger_{k, +z} b_{k, +z} \ket{p, (+z)} = 0$ so that all the quarks in the longitudinally-polarized nucleon have their spins polarized antiparallel to the spin of the nucleon.  This is simply a reflection of the conservation of helicity in the massless limit of Yukawa theory.  Additionally, we emphasize in \eqref{DISS-DECOMP6} that the Sivers function $f_{1T}^{\bot q}$ vanishes to order $G^2$ in the diquark model.  This is a manifestation of the symmetries discussed in \eqref{spinors-squared} in which transverse-spin dependence cannot appear at Born level in a quantum process.


 \section{The Sivers Function}
\label{sec-Sivers}

The key ideas introduced in this Chapter are all embodied in the quark Sivers function $f_{1T}^{\bot q}$.  In this Section, we will analyze in detail the sign reversal of the Sivers function between semi-inclusive deep inelastic scattering and the Drell-Yan process, performing explicit calculations within the diquark model \eqref{DISS-DECOMP4}.  This detailed analysis is original work which is considerably more technical than previous Sections of this Chapter and is beyond the level of a basic introduction.  Nonetheless, it is important to see the general statements made previously verified in an explicit calculation.  In this Section we follow closely our paper \cite{Brodsky:2013oya}.

 \subsection{The SIDIS / DY Sign-Flip Relation}
 \label{subsec-sign_flip}

We can extract the Sivers function from \eqref{DISS-DECOMP2} by anti-symmetrizing with respect to either the transverse spin $\ul{S}$ or the transverse momentum $\ul{k}$:
\begin{align}
 \label{DISS-SIVERS1}
 \frac{(\ul{S} \times \ul{k})}{m_N} f_{1T}^{\bot q} (x, k_T) &= \frac{1}{4} \Tr \left[\Phi^C (x, \ul{k}; p, \ul{S}) \, \gamma^+ \right] - (\ul{S} \rightarrow -\ul{S}) \\ \nonumber
 &= \frac{1}{4} \int \frac{d^{2-}r}{(2\pi)^3} e^{i k \cdot r} \bra{p \ul{S}} \psibar(0) \, \gamma^+
 U_C [0,r] \, \psi(r)
 \ket{p \ul{S}} - (\ul{S} \rightarrow -\ul{S}),
\end{align}
where the two-dimensional cross-product employed in $(\ul{S} \times \ul{k})$ is defined in \eqref{tcross}.  From the expression for the SIDIS cross-section \eqref{SIDIS18}, we see that the single transverse spin asymmetry (STSA) \eqref{AN1} of quark production in SIDIS is proportional to the Sivers function.  The Sivers function is thus the partonic analog of STSA, reflecting the transverse spin asymmetry of unpolarized quarks within the TMD parton distribution of a transversely-polarized nucleon.

As discussed in Sec.~\ref{subsubsec-STSA}, the asymmetry is odd under ``naive'' time-reversal ($PT$), so let us examine the transformation of \eqref{DISS-SIVERS1} under the combination $PT$ of both parity inversion and time reversal, which has the following transformation properties:
\begin{align}
 \label{DISS-SIVERS2}
 (PT) \, \psi(x^\mu) \, (PT)^\dagger &= \gamma^0 \gamma^1 \gamma^3 \, \psi(-x^\mu) \\ \nonumber
 (PT) \, \psibar(x^\mu) \, (PT)^\dagger &= - \psibar(-x^\mu) \, \gamma^0 \gamma^1 \gamma^3  \\ \nonumber
 (PT) \, \ket{p, \ul{S}} &= e^{i \phi} \ket{p, -\ul{S}} \\ \nonumber
 (PT) \, const \, (PT)^\dagger &= (const)^* \\ \nonumber
 \bra{f} (PT)^\dagger \, \hat{O} \, (PT) \ket{i} &= \bra{(PT) f} \hat{O} \ket{(PT) i}^*.
\end{align}
Loosely speaking, parity inversion changes the direction of momenta but leaves pseudovectors like spins unchanged, while time reversal changes the direction of both momenta and spins.  Thus their product $PT$ leaves momenta unchanged but flips the direction of the spin.  The last two properties in \eqref{DISS-SIVERS2} are consequences of the ``anti-linearity'' of time reversal, which introduces additional complex conjugation.

First, let us consider the Sivers function at lowest order in $\alpha_S$ which has a quark density interpretation; this amounts to neglecting the gauge link $U_C [0,r]$ in \eqref{DISS-SIVERS1}.  Inserting $(PT)^\dagger (PT)$ between the factors in the matrix element and using the transformation rules \eqref{DISS-SIVERS2} gives
\begin{align}
 \label{DISS-SIVERS3}
 \bra{p \ul{S}} \psibar(0) \gamma^+ \psi(r) \ket{p\ul{S}} &=
 \bra{p \ul{S}} (PT)^\dagger \left[ (PT) \psibar(0) (PT)^\dagger \right] \left[(PT) \gamma^+ (PT)^\dagger \right] \\ \nonumber &\times \left[(PT) \psi(r) (PT)^\dagger \right] (PT) \ket{p\ul{S}}
 \\ \nonumber &= 
 - \bra{p, - \ul{S}} \psibar(0) \, \gamma^0 \gamma^1 \gamma^3 (\gamma^+)^* \gamma^0 \gamma^1 \gamma^3 
 \, \psi(-r) \ket{p, -\ul{S}}^* 
 \\ \nonumber &=
 + \bra{p, - \ul{S}} \psibar(0) \, \gamma^+ \, \psi(-r) \ket{p, -\ul{S}}^* 
 \\ \nonumber &=
 + \bra{p, - \ul{S}} \psibar(-r) \, \gamma^+ \, \psi(0) \ket{p, -\ul{S}}
 \\ \nonumber \bra{p \ul{S}} \psibar(0) \gamma^+ \psi(r) \ket{p\ul{S}} &=
 + \bra{p, - \ul{S}} \psibar(0) \, \gamma^+ \, \psi(r) \ket{p, -\ul{S}}.
\end{align}
When substituted back into \eqref{DISS-SIVERS1}, this implies that the Sivers function vanishes at this level of accuracy:
\begin{align}
 \label{DISS-SIVERS4}
 f_{1T}^{\bot q} = 0 \;\; + \:\: \ord{\alpha_s}.
\end{align}
The vanishing of the Sivers function at the partonic level is a consequence of the time-reversal invariance of QCD \cite{Collins:1992kk} and is equivalent to our derivation in Sec.~\ref{subsubsec-STSA} of the vanishing of the single transverse spin asymmetry at Born level.

\begin{figure}
 \centering
 \includegraphics[width=0.9\textwidth]{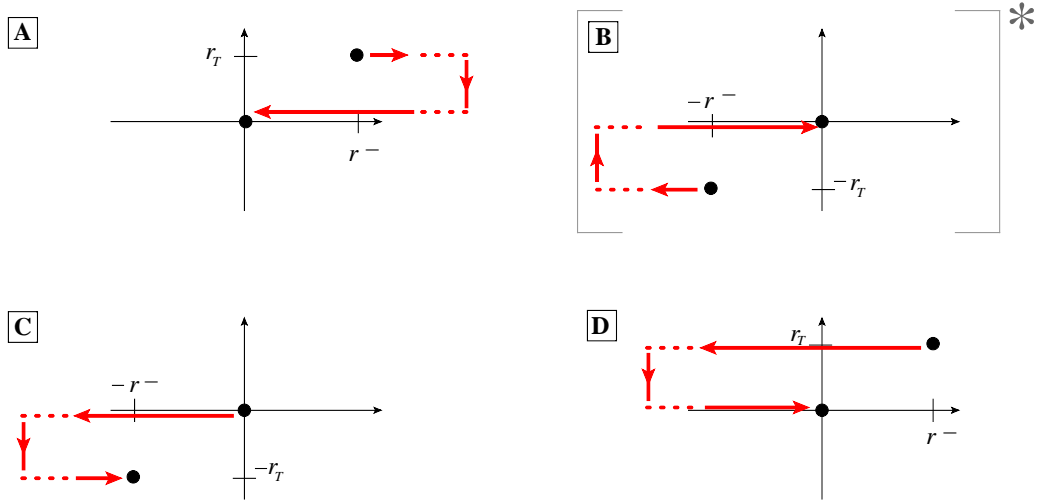}
 \caption{Transformation of the gauge link $U_C[0,r]$ under the manipulations which impose $PT$ invariance on the Sivers function.  (A) Gauge link $U_{FSI} [0,r]$ of \eqref{DISS-GAUGE5} representing final-state interactions.  (B) Under application of $PT$, the spacetime endpoints of the gauge link are reflected and a complex conjugate is introduced, but the direction of color flow toward the origin remains the same.  (C) Hermitian conjugation reverses the direction of the gauge link.  (D) A shift of coordinates returns the endpoints to their original positions, resulting in the initial-state interaction gauge link $U_{ISI} [0,r]$ of \eqref{DISS-GAUGE6}.}
 \label{fig-GaugeLinkPT}
\end{figure}

On the other hand, when we go beyond the Born level to include the gauge link $U_C[0,r]$, we introduce another variable into the time-reversal transformation.  Consider the transformation of a future-pointing gauge link $U_{FSI} [0,r]$ of \eqref{DISS-GAUGE5}.  A $PT$ transformation reflects the spacetime endpoints of each segment of the gauge link, but leaves the direction of color flow unchanged; it also complex-conjugates the gauge factor due to the anti-linearity of time reversal:
\begin{align}
 \label{DISS-SIVERS5}
 (PT)^\dagger \, U_{FSI} [0,r] \, (PT) = U_{ISI}^* [0,-r].
\end{align}
The result, illustrated in Fig.~\ref{fig-GaugeLinkPT}~(B), is a ``light-cone staple'' flowing from the reflected endpoint $-r$ to past light-cone infinity and back to the origin.  Repeating the steps of \eqref{DISS-SIVERS3} also applies Hermitian conjugation and a shift of coordinates; these steps, illustrated in Fig.~\ref{fig-GaugeLinkPT}~(C-D), transform the gauge link into the past-pointing gauge link of \eqref{DISS-GAUGE6}.
Thus the extension of \eqref{DISS-SIVERS3} beyond the lowest order gives
\begin{align}
 \nonumber
 \bra{p \ul{S}} \psibar(0) \gamma^+ U_{FSI}[0,r] \psi(r) \ket{p\ul{S}} &=
 \bra{p, - \ul{S}} \psibar(0) \, \gamma^+ \, \left[ (PT) U_{FSI}[0,r] (PT)^\dagger \right] 
 \psi(-r) \ket{p, -\ul{S}}^* 
 \\ \nonumber &=
 + \bra{p, - \ul{S}} \psibar(0) \, \gamma^+ \, U_{ISI}^*[0,-r] \psi(-r) \ket{p, -\ul{S}}^* 
 \\ \nonumber &=
 + \bra{p, - \ul{S}} \psibar(-r) \, \gamma^+ \, U_{ISI}[-r,0] \psi(0) \ket{p, -\ul{S}}
 \\ \label{DISS-SIVERS5a} \bra{p \ul{S}} \psibar(0) \gamma^+ U_{FSI}[0,r] \psi(r) \ket{p\ul{S}} &=
 + \bra{p, - \ul{S}} \psibar(0) \, \gamma^+ U_{ISI}[0,r] \, \psi(r) \ket{p, -\ul{S}}.
\end{align}
When applied to the Sivers function \eqref{DISS-SIVERS1}, this implies that
\begin{align}
 \nonumber
 \frac{(\ul{S} \times \ul{k})}{m_N} \left[f_{1T}^{\bot q} (x, k_T) \right]_{FSI} &= \frac{1}{4} \int \frac{d^{2-}r}{(2\pi)^3} e^{i k \cdot r} \bra{p \ul{S}} \psibar(0) \, \gamma^+
 U_{FSI} [0,r] \, \psi(r) \ket{p \ul{S}} - (\ul{S} \rightarrow -\ul{S})
 \\ \nonumber & \!\!\!\! =
 \frac{1}{4} \int \frac{d^{2-}r}{(2\pi)^3} e^{i k \cdot r} \bra{p, - \ul{S}} \psibar(0) \, \gamma^+
 U_{ISI} [0,r] \, \psi(r) \ket{p, - \ul{S}} - (\ul{S} \rightarrow -\ul{S})
 \\ \label{DISS-SIVERS6} &=
 - \frac{(\ul{S} \times \ul{k})}{m_N} \left[f_{1T}^{\bot q} (x, k_T) \right]_{ISI}
\end{align}
and therefore that the Sivers function itself switches sign between processes with a future-pointing contour and a past-pointing contour.  In particular, this implies a precise, measurable sign-flip relation between the Sivers functions measured in semi-inclusive deep inelastic scattering (SIDIS) and the Drell-Yan process (DY) \cite{Collins:2002kn}:
\begin{align}
 \label{DISS-SIVERS7}
 \left[ f_{1T}^{\bot q} \right]_{SIDIS} = - \left[ f_{1T}^{\bot q} \right]_{DY}.
\end{align}

To understand the interplay between STSA, the Sivers function, time reversal, and the SIDIS/DY sign flip, let us return to the diquark model defined in \eqref{DISS-DECOMP4}.  In the following sections, we will explicitly calculate the spin-dependent part of the SIDIS and DY cross-sections and analyze the manner in which the asymmetry arises from the imaginary part \eqref{ImPart3}.  While the analysis of SIDIS is straightforward, for DY we will find that the imaginary part responsible for the asymmetry is not exactly the same as in SIDIS, suggesting that the sign-flip relation \eqref{DISS-SIVERS7} is not exact.  Instead, we will see that the SIDIS/DY sign-flip holds to leading-twist accuracy and that subleading corrections enter which suggest the breakdown of the sign flip when $Q^2$ is not sufficiently large.


\subsection{SIDIS Sivers Function in the Diquark Model}
\label{SIDISF}

To begin, let us establish the kinematics.  Following
\cite{Brodsky:2002cx}, we work in the Drell-Yan-West frame which is
collinear to the nucleon $(\ul{p}=\ul{0})$ and boosted such that
$q^+ =0$ exactly.  In this frame, then, the photon's virtuality comes
from its transverse components: $Q^2 = q_T^2$.  We define the
longitudinal momentum fraction exchanged in the $t$-channel as $\Delta
\equiv r^+/ p^+$.  Then momentum conservation and the on-shell
conditions for the nucleon, quark, and diquark fix $r^-$ and $q^-$:
\begin{eqnarray}
\label{MDISkin}
r^- &=& p^- - (p-r)^- = \frac{m_N^2}{p^+} - \frac{r_T^2 + \lambda^2}{(1-\Delta)p^+} \\ \nonumber
q^- &=& (q+r)^- - r^- = \frac{(\ul{q}+\ul{r})_T^2}{\Delta p^+} - r^- 
\approx \frac{Q^2 + 2 (\ul{q} \cdot \ul{r})}{\Delta p^+} + \mathcal{O}\left(\frac{\bot^2}{p^+}\right).
\end{eqnarray}
These kinematics can be summarized as
\begin{align}
\label{MDISkin2}
p^\mu &= \left( p^+ \, , \, \frac{m_N^2}{p^+} \, , \, \ul{0} \right)
\\ \nonumber q^\mu &= \left( 0 \, , \,
  \frac{(\ul{q}+\ul{r})_T^2}{\Delta p^+} - \frac{m_N^2}{p^+} +
  \frac{r_T^2 + \lambda^2} {(1-\Delta)p^+} \, , \, \ul{q}
\right) \\ \nonumber r^\mu &= \left( \Delta p^+ \, , \,
  \frac{m_N^2}{p^+} - \frac{r_T^2 + \lambda^2}{(1-\Delta)p^+} \, ,
  \, \ul{r} \right).
\end{align}

When it is necessary to approximate the kinematics, we
will work in the limit 
\begin{align}\label{approx}
  s, Q^2 , q_T^2 \gg m_N^2 , \lambda^2, r_T^2, k_T^2
\end{align}
corresponding to Bjorken kinematics in which $s$ and $Q^2$ are large, but their ratio is fixed and $\ord{1}$.

\begin{figure}
\centering
\includegraphics[width=12cm]{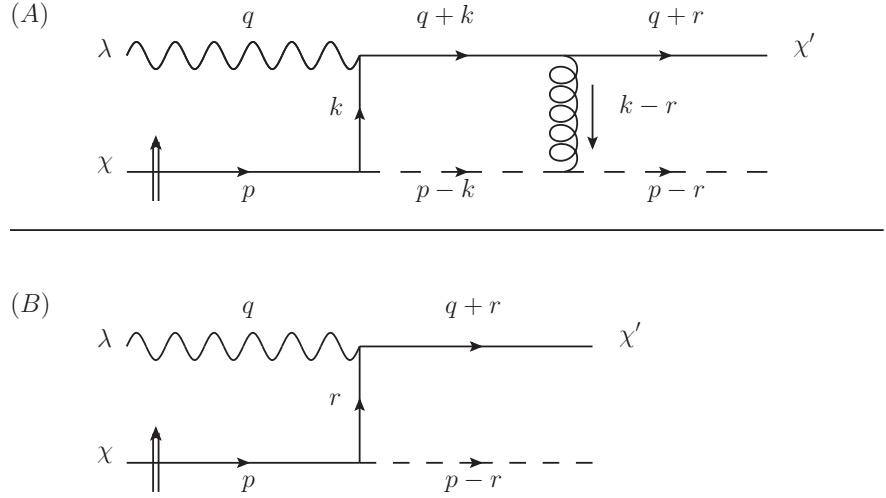}
\caption{Diagrams for the $\gamma^* + p^\uparrow \rightarrow q + X$
  SIDIS amplitude at one-loop order (A) and tree-level (B). The
  incoming solid line denotes the transversely polarized nucleon, which
  splits into a quark (outgoing solid line) and a diquark (dashed
  line).}
\label{figDIS}
\end{figure}

With these kinematics, we can write down the one-loop amplitude shown
in Fig.~\ref{figDIS}~(A) as
\begin{align}
\label{MA1DIS}
\mathcal{A}_1^{DIS} &= -i g^2 e_f G C_f \int \frac{d^4 k}{(2\pi)^4} \\ \nonumber
&\times \frac{\ubar{\chi'}(q+r) (2 \slashed{p} - \slashed{k} - \slashed{r})(\slashed{q} + \slashed{k}) 
  \slashed{\epsilon}_\lambda \slashed{k} U_\chi (p)} {[k^2 + i\epsilon] \, [(q+k)^2 + i\epsilon] \, [(k-r)^2 + i\epsilon] \, 
  [(p-k)^2 - \lambda^2 + i\epsilon]}
\\ \nonumber
&=
\frac{i g^2 e_f G C_F}{2 (2\pi)^4 (p^+)^3} \int \frac{dx \, dk^- \, d^2k}{x^2 (x-\Delta) (1-x)} 
\\ \nonumber  & \!\!\!\!\!\! \times
\frac{\ubar{\chi'}(q+r) (2\slashed{p}-\slashed{k}-\slashed{r})(\slashed{q}+\slashed{k})\slashed{\epsilon}_\lambda \slashed{k} U_\chi(p)}{ \left[ k^- - \frac{k_T^2 - i\epsilon}{x p^+} \right] \left[k^- + q^- - 
    \frac{ (\ul{q}+\ul{k})_T^2-i\epsilon}{x p^+} \right] \left[k^- - r^- - \frac{(\ul{k}-\ul{r})_T^2-i
      \epsilon}{(x-\Delta)p^+} \right] \left[k^- - p^- + \frac{k_T^2 + \lambda^2 - i\epsilon}{(1-x)p^+} 
  \right]} ,
\end{align}
where the longitudinal momentum fraction in the loop is $k^+ \equiv x
p^+$ and $C_F = (N_c^2 -1)/2 N_c$ is the Casimir operator in the
fundamental representation.  Similarly, the tree-level amplitude shown
in Fig. \ref{figDIS} (B) is
\begin{align}
\label{MA0DIS}
\mathcal{A}_0^{DIS} = \frac{e_f G}{r^2} \ubar{\chi'}(q+r)
\slashed{\epsilon}_\lambda \slashed{r} U_\chi(p) .
\end{align}

The lowest-order contribution to the spin-difference amplitude squared $\Delta \left|\mathcal{A}\right|^2$
defined in \eqref{AN2} comes from the overlap of these diagrams and, in particular, the
imaginary part of the denominators (cf. \eq{ImPart3}):
\begin{align}
\label{MADIS}
\Delta \left| \mathcal{A}_{DIS} \right|^2 &= 2i 
\left[ \frac{g^2 e_f^2 G^2 C_F}{2(2\pi)^4 r^2 (p^+)^3} \right]
\int \frac{dx \, d^2k}{x^2 (x-\Delta) (1-x)} \, \mathrm{Im} 
\left\{ \int dk^- \frac{i} {\left[k^- - \frac{k_T^2-i\epsilon}{x p^+} \right] } 
\right. \\ \nonumber &\times \left.
\frac{1} 
{
 \left[k^- + q^- - \frac{(\ul{q}+\ul{k})_T^2-i\epsilon}{x p^+} \right] 
 \left[k^- - r^- - \frac{(\ul{k}-\ul{r})_T^2 - i\epsilon}{(x-\Delta)p^+} \right] 
 \left[k^- - p^- + \frac{k_T^2 + \lambda^2 - i\epsilon}{(1-x)p^+} \right]
}
\right\} \\ \nonumber & \!\!\!\!\! \times \sum_{\chi',\lambda} \left[
\ubar{\chi}(p) \slashed{r} \slashed{\epsilon}^*_\lambda U_{\chi'}(q+r) \ubar{\chi'}(q+r) (2 \slashed{p} - 
\slashed{k} - \slashed{r}) (\slashed{q} + \slashed{k}) \slashed{\epsilon}_\lambda \slashed{k} U_{\chi}(p) 
- (\chi \rightarrow - \chi) \right].
\end{align}
Since we are considering a single-spin asymmetry, we will need to impose the condition that the incident virtual photon (and hence, the lepton that emitted it) is unpolarized.  As in \eqref{DISS-TMD8}, this reduces the interaction of the virtual photon with the active quark to a $\gamma^+$ vertex, which projects out the Sivers function via \eqref{DISS-DECOMP2}.  Thus we now sum over the spin of the outgoing quark and use
\begin{align}
  \label{eq:polsum}
  \sum_\lambda \epsilon^{* \, \mu}_\lambda (q) \, \epsilon^\nu_\lambda
  (q) \rightarrow - g^{\mu\nu},
\end{align}
obtaining
\begin{eqnarray}
\label{MADIS2}
\Delta \left| \mathcal{A}_{DIS} \right|^2 &=& \frac{2i g^2 e_f^2 G^2 C_F}{(2\pi)^4 r^2 (p^+)^3}
\int \frac{dx \, d^2k}{x^2 (x-\Delta)(1-x)} \, \mathcal{I} \,
\\ \nonumber &\times&
 \bigg[ \ubar{\chi}(p) \slashed{r} (\slashed{q} + \slashed{k}) (2 \slashed{p} -
\slashed{k} - \slashed{r}) (\slashed{q} + \slashed{r}) \slashed{k} U_\chi(p) \, - \,
(\chi \rightarrow -\chi) \bigg]
\end{eqnarray}
where the imaginary part that is essential for generating the
asymmetry comes from the expression
\begin{align}
\label{MIDIS}
\mathcal{I} &\equiv \mathrm{Im} \! \left\{ \!  \int \! \! \frac{i \;
    dk^-}{ \left[k^- - \frac{k_T^2-i\epsilon}{x p^+} \right] \!
    \!  \left[k^- + q^- - \frac{(\ul{q}+\ul{k})_T^2-i\epsilon}{x
        p^+} \right] \!  \!  \left[k^- - r^- -
      \frac{(\ul{k}-\ul{r})_T^2-i\epsilon}{(x-\Delta)p^+} \right] } \right. \! 
\\ \nonumber &\times \left.		
 \frac{1}{\left[k^- - p^- + \frac{k_T^2+\lambda^2-i\epsilon}{(1-x)p^+} \right]}				
\right\} \! .
\end{align}

Notice that the numerator in \eqref{MADIS2} containing the Dirac matrix
element is $k^-$-dependent, such that the $d k^-$ integration
contained in $\mathcal{I}$ from \eq{MIDIS} applies to it
too. Superficially the numerator of \eqref{MADIS2} could
scale as $(k^-)^3$ at large $k^-$, which would endanger convergence;
however, it actually only scales as $(k^-)^2$ since $(\gamma^+)^2 =
0$.  Thus the $k^-$ integral scales at most as $dk^-/(k^-)^{2}$, which
converges and allows us to close the contour in either the upper or
the lower half-plane.

In addition, we will demonstrate below that in the kinematic limit at
hand given by \eq{approx} the leading contribution to the Dirac matrix
element in \eq{MADIS2} is, in fact, $k^-$-independent. We will,
therefore, proceed under the assumption that this is the case and that
all the $k^-$ dependence in \eqref{MADIS2} is contained in the
integrand of \eq{MIDIS}, evaluating the integration in \eqref{MIDIS}
separately.

The imaginary part in \eqref{MIDIS} comes from the denominators, which
corresponds to putting two of the loop propagators on shell: one
occurs from the residue of the $dk^-$ integral and the other occurs by
taking the imaginary part.  However, there are strong kinematic constraints
that restrict which combinations of propagators can go on-shell
simultaneously.  We are working in the limit of massless quarks, and
$1 \leftrightarrow 2$ processes for on-shell massless particles are
forbidden by four-momentum conservation; cuts corresponding to such
processes will explicitly be impossible to put on shell.  Other cuts
correspond to spontaneous nucleon decay; nucleon stability against decay
through various channels must be imposed by hand, resulting in
kinematic constraints on the masses of the nucleon and the scalar.

There are four poles to the $dk^-$ integral \eqref{MIDIS}, labeled
below as \ding{172} - \ding{175}.  Depending on the hierarchy of the
longitudinal momentum fractions $x$ and $\Delta$, these poles may be
located either above or below the real $k^-$ axis.  Since the outgoing
quark and scalar are on-shell, we have $(q+r)^+ = \Delta p^+ > 0$ and
$(p-r)^+ = (1-\Delta)p^+ > 0$ so that $0<\Delta<1$.  This allows us to
write four distinct kinematic regimes in which to classify the poles:
$(x < 0 < \Delta < 1)$, $(0 < x < \Delta < 1)$, $(0 < \Delta < x <
1)$, and $(0 < \Delta < 1 < x)$.  The classification of the four pole
locations as above or below the real $k^-$ axis for each of these
regimes is listed in Table \ref{MDIStable}.

\begin{table}[ht]
\centering
\begin{tabular}{|cl|c|c|c|c|}
 \hline &Pole & $x<0$ & $0<x<\Delta<1$ & $0<\Delta<x<1$ & $x>1$ \\ \hline
 \ding{172}& $k^- = \frac{k_T^2-i\epsilon}{xp^+}$ & above & below & below & below \\
 \ding{173}& $k^- = - q^- + \frac{(\ul{q}+\ul{k})_T^2-i\epsilon}{xp^+}$ & above & below & below & below \\
 \ding{174}& $k^- = r^- + \frac{(\ul{k}-\ul{r})_T^2-i\epsilon}{(x-\Delta)p^+}$ & above & above & below &
  below \\
 \ding{175}& $k^- = p^- - \frac{k_T^2 + \lambda^2 -i\epsilon}{(1-x)p^+}$ & above & above & above &
  below \\ \hline \hline
 &Contribution: & $0$ & Case A & Case B & $0$ \\ \hline
\end{tabular}
\caption{Table classifying the pole locations of \eqref{MIDIS} as lying either above or below the $\mathrm{Re} \, k^-$ axis.}
\label{MDIStable} 
\end{table}

For $x<0$ or $x>1$, all the poles fall on the same side of the
$\mathrm{Re}\, k^-$ axis, so that we can close the contour in the
other direction and get zero contribution.  The physical region
corresponds to $0<x<1$, and there are two distinct time-orderings of
the diagram, $x<\Delta$ and $x>\Delta$.  We examine these two cases
below.

\begin{figure}[htb]
\centering
\includegraphics[width=0.9\textwidth]{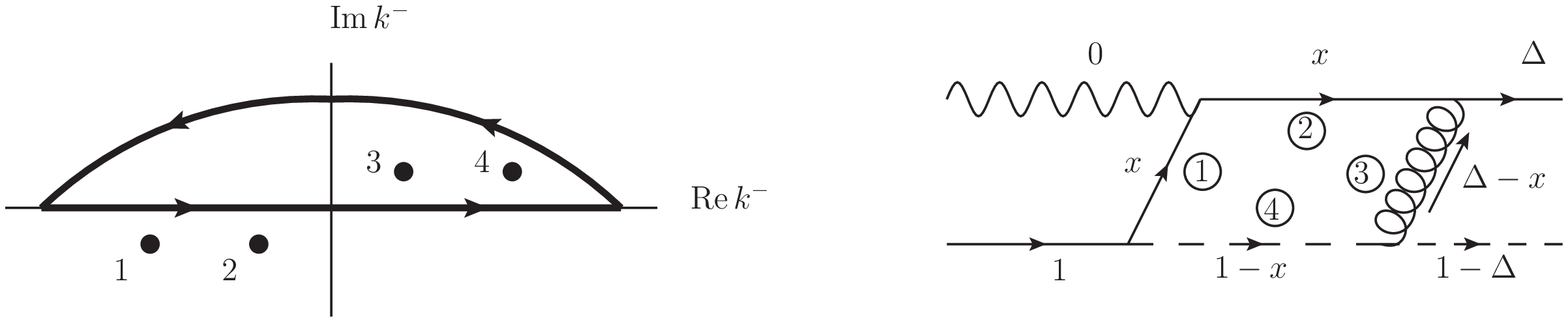}
\caption{Illustration of the poles (left) and
  corresponding time-ordered diagram (right) of \eqref{MIDIS} for the
  kinematic regime Case A: $0<x<\Delta<1$.  We choose to close the
  contour in the upper half-plane, enclosing the poles \ding{174} and
  \ding{175}.  Note that the placement of the poles is only schematic,
  indicating the sign of their imaginary part; the placement on the
  real axis has no significance.}
\label{MDISA} 
\end{figure}

\vspace{0.5cm}

For Case A: $0<x<\Delta<1$, we can close the contour in the upper
half-plane, enclosing the poles \ding{174} and \ding{175}, as in
Fig. \ref{MDISA}.  Let us consider the possible contributions to
\eqref{MIDIS} from the residue and imaginary parts of the various
poles.

\begin{itemize}
 \item \underline{Res[\ding{174}] Im[\ding{173}]: Kinematically Prohibited}
\end{itemize}

This term would yield a contribution of
\begin{eqnarray}
\label{Mcut1}
\mathcal{I} &=& \frac{-2\pi \, \mathrm{Im} \left\{ \frac{1}{r^- + q^- - \frac{(\ul{k}-\ul{r})_T^2}
{(\Delta-x)p^+} -\frac{(\ul{q}+\ul{k})_T^2}{xp^+} +i\epsilon} \right\}
}{\left[r^- - \frac{(\ul{k}-\ul{r})_T^2}{(\Delta-x)p^+} - \frac{k_T^2}
{xp^+} \right] \left[ r^- - p^- - \frac{(\ul{k}-\ul{r})_T^2}{(\Delta-x)p^+}+\frac{k_T^2+\lambda^2}
{(1-x)p^+} \right]} 
\\ \nonumber &=&
\frac{+2\pi^2 \, \delta\left[ r^- + q^- - \frac{(\ul{k}-\ul{r})_T^2} {(\Delta-x)p^+} -\frac{(\ul{q}+
\ul{k})_T^2}{xp^+} \right]} {\left[r^- - \frac{(\ul{k}-\ul{r})_T^2}{(\Delta-x)p^+} - \frac{k_T^2}
{xp^+} \right] \left[ r^- - p^- - \frac{(\ul{k}-\ul{r})_T^2}{(\Delta-x)p^+}+\frac{k_T^2+\lambda^2}
{(1-x)p^+} \right]}
\\ \nonumber \mathcal{I} &=&
\frac{+2\pi^2 \frac{x\Delta}{\Delta-x} p^+ \, \delta \left[ \left(
\ul{q}+\frac{\Delta}{\Delta-x}\ul{k}-\frac{x}{\Delta-x}\ul{r} \right)_T^2 \right]}
{\left[r^- - \frac{(\ul{k}-\ul{r})^2}{(\Delta-x)p^+} - \frac{k_T^2}
{xp^+} \right] \left[ r^- - p^- - \frac{(\ul{k}-\ul{r})_T^2}{(\Delta-x)p^+}+\frac{k_T^2+\lambda^2}
{(1-x)p^+} \right]},
\end{eqnarray}
but the argument of the delta function is positive definite, so it
cannot be satisfied; this cut is kinematically prohibited because it
corresponds to a 2 $\rightarrow$ 1 massless, on-shell process.

\begin{itemize}
 \item \underline{Res[\ding{174}] Im[\ding{172}]: Nucleon Decay}
\end{itemize}

Similarly, this cut would yield a contribution of
\begin{eqnarray}
\label{Mcut2}
\mathcal{I} &=& 
\frac{ +2\pi^2 \, \delta 
\left[ 
 r^- - \frac{(\ul{k}-\ul{r})_T^2}{(\Delta-x)p^+} - \frac{k_T^2}{xp^+} 
\right] }{
\left[
 r^- + q^- - \frac{(\ul{k}-\ul{r})_T^2}{(\Delta-x)p^+} - \frac{(\ul{q}+\ul{k})_T^2}{xp^+}
\right] 
\left[
 r^- - p^- - \frac{(\ul{k}-\ul{r})_T^2}{(\Delta-x)p^+} + \frac{k_T^2 + \lambda^2}{(1-x)p^+}
\right]}
\\ \nonumber \mathcal{I} &\propto&
\delta \bigg[ 
 -x(\Delta-x)\left(\lambda^2-(1-\Delta)m_N^2\right) - x(\Delta-x)r_T^2 - x(1-\Delta)(\ul{k}-\ul{r})_T^2
 \\ \nonumber &-& (1-\Delta)(\Delta-x)k_T^2
\bigg].
\end{eqnarray}
All of the terms inside the delta-function are negative definite except
for the first one, so we can impose the stability of the nucleon by
requiring that
\begin{equation}
\label{Mproton1}
\lambda^2 - (1-\Delta)m_N^2 > 0.
\end{equation}

Similarly, we can easily identify the combinations of poles that would be kinematically prohibited or would correspond to nucleon decay.

\begin{itemize}
 \item \underline{Res[\ding{174}] Im[\ding{175}]: Kinematically Prohibited}
 \item \underline{Res[\ding{175}] Im[\ding{174}]: Kinematically Prohibited}
\end{itemize}

\begin{itemize}
\item \underline{Res[\ding{175}] Im[\ding{172}]: Nucleon Decay}
\end{itemize}

This cut corresponds to nucleon decay through a different channel,
yielding a contribution of
\begin{eqnarray}
\label{Mcut5}
\mathcal{I} &=& 
\frac{ +2\pi^2 \, \delta 
\left[ 
 p^- - \frac{k_T^2+\lambda^2}{(1-x)p^+} - \frac{k_T^2}{xp^+}
\right] }{
\left[
 p^- + q^- - \frac{k_T^2+\lambda^2}{(1-x)p^+} - \frac{(\ul{q}+\ul{k})_T^2}{xp^+}
\right] 
\left[
 p^- - r^- - \frac{k_T^2 + \lambda^2}{(1-x)p^+} + \frac{(\ul{k}-\ul{r})_T^2}{(\Delta - x)p^+}
\right]}
\\ \nonumber \mathcal{I} &\propto&
\delta \bigg[ 
 -x \left(\lambda^2 - (1-x) m_N^2 \right) - k_T^2
\bigg].
\end{eqnarray}
To prevent nucleon decay through this channel, we need to impose the
slightly different condition
\begin{equation}
\label{Mproton2}
\lambda^2 - (1-x) m_N^2 > 0
\end{equation}

\begin{itemize}
 \item \underline{Res[\ding{175}] Im[\ding{173}]: Legal Cut}
\end{itemize}

This combination is the only legal cut of the four denominators that
can be put on-shell simultaneously.  This contribution is
\begin{eqnarray}
\label{Mcut6}
\mathcal{I} &=& 
\frac{ +2\pi^2 \, \delta 
\left[ 
 p^- + q^- - \frac{k_T^2 + \lambda^2}{(1-x)p^+} - \frac{(\ul{q}+\ul{k})_T^2}{xp^+}
\right] }{
\left[
 p^- - \frac{k_T^2+\lambda^2}{(1-x)p^+} - \frac{k_T^2}{xp^+}
\right] 
\left[
 p^- - r^- - \frac{k_T^2+\lambda^2}{(1-x)p^+} + \frac{(\ul{k}-\ul{r})_T^2}{(\Delta-x)p^+}
\right]}.
\end{eqnarray}
Expanding the argument of the delta function and keeping terms of
order $\mathcal{O}\left(\frac{\bot}{Q}\right)$ gives
\begin{align}\label{delta_full}
  \delta \left[ p^- + q^- - \frac{k_T^2 + \lambda^2}{(1-x)p^+} -
    \frac{(\ul{q}+\ul{k})_T^2}{xp^+} \right] \approx
  \frac{\Delta^2 p^+}{Q^2} \, \delta \left[ x - \left(1+ 2
      \frac{\ul{q} \cdot ( \ul{k}-\ul{r})}{Q^2} \right) \Delta
  \right].
\end{align}
The delta function sets $x \approx \Delta$ to leading order, but the
singularity of the delta function only falls within the kinematic
region of Case A, $0<x<\Delta<1$ if
\begin{equation}
\nonumber
\ul{q} \cdot (\ul{k} - \ul{r}) <0 ,
\end{equation}
which is restricted to only half of the total phase space of the
$d^2k$ integral.  As we will see, Case B complements this integral
with the other half of the phase space.  With this caveat, we can
write a final expression for the imaginary part as
\begin{align}
\label{MIDIS22}
\mathcal{I} = \frac{2\pi^2\Delta^2p^+}{Q^2} \frac{\delta \left[ x -
    \left(1+ 2 \frac{\ul{q} \cdot ( \ul{k}-\ul{r})}{Q^2} \right)
    \Delta \right]}{ \left[ p^- -
    \frac{k_T^2+\lambda^2}{(1-x)p^+} - \frac{k_T^2}{xp^+}
  \right] \left[ p^- - r^- - \frac{k_T^2+\lambda^2}{(1-x)p^+} +
    \frac{(\ul{k}-\ul{r})_T^2}{(\Delta-x)p^+} \right]}.
\end{align}

\vspace{0.5cm}

\begin{figure}[ht]
\centering
\includegraphics[width=\textwidth]{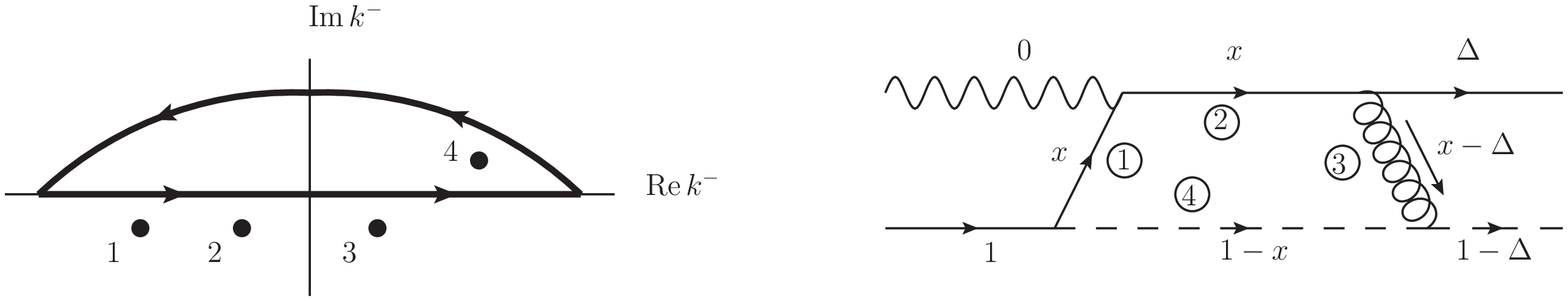}
\caption{Illustration of the poles (left) and corresponding
  time-ordered diagram (right) of \eqref{MIDIS} for the kinematic
  regime Case B: $0<\Delta<x<1$.  We choose to close the contour in
  the upper half-plane, enclosing only the pole \ding{175}.  Note that
  the placement of the poles is only schematic, indicating the sign of
  their imaginary part; the placement on the real axis has no
  significance.}
\label{MDISB} 
\end{figure}

For Case B: $0 < \Delta < x < 1$, we can close the contour in the
upper half-plane, enclosing only the pole \ding{175}, as in
Fig. \ref{MDISB}.  Again, we will consider the various contributions
to \eqref{MIDIS} from the residue and imaginary part of the various
poles.

\begin{itemize}
 \item \underline{Res[\ding{175}] Im[\ding{172}]: Nucleon Decay}
 \item \underline{Res[\ding{175}] Im[\ding{174}]: Kinematically Prohibited}
 \item \underline{Res[\ding{175}] Im[\ding{173}]: Legal Cut}
\end{itemize}

Again, this is the only combination of propagators that can be put on
shell simultaneously.  The expression is the same as in \eqref{Mcut6}
but with $x>\Delta$.  This means that the delta function
\begin{eqnarray}
  \nonumber
  \delta\left[
    x- \left(1+2 \frac{\ul{q}\cdot(\ul{k}-\ul{r})}{Q^2}  \right) \, \Delta
  \right]
\end{eqnarray}
has its singularity within the kinematic window of Case B, $0 < \Delta
< x < 1$, if
\begin{eqnarray}
\nonumber
\ul{q}\cdot(\ul{k}-\ul{r}) >0.
\end{eqnarray}

Thus Case B gives rise to the same expression \eqref{MIDIS22} at
leading order, but with validity in the complementary region of the
$d^2k$ phase space; the expression \eqref{MIDIS22} is thus valid
for all $\ul{k}$ and is a complete evaluation of \eqref{MIDIS}.

Substituting this expression back into \eqref{MADIS2} and integrating
over the delta function which sets $x \approx \Delta$ gives
\begin{align}
\label{MADIS3}
\Delta \left|\mathcal{A}_{DIS}\right|^2 & = -i g^2 e_f^2 G^2 C_F
\left(\frac{\Delta(1-\Delta)}{Q^2 (r_T^2 +a^2)} \right) \int
\frac{d^2k}{(2\pi)^2} \\ \nonumber & \times \,
\frac{\ubar{\chi}(p) \slashed{r} (\slashed{q}+\slashed{k}) (2
  \slashed{p}-\slashed{k}-\slashed{r}) (\slashed{q}+\slashed{r})
  \slashed{k} U_\chi (p) - (\chi \rightarrow
  -\chi)}{(\ul{k}-\ul{r})_T^2 (k_T^2 + a^2)},
\end{align}
where the mass parameter $a^2$ that regulates the infrared divergence
\begin{equation}
\label{Ma2}
a^2 \equiv \Delta \left( \lambda^2 - (1-\Delta)m_N^2 \right) > 0
\end{equation}
is ensured to be positive definite by the stability conditions
\eqref{Mproton1} and \eqref{Mproton2}, and we have used
\begin{equation}
\label{Mr2}
r^2 = \Delta \left( m_N^2 - \frac{r_T^2 + \lambda^2}{1-\Delta} \right) - r_T^2 = 
 - \frac{r_T^2 + a^2}{1-\Delta}.
\end{equation}
Note that making this cut has fixed the loop momentum $k^\mu$ to be
\begin{equation}
\label{MkDIS}
k^\mu = \left(\Delta p^+ \, , \, \frac{m_N^2}{p^+} - \frac{k_T^2+\lambda^2}{(1-\Delta)p^+} \, , \, 
 \ul{k} \right).
\end{equation}

Next we need to evaluate the numerator of \eqref{MADIS3} by computing
the difference between the matrix elements:
\begin{eqnarray}
\label{NDIS1}
N_{DIS} = \ubar{\chi}(p) \slashed{r} (\slashed{q} + \slashed{k}) (2 \slashed{p} - \slashed{k} - \slashed{r}) (\slashed{q} + \slashed{r}) \slashed{k} U_\chi(p) - (\chi \rightarrow -\chi).
\end{eqnarray}
The momenta involved in this spinor product obey the scale hierarchy
\begin{align}
\label{Nscale}
\overbrace{p^+ , r^+, k^+, q^-, q_T}^{\mathcal{O}(Q)} \, \gg \,
\overbrace{r_T, k_T, m_N, \lambda}^{\mathcal{O}(\bot)} \,
\gg \, \overbrace{p^- , r^-, k^-}^{\mathcal{O}\left( \bot^2/Q \right)},
\end{align}
with the dominant power-counting of the spin-dependent part of the
Dirac matrix element being $\mathcal{O} (Q^4 \bot^2)$. Evaluation of
$N_{DIS}$ in \eq{NDIS1} in the kinematics of \eq{Nscale} is somewhat
involved: after some algebra one can show that there are three classes
of Dirac structures that give a contribution of the leading order,
$\mathcal{O} (Q^4 \bot^2)$; all three involve taking the
$\mathcal{O}(Q)$ momenta from the middle three gamma matrices:
\begin{align}
\label{NDIS2}
N_{DIS} &= \frac{1}{8} \left[ (2p^+ - k^+ - r^+)(q^-)^2 \right] \,
\ubar{\chi}(p) \, \slashed{r} \gamma^+ \gamma^- \gamma^+ \slashed{k} \, U_\chi(p) - 
 (\chi \rightarrow - \chi)
\\ \nonumber &=
\left[(1-\Delta) p^+ (q^-)^2 \right] \,
\ubar{\chi}(p) \, \slashed{r} \gamma^+ \slashed{k} \, U_\chi(p) -  (\chi \rightarrow - \chi).
\end{align}
The three variations consist of taking $\gamma^-$ for both
$\slashed{r}$ and $\slashed{k}$, taking $\gamma^-$ for one and
$\gamma_\bot$ for the other, or taking $\gamma_\bot$ for both.

In the first case, if we take $\gamma^-$ for both $\slashed{r}$ and
$\slashed{k}$, we obtain
\begin{eqnarray}
\label{NDIS3}
\ubar{\chi}(p) \, \slashed{r} \gamma^+ \slashed{k} \, U_\chi(p) &\rightarrow&
\frac{1}{4} \Delta^2 (p^+)^2 \, \ubar{\chi}(p) \, \gamma^- \gamma^+ \gamma^- \, U_\chi(p)
\\ \nonumber &=&
\Delta^2 (p^+)^2 \, \ubar{\chi}(p) \, \gamma^- \, U_\chi(p) ,
\end{eqnarray}
but $\ubar{\chi}(p) \, \gamma^- \, U_\chi(p) = 2 p^-$ is
spin-independent and cannot generate the asymmetry.
Similarly, if we take $\gamma_\bot$ for both $\slashed{r}$ and
$\slashed{k}$, we obtain
\begin{eqnarray}
\label{NDIS4}
\ubar{\chi}(p) \, \slashed{r} \gamma^+ \slashed{k} \, U_\chi(p) &\rightarrow& 
r_\bot^i k_\bot^j \ubar{\chi}(p) \, \gamma_\bot^i \gamma^+ \gamma_\bot^j \, U_\chi(p),
\end{eqnarray}
but $\ubar{\chi}(p) \, \gamma_\bot^i \gamma^+ \gamma_\bot^j \,
U_\chi(p) = 2p^+ \delta^{ij}$ is also spin-independent and cannot
generate the asymmetry.
However, if we take one each of $\gamma_\bot$ and $\gamma^-$, we
obtain
\begin{eqnarray}
\label{NDIS5}
\ubar{\chi}(p) \, \slashed{r} \gamma^+ \slashed{k} \, U_\chi(p) &\rightarrow& 
-\frac{1}{2} \Delta p^+ \bigg[
r_\bot^i \ubar{\chi}(p) \, \gamma_\bot^i \gamma^+ \gamma^- \, U_\chi(p)
\\ \nonumber &+&
k_\bot^i \ubar{\chi}(p) \, \gamma^- \gamma^+ \gamma_\bot^i \, U_\chi(p)
\bigg].
\end{eqnarray}
We can further simplify this expression by using the Dirac equation
\begin{equation}
\label{MDirac}
0 = (\slashed{p}-m_N) U_\chi(p) = \left[\frac{1}{2}p^+ \gamma^- + \frac{m_N^2}{2p^+} \gamma^+ - m_N \right]
U_\chi(p)
\end{equation}
to rewrite the action of $\gamma^-$ in terms of $\gamma^+$ and $m_N$.  Since $(\gamma^+)^2 = 0$, this simplifies \eqref{NDIS5} to
\begin{eqnarray}
\label{NDIS6}
\ubar{\chi}(p) \, \slashed{r} \gamma^+ \slashed{k} \, U_\chi(p) &\rightarrow& 
- m_N \Delta (k_\bot^i - r_\bot^i) \, \ubar{\chi}(p) \, \gamma^+ \gamma_\bot^i \, U_\chi(p),
\end{eqnarray}
and $\ubar{\chi}(p) \, \gamma^+ \gamma_\bot^i \, U_\chi(p) = 2 i \chi
p^+ \delta^{i2}$, which is spin-dependent and generates the asymmetry.
Altogether this gives
\begin{eqnarray}
\label{NDIS7}
N_{DIS} &=& -2i\chi \Delta (1-\Delta) (p^+)^2 (q^-)^2 m_N (k_\bot^{2} - r_\bot^{2}) - (\chi \rightarrow -\chi)
\\ \nonumber &=&
-4 i \left( \frac{1-\Delta}{\Delta} \right) Q^4 m_N (k_\bot^{2} - r_\bot^{2}),
\end{eqnarray}
so that the spin-difference matrix element is pure imaginary, as was
proved in \eqref{spinors5}.

Substituting this result back into \eqref{MADIS3} gives (cf. Eq.~(31)
in \cite{Boer:2002ju}\footnote{As noted in Ref. \cite{Burkardt:2003je}, 
there should be an additional overall minus sign in
front of Eq. (21) of Ref. \cite{Brodsky:2002cx},
also in front of Eq. (31) of Ref. \cite{Brodsky:2002rv}
and Eqs. (31,33,36) of Ref. \cite{Boer:2002ju}.
})
\begin{align}
  \nonumber \Delta \left|\mathcal{A}_{DIS}\right|^2 & = -4 g^2 e_f^2
  G^2 C_F \left(\frac{(1-\Delta)^2 Q^2 m_N}{r_T^2 +a^2} \right)
  \int \frac{d^2k}{(2\pi)^2} \frac{k_\bot^{2} - r_\bot^{2}}
  {(\ul{k}-\ul{r})_T^2 \, (k_T^2 + a^2)} \\ \label{MADIS4} & =
  +\frac{g^2 e_f^2 G^2 C_F}{\pi} (1-\Delta)^2 \frac{Q^2 m_N
    r_\bot^{2}}{r_T^2 (r_T^2 + a^2)} \ln
  \left(\frac{r_T^2 + a^2}{a^2} \right) ,
\end{align}
where the $d^2k$ integral is performed using Feynman parameters obtaining
\begin{align}
  \label{eq:integral2}
  \int \frac{d^2k}{(2\pi)^2} \frac{k_\bot^{2} - r_\bot^{2}}
  {(\ul{k}-\ul{r})_T^2 \, (k_T^2 + a^2)} = - \frac{1}{4 \, \pi} \,
  \frac{r_\bot^{2}}{r_T^2} \, \ln \left(\frac{r_T^2 +
      a^2}{a^2} \right).
\end{align}
\eq{MADIS4} is the final expression for the spin-difference amplitude
squared for deep inelastic scattering.  This expression for $\Delta |\mathcal{A}|^2$ corresponds to the first nonzero contribution to Sivers function in the diquark model \eqref{DISS-DECOMP6} of \cite{Boer:2002ju}
\begin{align}
 \label{diquarkSivers}
 \left[f_{1T}^{\bot q} (\Delta,r_T) \right]_{SIDIS} = - \frac{g^2 G^2 C_F}{4(2\pi)^4} \Delta(1-\Delta) \frac{m_N^2}{r_T^2 (r_T^2 + a^2)} \ln \left(\frac{r_T^2 + a^2}{a^2}\right).
\end{align}

\begin{figure}
\centering
\includegraphics[width=0.7\textwidth]{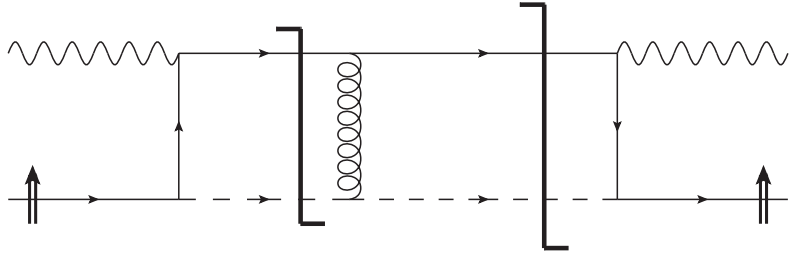}
\caption{Diagrammatic representation of the origin of complex phase
  leading to the single-spin asymmetry in SIDIS. The longer cut
  denotes the final state of the process, while the shorter cut
  demonstrates the origin of the phase needed for the asymmetry.}
\label{DIS-DY2} 
\end{figure}

Let us stress once again that the asymmetry in the
SIDIS case arises from the contribution of the diagram in \fig{figDIS}
(A) with the $(q+k)$- and $(p-k)$-lines (corresponding to the lines
labeled \ding{175} and \ding{173}) which are put on mass-shell. It is this and
only this contribution that gives the imaginary phase needed for the
asymmetry in SIDIS. This fact becomes more apparent if we diagrammatically
represent putting the $(q+k)$- and $(p-k)$-lines on mass-shell by a
cut, as shown in \fig{DIS-DY2}. In \fig{DIS-DY2} we show the interference term
which we have just calculated, with the longer cut representing the
true final state of the process, and the shorter vertical cut line
representing the imaginary phase generating the asymmetry. The shorter
cut follows the standard Cutkosky rules \cite{Cutkosky:1960sp}, with
the caveat stressed above that it should
not be applied to the spinor matrix element; that is, the shorter cut
applies to the denominators of the propagators only, as if we are
evaluating the diagram in a scalar field theory. Using the Cutkosky
rules one can clearly see that this is the only way the shorter cut
line can be placed in the diagram, as all other cuts would lead to
various prohibited $1 \to 2$ or $2 \to 1$ processes, including nucleon
decay.  Thus \fig{DIS-DY2} demonstrates that the imaginary phase
needed for the single-spin asymmetry arises only in diagrams where it is
possible to place a second cut. We will make use of this result in
the analysis of the Drell-Yan process below.


\subsection{Drell-Yan Sivers Function in the Diquark Model}
\label{sec:DYF}

\begin{figure}[htb]
\centering
\includegraphics[width= 0.8 \textwidth]{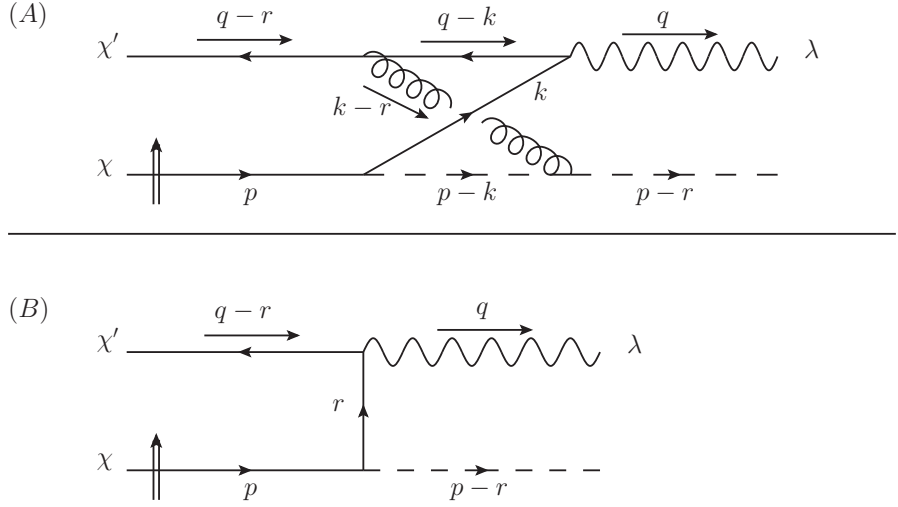}
\caption{Diagrams for the $\overline{q} + p^\uparrow \rightarrow
  \gamma^* + X$ DY amplitude at one-loop order (A) and tree-level
  (B). The incoming nucleon and anti-quark are denoted by the lower and
  upper solid lines correspondingly, with the outgoing diquark denoted
  by the dashed line.}
\label{figDY}
\end{figure}

We now perform a similar calculation for the Drell-Yan
process in the same diquark model considered above for deep inelastic
scattering.  We will consider the scattering of an antiquark on a
transversely-polarized nucleon with transverse spin eigenvalue $\chi$
that produces a virtual photon, which then decays into a dilepton pair
with invariant mass $q^2=Q^2$.  This process is shown in
Fig. \ref{figDY} at the level of virtual photon production:
$\overline{q} + p^\uparrow \rightarrow \gamma^* + X$.

Following \cite{Brodsky:2002rv}, we work in a generic frame collinear
to the nucleon ($\ul{p}=\ul{0}$).  We define the longitudinal
momentum fraction of the photon to be $\beta \equiv q^+ / p^+$ and the
momentum fraction exchanged in the $t$-channel to be $\Delta \equiv r^+
/ p^+$.  As before, four-momentum conservation and the on-shell
conditions fix $r^-$ and $q^-$ to be 
\begin{eqnarray}
\label{MDYkin}
r^- &=& p^- - (p-r)^- = \frac{m_N^2}{p^+} - \frac{r_T^2 + \lambda^2}{(1-\Delta)p^+} \\ \nonumber
q^- &=& (q-r)^- + r^- = \frac{(\ul{q}-\ul{r})_T^2}{(\beta-\Delta)p^+} + r^- \approx
 \frac{q_T^2 - 2 \ul{q}\cdot\ul{r}}{(\beta-\Delta)p^+} + \mathcal{O}\left(\frac{\bot^2}{p^+}
\right).
\end{eqnarray}
In this frame, the virtual photon's large invariant mass $Q^2$ comes
in part from its transverse components and in part from its longitudinal
components:
\begin{eqnarray}
\label{MQ2DY}
Q^2 \equiv q^2 = \beta p^+ q^- - q_T^2 \approx \left(\frac{\Delta}{\beta-\Delta}\right) q_T^2 +
\mathcal{O}\left(\frac{\bot}{Q}\right).
\end{eqnarray}
This allows us to approximate $q^-$ as
\begin{eqnarray}
\label{Mq-DY}
q^- \approx \frac{Q^2}{\Delta p^+} + \mathcal{O}\left(\frac{Q \bot}{p^+}\right),
\end{eqnarray}
which agrees with the corresponding expression \eqref{MDISkin} for DIS
to leading order in $Q^2$. The kinematics can be summarized as
\begin{samepage}
\begin{align}
\label{MDYkin2}
p^\mu &= \left( p^+ \, , \, \frac{m_N^2}{p^+} \, , \, \ul{0} \right)
\\ \nonumber q^\mu &= \left( \beta p^+ \, , \,
  \frac{(\ul{q}-\ul{r})_T^2}{(\beta-\Delta)p^+} + \frac{m_N^2}{p^+} -
  \frac{r_T^2+\lambda^2}{(1-\Delta)p^+} \, , \, \ul{q} \right)
\\ \nonumber r^\mu &= \left( \Delta p^+ \, , \, \frac{m_N^2}{p^+} -
  \frac{r_T^2 + \lambda^2}{(1-\Delta)p^+} \, , \, \ul{r}
\right).
\end{align}
\end{samepage}
Notice that, in this frame, the on-shell conditions for the antiquark,
scalar, and dilepton pair imply that $(\beta-\Delta) > 0 , \beta > 0 ,
$ and $(1-\Delta)>0$.  Additionally, to leading order, the positivity
constraint on $q^-$ \eqref{Mq-DY} implies that $\Delta > 0$, and we
can choose 
our frame such that $\beta < 1$, although this is not strictly
necessary.  Altogether, this gives the hierarchy of the fixed scales to
be $0 < \Delta < \beta < 1$.

With these kinematics, we can evaluate the one-loop amplitude shown in
Fig. \ref{figDY} (A) as
\begin{align}
\label{MA1DY}
\mathcal{A}_1^{DY} &= \frac{i g^2 e_f G C_F}{(2\pi)^4} \int d^4{k} \frac{ \vbar{\chi'}(q-r) (2\slashed{p} 
- \slashed{k} - \slashed{r})(\slashed{k}-\slashed{q})\slashed{\epsilon}^*_\lambda \slashed{k} U_\chi(p)}
{\left[k^2 + i\epsilon \right] \left[(k-q)^2 + i\epsilon \right] \left[(k-r)^2 + i \epsilon \right]
\left[(p-k)^2 - \lambda^2 + i\epsilon \right]}
\\ \nonumber &=
\frac{-i g^2 e_f G C_F}{2(2\pi)^4 (p^+)^3} \int \frac{dx \, dk^- \, d^2k}{x(x-\beta)(x-\Delta)(1-x)} 
\\ \nonumber &\!\!\!\!\! \times
\frac{\vbar{\chi'}(q-r) (2\slashed{p} - \slashed{k} - \slashed{r}) (\slashed{k} - \slashed{q})
\slashed{\epsilon}_\lambda^* \slashed{k} U_\chi(p)}{ \left[ k^- - \frac{k_T^2-i\epsilon}{xp^+} \right]
\left[k^- - q^- - \frac{(\ul{k}-\ul{q})_T^2-i\epsilon}{(x-\beta)p^+}\right]\left[k^- - r^- - \frac{(\ul{k}-\ul{r})_T^2-i\epsilon}{(x-\Delta)p^+}\right]\left[k^- - p^- + \frac{k_T^2 + \lambda^2 - i\epsilon}{(1-x)p^+}\right]},
\end{align}
where $x \equiv k^+ / p^+$ is the longitudinal momentum fraction in
the loop.  Similarly, the tree-level amplitude shown in
Fig. \ref{figDY} (B) is
\begin{align}
\label{MA0DY}
\mathcal{A}_0^{DY} = - \frac{e_f G}{r^2} \vbar{\chi'}(q-r)
\slashed{\epsilon}_\lambda^* \slashed{r} U_\chi(p).
\end{align}
This allows us to calculate the spin-difference amplitude squared
following \eqref{ImPart3} as
\begin{align}
\nonumber
\Delta |\mathcal{A}_{DY}|^2 &= 2i \left[\frac{g^2 e_f^2 G^2 C_F}{2(2\pi)^4(p^+)^3 r^2}\right] \int \frac{dx 
\, d^2k} {x(x-\beta)(x-\Delta)(1-x)} \mathrm{Im}\left\{\int dk^- \frac{i}{\left[k^- - 
\frac{k_T^2-i\epsilon}{xp^+}\right]} \right.
\\ \label{MADY} &\!\!\!\!\! \times \left.
\frac{1}{\left[k^- - q^- - \frac{(\ul{k}-\ul{q})_T^2-i\epsilon}{(x-\beta)p^+} \right] \left[k^- - r^-
- \frac{(\ul{k}-\ul{r})_T^2-i\epsilon}{(x-\Delta)p^+} \right] \left[k^- - p^- + \frac{k_T^2 +
\lambda^2 - i\epsilon}{(1-x)p^+} \right]} \right\}
\\ \nonumber &\!\!\!\!\! \times
\sum_{\chi' , \lambda} \left[ \ubar{\chi}(p) \, \slashed{r} \slashed{\epsilon}_\lambda V_{\chi'}(q-r) \vbar{\chi'}(q-r) (2\slashed{p} - \slashed{k} - \slashed{r}) (\slashed{k} - \slashed{q}) \slashed{\epsilon}_\lambda^*
\slashed{k} \, U_\chi(p) - (\chi \rightarrow -\chi) \right]
\end{align}
where we sum over the spin of the incoming antiquark and use
\eq{eq:polsum}.  Performing these sums and simplifying the result gives
\begin{align}
\label{MADY2}
\Delta |\mathcal{A}_{DY}|^2 & = \frac{2 i g^2 e_f^2 G^2 C_F}{(2\pi)^4
  r^2 (p^+)^3} \int \frac{dx \, d^2k} {x(x-\beta)(x-\Delta)(1-x)} \,
\mathcal{I} \\ \nonumber & \times \, \left[ \ubar{\chi}(p) \,
  \slashed{r} (\slashed{k} - \slashed{q}) (2\slashed{p} - \slashed{k}
  - \slashed{r}) (\slashed{q}-\slashed{r}) \slashed{k} \, U_\chi(p) -
  (\chi \rightarrow -\chi) \right] ,
\end{align}
where the imaginary part necessary for the asymmetry is generated by
\begin{align}
\label{MIDY}
\mathcal{I} &\equiv \mathrm{Im} \! \left\{ \!  \int \! \! \frac{i \;
    dk^-}{ \left[k^- - \frac{k_T^2-i\epsilon}{xp^+} \right] \! \!
    \left[k^- - q^- -
      \frac{(\ul{k}-\ul{q})_T^2-i\epsilon}{(x-\beta)p^+} \right] \!
    \! \left[k^- - r^- -
      \frac{(\ul{k}-\ul{r})_T^2-i\epsilon}{(x-\Delta)p^+} \right] } \right.
\\ \nonumber &\times 
\left. \frac{1}{\left[k^- - p^- + \frac{k_T^2 + \lambda^2-i\epsilon}{(1-x)p^+} \right]} \right\} \! .
\end{align}
As before, the imaginary part \eqref{MIDY} corresponds to putting two
of the loop propagators on-shell simultaneously: one from performing
the $k^-$ integral and another from taking the imaginary part.  The
propagators that can be simultaneously put on-shell are strongly
constrained by the kinematics and by the requirement of nucleon
stability. 

In Table \ref{MDYtable} we classify the four poles \ding{172} - \ding{175} of
this expression as lying either above or below the $\mathrm{Re} \,
k^-$ axis for the five distinct kinematic regimes:
$(x<0<\Delta<\beta<1)$, $(0<x<\Delta<\beta<1)$,
$(0<\Delta<x<\beta<1)$, $(0<\Delta<\beta<x<1)$, and
$(0<\Delta<\beta<1<x)$.  As before all the poles lie to one side of
the real axis unless $0<x<1$, so there are three distinct cases to
evaluate, each of which corresponds to a particular time-ordering of
the diagram.  We consider each of these cases below.
%
\begin{table}[ht]
\centering
\begin{tabular}{|cl|c|c|c|c|c|}
 \hline &Pole & $x<0$ & $0<x<\Delta$ & $\Delta<x<\beta$ & $\beta<x<1$ & $x>1$ 
  \\ \hline
 \ding{172}& $k^- = \frac{k_T^2-i\epsilon}{xp^+}$ & above & below & below & below & below \\
 \ding{173}& $k^- = q^- + \frac{(\ul{k}-\ul{q})_T^2-i\epsilon}{(x-\beta)p^+}$ & above & above & above & 
  below & below \\
 \ding{174}& $k^- = r^- + \frac{(\ul{k}-\ul{r})^2-i\epsilon}{(x-\Delta)p^+}$ & above & above & below &
  below & below \\
 \ding{175}& $k^- = p^- - \frac{k_T^2 + \lambda^2 -i\epsilon}{(1-x)p^+}$ & above & above & above &
  above & below \\ \hline \hline
 &Contribution: & $0$ & Case A & Case B & Case C & $0$ \\ \hline
\end{tabular}
\caption{\label{MDYtable} Table classifying the pole locations of \eqref{MIDY} as lying either above or below the $\mathrm{Re} \, k^-$ axis.}
\end{table}
%

\vspace{0.5cm}

\begin{figure}[htb]
\centering
\includegraphics[width=\textwidth]{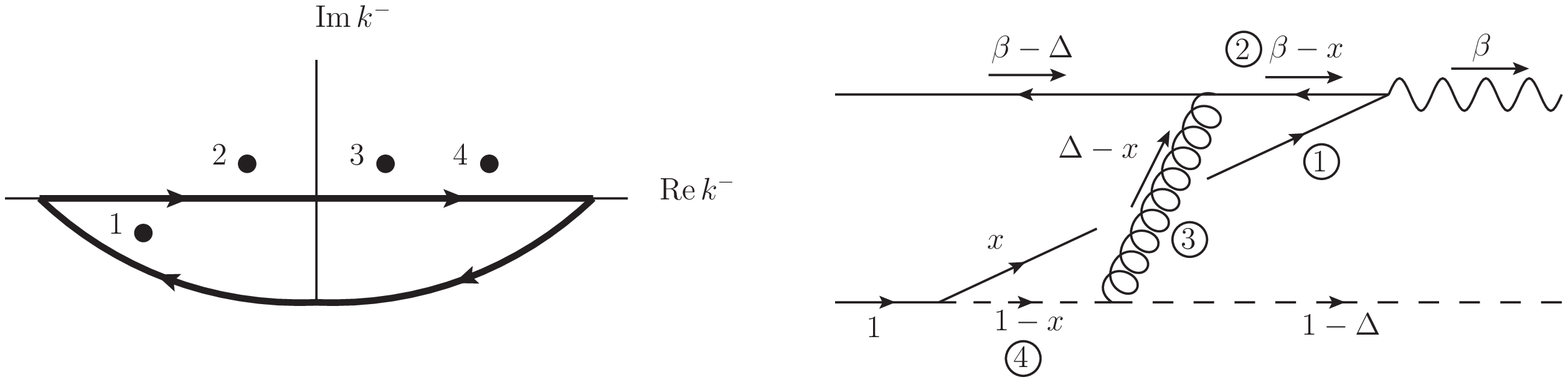}
\caption{\label{MDYA} Illustration of the poles (left) and
  corresponding time-ordered diagram (right) of \eqref{MIDY} for the
  kinematic regime Case A: $0<x<\Delta<\beta<1$.  We choose to close
  the contour in the lower half-plane, enclosing only the pole
  \ding{172}.  Note that the placement of the poles is only schematic,
  indicating the sign of their imaginary part; the placement on the
  real axis has no significance.}
\end{figure}

For Case A: $0<x<\Delta<\beta<1$, we choose to close the contour in
the lower half-plane, enclosing only the pole \ding{172}, as shown in
Fig. \ref{MDYA}.  Let us consider the possible contributions to
\eqref{MIDY} from the residue and imaginary parts of the various
poles.

\begin{itemize}
 \item \underline{ Res[\ding{172}] Im[\ding{175}]: Nucleon Decay }
 \item \underline{ Res[\ding{172}] Im[\ding{174}]: Nucleon Decay }
 \item \underline{ Res[\ding{172}] Im[\ding{173}]: Legal Cut }
\end{itemize}
This corresponds to the only legal cut of the diagram as shown in
Fig. \ref{MDYA}; it is permitted because it corresponds to a $2
\rightarrow 1$ process in which the two massless quarks become a
single ``massive'' time-like photon with ``mass'' $Q$.  Equivalently,
we can recognize that the subsequent leptonic decay of the time-like
virtual photon makes this cut correspond to a massless, on-shell $2
\rightarrow 2$ scattering process, which is allowed.  This cut makes a
contribution of
\begin{eqnarray}
\label{Mcut10}
\mathcal{I} &=& \frac{+2\pi^2 \delta
 \left[
  \frac{k_T^2}{xp^+} - q^- + \frac{(\ul{k}-\ul{q})_T^2}{(\beta-x)p^+}
 \right] }{
 \left[
  \frac{k_T^2}{xp^+} - r^- + \frac{(\ul{k}-\ul{r})_T^2}{(\Delta-x)p^+}
 \right]
 \left[
  \frac{k_T^2}{xp^+} - p^- + \frac{k_T^2 + \lambda^2}{(1-x)p^+}
 \right]},
\end{eqnarray}
and
\begin{align}
  \nonumber \delta\left[\frac{k_T^2}{xp^+} - q^- +
    \frac{(\ul{k}-\ul{q})_T^2}{(\beta-x)p^+}\right] \approx
  \frac{\Delta(\beta-\Delta)p^+}{Q^2} \ \delta \left[x - \left(1 + 2
      \, \frac{\ul{q} \cdot (\ul{k} - \ul{r})} {Q^2} \right)
    \Delta \right].
\end{align}
As usual, the $\delta$-function sets $x \approx \Delta$, but the
singularity only falls within the kinematic window of Case A
$(x<\Delta)$ for $\ul{q} \cdot (\ul{k}-\ul{r})<0$.  As with DIS,
this half of the $d^2k$ phase space will be complemented by an equal
contribution for Case B $\Delta < x < \beta$.  Thus, the legal cut
gives
\begin{align}
\label{MIDY22}
\mathcal{I} = \frac{2\pi^2 \Delta (\beta - \Delta)p^+}{Q^2}
\frac{\delta \left[x - \left(1 + 2 \, \frac{\ul{q} \cdot (\ul{k} -
        \ul{r})} {Q^2} \right) \Delta \right]}{ \left[
    \frac{k_T^2}{xp^+}-r^- +
    \frac{(\ul{k}-\ul{r})_T^2}{(\Delta-x)p^+} \right] \left[
    \frac{k_T^2}{xp^+} - p^- + \frac{k_T^2 +
      \lambda^2}{(1-x)p^+} \right] }
\end{align}

\vspace{0.5cm}
%
\begin{figure}
\centering
\includegraphics[width=\textwidth]{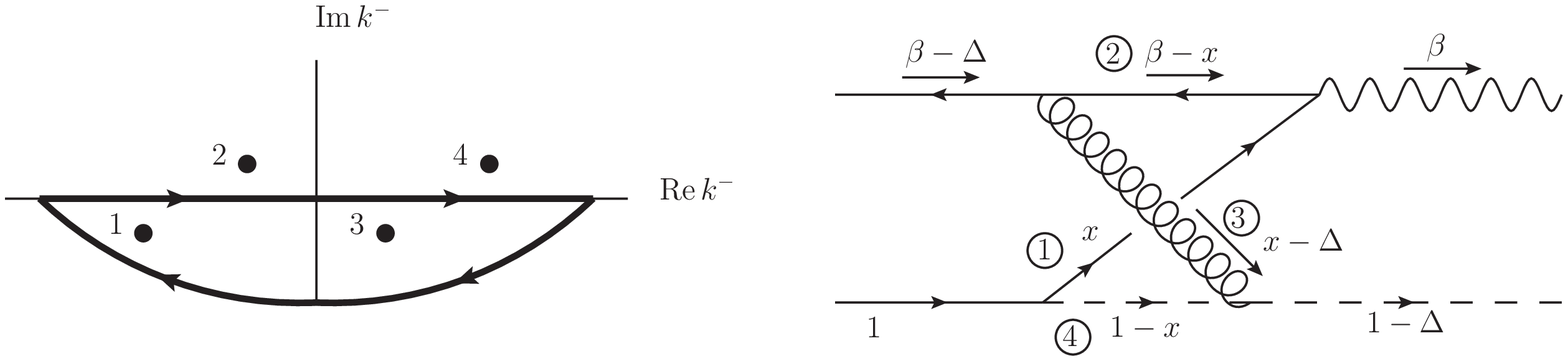}
\caption{\label{MDYB} Illustration of the poles (left) and
  corresponding time-ordered diagram (right) of \eqref{MIDY} for the
  kinematic regime Case B: $0<\Delta<x<\beta<1$.  We choose to close
  the contour in the lower half-plane, enclosing the poles \ding{172}
  and \ding{174}.  Note that the placement of the poles is only
  schematic, indicating the sign of their imaginary part; the
  placement on the real axis has no significance.}
\end{figure}
%
For Case B: $0<\Delta<x<\beta<1$, we close the contour in the lower
half-plane, enclosing the poles \ding{172} and \ding{174}, as shown in
Fig. \ref{MDYB}.  Let us consider the possible contributions to
\eqref{MIDY} from the residue and imaginary parts of the various
poles.
\begin{itemize}
 \item \underline{ Res[\ding{172}] Im[\ding{175}]: Nucleon Decay}
\end{itemize}

\begin{itemize}
 \item \underline{ Res[\ding{172}] Im[\ding{174}] + Res[\ding{174}] Im[\ding{172}]: False Pole (Cancels)}
\end{itemize}
Evaluating Res[\ding{172}] Im[\ding{174}] would give a contribution of
\begin{eqnarray}
\label{Mcut11}
\mathcal{I}_1 &=& \frac{+2\pi \, \mathrm{Im} \left\{ \frac{1}{
  \frac{k_T^2-i\epsilon}{xp^+} - r^- - \frac{(k-r)_T^2-i\epsilon}{(x-\Delta)p^+}
 } \right\} }{
 \left[
  \frac{k_T^2}{xp^+}-q^- + \frac{(k-q)_T^2}{(\beta-x)p^+}
 \right]
 \left[
  \frac{k_T^2}{xp^+} - p^- + \frac{k_T^2 + \lambda^2}{(1-x)p^+}
 \right]}
\\ \nonumber &=& \frac{\pm 2\pi^2 \delta
 \left[
  \frac{k_T^2}{xp^+} - r^- - \frac{(k-r)_T^2}{(x-\Delta)p^+}
 \right] }{
 \left[
  \frac{k_T^2}{xp^+}-q^- + \frac{(k-q)_T^2}{(\beta-x)p^+}
 \right]
 \left[
  \frac{k_T^2}{xp^+} - p^- + \frac{k_T^2 + \lambda^2}{(1-x)p^+}
 \right]},
\end{eqnarray}
where the sign ambiguity of the $i\epsilon$ components indicates the
presence of a false pole.  Whatever the sign of \eqref{Mcut11}, it is
exactly canceled by the contribution of Res[\ding{174}] Im[\ding{172}]:
\begin{eqnarray}
\label{Mcut12}
\mathcal{I}_2 &=& \frac{+2\pi \, \mathrm{Im} \left\{ \frac{1}{
  r^- + \frac{(k-r)_T^2-i\epsilon}{(x-\Delta)p^+} - \frac{k_T^2-i\epsilon}{xp^+}
 } \right\} }{
 \left[
  r^- - q^- + \frac{(k-r)_T^2}{(x-\Delta)p^+} + \frac{(k-q)_T^2}{(\beta-x)p^+}
 \right]
 \left[
  r^- - p^- + \frac{(k-r)_T^2}{(x-\Delta)p^+} + \frac{k_T^2 + \lambda^2}{(1-x)p^+}
 \right]}
\\ \nonumber &=& \frac{\mp 2\pi^2 \delta
 \left[
  r^- + \frac{(k-r)_T^2}{(x-\Delta)p^+} - \frac{k_T^2}{xp^+}
 \right] }{
 \left[
  r^- - q^- + \frac{(k-r)_T^2}{(x-\Delta)p^+} + \frac{(k-q)_T^2}{(\beta-x)p^+}
 \right]
 \left[
  r^- - p^- + \frac{(k-r)_T^2}{(x-\Delta)p^+} + \frac{k_T^2 + \lambda^2}{(1-x)p^+}
 \right]}.
\end{eqnarray}
Thus $\mathcal{I}_1 + \mathcal{I}_2 = 0$, so that this cut is prohibited.

\begin{itemize}
 \item \underline{ Res[\ding{174}] Im[\ding{173}]: Kinematically Prohibited }
 \item \underline{ Res[\ding{172}] Im[\ding{173}]: Legal Cut }
\end{itemize}
Again, this is the only legal cut of the diagram in Fig.~\ref{MDYB}.
The expression is the same as in \eqref{Mcut10} from Case A, but with
$x > \Delta$.  This means that the delta function
\begin{eqnarray}
  \nonumber
  \delta \left[ x - \left( 1 + 2 \, \frac{\ul{q} \cdot (\ul{k} - \ul{r})}{Q^2} \right) \Delta \right]
\end{eqnarray}
has its singularity within the kinematic window of Case B,
$0<\Delta<x<\beta<1$, if $\ul{k} \cdot (\ul{q}-\ul{r}) > 0$.
Hence we again recover \eqref{MIDY22}, but with validity in the other
half of the $d^2k$ phase space.  Cases A and B thus complement each
other, and we will show that Case C does not make any contribution to
\eqref{MIDY}.

\vspace{0.5cm}

\begin{figure}
\centering
\includegraphics[width=\textwidth]{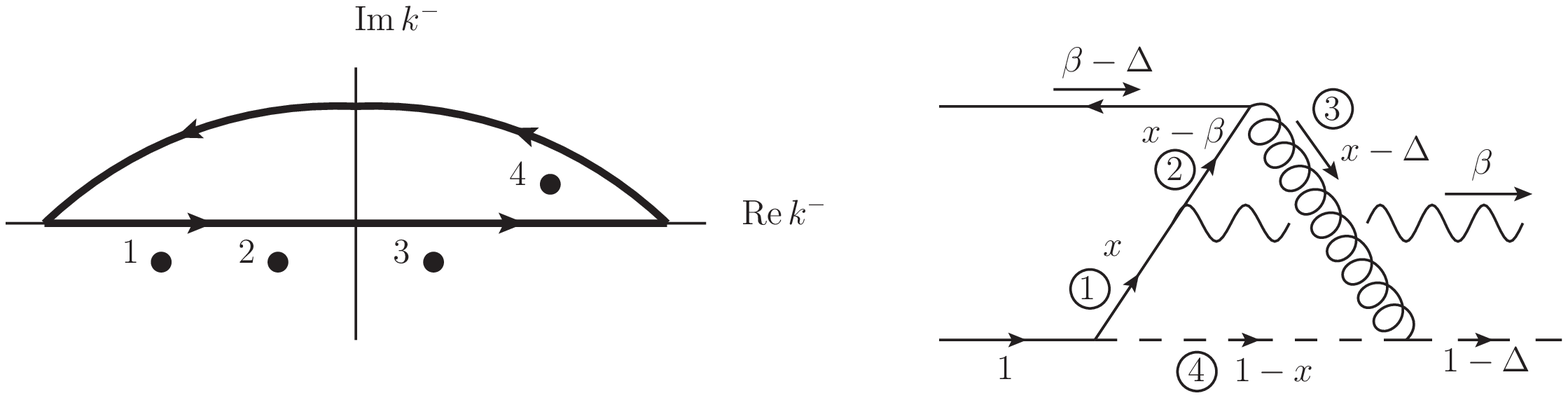}
\caption{\label{MDYC} Illustration of the poles (left) and
  corresponding time-ordered diagram (right) of \eqref{MIDY} for the
  kinematic regime Case C: $0<\Delta<\beta<x<1$.  We choose to close
  the contour in the upper half-plane, enclosing only the pole
  \ding{175}.  Note that the placement of the poles is only schematic,
  indicating the sign of their imaginary part; the placement on the
  real axis has no significance.}
\end{figure}
%
For Case C: $0<\Delta<\beta<x<1$, we choose to close the contour in
the upper half plane, enclosing only the pole \ding{175}, as
illustrated in Fig. \ref{MDYC}.  We demonstrate below that there is no
viable cut for this time-ordering of the process.

\begin{itemize}
 \item \underline{ Res[\ding{175}] Im[\ding{172}]: Nucleon Decay }
 \item \underline{ Res[\ding{175}] Im[\ding{174}]: Kinematically Prohibited }
 \item \underline{ Res[\ding{175}] Im[\ding{173}]: Nucleon Decay }
\end{itemize}

Thus there is no viable cut of the diagram for the kinematics of
Case C, and this case makes no contribution to the asymmetry.
Therefore \eqref{MIDY22} gives the complete expression for the
imaginary part and is our final result.

Substituting \eqref{MIDY22} back into \eqref{MADY2} and integrating over the
delta function sets $x \approx \Delta$, giving
\begin{align}
\label{MADY3}
\Delta |\mathcal{A}_{DY}|^2 &= -i g^2 e_f^2 G^2 C_F \left(
  \frac{\Delta (1-\Delta)}{Q^2 (r_T^2 + a^2)} \right) \int
\frac{d^2k}{(2\pi)^2} \\ \nonumber &\times \frac{\ubar{\chi}(p) \,
  \slashed{r} (\slashed{k}-\slashed{q})
  (2\slashed{p}-\slashed{k}-\slashed{r}) (\slashed{q}-\slashed{r})
  \slashed{k} \, U_\chi(p) - (\chi \rightarrow -\chi)} {(\ul{k} -
  \ul{r})_T^2 \, (k_T^2 + a^2)},
\end{align}
where we have again employed \eqref{Ma2} and \eqref{Mr2}.  In performing the
longitudinal integrals, we have fixed the loop momentum $k^\mu$ to be
\begin{equation}
\label{MkDY}
k^\mu = \left(\Delta p^+ \, , \, \frac{k_T^2}{\Delta p^+} \, , \, \ul{k} \right) .
\end{equation}

Comparison of \eqref{MADIS3} with \eqref{MADY3} shows that the only
difference between the two processes occurs in the numerators, rather
than in the denominators.  The essential difference in the numerators
is the reversal of the intermediate (anti)quark propagator from
$\slashed{q}+\slashed{k}$ in deep inelastic scattering to $\slashed{k}
- \slashed{q}$ in the Drell-Yan process.  We will return to this point
later in the analysis of the results.


Next we need to evaluate the spin-difference matrix element appearing
in the numerator of \eqref{MADY3}:
\begin{equation}
\label{NDY1}
N_{DY} = \ubar{\chi}(p) \, \slashed{r} (\slashed{k}-\slashed{q}) (2\slashed{p}-\slashed{k}-\slashed{r})
 (\slashed{q}-\slashed{r}) \slashed{k} \, U_\chi(p) - (\chi \rightarrow -\chi).
\end{equation}
The momenta obey the same scale hierarchy \eqref{Nscale} as in deep
inelastic scattering, with the addition of $q^+$ as a scale at
$\mathcal{O}(Q)$ in our frame for Drell-Yan.  The other momenta can
differ from their values in DIS by factors of $\mathcal{O}(1)$, but
the power-counting is the same.  Again, the dominant power-counting of
the matrix element is $\mathcal{O}(Q^4 \bot^2)$, which only arises
from taking
\begin{eqnarray}
\label{NDY2}
(\slashed{k}-\slashed{q}) (2\slashed{p}-\slashed{k}-\slashed{r}) (\slashed{q} - \slashed{r}) &\rightarrow&\
- \frac{1}{8} (q^-)^2 (2p^+ - k^+ - r^+) \, \gamma^+ \gamma^- \gamma^+
\\ \nonumber
&=&
-(1-\Delta) (p^+)(q^-)^2 \gamma^+
\end{eqnarray}
so that
\begin{eqnarray}
\label{NDY3}
N_{DY} = - \left[(1-\Delta) p^+ (q^-)^2 \right] \, \ubar{\chi}(p) \, \slashed{r} \gamma^+ \slashed{k} \,
 U_\chi(p) - (\chi \rightarrow -\chi).
\end{eqnarray}
Comparing \eqref{NDY3} with \eqref{NDIS2}, we see that
\begin{equation}
\label{NDY4}
N_{DY} = - N_{DIS}
\end{equation}
to leading order in $Q$, so we can immediately write the numerator for
Drell-Yan using \eqref{NDIS7} as
\begin{equation}
\label{NDY5}
N_{DY} = +4i \left( \frac{1-\Delta}{\Delta} \right) Q^4 m_N (k_\bot^{2} - r_\bot^{2}).
\end{equation}
Substituting this back into \eqref{MADY3} yields the same transverse
momentum integral as in DIS, which we can evaluate using Feynman
parameters to obtain the final answer
\begin{align}
\label{MADY4}
\Delta |\mathcal{A}_{DY}|^2 & = -\frac{g^2 e_f^2 G^2 C_F}{\pi}
(1-\Delta)^2 \frac{Q^2 M r_\bot^{2}}{ r_T^2 (r_T^2 + a^2)}
\, \ln \left(\frac{r_T^2 + a^2}{a^2} \right) \\ \nonumber & = -
\Delta |\mathcal{A}_{DIS}|^2.
\end{align}
Thus we conclude that the spin-difference amplitude squared from the
Drell-Yan process is exactly the negative of that from deep
inelastic scattering, \eqref{MADIS4}, and hence the same is true for the Sivers functions: 
\cite{Sivers:1989cc,Sivers:1990fh,Collins:2002kn,Belitsky:2002sm}
\begin{align}
 \label{signflip}
 \left[f_{1T}^{\bot q}(\Delta, r_T) \right]_{DY} = - 
 \left[f_{1T}^{\bot q}(\Delta, r_T) \right]_{SIDIS}.
\end{align}
Note that the relation in \eq{MADY4} is only valid if one writes
the spin-difference amplitudes in terms of $Q^2$ and $\Delta = x_{F}$,
as is proper for a TMD parton distribution function like the Sivers function.

To obtain the single-spin asymmetry $A_N$ one needs to divide $\Delta
|\mathcal{A}|^2$ for DY and SIDIS by twice the unpolarized amplitude
squared (averaged over the incoming nucleon polarizations), as follows
from \eq{AN1}. Both in the SIDIS and DY cases the unpolarized
amplitude squared is dominated by the Born-level processes, with the
amplitudes given in Eqs.~\eqref{MA0DIS} and \eqref{MA0DY}
correspondingly. One can easily show that the squares of those
amplitudes, averaged over the nucleon polarizations, are, in fact,
equal, such that \eq{MADY4} leads to
\cite{Collins:2002kn,Brodsky:2002rv}
\begin{align}
  \label{eq:ANDISDY}
  A_N^{DY} = - A_N^{DIS}.
\end{align}

This conclusion \eqref{eq:ANDISDY} was reached by reducing the SIDIS process to $\gamma^* +
p^\uparrow \to q + X$ scattering and summing over polarizations of
the incoming virtual photon. This is not an exact representation of
the physical SIDIS process, since we have to convolute the hadronic
interaction part of the diagram with a lepton tensor coming from the
electron-photon interactions as in \eqref{SIDIS1}. Likewise, for the Drell-Yan process we
have replaced the second hadron by an antiquark, reducing it to the
$q + p^\uparrow \to \gamma^* + X$ scattering.  Replacing the $l^+
l^-$-pair by a time-like photon is also an approximation, true up to
an overall multiplicative factor which can be obtained by integrating out the dilepton pair.

\begin{figure}
\centering
\includegraphics[width=0.7\textwidth]{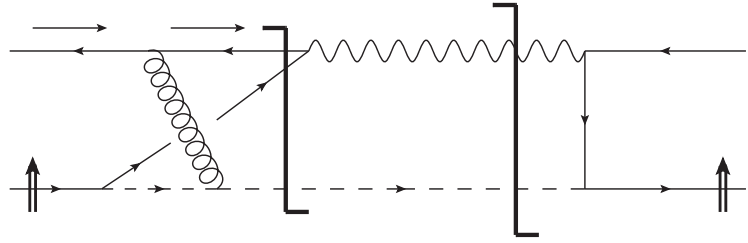}
\caption{Diagrammatic representation of the origin of complex phase
  leading to the single-spin asymmetry in the Drell-Yan process. The
  longer cut denotes the final state of the process, while the shorter
  cut demonstrates the origin of the phase needed for the asymmetry.}
\label{DIS-DY1} 
\end{figure}

It is interesting to investigate the diagrammatic origin of the
sign-flip in Eqs.~\eqref{MADY4} and \eqref{eq:ANDISDY}. To do that we
consider the diagram contributing to the single-spin asymmetry in the
Drell-Yan process shown in \fig{DIS-DY1}. As shown above, the asymmetry in the Drell-Yan case
arises due to putting the $(q-k)$- and $k$-lines in \fig{figDY} (A)
(corresponding to lines \ding{172} and \ding{173} in Figs.~\ref{MDYA}
and \ref{MDYB}) on mass-shell: this is illustrated in \fig{DIS-DY1} by
the second (shorter) cut, in analogy to \fig{DIS-DY2}. Comparing
Figures \ref{DIS-DY1} and \ref{DIS-DY2}, we see that the minus sign in
Eqs.~\eqref{MADY4} and \eqref{eq:ANDISDY} arises due to the
replacement of the outgoing eikonal quark in \fig{DIS-DY2} by the
incoming eikonal antiquark in \fig{DIS-DY1}: this is in complete
analogy with the derivation \eqref{DISS-SIVERS7} of the sign flip
due to the time-reversal of the gauge links \cite{Collins:2002kn} 
(see also \cite{Belitsky:2002sm}).

However, a closer inspection of Figures \ref{DIS-DY2} and
\ref{DIS-DY1} reveals that the cuts generating the complex phase
appear to be different: in \fig{DIS-DY2} the (shorter) cut crosses the
struck quark and the diquark lines, while in \fig{DIS-DY1} the
(shorter) cut crosses the anti-quark line and the line of the quark in
the nucleon wave function. While we have already identified the
outgoing quark/incoming antiquark duality in SIDIS vs.\ DY as
generating the sign flip, the fact that in the nucleon wave function
the diquark is put on mass shell in SIDIS and the quark is put on mass
shell in DY makes one wonder why the absolute magnitudes of the
Sivers functions in \eq{signflip} are equal. After all, different cuts may
lead to different contributions to the magnitudes of the asymmetry.
 
\begin{figure}[htb]
\centering
\includegraphics[width=0.3\textwidth]{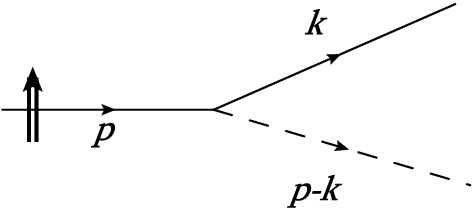}
\caption{Polarized nucleon splitting into a quark and a diquark, which
  is a part of the diagrams in both Figs.~\protect\ref{DIS-DY2} and
  \protect\ref{DIS-DY1}.}
\label{splitting} 
\end{figure}

In the diagrams at hand the origin of
the equivalence of the shorter cuts in Figs.~\ref{DIS-DY2} and
\ref{DIS-DY1} is as follows. Consider the splitting of a polarized
nucleon into a quark and a diquark as shown in \fig{splitting}: this
subprocess is common to both diagrams in Figs.~\ref{DIS-DY2} and
\ref{DIS-DY1}. The essential difference between Figs.~\ref{DIS-DY2}
and \ref{DIS-DY1} that we are analyzing is in the fact that in
\fig{DIS-DY2} the diquark is on mass shell, while in \fig{DIS-DY1} the
quark is on mass shell.

Concentrating on the denominators of the quark and diquark propagators
in \fig{splitting} we shall write for the SIDIS case of \fig{DIS-DY2} (quark
is off mass shell, diquark is on mass shell)
\begin{align}
  \label{eq:denom1}
  \frac{1}{k^2} \, \delta \left( (p-k)^2 - \lambda^2 \right) =
  \frac{-1}{p^+ \, (k_T^2 + a^2)} \, \delta \left( k^- -
    \frac{m_N^2}{p^+} + \frac{k_T^2 + \lambda^2}{(1-\Delta)
      \, p^+} \right) \approx \frac{-1}{p^+ \, (k_T^2 +
    a^2)} \, \delta (k^-),
\end{align}
where we have used Eqs.~\eqref{MDISkin2}, \eqref{Nscale}, and
\eqref{Ma2} along with $x \approx \Delta$, and, in the last step,
neglected all ${\cal O} (\perp^2/Q)$ terms inside the delta-function
since the numerator of the diagram does not depend in the exact value
of $k^-$ as long as it is small.

A similar calculation for the Drell-Yan process from \fig{DIS-DY1}
(quark is on mass shell, diquark is off mass shell in \fig{splitting})
employing Eqs.~\eqref{MDYkin2} and \eqref{Nscale} leads to
\begin{align}
  \label{eq:denom2}
  \frac{1}{(p-k)^2 - \lambda^2} \, \delta \left( k^2 \right) =
  \frac{-1}{p^+ \, (k_T^2 + a^2)} \, \delta \left( k^- -
    \frac{k_T^2}{\Delta \, p^+} \right) \approx
  \frac{-1}{p^+ \, (k_T^2 + a^2)} \, \delta (k^-).
\end{align}
We see that although the two contributions in Eqs.~\eqref{eq:denom1} and
\eqref{eq:denom2} are, in general, different, in the kinematics
\eqref{Nscale} they are apparently equivalent, leading to two
different cuts in Figs.~\ref{DIS-DY2} and \ref{DIS-DY1} giving the
same magnitude of the Sivers function.

\begin{figure}[htb]
\centering
\includegraphics[width=0.7\textwidth]{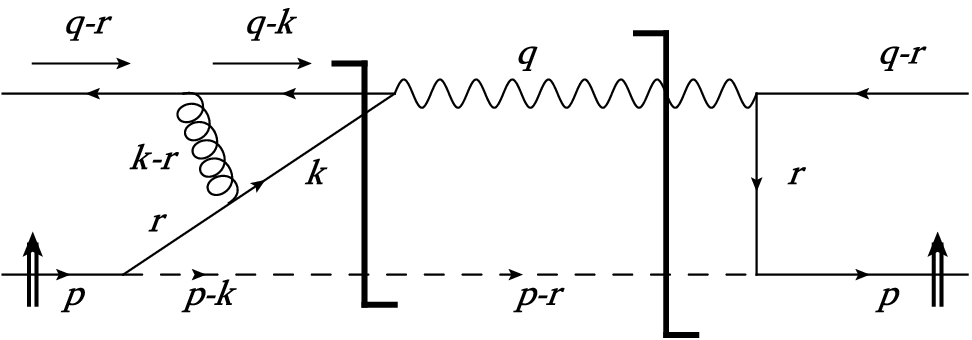}
\caption{The potential contribution to the asymmetry in DY coming from
  the quark-photon vertex correction.}
\label{DY2_fig}
\end{figure}

For completeness, let us note that, in the framework of the diquark
model at hand, there is another diagram in the Drell-Yan process which
at first glance contains both the spin-dependence and a complex phase
needed to generate the single-spin asymmetry. The diagram is shown in
\fig{DY2_fig} with its contribution to the single-spin asymmetry denoted
by the double-cut notation of Figs.~\ref{DIS-DY2} and
\ref{DIS-DY1}. The potential contribution to the asymmetry arises due
to a phase generated by the correction to the quark-photon vertex in
\fig{DY2_fig}. Note that an analogous graph cannot give an imaginary part
in the case of SIDIS, since there the virtual photon is space-like.

We will also demonstrate  that the contribution of the diagram in
\fig{DY2_fig} to the single-spin asymmetry is zero. To do this one needs
to evaluate the numerator of this diagram (minus the spin-flip term):
\begin{align}
  \label{eq:num_zero}
  & \sum_\lambda \ubar{\chi}(p) \, \slashed{r}
  \slashed{\epsilon}_\lambda (\slashed{q}-\slashed{r}) \gamma^\mu
  (\slashed{q} - \slashed{k}) \slashed{\epsilon}^*_\lambda \slashed{k}
  \gamma_\mu \slashed{r} \, U_\chi(p) - (\chi \rightarrow -\chi)
  \notag \\ & = - 8 \, k \cdot (q-r) \, \ubar{\chi}(p) \, \slashed{r}
  (\slashed{q} - \slashed{k}) \slashed{r} \, U_\chi(p) - (\chi
  \rightarrow -\chi) \notag \\ & = - 8 \, k \cdot (q-r) \, \left[ 2 \,
    r \cdot (q-k) \, \ubar{\chi}(p) \, \slashed{r} \, U_\chi(p) - r^2
    \, \ubar{\chi}(p) \, (\slashed{q} - \slashed{k}) \, U_\chi(p)
  \right] - (\chi \rightarrow -\chi) \nonumber \\ &= 0.
\end{align}
The zero answer results from the fact that, as can be checked
explicitly, forward Dirac matrix elements of transverse spinors with a
single gamma-matrix, i.e. expressions like $\ubar{\chi}(p) \,
\gamma^\mu \, U_\chi(p)$, are $\chi$-independent. Hence, the diagram
in \fig{DY2_fig} does not contribute to the asymmetry.  In fact, the
second line of \eq{eq:num_zero} is proportional to the square of the
Born term from \fig{figDY} (B): as we have shown,
the square of the Born diagram cannot lead to a non-zero single-spin
asymmetry.

Finally, let us point out that in the calculation of the asymmetries
in both SIDIS and DY, we have neglected diagrams in which the virtual
photon couples to either the nucleon or the scalar diquark instead of
the (anti)quark. These diagrams are necessary to ensure gauge
invariance, but they are suppressed by powers of $\bot / Q$, which
allowed us to neglect them.



\subsection{The Physical Picture: QCD Lensing}
\label{sec:Lensing}

From the general considerations of time reversal symmetry and an explicit calculation using the diquark model \eqref{DISS-DECOMP4}, we find that a fundamental prediction of QCD within the transverse-momentum paradigm of hadronic structure is that the Sivers function \eqref{DISS-SIVERS1} as measured in SIDIS should have equal magnitude and opposite sign from the Sivers function measured in DY \eqref{DISS-SIVERS7}.  The fact that the Sivers function is nonzero at all \eqref{DISS-SIVERS4} is evidence that what these experiments measure is not simply a density of quarks in the nucleon, but is modified by the presence of initial- or final-state interactions.  The idea that the Sivers function measured in SIDIS and DY reflects a warping or distortion of the quark densities motivates the physical picture of ``QCD lensing'' \cite{Brodsky:2002cx, Brodsky:2002rv} in analogy with the phenomenon of gravitational lensing.

\begin{figure}
\centering
\includegraphics[width=0.7\textwidth]{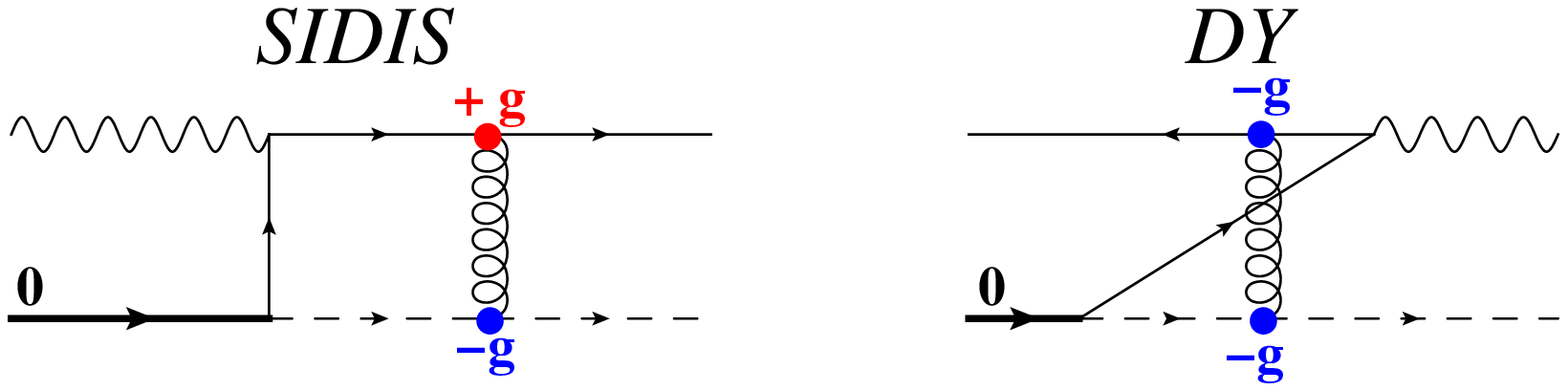}
\caption{The ``lensing'' interactions in SIDIS and DY as the origin of the Sivers function.  The final-state interaction in SIDIS is attractive, while the initial state interaction in DY is repulsive.  The sign of the potential is fully determined by conservation of color-charge.}
\label{myrelsign}
\end{figure}

Consider the representation of SIDIS and DY in the diquark model \eqref{DISS-DECOMP4} as visualized in 
Fig.~\ref{myrelsign}.  In SIDIS the color-neutral nucleon splits into a quark with charge $(+g)$ and a diquark with charge $(-g)$; the final-state interaction between these two particles thus corresponds to an attractive force which deflects the momentum of the outgoing quark and generates the asymmetry.  For DY, on the other hand, the incoming antiquark has charge $(-g)$ and undergoes an initial-state interaction with the diquark of charge $(-g)$; this corresponds to a repulsive force which deflects the antiquark and again generates the asymmetry.  The interpretation in terms of lensing, depicted in Fig.~\ref{myrelsign}, is that the SIDIS / DY sign-flip reflects this reversal of the deflection from an attractive force versus a repulsive one.  This simple interpretation should be used with care, however, as it would seem to imply that the rescattering diagram squared is responsible for the asymmetry.  Instead, it describes the quantum interference between the rescattering diagram and the Born-level diagram as calculated in \eqref{ImPart2}.  This lensing mechanism depends on the sensitivity of the initial- or final-state interactions to the total color charge of the nucleon remnants represented by the diquark.  Another competing mechanism due to orbital angular momentum and nuclear shadowing resulting from uncorrelated color fields is discussed in Chapter~\ref{chap-MVspin}.

\chapter{The Saturation Paradigm of High-Energy Hadronic Structure}
\label{chap-CGC}

In Chapter~\ref{chap-TMD}, we discussed the quark and gluon degrees of freedom that can be probed by deep inelastic scattering (DIS) and the Drell-Yan process.  Deep inelastic scattering, for example, is described by a hard momentum scale $Q^2$ (the exchanged photon's virtuality) and the Bjorken scaling variable 
\begin{align}
 \label{xB1}
 x \equiv \frac{Q^2}{2 p \cdot q} = \frac{Q^2}{s + Q^2 - m_N^2} \approx \frac{Q^2}{s + Q^2}
\end{align}
with $p$ the momentum of the nucleon, $q$ the momentum of the photon, and $s \equiv (p+q)^2$ the photon-nucleon center-of-mass energy squared.  The virtuality $Q^2$ sets the transverse scale of the photon-nucleon interaction (as can be seen in the Drell-Yan-West frame of Sec.~\ref{SIDISF}), while the center-of-mass energy $s$ sets the longitudinal scale.  From \eqref{xB1}, we see that the Bjorken scaling variable $x$ fixes the ratio of $s$ to $Q^2$, so that $x$ can stand in for $s$ as a measure of the longitudinal extent of the photon-nucleon interaction.  

Let us compare the longitudinal coherence length (ie, wavelength) $\ell_\gamma$ of the virtual photon to the longitudinal size of the nucleon as a function of $x$.  For convenience, we will work in the photon-nucleon center-of-mass frame, with the virtual photon traveling along the light-cone minus axis and the nucleon traveling along the light-cone plus axis:
\begin{align}
 \label{xB2}
 p^\mu &= \left( p^+ , \frac{m_N^2}{p^+} , \ul{0} \right) \\ \nonumber
 q^\mu &= \left( - \frac{Q^2}{q^-} , q^- , \ul{0} \right) \\ \nonumber
 s &\approx p^+ q^- .
\end{align}
The longitudinal size of the nucleon in this frame is Lorentz-contracted by a large boost factor, $L^- \sim (m_N / p^+) \, R_N$, with its size in the rest frame roughly set by the mass: $R_N \sim 1 / m_N$.  The coherence length of the virtual photon in the light-cone minus direction is set by its light-cone plus momentum: $\ell_\gamma^- \sim 1 / |q^+|$.  The ratio of these two length scales is
\begin{align}
 \label{xB3}
 \frac{\ell_\gamma^-}{L^-} \sim \frac{1}{m_N R_N} \frac{p^+}{|q^+|} \sim \frac{p^+ q^-}{Q^2} \sim \frac{1-x}{x} .
\end{align}
This demonstrates that, in the \textit{Bjorken kinematics} of Chapter~\ref{chap-TMD}, where $x \sim \ord{1}$, the coherence length of the virtual photon can be less than or equal to the longitudinal size of the nucleon; thus the photon can resolve individual partons within the nucleon.  

But on the other hand, when $x$ becomes small, $x \ll 1$, the coherence length of the virtual photon becomes large, potentially even larger than the longitudinal size of the nucleon.  From \eqref{xB1} we see that, for fixed $Q^2$, small-$x$ corresponds to large center-of-mass energy $s$; the kinematic limit in which the center-of-mass energy is the dominant scale is known as the \textit{Regge limit}.  The long coherence length of the photon in the Regge limit means that, if there are multiple partons at a given transverse position within the nucleon, the virtual photon will interact with them all coherently rather than resolving a single parton.  Such configurations are unlikely for dilute targets like a nucleon, but can play an important role when the density of partons per unit transverse area becomes large.  Thus, one would naturally expect the physics of DIS on a dense target in the Regge limit to be very different from the physics of the Bjorken limit considered in Chapter~\ref{chap-TMD}.  

In this Chapter, we study the effects of high transverse densities in the Regge limit.  We will make extensive use of a target system for which the density of partons per unit transverse area is large: a heavy nucleus.  The large number $A$ of nucleons in a heavy nucleus provides an external parameter which we will use to systematically re-sum the effects of these high densities.  First, we re-examine deep inelastic scattering in Regge kinematics, finding that a different channel for the interaction of the virtual photon with the target becomes dominant when $x \ll 1$.  Then we apply this to study DIS on a heavy nucleus, re-summing the leading high-density effects which are enhanced by a power of the large parameter $A$.  By comparing this result with the solution of the classical Yang-Mills equations for a heavy nucleus, we will show that the leading high-density effects in DIS are actually dominated by the classical gluon fields of the nucleus.  Finally, we will discuss the role of quantum evolution corrections at very small $x$, which drive up the transverse density dynamically through a cascade of gluon bremsstrahlung.  Together, these considerations paint a picture of high-energy, high-density physics known as the \textit{color-glass-condensate} (CGC) in which the classical gluon fields become the dominant degrees of freedom.


\section{Dipole Channel of DIS on a Heavy Nucleus}
\label{sec-dipole_DIS}

\subsection{Regge Kinematics and the Dipole Channel}
\label{subsec-Regge}

\begin{figure}
 \centering
 \includegraphics[width=0.85\textwidth]{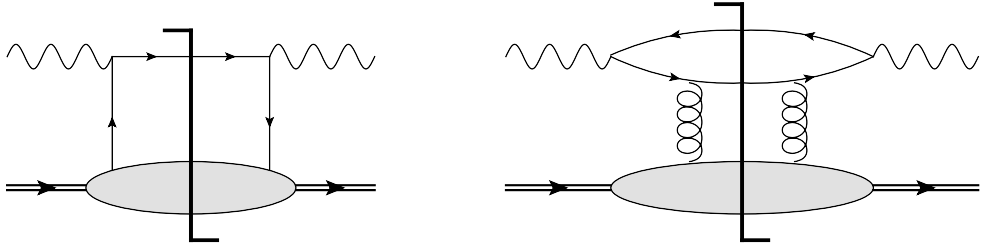}
 \caption{The leading-order diagrams for DIS in the Bjorken limit (left panel) and Regge limit (right panel).  The ``handbag diagram'' on the left leads to a cross-section that scales as $\ord{\alpha_{EM} \, \alpha_s \, x^1}$ and is suppressed at small-$x$, whereas the ``dipole diagram'' on the right leads to a cross-section that scales as $\ord{\alpha_{EM} \, \alpha_s^2}$ and dominates in the Regge limit.}
 \label{fig-Regge_DIS}
\end{figure}

In Regge kinematics $s \gg Q^2$, \eqref{xB1} simplifies to
\begin{align}
 \label{Regge1}
 x \approx \frac{Q^2}{s} \ll 1
\end{align}
so that powers of $x$ in a cross-section become suppression factors.  Regge \cite{Regge:1959a, Regge:1960a} derived a simple rule at high energy for determining the scaling of a cross-section with the center-of-mass energy $s$ based on the spin of the particle being exchanged in the $t$-channel.  For a particle with spin $j$ being exchanged $n$ times in the amplitude (and another $n$ in the complex-conjugate amplitude), the cross-section scales with the energy $s$ (and hence, through \eqref{Regge1}, with $x$) as
\begin{align}
 \label{Regge2}
 \sigma \sim s^{(j-1) 2n} \sim x^{-(j-1) 2n} .
\end{align}
A scattering process that exchanges a single quark in the amplitude ($j=\tfrac{1}{2} , n = 1$) scales as $s^{-1}$ and hence $x^1$.  These processes, which include the lowest-order ``handbag'' amplitude for DIS in the Bjorken limit (Fig.~\ref{fig-Regge_DIS}, left panel; c.f. Fig.~\ref{figDIS2}), are thus suppressed at small-$x$.  On the other hand, a scattering process which exchanges a single gluon in the amplitude ($j=1, n=1$) scales as $s^0$ and hence $x^0$.  Gluon exchange, therefore, is not $x$-suppressed at high-energies; in fact, the exchange of any number $n$ of gluons still scales as $x^0$ at small-$x$.

This consideration allows us to identify the dominant channel for DIS in the Regge limit.  In the Bjorken limit of Chapter~\ref{chap-TMD}, the lowest-order diagram for DIS is the ``handbag'' diagram shown in the left panel of Fig.~\ref{fig-Regge_DIS} (c.f. Figs.~\ref{figDIS2} , \ref{figDIS3}).  Counting powers of the electromagnetic and strong coupling (there must be at least one power of $\alpha_s$ to generate the quarks in the distribution) as well as powers of $x$ using \eqref{Regge2}, we find this ``knockout'' process scales as $\ord{\alpha_{EM} \, \alpha_s \, x^1}$.  Thus this channel for the interaction of the virtual photon with the target is suppressed as we approach the Regge limit $x \ll 1$.  On the other hand, the long coherence length of the photon \eqref{xB3} at small-$x$ also means that its quantum fluctuations inherit a long lifetime.  Thus, an alternative to the ``knockout'' process of DIS is for the virtual photon to fluctuate into a quark-antiquark dipole; the dipole can then interact with the target through the exchange of gluons, as shown in the right panel of Fig.~\ref{fig-Regge_DIS}.  From \eqref{Regge2} we see that this process scales as $\ord{\alpha_{EM} \, \alpha_s^2}$: it is not suppressed at small-$x$, although it contains an extra power of the strong coupling compared to the ``handbag'' diagram.  Thus, when $x \ll \alpha_s \ll 1$, deep inelastic scattering is dominated by the fluctuation of the photon into a long-lived $q \bar q$ dipole and its subsequent QCD scattering on the target.
\footnote{Of course, a similar process could occur consisting of a dipole fluctuation and \textit{photon} exchanges with the target.  The QCD process considered here dominates over the QED one simply because of the larger coupling.}

We can therefore express the DIS cross-section in the Regge limit as a convolution of two ingredients: the QED light-cone wave functions describing the fluctuation of the virtual photon into the dipole, and the QCD scattering cross-section for this dipole in the field of the target \cite{Kovchegov:2012mbw}.  It is especially convenient to formulate this convolution of the wave functions and dipole cross-section in transverse coordinate space; because of the extreme Lorentz contraction in the Regge limit, there is little opportunity for the dipole to drift apart during the brief interaction with the target.  In the rest frame of the target, the change in the transverse dipole size $r_T$ during the interaction would be $\Delta r_T = v_T \, \Delta t = (k_T / E) L$, where $(k_T / E)$ is the relative velocity of the quark and antiquark and $\Delta t = L / c$ is the time needed to traverse the length $L$ of the target.  Noting that the relative momentum $k_T$ is Fourier conjugate to the dipole separation $r_T$ and that the transverse momentum provided by the virtual photon is of the order of $Q$, we have $k_T \sim 1/r_T \sim Q$.  In our highly boosted frame, the relevant ratio of distance to energy is $L/E \rightarrow L^- / q^-$, giving
\begin{align}
 \label{Regge3}
 \Delta r_T &= k_T \frac{L^-}{q^-} = x  \, \frac{k_T}{Q^2} \, p^+ L^- \sim x \, \frac{1}{Q} 
 \\ \nonumber
 \frac{\Delta r_T}{r_T} & \sim x \ll 1 ,
\end{align}
where we have used \eqref{xB2} and \eqref{Regge1}, and $p^+ L^- \sim m_N R_N \sim 1$.  Thus the fractional change in the transverse size of the dipole is negligible during the scattering process; when the wave functions and dipole scattering cross-section are Fourier-transformed to transverse coordinate space, they will therefore be diagonal in $r_T$.

Using the standard rules of light-cone perturbation theory (LCPT) and the associated two-particle phase space in the conventions of \cite{Kovchegov:2012mbw}, we can write the total DIS cross-section as
\begin{align}
 \label{Regge4}
 \sigma_{tot}^{\gamma^*} = \int \frac{d^2 r \, dz}{2(2\pi) z (1-z)} \left( \left| \Psi_T (\ul r , z) 
 \right|^2 + \left| \Psi_L (\ul r , z) \right|^2 \right) \sigma_{tot}^{dip} (\ul r)
\end{align}
where $z, (1-z)$ are the fractions of $q^-$ carried by the quark (antiquark), $\Psi_L$ and $\Psi_T$ are the light-cone wave functions the fluctuation into a quark/antiquark dipole from longitudinally- and transversely-polarized photons, respectively, and $\sigma_{tot}^{dip} (\ul r)$ is the dipole scattering cross-section on the target.  The momentum-space light-cone wave functions $\Psi_{T,L}$ are given directly from the LCPT rules as
\begin{align}
 \label{Regge5}
 \Psi_{T,L} (\ul k , z) = e Z_f \frac{z (1-z)}{k_T^2 + m_f^2 + Q^2 \, z (1-z)} \,
 \ubar{\sigma}(k) \slashed{\epsilon}^\lambda_{T, L} V_{\sigma '} (q-k) ,
\end{align}
where $Z_f$ and $m_f$ are the fractional charge and mass of the quark flavor $f$ produced by the splitting, $\epsilon_{T, L}^\lambda$ are the polarization vectors of the virtual photon, and $\sigma, \sigma '$ are the spins of the quark and antiquark.  Fourier-transforming these wave functions to transverse coordinate space and squaring them is straightforward, yielding
\begin{gather}
 \begin{aligned}
 \label{dipoleWF}
 \left| \Psi_T (\ul r , z) \right|^2 &= \frac{2 N_c \alpha_{EM}}{\pi} \, z (1-z) \,
 \sum_f Z_f^2 \bigg\{ a_f^2 [z^2 + (1-z)^2] K_1^2 (r_T a_f) 
 \\ &+ m_f^2 K_0^2 (r_T a_f) \bigg\}
 \\
 \left| \Psi_L (\ul r , z) \right|^2 &= \frac{2 N_c \alpha_{EM}}{\pi} \sum_f Z_f^2 \left\{ 
 4 Q^2 z^3 (1-z)^3 K_0^2 (r_T a_f) \right\}
 \end{aligned}
\end{gather}
where $a_f^2 \equiv Q^2 \, z(1-z) + m_f^2$.

We can further simplify \eqref{Regge4} by using the optical theorem to rewrite the dipole cross-section $\sigma_{tot}^{dip}$ in terms of the imaginary part of the dipole forward-scattering amplitude:
\begin{align}
 \label{Regge6}
 \sigma_{tot}^{dip} = 2 \, \mathrm{Im} \, A^{dip}_{fwd} \equiv 2 N,
\end{align}
where $A^{dip}_{fwd}$ is the dipole forward scattering amplitude ($T$-matrix) that has been rescaled by a factor of $2 s$ and $N$ is its imaginary part.  The amplitude $N$ describes the $2 \rightarrow 2$ process of the dipole scattering on the target, and its arguments can be expressed in terms of the center-of-mass energy $s$ and a momentum transfer vector $\Delta k^\mu$, together with the internal degrees of freedom of the dipole like $\ul r$ and $z$.  As we argued in \eqref{Regge3}, the dipole separation $\ul r$ does not change during the interaction with the target, and in the high-energy scattering considered here, the dipole cross-section depends on the total center-of-mass energy $s$ but not the distribution $z$ between the quark and antiquark.  For the forward amplitude considered in \eqref{Regge6}, the net momentum transfer $\Delta k$ between the dipole and target is zero; when the transverse momentum transfer is Fourier transformed into the impact parameter $\ul b$
\begin{align}
 \label{bint}
 N (\ul{\Delta k}) = \int d^2 b \, e^{- i \ul{\Delta k} \cdot \ul{b}} N(\ul{b}) ,
\end{align}
setting $\ul{\Delta k}=0$ corresponds to integrating over all impact parameters $d^2 b$.  Thus we write
\begin{align}
 \label{Regge7}
 \sigma_{tot}^{dip} = 2 \int d^2 b \, N(\ul r , \ul b , s),
\end{align}
and hence
\begin{align}
 \label{Regge8}
 \sigma_{tot}^{\gamma^*} = \int \frac{d^2 r \, d^2 b \, dz}{(2\pi) z (1-z)} \left( \left| \Psi_T 
 (\ul r , z) \right|^2 + \left| \Psi_L (\ul r , z) \right|^2 \right) N (\ul r , \ul b , s) .
\end{align}
%

\begin{figure}
 \centering
 \includegraphics[width=0.4\textwidth]{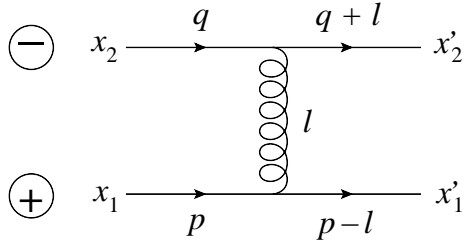}
 \caption{Quark-quark scattering at high energies.  The interaction is dominated by the exchange of ``Glauber gluons'' whose longitudinal momenta $\ell^+ , \ell^-$ are negligibly small.  }
 \label{fig-quark_scatter}
\end{figure}

Eq.~\ref{Regge8} expresses the direct connection between the cross-section $\sigma^{\gamma^*}$ for deep inelastic scattering in the Regge limit and the high-energy scattering amplitude $N (\ul r, \ul b , s)$ of a $q \bar q$ dipole on a target.  As illustrated in Fig.~\ref{fig-Regge_DIS}, the interaction of a dipole at high energies is driven by the exchange of gluons.  The most fundamental element of such processes is the elementary scattering of two quarks in the high-energy Regge limit, as illustrated in Fig~\ref{fig-quark_scatter}.  Let us evaluate this diagram using ordinary Feynman perturbation theory in the covariant (Lorenz) gauge $\partial_\mu A^\mu = 0$ (equivalent to the Feynman gauge):
\begin{align}
 \label{eik1}
 \mathcal{M} = \frac{g^2}{\ell^2} \, T^a_{i' i} T^a_{j' j} \, \left[ \ubar{\sigma_1 '} (p-\ell) 
 \gamma_\mu U_{\sigma_1} (p) \right]  \left[ \ubar{\sigma_2 '} (q+\ell) \gamma^\mu U_{\sigma_2} (q) \right]
\end{align}
where the spins $\sigma_1 ,  \sigma_1 '$ and colors $i , i'$ refer to the quark $p$ traveling with high energy along the light-cone plus axis, and similarly for $\sigma_2 , \sigma_2 ' ; j , j'$ describing the quark $q$ traveling with high energy along the light-cone minus axis.  The momenta of the incident on-shell quarks are
\begin{align}
 \label{eik2}
 p^\mu &= \left( p^+ , \frac{m^2}{p^+} , \ul{p} \right) \\ \nonumber
 q^\mu &= \left( \frac{m^2}{q^-} , q^- , \ul{q} \right) ,
\end{align}
and the on-shell conditions for the outgoing quarks constrain the light-cone momenta of the exchanged gluon:
\begin{align}
 \label{eik3}
 \ell^+ &= (q+\ell)^+ - q^+ = \frac{(\ul{q} + \ul{\ell})_T^2 + m^2}{q^- + \ell^-} - \frac{q_T^2 + m^2}{q^-} \sim \frac{\bot^2}{q^-} \\ \nonumber
 \ell^- &= p^- - (p-\ell)^- = \frac{p_T^2 + m^2}{p^+} - \frac{(\ul{p} - \ul{\ell})_T^2 + m^2}{p^+ - \ell^+} \sim \frac{\bot^2}{p^+} .
\end{align}
High-energy scattering is thus dominated by the exchange of gluons, sometimes referred to as ``Glauber gluons,'' which have negligible light-cone momenta but finite transverse momenta:
\begin{align}
 \label{eik4}
 \ell^\mu \approx \left( 0^+ , 0^- , \ul{\ell} \right).
\end{align}
Thus $\ell^2 \approx - \ell_T^2$ for a Glauber gluon.  

A corollary of the Gordon identity (see, e.g. \cite{Peskin:1995ev}),
\begin{align}
 \label{eik5}
 \ubar{\sigma '}(k) \gamma^\mu U_\sigma (k) = 2 k^\mu \delta_{\sigma \sigma'}
\end{align}
shows that the spinor products of \eqref{eik1} are dominated in the kinematics \eqref{eik2} by the large light-cone momenta $p^+ , q^-$ of the associated quarks:
\begin{align}
 \label{eik6}
 \left[ \ubar{\sigma_1 '} (p-\ell) \gamma_\mu U_{\sigma_1} (p) \right]  
 \left[ \ubar{\sigma_2 '} (q+\ell) \gamma^\mu U_{\sigma_2} (q) \right] &\approx 
 \frac{1}{2} \left[ 2 p^+ \delta_{\sigma_1 \sigma_1'} \right] \left[ 2 q^- \delta_{\sigma_2 \sigma_2 '} \right] 
 \\ \nonumber &= 2 p^+ q^- \delta_{\sigma_1 \sigma_1'} \delta_{\sigma_2 \sigma_2 '} .
\end{align}
Collecting these results, we evaluate \eqref{eik1} as
\begin{align}
 \label{eik7}
 \mathcal{M} = - g^2 \, T^a_{i' i} T^a_{j' j} \, \delta_{\sigma_1 \sigma_1'} \delta_{\sigma_2 \sigma_2'} \, \frac{2 s}{\ell_T^2},
\end{align}
where $s \equiv (p+q)^2 \approx p^+ q^-$ is the center-of-mass energy squared of the collision.  It is convenient to define an energy-rescaled amplitude
\begin{align}
 \label{eik8}
 A \equiv \frac{\mathcal{M}}{2 s} = - g^2 \, T^a_{i' i} T^a_{j' j} \, \delta_{\sigma_1 \sigma_1'} \delta_{\sigma_2 \sigma_2'} \, \frac{1}{\ell_T^2} ,
\end{align}
so that the scattering is energy independent at lowest order.  As we argued in \eqref{Regge3}, it is convenient to express high-energy scattering amplitudes in coordinate space; to illustrate this, let us Fourier-transform \eqref{eik8}:
\begin{align}
 \label{eik9}
 \tilde{A} &\equiv \int \frac{d^2 p}{(2\pi)^2} \frac{d^2 q}{(2\pi)^2} \frac{d^2 \ell}{(2\pi)^2} 
 \, e^{i \ul{p} \cdot (\ul{x}_1' - \ul{x}_1)} \, e^{i \ul{q} \cdot (\ul{x}_2' - \ul{x}_2)} \, e^{i \ul{\ell} \cdot (\ul{x}_2' - \ul{x}_1')} \, A (\ul{p},\ul{q},\ul{\ell})
 \\ \nonumber &=
 \delta^2 (\ul{x}_1' - \ul{x}_1) \, \delta^2 (\ul{x}_2' - \ul{x}_2) \left[ - g^2 \, T^a_{i' i} T^a_{j' j} \delta_{\sigma_1 \sigma_1'} \delta_{\sigma_2 \sigma_2'} \int \frac{d^2 \ell}{(2\pi)^2} e^{i \ul{\ell} \cdot (\ul{x_2'}
 - \ul{x_1'})} \frac{1}{\ell_T^2} \right]
 \\ \nonumber &=
 \delta^2 (\ul{x}_1' - \ul{x}_1) \, \delta^2 (\ul{x}_2' - \ul{x}_2) \,
 \left[ - g^2 \, T^a_{i' i} T^a_{j' j} \, \delta_{\sigma_1 \sigma_1'} \delta_{\sigma_2 \sigma_2'} \, \frac{1}{2\pi} \ln \frac{1}{|\ul{x_2'} - \ul{x_1'}|_T \Lambda} \right]
\end{align}
where $\Lambda$ is an infrared cutoff.

We see from the delta functions in \eqref{eik9} that transverse coordinates of the quarks are unchanged by the scattering $(\ul{x}_1 = \ul{x}_1' , \ul{x}_2 = \ul{x}_2')$ as anticipated by \eqref{Regge3}, which again reflects the instantaneous nature of high-energy scattering.  The only nontrivial integral performed in the Fourier transform \eqref{eik9} is over the transverse momentum carried by the Glauber gluon, which generates a logarithm in the transverse separation.  This logarithm reflects the familiar example from classical electrodynamics of the logarithmic potential of an infinite wire of charge.  Since the deflection of the color-charges is negligible in the high-energy limit, this is indeed the situation realized by quark-quark scattering in Regge kinematics, as we will see again explicitly in the classical calculation of Sec.~\ref{subsec-MV}.  The physical situation realized by this example - instantaneous interaction of high-energy particles that preserves transverse coordinates - is referred to as \textit{eikonal scattering}; the corresponding high-energy limit in which only the leading behavior with energy is kept is the \textit{eikonal approximation}.  

\begin{figure}
 \centering
 \includegraphics[width=0.5\textwidth]{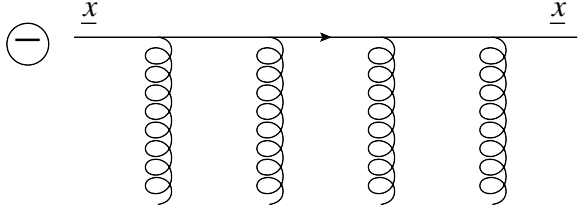}
 \caption{Illustration of the propagation of a high-energy quark in a background field of Glauber gluons represented by a Wilson line.  The eikonal kinematics demonstrated by a single scattering \eqref{eik9} reduce the interaction to a net color rotation with the total phase given by the path-ordered exponential \eqref{Wilson1}.}
 \label{fig-Wlinedefn}
\end{figure}

If a high-energy quark moving along the light-cone minus axis interacts with many such Glauber gluons as in Fig.~\ref{fig-Wlinedefn}, the transverse coordinate $\ul{x}$ and the eikonal momentum $q^-$ will still be unchanged.  The only effect is a net $SU(N_c)$ color rotation performed by a sequential interaction with the gauge fields $A^{\mu a} T^a$ in the order they are encountered.  The eikonal kinematics make it easy to re-sum this total color rotation into the form of a \textit{Wilson line} given by the path-ordered exponential \cite{Wilson:1974sk}
\begin{align}
 \label{Wilson1}
 V(\ul{x}, x_i^- , x_f^-) = \mathcal{P} \, \exp\left[ \frac{i g}{2} \int_{x_i^-}^{x_f^-} d x^- \,
 A^{+ a}(\ul{x},x^-) \, T^a \right]
\end{align}
where the quark propagates along the light-cone minus axis from $x_i^-$ to $x_f^-$, $\mathcal{P}$ stands for path-ordering, and $T^a$ are the $SU(N_c)$ generators in the fundamental representation.  The straight-line segment along the minus light-cone \eqref{Wilson1} is a special case of the more general integral along an arbitrary contour $C$, such as the gauge-link defined in \eqref{DISS-GAUGE2}.  Analogous Wilson lines can be defined for the propagation of eikonal gluons, which couple to the $SU(N_c)$ generators in the adjoint representation.

\subsection{Eikonal Scattering of a Color Dipole}
\label{subsec-dipole}

\subsubsection{$\gamma^* \gamma^*$ Scattering}

\begin{figure}
 \centering
 \includegraphics[width=0.85\textwidth]{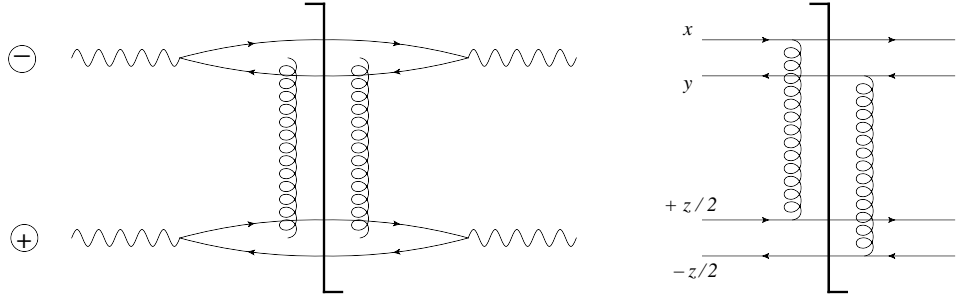}
 \caption{Scattering of two virtual photons by their fluctuation into $q \bar q$ dipoles and the exchange of gluons.  The disconnected gluon lines in the left panel represent a summation over all possible attachments of those gluons to the quark or antiquark in each dipole.  One such diagram is shown in the right panel, along with coordinates denoting the transverse positions of the quarks and antiquarks.}
 \label{fig-dipole_dipole}
\end{figure}

We would ultimately like to use \eqref{Regge4} to calculate the DIS cross-section on a heavy nucleus to understand the effects of longitudinal coherence introduced by the Regge limit.  The first step toward this goal is to build upon the quark-quark scattering amplitude \eqref{eik8} to construct the cross-section for the eikonal scattering of two dipoles.  A useful example to put this process in context is the scattering of two virtual photons by their fluctuation into dipoles, as illustrated in Fig.~\ref{fig-dipole_dipole}; this process is essentially DIS on a target which is also a dipole.
%
%
Thus we can use \eqref{Regge4} and \eqref{Regge8} to write
\begin{align}
 \label{2gamma1}
 \hspace{-0.275cm}
 \sigma_{tot}^{\gamma^* \gamma^*} \!\! &= \!\! \int \frac{d^2 r_1 \, dz_1}{2(2\pi) z_1 (1-z_1)} 
 \frac{d^2 r_2 \, dz_2}{2(2\pi) z_2 (1-z_2)} \left| \Psi_{dip} (\ul{r_1} , z_1 ) \right|^2
 \left| \Psi_{dip} (\ul{r_2} , z_2 ) \right|^2 \sigma_{tot}^{dip \, dip} (\ul{r_1}, \ul{r_2})
\end{align}
with
\begin{align}
 \label{2gamma2}
 \sigma_{tot}^{dip \, dip} (\ul{r_1}, \ul{r_2}) = 2 \int d^2 b \, N^{dip \, dip} (\ul{r_1} , \ul{r_2} , \ul{b}) = \int d^2 b \, \frac{d\sigma^{dip \, dip}}{d^2 b} (\ul{r_1} , \ul{r_2} , \ul{b}) .
\end{align}

The dipole-dipole cross-section \eqref{2gamma2} consists of a number of sub-processes built from the eikonal quark-quark scattering amplitude \eqref{eik8}.  It is important to note that, from the point of view of the forward-scattering amplitude $N$, the exchange of a single $t$-channel gluon cannot contribute.  This is because in QCD the gluon carries color charge, so, unlike QED, the exchange of a single gauge boson changes the color state of the target and is thus necessarily non-forward.  The lowest-order contribution to the forward-scattering amplitude $N$ in QCD comes from the exchange of 2 gluons in a color-singlet state, which is equivalent to the square of the single-gluon exchange amplitude \eqref{eik8}.

We can construct the dipole-dipole cross-section in transverse coordinate space \eqref{2gamma2} by summing all of the $(2)^4 = 16$ possible diagrams that apply \eqref{eik9}.  One such diagram is shown in the right panel of Fig.~\ref{fig-dipole_dipole}; leaving off the color factor for now, we evaluate its contribution to be
\begin{align}
 \label{2gamma3}
 \bigg[ A A^* \bigg]_{Fig. \, \ref{fig-dipole_dipole}} &=
 \left( \frac{-g^2}{2\pi} \right)^2 \ln \frac{1}{|\ul{x} - \ul{z}/2|_T \Lambda} \, \ln \frac{1}{|\ul{y} + \ul{z}/2|_T \Lambda}
 \\ \nonumber &=
 \frac{ (-g^2)^2}{(2\pi)^4} \int \frac{d^2 \ell \, d^2 q}{\ell_T^2 \, q_T^2} \, 
 e^{i \ul{\ell} \cdot (\ul{x} - \ul{z}/2)} \, e^{-i \ul{q} \cdot (\ul{y} + \ul{z}/2)}
\end{align}
where we have utilized the inverse transform of \eqref{eik9} and the coordinate system is illustrated in the figure.  By starting with the coordinate-space logarithms and transforming back to momentum space in this way, we can absorb the coordinate differences between the various gluon attachments into Fourier factors.  Aside from this, the only difference among the 16 diagrams is the possible minus sign due to attaching a gluon to an antiquark rather than a quark.  For quark-quark or antiquark-antiquark scattering, the amplitude is given by \eqref{eik8}, while for quark-antiquark scattering, an extra minus sign is introduced.  

Tabulating the other 15 contributions as in \eqref{2gamma3} is tedious but straightforward.  The result can be expressed in the factorized form
\begin{align}
 \label{2gamma4}
 \sum_{diags} = \frac{g^4}{(2\pi)^4} \int \frac{d^2 \ell \, d^2 q}{\ell_T^2 \, q_T^2} \, &\left[
 e^{i \tfrac{1}{2} (\ul{\ell} - \ul{q}) \cdot {\ul z}} - 
 e^{i \tfrac{1}{2} (\ul{\ell} + \ul{q}) \cdot {\ul z}} - 
 e^{-i \tfrac{1}{2} (\ul{\ell} + \ul{q}) \cdot {\ul z}} + 
 e^{-i \tfrac{1}{2} (\ul{\ell} - \ul{q}) \cdot {\ul z}} \right]
 \\ \nonumber &\times \left[
 e^{i (\ul{\ell} - \ul{q}) \cdot \ul{x}} - 
 e^{i \ul{\ell} \cdot \ul{x}} \, e^{-i \ul{q} \cdot \ul{y}} - 
 e^{i \ul{\ell} \cdot \ul{y}} \, e^{-i \ul{q} \cdot \ul{x}} + 
 e^{i (\ul{\ell} - \ul{q}) \cdot \ul{y}} \right] .
\end{align}
To relate this back to the dipole-dipole cross-section at fixed impact parameter, let us define the average coordinate $\ul{b} \equiv \tfrac{1}{2} (\ul{x} + \ul{y})$ and relative coordinate $\ul{r} \equiv \ul{x} - \ul{y}$.  Then the factor $\exp[i (\ul{\ell} - \ul{q}) \cdot \ul{b}]$ is common to all of the exponentials in the second bracketed factor of \eqref{2gamma4}; pulling this common factor out gives
\begin{align}
 \nonumber
 \sum_{diags} &=
 \frac{g^4}{(2\pi)^4} \int \frac{d^2 \ell \, d^2 q}{\ell_T^2 \, q_T^2} \,
 e^{i (\ul{\ell} - \ul{q}) \cdot \ul{b}} \,  \left[
 e^{i \tfrac{1}{2} (\ul{\ell} - \ul{q}) \cdot {\ul z}} - 
 e^{i \tfrac{1}{2} (\ul{\ell} + \ul{q}) \cdot {\ul z}} - 
 e^{-i \tfrac{1}{2} (\ul{\ell} + \ul{q}) \cdot {\ul z}} + 
 e^{-i \tfrac{1}{2} (\ul{\ell} - \ul{q}) \cdot {\ul z}} \right]
 \\ \label{2gamma5} &\times
 \left[
 e^{i \tfrac{1}{2} (\ul{\ell} - \ul{q}) \cdot {\ul r}} - 
 e^{i \tfrac{1}{2} (\ul{\ell} + \ul{q}) \cdot {\ul r}} - 
 e^{-i \tfrac{1}{2} (\ul{\ell} + \ul{q}) \cdot {\ul r}} + 
 e^{-i \tfrac{1}{2} (\ul{\ell} - \ul{q}) \cdot {\ul r}}
 \right] .
\end{align}
Integrating over impact parameters as in \eqref{2gamma2} generates a delta function $\delta^2 (\ul{\ell} - \ul{q})$ that we can use to integrate over $d^2 q$, giving
\begin{align}
 \label{2gamma6}
 \int d^2 b \bigg[ \sum_{diags} \bigg] = \frac{g^4}{(2\pi)^2} \int \frac{d^2 \ell}{\ell_T^4} \,
 \left[ 2 - e^{i \ul{\ell} \cdot \ul{z}} - e^{- i \ul{\ell} \cdot \ul{z}} \right] \,
 \left[ 2 - e^{i \ul{\ell} \cdot \ul{r}} - e^{- i \ul{\ell} \cdot \ul{r}} \right] .
\end{align}
Finally, we need to re-insert the color factor; each dipole yields the same trace of the $SU(N_c)$ generators, and there is a conventional factor of $1 / N_c^2$ from averaging the colors of each dipole:
\begin{align}
 \label{2gamma7}
 \frac{1}{N_c^2} \, \Tr[T^a \, T^b] \, \Tr[T^a \, T^b] = \frac{1}{2 N_c^2} \, \delta^{a b} \, 
 \Tr[T^a T^b] = \frac{C_F}{2 N_c}.
\end{align}
where $C_F = (N_c^2 - 1)/2N_c$ is the quadratic Casimir of $SU(N_c)$ in the fundamental representation.  Combining this color factor with \eqref{2gamma6} and returning to the coordinate labels $\ul{r_1} , \ul{r_2}$ gives
\begin{align}
 \label{2gamma8}
 \sigma_{tot}^{dip \, dip} (\ul{r_1},\ul{r_2}) = 
 \frac{2 C_F}{N_c} \, \alpha_s^2 \, \int \frac{d^2 \ell}{\ell_T^4} \,
 \left[ 2 - e^{i \ul{\ell} \cdot \ul{r_1}} - e^{- i \ul{\ell} \cdot \ul{r_1}} \right] \,
 \left[ 2 - e^{i \ul{\ell} \cdot \ul{r_2}} - e^{- i \ul{\ell} \cdot \ul{r_2}} \right] .
\end{align}

The symmetric form of \eqref{2gamma8} is suggestive, with each factor in brackets describing the emission or absorption of gluons from all possible attachments to the quark and antiquark in the $\ul{r_1}$ and $\ul{r_2}$ dipoles.  Such factors describe the overall distribution of gluons emitted from or absorbed by the projectile and target dipoles, and the convolution in \eqref{2gamma8} splices these distributions together to assemble the scattering cross-section.  Motivated by this observation, we define the \textit{unintegrated gluon distribution} of the dipole and virtual photon as (see, e.g. \cite{Kovchegov:2012mbw})
\begin{align}
 \label{2gamma9}
 \phi^{dip}(\ul{r}, \ul {k}) &\equiv \frac{\alpha_s C_F}{\pi} \, \frac{1}{k_T^2} 
 \left[ 2 - e^{i \ul{k} \cdot \ul{r}} - e^{- i \ul{k} \cdot \ul{r}} \right] 
 \\ \nonumber
 \phi^{\gamma^*}( \ul{k}) &\equiv \int \frac{d^2 r \, dz}{2(2\pi) z (1-z)} \,
 \left| \Psi_{dip}(\ul{r}, z) \right|^2 \, \phi^{dip} (\ul{r}, \ul{k}) ,
\end{align}
which, together with \eqref{2gamma8} and \eqref{2gamma1}, gives the total photon-photon and dipole-dipole cross-sections the particularly simple form
\begin{align}
 \label{2gamma10}
 \sigma_{tot}^{dip \, dip} (\ul{r_1},\ul{r_2}) &= \frac{2 \pi^2}{N_c C_F} \, \int d^2 k \,
 \phi^{dip}_1 (\ul{r_1},\ul{k}) \, \phi^{dip}_2 (\ul{r_2},\ul{k})
 \\ \nonumber
 \sigma_{tot}^{\gamma^* \gamma^*} &= \frac{2 \pi^2}{N_c C_F} \, \int d^2 k \,
 \phi^{\gamma^*}_1 (\ul{k}) \, \phi^{\gamma^*}_2 (\ul{k}).
\end{align}
The extremely simple form of \eqref{2gamma10} in which the cross-section appears as just a convolution over the transverse momentum of the projectile and target gluon distributions is a feature known as \textit{$k_T$-factorization}.  Such $k_T$-factorization is known to occur in a variety of scattering and production cross-sections at high energy \cite{Kovchegov:2012mbw}, beyond just the lowest order calculation presented here.

Since the dipole wave functions of \eqref{dipoleWF} depend only on the size $r_T$ of the dipole and not its vector orientation $\ul{r}$, the relevant quantity is the angular-averaged gluon distribution
\begin{align}
 \label{2gamma11}
 \left\langle \phi^{dip} \right\rangle (r_T , k_T) \equiv \int \frac{d\varphi_r}{2\pi} \, \phi^{dip} (\ul{r}, \ul{k}) = \frac{\alpha_s C_F}{\pi} \, \frac{2}{k_T^2} \left[1 - J_0 (k_T r_T) \right]
\end{align}
which can be thought of as either the distribution of gluons emitted by a dipole or the interaction of a dipole with an external gluon field.  Consider two limits of this distribution, corresponding to large momenta $k_T r_T \gg 1$ or small momenta $k_T r_T \ll 1$; using the asymptotic forms of the Bessel function gives
\begin{align}
 \label{2gamma12}
 k_T r_T \gg 1 : &\hspace{0.5cm}
 \left\langle \phi^{dip} \right\rangle \sim \frac{\alpha_s C_F}{\pi} \, \left(\frac{2}{k_T^2} \right)
 \\ \nonumber
 k_T r_T \ll 1 : &\hspace{0.5cm}
 \left\langle \phi^{dip} \right\rangle \sim \frac{\alpha_s C_F}{\pi} \, \left(\frac{r_T^2}{2} \right) .
\end{align}
Thus in the lowest-order approximation considered here, high-momentum gluons emitted by a large dipole ($k_T r_T \gg 1$) are just a superposition of the quark and antiquark's individual $1/k_T^2$ gluon fields \eqref{eik7}, while a small dipole couples to long-wavelength external fields ($k_T r_T \ll 1$) proportional to its dipole moment squared $d^2 \sim \alpha_s r_T^2$.  

\subsubsection{Dipole DIS on a Single Nucleon}

\begin{figure}
 \centering
 \includegraphics[width=0.5\textwidth]{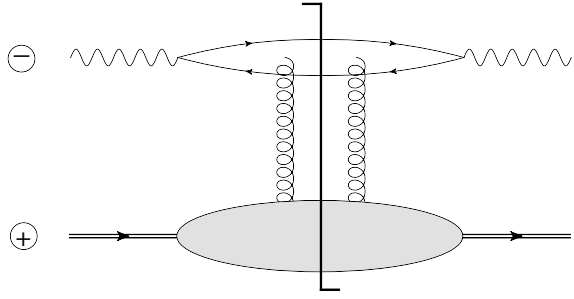}
 \caption{Dipole channel of DIS on a nucleon.  Analogous to the photon-photon / dipole-dipole scattering considered previously, the virtual photon generates a perturbatively small dipole that interacts with the gluon field of the nucleon.}
 \label{fig-dipole_nucleon}
\end{figure}

Now let us generalize these considerations to study the dipole channel of DIS on a nucleon, as illustrated in Fig.~\ref{fig-dipole_nucleon}.  With no external preferred direction to the nucleon or virtual photon, it is again the angular-averaged cross-section that contributes:
\begin{align}
 \label{dipN1}
 \left\langle \sigma_{tot}^{dip \, N} \right\rangle (r_T , R_T) = \frac{2\pi^2}{N_c C_F} \int d^2 k \,
 \big\langle \phi^{dip} \big\rangle (r_T , k_T) \, \big\langle \phi^N \big\rangle (R_T , k_T)
\end{align}
with the angular-averaged dipole field given by \eqref{2gamma11}.  Unlike the dipole-dipole scattering considered previously, here there is a clear separation of scales between the dipole projectile and the nucleon target.  The dipole is generated by the quantum fluctuation of a highly-virtual photon, with a characteristic size set by the virtuality, $r_T^2 \sim 1/ (Q^2 z (1-z) )$ \eqref{dipoleWF}.  The nucleon, on the other hand, has a size set by its mass $m_N$, or equivalently, by $\Lambda_{QCD}$ : $R_T^2 \sim 1 / \Lambda^2$.  Since $Q^2 \gg \Lambda^2$ for deep inelastic scattering, we see that the dipole is perturbatively small, scattering in the field of the much larger nucleon.  

The characteristic momentum scale of the nucleon's field is set by $k_T \sim 1/R_T$ so that the interaction with the dipole takes place well into the small-dipole, long-wavelength asymptotics $k_T r_T \sim r_T / R_T \ll 1$.  Thus we can use the asymptotics \eqref{2gamma12} for the dipole field $\langle \phi^{dip} \rangle$, and we must impose a corresponding UV cutoff $k_T < 1/r_T$ on the nucleon field $\langle \phi^N \rangle$ for consistency.  This allows us to simplify \eqref{dipN1} as
\begin{align}
 \label{dipN2}
 \left\langle \sigma_{tot}^{dip \, N} \right\rangle (r_T , R_T) &= \frac{2\pi^2}{N_c C_F} \, \big\langle \phi^{dip} \big\rangle (k_T \ll 1/r_T) \hspace{-0.4cm} \int\limits^{(k_T < 1/r_T)} \hspace{-0.4cm} d^2 k \, \big\langle \phi^N \big\rangle (R_T , k_T)
 \\ \nonumber &=
 \frac{\alpha_s \pi^2}{N_c} \, r_T^2 \hspace{-0.4cm} \int\limits^{(k_T^2 < 1/r_T^2)} \hspace{-0.4cm} d k_T^2 \, \big\langle \phi^N \big\rangle (R_T , k_T)
 \\ \nonumber \left\langle \sigma_{tot}^{dip \, N} \right\rangle (r_T , R_T) &\equiv
 \frac{\alpha_s \pi^2}{N_c} \, r_T^2 \, x G_N \left(\frac{1}{r_T^2}\right)
\end{align}
where $x G_N (1/r_T^2)$ is the integral giving the total gluon field of the nucleon with momenta less than the UV cutoff $k_T^2 < 1/r_T^2$.  Thus the small dipole is measuring the gluon field strength of the nucleon, with a coupling proportional to the square of the dipole size.  We can fully evaluate \eqref{dipN2} if we model the ``nucleon'' by a large dipole using \eqref{2gamma11},
\begin{align}
 \label{dipN3}
 x G_{dip} \left(\frac{1}{r_T^2}\right) &= \frac{2 \alpha_s C_F}{\pi} \int\limits_{0}^{1/r_T^2} \frac{d k_T^2}{k_T^2} \, \left[1 - J_0(k_T R_T) \right] 
 \\ \nonumber &=
 \frac{2 \alpha_s C_F}{\pi} \left\{ \int\limits_{0}^{1/R_T^2} \frac{d k_T^2}{k_T^2} \, \left[1 - J_0(k_T R_T) \right] + \int\limits_{1/R_T^2}^{1/r_T^2} \frac{d k_T^2}{k_T^2} - \int\limits_{1/R_T^2}^{1/r_T^2} \frac{d k_T^2}{k_T^2} \, J_0 (k_T R_T) \right\}
 \\ \nonumber &\approx
 \frac{2 \alpha_s C_F}{\pi} \ln \frac{R_T^2}{r_T^2}
 \\ \nonumber x G_{dip} \left(\frac{1}{r_T^2}\right) &\approx
 \frac{4 \alpha_s C_F}{\pi} \ln \frac{1}{r_T \Lambda} ,
\end{align}
where we have dropped the first and third integrals in the large braces because they are finite in the limit $r_T / R_T \rightarrow 0$.  Using this result in \eqref{dipN2} gives the explicit dipole-``nucleon'' cross-section as
\begin{align}
 \label{dipN4}
 \left\langle \sigma_{tot}^{dip \, dip} \right\rangle (r_T) = \frac{4\pi C_F \alpha_s^2}{N_c} \, r_T^2
 \ln\frac{1}{r_T \Lambda}
\end{align}
with $\Lambda$ now playing the role of an infrared cutoff.  For comparison, the evaluation with finite dipole sizes without the assuption $r_T \ll R_T$ is given by \cite{Kovchegov:2012mbw}
\begin{align}
 \label{dipdipfull}
 \left\langle \sigma_{tot}^{dip \, dip} \right\rangle (r_T) = \frac{4\pi C_F \alpha_s^2}{N_c} \,
 r_<^2 \left[ 1 + \ln\frac{r_>}{r_<} \right] ,
\end{align}
where $r_<$ $(r_>)$ is the minimum (maximum) of $r_T , R_T$.  Eq.~\eqref{dipN2}, together with \eqref{Regge4}, gives an evaluation of the dipole DIS cross-section on a nucleon target in terms of its gluon distribution $x G_N$, which can be calculated in the dipole model using \eqref{dipN4} or \eqref{dipdipfull}.

\subsection{Glauber-Gribov-Mueller Multiple Rescattering}
\label{subsec-GGM}

Now let us build up the dipole cross-section for interacting with a heavy nucleus having a large number $A \gg 1$ of nucleons.  The model presented here is known as the Glauber-Gribov-Mueller picture of multiple rescatterings in a heavy nucleus \cite{Glauber:1955qq, Glauber:1970jm, Franco:1965wi, Gribov:1968jf, Gribov:1968gs, Mueller:1989st}.  The nucleus is considered to be a large bag of $A$ nucleons with nuclear radius $R_A$ and nucleon number density $\rho_A$.  The typical number of nucleons in a given direction is $A^{1/3} \gg 1$, and the nucleus is considered to be dilute, with a finite longitudinal separation between the nucleons of order $\Delta x^3 \sim R_A \, A^{-1/3}$.  Despite the nucleus being dilute in a three-dimensional sense, the transverse number density of nucleons at a given impact parameter $\ul{b}$ (sometimes called the \textit{nuclear profile function})
\begin{align}
 \label{dipA1}
 T(\ul{b}) \equiv \int d b^3 \, \rho_A (\ul{b}, b^3)
\end{align}
is large, scaling with the nucleon number as $T(\ul{b}) \sim A^{1/3}$.  For a uniform sphere of constant density, for example, the nuclear profile function is
\begin{align}
 \label{Tbextra}
 T(\ul{b}) = 2 \rho_A \sqrt{R_A^2 - b_T^2} .
\end{align}

For small values of $A$, the interaction of the DIS dipole with the nucleus follows the same power-counting as for the scattering from a single nucleon, and the leading-order contribution to the DIS cross-section is just the $\ord{\alpha_s^2}$ 2-gluon exchange of Fig.~\ref{fig-dipole_nucleon}.  In this regime, the transverse density $T(\ul{b})$ is dilute and the effects of multiple scattering are negligible, so that the scattering cross-section on the nucleus is just the superposition of scattering on any of the $A$ nucleons individually.  But for a heavy nucleus with $A^{1/3} \gg 1$, the probability of the dipole to interact with multiple nucleons is enhanced.  Suppose the dipole interacts with one nucleon by the usual 2-gluon exchange; if it were to interact a second time with the same nucleon by another color-singlet 2-gluon exchange, the process would be suppressed by an additional factor of $\ord{\alpha_s^2}$.  However, if the second interaction occurs on a \textit{different} nucleon, there is an additional combinatoric enhancement due to the large number $\sim A^{1/3}$ of such nucleons available at a given impact parameter; this rescattering on an independent nucleon would contribute a factor of $\ord {\alpha_s^2 \, A^{1/3}}$ and therefore be less suppressed.  When the transverse density is large enough to offset the suppression from the coupling, $\alpha_s^2 \, A^{1/3} \sim \ord{1}$, then all such rescatterings become equally important and need to be re-summed.  This discussion highlights a general principle: because correlations between two (or more) nucleons are suppressed by powers of $A$, the dominant effect of the large nucleon number is to present the nucleus as a bag of \textit{uncorrelated, independent} nucleons.  

\begin{figure}
 \centering
 \includegraphics[width=0.8\textwidth]{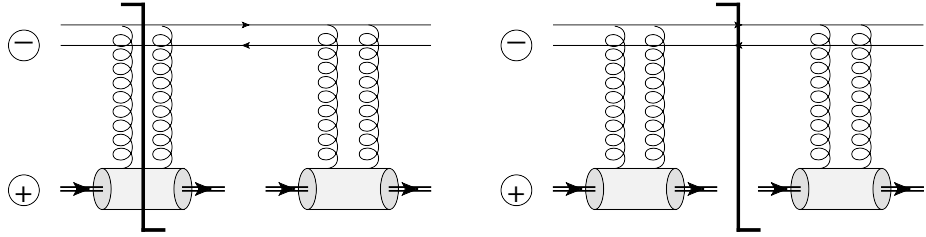}
 \caption{Two possible cuts of the forward-scattering amplitude for a dipole on two nucleons.  The cross-section \eqref{Regge6} is given in terms imaginary part of the forward-scattering amplitude, which is obtained by inserting a final-state cut in all possible ways \cite{Cutkosky:1960sp}.  For the two-gluon case, the only nontrivial cut is the one visualized in Fig.~\ref{fig-dipole_nucleon}, but with multiple scattering included, now many such cuts are possible; two of these cuts are visualized here.}
 \label{fig-Cuts}
\end{figure}

We would thus like to calculate the dipole-nucleus cross-section $\sigma_{tot}^{dip \, A} (r_T , \ul{b})$, or equivalently, the imaginary part of the forward-scattering amplitude $N(r_T , \ul{b})$, in terms of the dipole-nucleon cross-section \eqref{dipN2}.  Now that we are including rescattering on other nucleons, the relationship between $N$ and $\sigma_{tot}$ is considerably more complicated.  For the previous case when a total of two gluons were exchanged, the imaginary part of the two-gluon exchange was given by a unique final-state cut (see, e.g. Fig.~\ref{fig-dipole_dipole}) that separated the cross-section into the square of the one-gluon exchange amplitude which we calculated in \eqref{eik7}.  Since we now want to include multiple iterations of two-gluon exchange, there are many possible final-state cuts that can contribute (two such cuts are shown in Fig.~\ref{fig-Cuts}), and it is not at all clear \textit{a priori} that we will be able to reduce all the relevant diagrams down to just squares of single-gluon exchange; nonetheless, that is what we will now show.  

\begin{figure}
 \centering
 \includegraphics[width=0.8\textwidth]{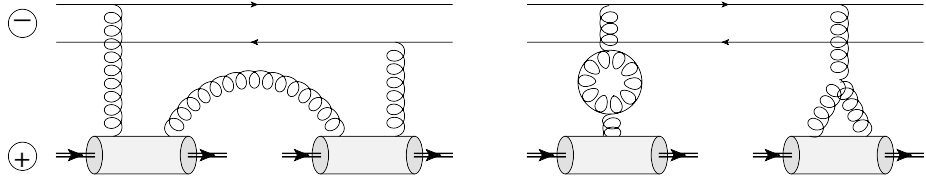}
 \caption{Examples of diagrams that do not contribute to the leading-$A$ calculation of the (imaginary part of) the forward-scattering amplitude $N$.  The left panel represents nucleon-nucleon correlations due to gluon exchange, which are sub-leading in $A$.  The right-panel shows two examples of color non-singlet exchanges which do not contribute to the forward amplitude at all.}
 \label{fig-NN1}
\end{figure}

Let us analyze in detail the rescatterings on two nucleons in the covariant gauge $\partial_\mu A^\mu = 0$; once we have identified the relevant contributions, it will be straightforward to extend this to an arbitrary number of nucleons.  To begin, let us emphasize that each contribution to $N(r_T , \ul{b})$ must be forward at the level of each nucleon.  Since each nucleon is color-neutral, this necessitates that the two gluons exchanged with each nucleon must be in a color-singlet configuration.  Secondly, we know that correlations between nucleons are suppressed by powers of $A$ and therefore do not contribute to this leading-order calculation.  These considerations allow us to neglect the diagrams shown in Fig.~\ref{fig-NN1}.

\begin{figure}
 \centering
 \includegraphics[width=0.85\textwidth]{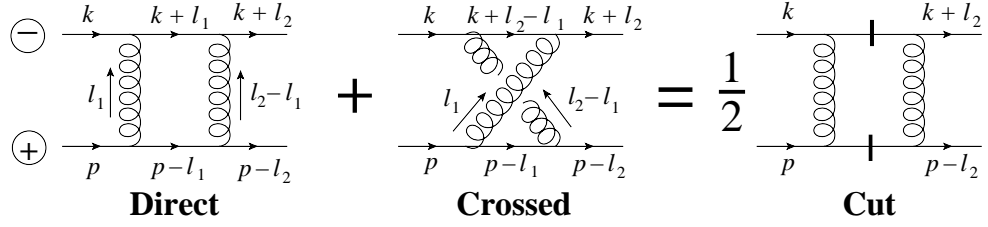}
 \caption{The combination of ``direct'' and ``crossed'' gluon exchange, as discussed in the text.  The sum of the two processes for color-singlet exchange in the high-energy limit has the remarkable feature of effectively putting the intermediate propagators on-shell, denoted by the thick vertical lines.  This enables many of the simplifications used in this Section to the dipole-nucleus forward-scattering amplitude.}
 \label{fig-NN2}
\end{figure}

Since taking the imaginary part no longer imposes a unique cut between the two gluons being exchanged to a given nucleon, we must consider new, distinct diagrams in which the two gluon propagators are ``crossed,'' as shown in the middle panel of Fig.~\ref{fig-NN2}.  Such crossed diagrams are only topologically distinct from the direct diagrams if the gluons interact with at least one fermion twice; the two gluons can even form a loop if they interact with the same particle in both the dipole and the nucleon.  To illustrate the properties of these graphs, we will analyze the case of quark-quark scattering shown in Fig.~\ref{fig-NN2}.  

Note that, since the two gluons are in a color-singlet state, the color factor of the direct and crossed diagrams is identical.  Kinematically, the upper quark line carries a large light-cone minus momentum $k^-$, while the bottom line carries a large light-cone plus momentum $p^+$, and the on-shell conditions for the outgoing quarks specify $\ell_2^+ = \ell_2^- = 0$ with eikonal accuracy.  Moreover, if we analyze the Dirac structure of the upper line carrying the large $k^-$ momentum in both cases,
\begin{align}
 \label{dipA2}
 D_{direct}(k) &\equiv 
 \ubar{}(k + \ell_2) \:\: \gamma^\mu (\slashed{k} + \slashed{\ell}_1 - m) \gamma^\nu \:\: U(k)
 \\ \nonumber &\approx 
 \ubar{}(k^-) \:\: \gamma^\mu (\tfrac{1}{2} k^- \gamma^+) \gamma^\nu \:\: U(k^-) 
 \\ \nonumber &\, \\ \nonumber
 D_{crossed}(k) &\equiv 
 \ubar{}(k + \ell_2) \:\: \gamma^\nu (\slashed{k} + \slashed{\ell}_2 - \slashed{\ell}_1 - m) 
 \gamma^\mu \:\: U(k)
 \\ \nonumber &\approx 
 \ubar{}(k^-) \:\: \gamma^\nu (\tfrac{1}{2} k^- \gamma^+) \gamma^\mu \:\: U(k^-) ,
\end{align}
we can see already from the Lorentz indices that the eikonal contribution will come from $\mu = \nu = -$, since the product $\ubar{}(k^-) \: \gamma^- \gamma^+ \gamma^- \: U(k^-) = 4 \ubar{}(k^-) \: \gamma^- \: U(k^-) = 8 k^-$ contributes a factor of the large $k^-$ momentum.  Thus the Dirac structure of the upper line is the same for the direct and crossed diagrams: $D_{direct}(k) = D_{crossed}(k)$ with eikonal accuracy.  The same reasoning holds for the lower line, which is dominated by the large $p^+$ momentum; these considerations show that in the eikonal limit, the color and Dirac structure of the direct and crossed diagrams are the same.  Thus, when we add the two diagrams together, the only factors not common to both come from the denominators of the intermediate quark lines $(k+\ell_1)$ vs $(k + \ell_2 - \ell_1)$.  The sum of these factors in the eikonal limit is given by
\begin{align}
 \label{dipA3}
 \frac{i}{(k + \ell_1)^2 - m^2 + i\epsilon} &+ \frac{i}{(k + \ell_2 - \ell_1)^2 - m^2 + i\epsilon}
 = \\ \nonumber &\approx
 \frac{i}{k^- \ell_1^+ - \bot^2 + i\epsilon} + \frac{i}{- k^- \ell_1^+ - \bot^2 + i\epsilon} 
 \\ \nonumber
 &= \frac{i}{k^-} \left[\frac{1}{\ell_1^+ - \bot^2/k^- + i\epsilon} - \frac{1}{\ell_1^+ + \bot^2/k^- - i\epsilon} \right] 
 \\ \nonumber
 &= \frac{i}{k^-} \left[\left( P.V.\frac{1}{\ell_1^+} - i \pi \, \delta\left(\ell_1^+ - \frac{\bot^2}{k^-}\right) \right) - \left( P.V.\frac{1}{\ell_1^+} + i \pi \, \delta\left(\ell_1^+ + \frac{\bot^2}{k^-} \right) \right) \right]
 \\ \nonumber
 &= \frac{i}{k^-} \bigg[ - 2\pi i \, \delta(\ell_1^+) \bigg]
\end{align}
where $P.V.$ is the principal value regularization.  Adding the direct and crossed diagrams has generated a delta function which effectively puts the intermediate quark propagator of the upper line on-shell: $k^- \ell_1^+ \approx (k+\ell_1)^2 \approx (k+\ell_2 - \ell_1)^2 \approx 0$.  We have emphasized in the arguments of the delta functions in \eqref{dipA3} that the two diagrams approach the on-shell limit $\ell_1^+ = 0$ from opposite directions, with positive $\ell_1^+ = \bot^2/k^- \rightarrow 0$ from the direct diagram and negative $\ell_1^+ = - \bot^2/k^- \rightarrow 0$ from the crossed diagram.  The two processes cover complementary halves of the phase space, with the sum of the two giving a symmetric approach to the on-shell point.  

For it to be kinematically possible to put the upper quark line on-shell, there are two additional constraints that must be satisfied: the light-cone momenta $(k+\ell_1)^-$ or $(k + \ell_2 - \ell_1)^-$ must be positive definite to correspond to physical on-shell particles, and the $t$-channel gluons must be spacelike ($\ell_1^+ \ell_1^- < 0)$.  
\footnote{Verification of these statements requires a careful treatment of the complete pole structure of the diagrams of Fig.~\ref{fig-NN2} in covariant gauge.  Such a detailed derivation would be a significant distraction from the treatment of dipole DIS on a nucleus considered here and thus, although an instructive example, it is left as an ``exercise for the reader.''}
These constraints imposed by picking up the pole $\delta(\ell_1^+)$ set the corresponding limits on the minus momentum: $-k^- < \ell_1^- < 0$ for the direct channel, and $0 < \ell_1^- < k^-$ for the crossed channel.  This explicitly sets finite, symmetric bounds $| \ell_1^- | < k^-$ on the range of integration of the loop momentum $\ell_1^-$ which, in the eikonal limit, is only contained in the intermediate propagator of the lower $(p-\ell_1)$ quark line.  This allows us to explicitly symmetrize the $\ell_1^-$ loop integral by adding $(\ell_1^- \rightarrow - \ell_1^-)$ and dividing by two; when applied to the lower eikonal $(p - \ell_1)$ propagator, this generates a second delta function:
\begin{align}
 \label{dipA4}
 \frac{i}{(p-\ell_1)^2 - m^2 + i\epsilon} &\approx \frac{i}{- p^+ \ell_1^- - \bot^2 + i\epsilon}
 \\ \nonumber &\rightarrow
 \frac{1}{2} \left[ \frac{i}{- p^+ \ell_1^- - \bot^2 + i\epsilon} + (\ell_1^- \rightarrow - \ell_1^-) \right]
 \\ \nonumber &=
 \frac{i}{2p^+} \left[ \frac{1}{\ell_1^- - \bot^2/p^+ + i\epsilon} - \frac{1}{\ell_1^- + \bot^2/p^+ - i\epsilon} \right]
 \\ \nonumber &=
 \frac{i}{2p^+} \bigg[ -2\pi i \delta(\ell_1^-) \bigg].
\end{align}

Eqs.~\eqref{dipA4} and \eqref{dipA3} are represented graphically by Fig.~\ref{fig-NN2}.  The sum of direct and crossed diagrams generates an effective cut of the intermediate state which puts the intermediate propagators on-shell, eliminating the apparent loop integrals and reducing the two-gluon scattering amplitude to the square of the one-gluon scattering amplitude \eqref{eik8}.  The associated factors of $1/p^+ k^- = 1/s$ normalize the second one-gluon scattering amplitude by scaling out the energy
as in \eqref{eik8}, and the factor of $\tfrac{1}{2}$ due to symmetrization is consistent with the Abramovsky-Gribov-Kancheli (AGK) cutting rules \cite{Abramovsky:1973fm}.  This considerable simplification, together with other effective cuts, will allow us to completely express the dipole-nucleus scattering in terms of the one-gluon scattering amplitude \eqref{eik8}.

\begin{figure}
 \centering
 \includegraphics[width=0.7\textwidth]{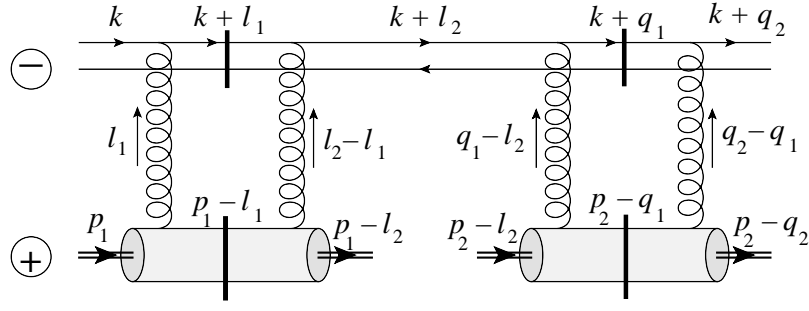}
 \caption{Sequential interaction by 2-gluon exchange with two different nucleons.  For each nucleon, the ``crossed'' diagrams of Fig.~\ref{fig-NN2} have been added, generating the effective cuts denoted by the thick vertical lines.  The only nontrivial momentum transfer is the momentum $\ell_2$ exchanged between one nucleon and another.}
 \label{fig-NN3}
\end{figure}

Now let us apply this result to analyze the sequential interaction of the dipole with two nucleons as shown in Fig.~\ref{fig-NN3}, each by the usual 2-gluon exchange.  By adding the crossed diagrams for each nucleon, we effectively put the intermediate states on-shell through Eqs.~\eqref{dipA3} and \eqref{dipA4}.  Together with the on-shell conditions for the external lines, this eliminates many of the longitudinal momentum components to eikonal order:
\begin{align}
 \label{dipA5}
 k^+ , \ell_1^+ , q_1^+ , q_2^+ &= 0 \\ \nonumber
 p_1^- , \ell_1^- , \ell_2^- , p_2^- , q_1^- , q_2^- &= 0 .
\end{align}
Aside from the eikonal momenta $k^- , p_1^+ , p_2^+$, the only nontrivial light-cone momentum is $\ell_2^+$, the longitudinal momentum exchanged between the two nucleons.  This momentum is Fourier-conjugate to the longitudinal separation $\Delta x^-$ between the nucleons, which is a small but finite number for the dilute Glauber nucleus considered here.  Thus it is natural to Fourier-transform over $\ell_2^+$ into longitudinal coordinate space $\Delta x^-$; in the eikonal limit the only factor that depends on $\ell_2^+$ is the quark propagator $(k+\ell_2)$ between the nucleons.  But when we perform the Fourier transform by contour integration,
\begin{align}
 \label{dipA6}
 \int \frac{d\ell_2^+}{2\pi} \, e^{-i \tfrac{1}{2} \ell_2^+ \Delta x^-} \, \frac{i}{(k+\ell_2)^2 - m^2 + i \epsilon} &\approx
 \int \frac{d\ell_2^+}{2\pi} \, e^{-i \tfrac{1}{2} \ell_2^+ \Delta x^-} \, \frac{i}{k^- \ell_2^+ + i \epsilon} 
 \\ \nonumber &=
 \frac{1}{k^-} \theta(\Delta x^-)
\end{align}
we need to close the contour below to enclose the pole and obtain a nonzero contribution.  This is only possible if $\Delta x^- > 0$ in the Fourier factor, so performing this transformation to coordinate space enforces a path-ordering of the nucleons along the $x^-$-axis, in addition to setting $\ell_2^+ = 0$ and putting the propagator between the two nucleons on-shell.  Thus, by going to $x^-$ coordinate space, we cut the last of the virtual propagators in Fig.~\ref{fig-NN3}, reducing the two-nucleon forward scattering amplitude down to four iterations of the on-shell-to-on-shell coordinate-space amplitudes \eqref{eik9} calculated previously.

\begin{figure}
 \centering
 \includegraphics[width=0.7\textwidth]{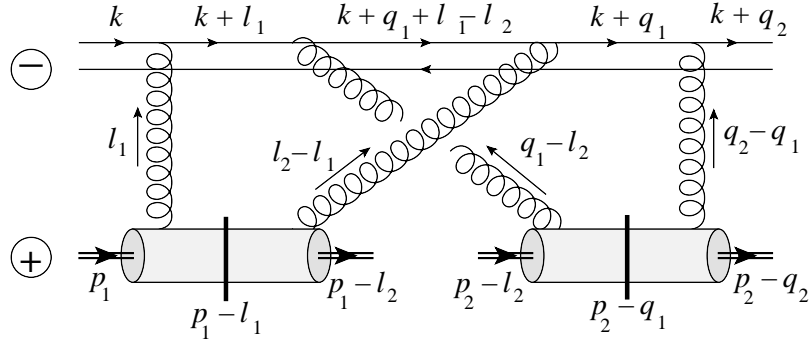}
 \caption{Scattering on two nucleons in which the multiple gluon exchanges have their $x^-$ ordering broken.  Again, the thick vertical lines denote effective cuts generated through the inclusion of crossed diagrams as in Fig.~\ref{fig-NN2}.}
 \label{fig-NN4}
\end{figure}

Finally, let us calculate the diagram of Fig.~\ref{fig-NN4} in which the $x^-$-ordering suggested by \eqref{dipA6} is broken.  We include the diagrams in which the attachments to the nucleons are crossed, generating the cuts indicated in the figure, but we leave the attachments to the dipole as is (to avoid a proliferation of gluon orderings).  The on-shell conditions for the external lines and the effective cuts of the nucleons eliminate many of the light-cone momenta,
\begin{align}
 \label{dipA7}
 k^+ , q_2^+ &= 0 \\ \nonumber
 p_1^- , \ell_1^- , \ell_2^- , p_2^- , q_1^- , q_2^- &=0 ,
\end{align}
leaving only the eikonal momenta $k^- , p_1^+ , p_2^+$, the longitudinal momentum transfer $\ell_2^+$ between nucleons, and the loop momenta $\ell_1^+ , q_1^+$.  The crossed $(\ell_2 - \ell_1)$ , $(q_1 - \ell_2)$ gluons in Fig.~\ref{fig-NN4} entangle the loop integration with the Fourier transform in $\ell_2^+$ we would like to perform.  But when we close the $\ell_2^+$ contour above and generate a fixed ordering of the nucleons as in \eqref{dipA6},
\begin{align}
 \label{dipA8}
 \int \frac{d\ell_1^+}{2\pi} &\, \frac{d q_1^+}{2\pi} \, \frac{d\ell_2^+}{2\pi} \, 
 e^{-i \tfrac{1}{2} \ell_2^+ \Delta x^-} \, 
  \frac{i}{k^- \ell_1^+ + i\epsilon} \:
  \frac{i}{k^-(q_1^+ + \ell_1^+ - \ell_2^+) + i\epsilon} \: 
	\frac{i}{k^- q_1^+ + i\epsilon} =
 \\ \nonumber &=
 \frac{1}{k^-} \int \frac{d\ell_1^+}{2\pi} \, \frac{d q_1^+}{2\pi} \, 
 e^{-i \tfrac{1}{2} (q_1^+ + \ell_1^+) \Delta x^-} \, 
  \frac{i}{k^- \ell_1^+ + i\epsilon} \, 
	\theta(-\Delta x^-) \, 
	\frac{i}{k^- q_1^+ + i\epsilon}
 \\ \nonumber &=
 \frac{\theta(-\Delta x^-)}{k^-} 
  \left[ \int \frac{d\ell_1^+}{2\pi} \, e^{+i \tfrac{1}{2} \ell_1^+ |\Delta x^-|} \, 
	\frac{i}{k^- \ell_1^+ + i\epsilon} \right]
	\left[ \int \frac{d q_1^+}{2\pi} \, e^{+i \tfrac{1}{2} q_1^+ |\Delta x^-|} \,
	\frac{i}{k^- q_1^+ + i\epsilon} \right]
 \\ \nonumber &= 0 ,
\end{align}
it becomes impossible to enclose the poles of the other propagators.  The inconsistency of the $x^-$-ordering forces this diagram to be zero.  

\begin{figure}
 \centering
 \includegraphics[width=0.6\textwidth]{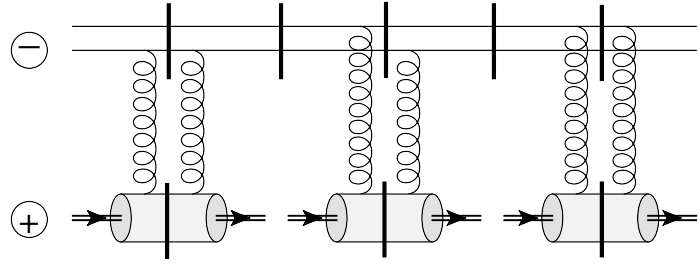}
 \caption{Diagrams contributing to the dipole-nucleus forward scattering amplitude at lowest order.  Crossed diagrams have been included, generating effective cuts in the middle of the 2-gluon exchanges, and the longitudinal momenta exchanged between nucleons has been Fourier-transformed, generating the effective cuts between the interaction with each nucleon.  The result is a sequence, ordered along the $x^-$-axis, that iterates the on-shell-to-on-shell 2-gluon exchange of Fig.~\ref{fig-dipole_nucleon}.}
 \label{fig-NN5}
\end{figure}

The two ingredients we have demonstrated above - the inclusion of crossed diagrams and the Fourier transform to $x^-$ coordinate space - can be used to decouple all such diagrams into the general structure illustrated in Fig.~\ref{fig-NN5} \cite{Mueller:1989st, Kovchegov:2012mbw}.  Going to $x^-$-space, together with the dilute nucleus approximation, yields a sequence of on-shell-to-on-shell interactions with individual nucleons, ordered in $x^-$.  And the summation of direct and crossed gluon exchange with a given nucleon as in Fig~\ref{fig-NN2} generates cuts between each gluon exchange, such that the dipole-nucleon interaction is given by the individual on-shell-to-on-shell cross-section \eqref{fig-dipole_nucleon}.  This has effectively reduced all interactions down to squares of the eikonal one-gluon exchange given by \eqref{eik8}.

\subsection{Nuclear Shadowing and Saturation}
\label{subsec-saturation}

When the number $\sim A^{1/3}$ of nucleons at a given impact parameter is large enough that $\alpha_s^2 A^{1/3} \sim \ord{1}$, we must re-sum the multiple interactions with all of these nucleons.  Since the individual dipole-nucleon cross-section \eqref{dipN2} is diagonal in transverse coordinate space, and since the interactions with each nucleon are ordered in $x^-$, the most direct way to re-sum the interaction with multiple nucleons is to formulate a differential equation in $(x^- , \ul{x})$ space.  Each successive time the dipole rescatters on another nucleon, the probability decreases that the dipole survives to contribute to the forward-scattering amplitude with the same kinematics; this effect is known as \textit{nuclear shadowing}.  This intuitive interpretation is most easily expressed in terms of the (forward matrix element of the) $S$-matrix, rather than $N$.  The two are related through the $T$-matrix,
\begin{align}
 \label{STN}
 T &\equiv \mathrm{Re} T + i \, N \\ \nonumber
 S &\equiv 1 + i \, T = 1 - N + i \, \mathrm{Re} T \approx 1 - N ,
\end{align}
where for high-energy QCD, the $T$-matrix is purely imaginary \cite{Low:1975sv, Nussinov:1975qb}.  Thus, defining a partial $S$-matrix $s(r_T , \ul{b}, b^-)$ as the probability of the dipole to survive up to a depth $b^-$, this probability decreases for every scattering on an additional nucleon.  The probability per unit $b^-$ for such a scattering to occur is given by the three-dimensional density of nucleons $\rho_A (\ul{b} , b^-)$ times the scattering cross-section $\sigma_{tot}^{dip \, N}$ of the dipole on that nucleon.  Thus the differential equation that describes the attenuation of the dipole $S$-matrix is
\begin{align}
 \label{dipA9}
 \frac{\partial}{\partial b^-} s(r_T , \ul{b}, b^-) = - \frac{1}{2} \rho_A (\ul{b} , b^-) \, 
 \sigma_{tot}^{dip \, N} (r_T , \ul{b}) \, s(r_T , \ul{b} , b^-),
\end{align}
where the factor of $1/2$ converts $\sigma_{tot}^{dip \, N}$ into its contribution to the forward-scattering amplitude, as in \eqref{Regge7}.  The initial condition to \eqref{dipA9} is $s=1$ at $b^- \rightarrow - \infty$, reflecting a $100\%$ survival probability before any scattering occurs. 

The solution to \eqref{dipA9} is just a simple exponential, 
\begin{align}
 \label{dipA10}
 S(r_T, \ul{b}) &\equiv \lim_{b^- \rightarrow \infty} s(r_T, \ul{b}, b^-) 
 \\ \nonumber &=
 \exp \left[- \frac{1}{2} \sigma_{tot}^{dip \, N} (r_T , \ul{b}) \int_{-\infty}^{\infty} d x^- 
 \rho_A (\ul{b} , x^-) \right]
 \\ \nonumber &=
 \exp \left[- \frac{1}{2} \sigma_{tot}^{dip \, N} (r_T , \ul{b}) \, T(\ul{b}) \right] ,
\end{align}
where $T(\ul{b})$, defined in \eqref{dipA1}, is the total density of nucleons per unit transverse area.  Re-expressing \eqref{dipA10} in terms of $N$ through \eqref{STN} and substituting the dipole-nucleon cross-section \eqref{dipN2} gives
\begin{align}
 \label{GGMsoln1}
 N(r_T, \ul{b}) = 1 - \exp \left[ - \frac{\alpha_s \pi^2}{2 N_c} \, T(\ul{b}) \, r_T^2 \, x G_N \left(\frac{1}{r_T^2}\right) \right] .
\end{align}
%
\begin{figure}
 \centering
 \includegraphics[width=0.5\textwidth]{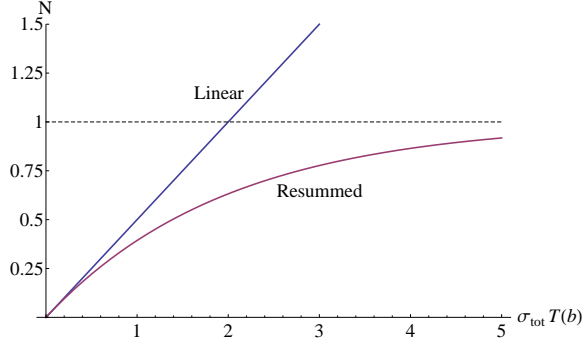}
 \caption{Linear and resummed forward-scattering amplitudes $N$.  The linear amplitude for scattering on a single nucleon can grow without bound, while the resummed amplitude is always bounded by unity.}
 \label{fig-GGMplot2}
\end{figure}
%
To make this expression more concrete, we can model the nucleons as dipoles, using $\sigma_{tot}^{dip \, dip}$ from \eqref{dipdipfull} in \eqref{dipA10}, which, in the small-dipole / large-dipole asymptotic limit \eqref{dipN4} gives
\begin{align}
 \label{GGMsoln2}
 N(r_T, \ul{b}) = 1 - \exp \left[ - \frac{2\pi C_F \alpha_s^2}{N_c} \, T(\ul{b}) \, r_T^2 \, \ln\frac{1}{r_T \Lambda} \right] .
\end{align}

Comparing the Glauber-Gribov-Mueller (GGM) multiple scattering amplitude \eqref{dipA10} with the single-nucleon scattering amplitude $N(r_T,\ul{b}) = \tfrac{1}{2} \sigma_{tot}^{dip \, N} (r_T) \, T(\ul{b})$, we see (as in Fig.~\ref{fig-GGMplot2}) that in the limit of low densities $\sigma_{tot}^{dip \, N} (r_T) \, T(\ul{b}) \ll 1$, the resummed amplitude reduces to the linear one.  However, at high densities, the linear amplitude can grow without bound, while the exponential form of \eqref{dipA10} gives an amplitude which \textit{saturates} to unity (see Fig.~\ref{fig-GGMplot2}).  The growth of the forward amplitude without bound would constitute a violation of unitarity \cite{Kovchegov:2012mbw}, reflecting a scattering probability greater than $100\%$.  Multiple scattering effects are thus important for the unitarization of the dipole-nucleus scattering cross-section.  

\begin{figure}
 \centering
 \includegraphics[width=0.5\textwidth]{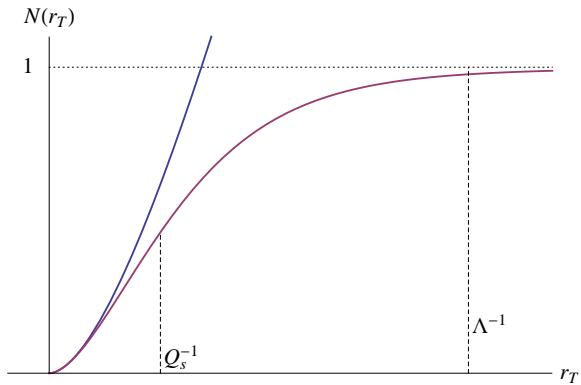}
 \caption{Linear and resummed forward-scattering amplitudes $N$ using the dipole-dipole cross-section \eqref{dipdipfull}.  As the size $r_T$ of the DIS dipole increases, it couples more strongly to the fields of the ``nucleon''.  In the linear amplitude, this causes $N$ to grow without bound, but in the resummed amplitude, the growth of $N$ is softened by the exponential attenuation due to multiple rescattering.  The momentum scale $Q_s$ which couples to $r_T$ in the exponential turns over the growth of $N$ on shorter scales than the nucleon size $\Lambda^{-1}$, suggesting that the color fields are now only coherent over smaller distances $Q_s^{-1} \ll \Lambda^{-1}$.}
 \label{fig-GGMplot}
\end{figure}

To further unfold the implications of the GGM formula, let us examine the dependence on the dipole size.  As we saw in \eqref{dipN4} and \eqref{dipdipfull}, the perturbatively small DIS dipole measures the strength of the gluon field of the nucleon, with a coupling proportional to the dipole moment squared $d^2 \sim \alpha_s r_T^2$ of the small dipole.  Thus, as $r_T$ increases, the strength of the scattering amplitude $N$ increases, reflecting the increased coupling of the dipole to the external field.  This is true as long as the dipole is small, interacting with a larger, nearly uniform color field.  For the general case of two dipoles interacting, \eqref{dipdipfull}, the growth of $N$ with $r_T$ is softened (although not cut off) when the small dipole becomes comparable in size to the large one (the ``nucleon'') at $r_T \Lambda \sim 1$.  However, in the resummed amplitudes \eqref{GGMsoln1} and \eqref{GGMsoln2}, the momentum scale that multiplies $r_T^2$ is different: it has been enhanced by the density factor $T(\ul b)$.  This momentum scale is known as the \textit{saturation scale} $Q_s$, and it can be either defined implicitly from \eqref{GGMsoln1} as the point at which the magnitude of the exponent becomes unity,
\begin{align}
 \label{Qsat1}
 Q_s^2 (\ul{b}) \equiv \frac{\alpha_s \pi^2}{2 N_c} \, T(\ul b) \, x G \left( Q_s^2(b) \right),
\end{align}
or explicitly from the dipole form \eqref{GGMsoln2} as
\footnote{For $Q_s^2$ from a single quark rather than a dipole, the value of $Q_s^2$ is half this.}
\begin{align}
 \label{Qsat2}
 Q_s^2 (\ul b) \equiv \frac{8\pi \alpha_s^2 C_F}{N_c} \, T(\ul{b}),
\end{align}
so that \eqref{GGMsoln2} becomes
\begin{align}
 \label{Qsat3}
 N(r_T, \ul{b}) = 1 - \exp \left[ - \frac{1}{4} \, r_T^2 \, Q_s^2(\ul{b}) \, \ln\frac{1}{r_T \Lambda} \right].
\end{align}

\begin{figure}
 \centering
 \includegraphics[width=0.5\textwidth]{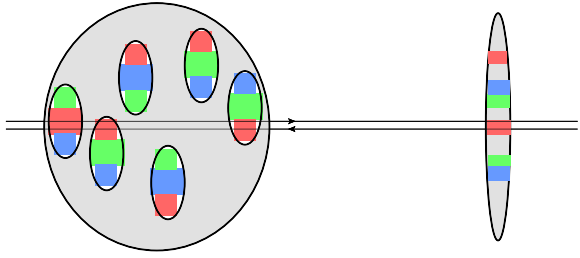}
 \caption{Illustration of the origin of a short length scale $1/Q_s \ll 1/\Lambda$ through multiple scattering.  A perturbatively small dipole scatters many times on nucleons whose color fields have typical size $1/\Lambda$, are independent of each other, and are randomly distributed in the $SU(N_c)$ color space.  Because of cancellation between these random colors, the ``color domains'' in the transverse plane (after integration over the longitudinal direction) are only correlated over shorter distances $1/Q_s$.  The higher the number of nucleons $A$, the greater this random cancellation, and the smaller the color domains become.}
 \label{fig-ColorDomains}
\end{figure}

As shown in Fig.~\ref{fig-GGMplot}, the growth of $N$ with $r_T$ turns over at $Q_s^{-1}$ rather than at $\Lambda^{-1}$ when the effects of multiple scattering have become important.  This suggests that, in the high transverse densities of a Lorentz-contracted nucleus, the transverse sizes over which the color fields are uniform are much smaller, of order $Q_s^{-1}$.  The color fields of many independent nucleons in the large-$A$ limit are randomly distributed in color space, so that the superposition of these fields seen by the DIS dipole tends to dilute the color fields of the nucleons.  Thus the effect of high transverse densities is to shorten the size of the correlated ``color domains'' in the transverse plane from $\Lambda^{-1}$ to $Q_s^{-1}$, as visualized in Fig.~\ref{fig-ColorDomains}.  This is another manifestation of nuclear shadowing through multiple rescattering effects.  Since $T(\ul{b}) \sim \Lambda^2 \, A^{1/3}$, we see that $Q_s^2$ scales parametrically as $\alpha_s^2 \, A^{1/3}$; it is a dynamical scale generated by the resummation of coherent multiple rescattering.

Finally, let us note that, despite the apparent quantum loops generated by multiple gluon exchange with the nucleus, in the resummed diagrams of Fig~\ref{fig-NN5} these loops have all been cut.  Thus our summation of the leading-$A$ contributions contains no genuine quantum corrections; it therefore appears that the Glauber-Gribov-Mueller formula \eqref{GGMsoln1} and the emergence of the saturation scale \eqref{Qsat2} are \textit{classical} effects.  If the leading-$A$ effects considered here are truly classical in origin, we should be able to recover them by analyzing the classical equations of motion of QCD; this is the approach we pursue in the next Section.


\section{The McLerran-Venugopalan Model: Classical Gluon Fields of a Heavy Nucleus}
\label{sec-classical_fields}

\subsection{Yang-Mills Equations for High-Energy Point Charges}
\label{subsec-MV}

Motivated by the Glauber-Gribov-Mueller formula \eqref{dipA10} for deep inelastic scattering on large nuclei, we wish to consider the small-$x$ gluon wave function of an ultra-relativistic heavy nucleus in the classical limit.  This approach was pioneered in \cite{McLerran:1994vd , McLerran:1993ni , McLerran:1993ka} and is known as the \textit{McLerran-Venugopalan (MV)} model of a heavy nucleus. Such a wave function $A^{a \, \mu}$ would be a solution of the classical Yang-Mills equation 
\begin{align}
 \label{eq-EOM}
 &\left(\mathcal{D}_\mu F^{\mu \nu}\right)^a \equiv \partial_\mu F^{\mu \nu a} + g f^{a b c} A_\mu^b 
 F^{\mu \nu c} = - J^{\nu a}
 \\ \nonumber \, \\ \nonumber
 & \partial_\mu \left[\partial^\mu A^{\nu a} - \partial^\nu A^{\mu a} + g f^{a b c} A^{\mu b} A^{\nu c}
 \right] + g f^{a b c} A_\mu^b \left[ \partial^\mu A^{\nu c} - \partial^\nu A^{\mu c} + g f^{c d e}
 A^{\mu d} A^{\nu e} \right] = - J^{\nu a}
\end{align}
for a nucleus under Regge kinematics.  When the nucleus is boosted to high energy in the center-of-mass frame considered in Sec.~\ref{subsec-GGM}, the color current $J^{\nu a}$ is purely in the light-cone plus direction.  For simplicity, let us consider the nucleus to be a superposition of point charges with positions $(\ul{x}_n , x_n^-)$ moving along the light-cone plus axis.  Then the current is given by
\begin{align}
 \label{eq-MV current}
 J^{\nu a} (\ul{x}, x^+ , x^-) &= \delta^{\nu +} \: \rho^a (\underline x , x^-) \\ \nonumber
 &= \delta^{\nu +} \, 2 g \sum_n T^a_n \, \delta(x^- - x_n^-) \, \delta^2(\ul{x} - \ul{x}_n)
\end{align}
where the factor of 2 arises from the choice of metric.  The color quantum number $a$ of the current is carried by the fundamental $SU(N_c)$ generators $T_n^a$, reflecting the color field radiated from the $n^{th}$ charge; thus the matrix $(T_n^a)_{j i}$ acts on the internal color-space of the $n^{th}$ charge, rotating its fundamental color quantum number from $i$ to $j$.  This is to be distinguished from ``external'' color matrices $T^a$ which are often contracted with quantities such as $A^{\nu a}$ to form color matrices $\hat{A}^{\nu} \equiv A^{\nu a} T^a$ with definite gauge-transformation properties.

Our goal is to solve \eqref{eq-EOM} for the source current given by \eqref{eq-MV current} and analyze the results to see if the high transverse densities due to the boost generate saturation effects analogous to the Glauber-Gribov-Mueller calculation \eqref{GGMsoln1}.  To solve \eqref{eq-EOM}, let us first consider the limit of parametrically weak coupling $g \ll 1$, so that nonlinear effects involving the self-interaction of the fields $A^\mu A^\nu$ are suppressed.  The weak-field limit is also the Abelian limit of the theory, since the nonlinear terms $A^\mu A^\nu$ are dropped, simplifying the  field-strength tensor to $F^{\mu \nu a} \approx \partial^\mu A^{\nu a} - \partial^\nu A^{\mu a}$.  In this limit, the Yang-Mills equations reduce down to just the linear Maxwell's equations
\begin{align}
 \label{YM1}
 \partial_\mu F^{\mu \nu a} = - \delta^{\nu +} \, 2 g \sum_n T^a_n \, \delta(x^- - x_n^-) \, \delta^2(\ul{x} - \ul{x}_n) .
\end{align}
Fixing the gauge to be the covariant gauge $\partial_\mu A^{\mu a} = 0$
\footnote{This is also known as the Lorenz gauge, and it is equivalent to the Feynman gauge.  When we construct the solution, we will see that it satisfies $A^- = 0$ and thus is also equivalent to the ``light-cone gauge of the projectile.''}
eliminates the term $\partial_\mu \partial^\nu A^{\mu a}$ coming from the left-hand side, leaving
\begin{align}
 \label{YM2}
 &\partial_\mu \partial^\mu A^{+ a}(\ul{x}, x^+ , x^-) = - 2 g \sum_n T^a_n \, \delta(x^- - x_n^-) \, \delta^2(\ul{x} - \ul{x}_n) \\ \nonumber
 &\partial_\mu \partial^\mu A^{\nu a} = 0 \hspace{1.5cm} \nu = -, \bot.
\end{align}
We note that the transverse and minus components can be trivially solved by $A^- = A_\bot^i = 0$, and we solve the plus equation as follows.  Since the source is $x^+$-independent, the solution $A^{\mu a}$ should be as well; this realization means that the action of $\partial^- = 2 \partial_+$ coming from the D'Alembertian $\partial_\mu \partial^\mu = \partial^+ \partial^- - \nabla_T^2$ vanishes.  This reduces the left-hand side to $- \nabla_T^2 A^{+ a} (\ul{x}, x^-)$, with the only $x^-$-dependence in the source coming from a set of delta functions.  Since the dynamics are now entirely transverse, the solution $A^{\mu a}$ must also reflect this localization in $x^-$.  Thus we define
\begin{align}
 \label{YM3}
 A^{+ a} (\ul{x} , x^-) \equiv \sum_n \delta(x^- - x_n^-) \alpha^a_n (\ul{x}), 
\end{align}
which reduces \eqref{YM2} down to the 2-dimensional Poisson equation
\begin{align}
 \label{YM4}
 \nabla_T^2 \alpha_n^a (\ul{x}) = + 2 g T^a_n \delta^2 (\ul{x} - \ul{x}_n).
\end{align}
This 2D Poisson equation is exactly the same as that for the electrostatic potential from an infinite line of electric charge, with the solution given in any elementary textbook: a logarithmic potential.  Writing down the general form of the solution for $\alpha_n^a (\ul{x})$ and applying the Laplacian gives
\begin{align}
 \label{YM5}
 &\alpha_n^a (\ul{x}) = c_{n}^a \ln (|\ul{x} - \ul{x}_n|_T \Lambda) \\ \nonumber
 &\nabla_T^2 \alpha_n^a = c_{n}^a \partial_\bot^i \left( \frac{(x-x_n)_\bot^i}{|\ul{x}-\ul{x}_n|_T^2} \right) = c_{n}^a \left( 2\pi \delta^2 (\ul{x} - \ul{x}_n) \right).
\end{align}
Finally, comparing \eqref{YM5} with \eqref{YM4} gives the general solution to \eqref{YM1} in the weak-coupling limit as
\begin{align}
 \label{Acov}
 & A_{cov}^{+ a} (\ul{x} , x^+ , x^-) = \frac{g}{\pi} \sum_n T_n^a \delta(x^- - x_n^-) \ln ( |\ul{x} - \ul{x}_n|_T \Lambda) \\ \nonumber
 & A_{cov}^{- a} = A_{cov}^{i a} = 0,
\end{align}
where we have included a subscript to emphasize that this is the solution in covariant gauge.  The picture of the gluon field in this gauge is of a ``shockwave'' localized in $x^-$ that radiates out in the transverse direction with a logarithmic potential.  The fact that we recover an elementary result from classical electrodynamics is not surprising, since the assumptions that led to \eqref{Acov} reduced the Yang-Mills equations \eqref{eq-EOM} down to Maxwell's equations \eqref{YM1}.  However, when we relax the weak-field assumption and examine the full nonlinear equations, we find that \eqref{Acov} is \textit{still the solution} of the full Yang-Mills equations, because all of the nonlinear terms in \eqref{eq-EOM} in this gauge are proportional to either 
\begin{align}
 \label{YM6}
 A_\mu A^\mu = 0 \hspace{1.5cm} \mathrm{or} \hspace{1.5cm} 
 A^\mu \partial_\mu A^+ \sim A^+ \frac{\partial}{\partial x^+} A^+ = 0.
\end{align}
Therefore \eqref{Acov} is the full solution to \eqref{eq-EOM} in covariant gauge, retaining the simple physical interpretation carried over from classical electrodynamics.

However, it is difficult to directly interpret the gauge field \eqref{Acov} in terms of gluons because, as we saw in \eqref{gluonTMD3}, a gauge-invariant definition of gluon number invokes the field $A^{\mu a}_{LC}$ in the light-cone gauge $A_{LC}^+=0$.  Therefore, to interpret \eqref{Acov} we would like to perform a gauge transformation to find the corresponding field in the light-cone gauge.  The gauge transformation takes the form of an $SU(N_c)$ color rotation $S \sim \exp[i \theta^a T^a]$ that rotates the color quantum numbers of the fermions as in \eqref{DISS-GAUGE1}.  If this is the transformation that takes the covariant gauge into the light-cone gauge, then it applies to the gluon field $\hat{A}^\mu \equiv A^{\mu a} \: T^a$ and the field strength tensor $\hat{F}^{\mu \nu} \equiv F^{\mu \nu a} \: T^a$ as
\begin{align}
 \label{SUN1}
 \hat{A}_{LC}^\mu &= S \: \hat{A}_{cov}^\mu \: S^{-1} - \frac{i}{g} \left( \partial^\mu S \right) S^{-1} 
 \\ \nonumber
 \hat{F}_{LC}^{\mu \nu} &= S \: \hat{F}_{cov}^{\mu \nu} \: S^{-1} ,
\end{align}
where it is to be emphasized that the color matrices $T^a$ used in formulating these transformations are ``external'' and operate in a different color space than the ``internal'' color matrices $T_n^a$ that act on the fermions.

We can determine the form of the light-cone gauge fields $A_{LC}^{\mu a}$ most straightforwardly by going through the field-strength tensor as an intermediary.  In the covariant gauge \eqref{Acov}, there is only one nonzero component of the field-strength tensor: $F_{cov}^{+ i a}$.  Since the gauge transformation \eqref{SUN1} only multiplies this by a color factor, this remains the only nonzero component in the light-cone gauge.  In both gauges, these tensors are uniquely determined by one component of the gauge field, so we can use \eqref{SUN1} to immediately write
\begin{align}
 \label{SUN2}
  F_{LC}^{+ i a} = \partial^+ A_{LC}^{i a} = 2 \frac{\partial}{\partial x^-} A_{LC}^{i a}
=  \left[ S \: \hat{F}_{cov}^{+ i} \: S^{-1} \right]^a =  \left[ S \left( \frac{\partial}{\partial x^i} \hat{A}_{cov}^+ \right) S^{-1} \right]^a .
\end{align}
By integrating this expression over $x^-$, we can solve for the light-cone field $A_{LC}^{i a}$, which is necessarily the only nonzero component in the light-cone gauge,
\begin{align}
 \label{SUN3}
 &A_{LC}^{i a} (\ul{x} , x^-) = \frac{1}{2} \int\limits_{-\infty}^{x^-} db^- \, \left[ S(\ul{x}, b^-) \, 
 \left( \frac{\partial}{\partial x^i} \hat{A}_{cov}^+ (\ul{x},b^-) \right) S^{-1} (\ul{x},b^-) \right]^a
 \\ \nonumber
 &A_{LC}^{+ a} = A_{LC}^{- a} = 0 
\end{align}
where we have chosen the boundary condition at infinity $A_{LC}^{i a} (x^- \rightarrow -\infty) = 0$ to fix the residual gauge freedom.  Using the covariant-gauge field \eqref{Acov}, we find
\begin{align}
 \nonumber
 A_{LC}^{i a} (\ul{x} , x^-) &= \frac{g}{2\pi} \int\limits_{-\infty}^{x^-} db^- \, 
 \left[ S(\ul{x}, b^-) \, \left( \sum_n T_n^b \, \delta(b^- - x_n^-) \, 
 \frac{(x-x_n)_\bot^i}{|\ul{x}-\ul{x}_n|_T^2} \right) T^b \, S^{-1}(\ul{x}, b^-) \right]^a
 \\ \label{ALC} &=
 \frac{g}{2\pi} \sum_n \theta(x^- - x_n^-) \, \frac{(x-x_n)_\bot^i}{|\ul{x}-\ul{x}_n|_T^2} \, 
 T_n^b \otimes \left[ S(\ul{x}, x_n^-) \, T^b \, S^{-1}(\ul{x}, x_n^-) \right]^a
\end{align}
where we have used the fact that the gauge transformation \eqref{SUN1} acts on the external color matrix, not on the quark color matrices.  Apart from the net color rotation provided by $S$, the main difference between the light-cone field $A_{LC}^i$ in \eqref{ALC} and the covariant field $A_{cov}^+$ is that the light-cone fields have the structure of a theta function instead of a delta function.  The color fields emitted by the point charges flow ``downstream,'' affecting everything at $x^-$ values greater than the value at the source.  The light-cone fields also decay as $1 / |\ul{x} - \ul{x}_n|_T$, unlike the covariant-gauge fields that are logarithmic in the distance.

To fully specify the light-cone field \eqref{ALC}, we need to determine the precise form of the color rotation $S$.  This can be done by defining $\hat{A}_{LC}^+ = 0$ in \eqref{SUN1}, which we can rewrite as
\begin{align}
 \label{SUN4}
 \frac{\partial}{\partial x^-} S(\ul{x}, x^-) = \frac{- i g}{2} \, S(\ul{x}, x^-) \, \hat{A}_{cov}^+ (\ul{x}, x^-) .
\end{align}
If $S$ and $\hat{A}_{cov}^+$ were scalars instead of matrices, the solution to this equation would simply be an exponential.  Since they are matrices, however, the order of the factors matters and the solution is a bit more subtle.  To see the form of the solution, consider iterating \eqref{SUN4} twice:
\begin{align}
 \label{SUN5}
 S(\ul{x}, x^- + \delta x_1^- + \delta x_2^-) &= S(\ul{x} , x^- + \delta x_1^-) + \delta x_2^- 
 \frac{\partial}{\partial x^-} S(\ul{x} , x^- + \delta x_1^-)
 \\ \nonumber &=
 S(\ul{x}, x^- + \delta x_1^-) \left[ 1 - \frac{i g}{2} \delta x_2^- \hat{A}_{cov}^+ (\ul{x}, x^- +
 \delta x_1^-) \right]
 \\ \nonumber &=
 \left[ S(\ul{x}, x^-) + \delta x_1^- \frac{\partial}{\partial x^-} S(\ul{x}, x^-) \right]
 \left[ 1 - \frac{i g}{2} \delta x_2^- \hat{A}_{cov}^+ (\ul{x}, x^- + \delta x_1^-) \right]
 \\ \nonumber &=
 S(\ul{x}, x^-) \left[ 1 - \frac{i g}{2} \delta x_1^- \hat{A}_{cov}^+ (\ul{x}, x^-) \right]
 \left[ 1 - \frac{i g}{2} \delta x_2^- \hat{A}_{cov}^+ (\ul{x}, x^- + \delta x_1^-) \right] .
\end{align}
We see that every iteration introduces another copy of the factor in brackets, evaluated at different $x^-$ points.  The largest values in $x^-$ are farthest to the right, with successively smaller values of $x^-$ on the left.  This expression has the form of a path-ordered exponential, with the integral in the exponent starting from $x^-$ and traveling out toward $-\infty$ if we were to continue iterating \eqref{SUN5}.  The only other subtlety is that the infinitesimal step length $\delta x$ in the usual form of the path-ordered exponential is $\delta x = (x_f - x_i) / N$ with $N$ the discretization number.  In \eqref{SUN5} the step lengths $\delta x_1^- , \delta x_2^-$ are moving toward the initial point $\delta x^- \sim (x_i^- - x_f^-) / N$; thus we introduce an extra minus sign in the factors in brackets when we put this into the standard form of the path-ordered exponential.  The resulting solution is
\begin{align}
 \label{SUN6}
 S(\ul{x}, x^-) = \mathcal{P} \exp\left[ + \frac{i g}{2} \int\limits_{x^-}^{-\infty} d b^- \hat{A}_{cov}^+ (\ul{x}, b^-) \right]
\end{align}
for the color factor which transforms from covariant gauge to light-cone gauge.  Inserting the specific form of the covariant-gauge field \eqref{Acov} gives
\begin{align}
 \label{SUN7}
 S(\ul{x}, x^-) &= \mathcal{P} \exp\left[ \frac{i g^2}{2\pi} \int\limits_{x^-}^{-\infty} db^- \left( \sum_n
 T^a \otimes T_n^a \delta(b^- - x_n^-) \ln(|\ul{x} - \ul{x}_n|_T \Lambda) \right) \right]
 \\ \nonumber &=
 \mathcal{P} \exp\left[ \frac{-i g^2}{2\pi} T^a \otimes \sum_n T_n^a \theta(x^- - x_n^-) 
 \ln(|\ul{x} - \ul{x}_n|_T \Lambda) \right] ,
\end{align}
where an extra minus sign is generated by picking up the delta function in an integral running in the negative direction.  Since each of the color matrices $T_n^a$, as well as the external color matrix $T^a$ all operate in different color spaces, they all commute with each other.  Thus the summation in the exponent can be converted into an ordinary product of individual exponential factors, and the path-ordering operator arranges these factors with the terms corresponding to $x^-$ on the right:
\begin{align}
 \label{SUN8}
 S(\ul{x}, x^-) &= \mathcal{P} \prod_n \exp\left[ \frac{-i g^2}{2\pi} T^a \otimes T_n^a \theta(x^- - x_n^-) 
 \ln(|\ul{x} - \ul{x}_n|_T \Lambda) \right]
 \\ \nonumber &\equiv
 \prod\limits_{-\infty \cdots x^-} \!\!\!\!\!\!{}_n \: \exp\left[ \frac{-i g^2}{2\pi} T^a \otimes T_n^a 
 \theta(x^- - x_n^-) \ln(|\ul{x} - \ul{x}_n|_T \Lambda) \right]
\end{align}
where the notation of the product in the last line emphasizes the ordering of the factors in $n$ from left to right with increasing $x^-$.  The specific color rotation \eqref{SUN8}, together with the spacetime structure given in \eqref{ALC} fully specifies the gluon field $A^{\mu a}_{LC}$ in the light-cone gauge which is the solution of the classical Yang-Mills equations \eqref{eq-EOM} for the source \eqref{eq-MV current}.

\subsection{The Non-Abelian Weizs\"acker-Williams Field}
\label{subsec-WW}

The Weizs\"acker-Williams ``equivalent photon approximation'' \cite{Jackson:1975a, Peskin:1995ev} in QED describes the classical radiation of effectively on-shell photons from a high-energy source.  Its generalization to QCD is the \textit{non-Abelian Weizs\"acker-Williams field} $\phi^{WW}$ \cite{Kovchegov:1997pc} of gluon radiation, defined as
\begin{align}
 \label{WW1}
 \phi^{WW}(x,k_T) &\equiv \frac{\pi}{2(2\pi)^3} \sum_{\lambda}^{phys} \, \left\langle a_{k \lambda}^{\dagger a} \, a_{k \lambda}^a \right\rangle_A 
 \\ \nonumber &=
 \frac{\pi}{2(2\pi)^3} \sum_{\lambda}^{phys} \, \int d^{2-}b \, \rho_A(\ul{b},b^-) \,
 \bra{A(p,b)} a_{k \lambda}^{\dagger a} \, a_{k \lambda}^a \ket{A(p,b)} .
\end{align}
The Weizs\"acker-Williams field counts the number $\frac{dN}{dk_T^2 \, dy}$ of gluons produced per unit $k_T^2$, per unit rapidity in the classical state of a nucleus $A$.  The factor of $\pi$ arises from the integration over the azimuthal direction $d^2 k = k_T d k_T d\phi_k = \pi d k_T^2$; the lack of dependence of \eqref{WW1} on the direction $\phi_k$ implicitly assumes an unpolarized state.  The averaging in the second line is over the three-dimensional positions of charges in the nucleus.

The definition \eqref{WW1} in terms of a gluon number density with fixed transverse momentum recalls the definition of transverse-momentum-dependent parton distribution functions (TMD's) from Chapter~\ref{chap-TMD}.  Indeed, if we compare \eqref{gluonTMD7} with \eqref{WW1}, we see that same operator appearing in the Weizs\"acker-Williams field occurs as a specific projection of the gluon field correlator \eqref{gluonTMD3}:
\begin{align}
 \label{WW2}
 f_1^g (x, k_T) &\equiv \frac{1}{2} \sum_S \left( \tilde{\Phi}^{\mu \nu}(x,\ul{k}; S) \, 
 (- g_{T \, \mu\nu}) \right)
 \\ \nonumber &=
 \frac{1}{2(2\pi)^3} \frac{1}{2 p^+ \mathcal{V}^-} \frac{1}{x} \sum_\lambda^{phys} \bra{p} a_{k\lambda}^{\dagger a} \, a_{k \lambda}^a \ket{p},
\end{align}
where $f_1^g$, defined in \eqref{gluonTMD8}, is the unpolarized gluon TMD and the factor of $1/ 2 p^+ \mathcal{V}^-$ plays the role of the density in the plane wave state $\ket{p}$ as in \eqref{newTMD4}.  Note that \eqref{WW2} only holds at the lowest (classical) order in which the gauge link \eqref{DISS-GAUGE2} can be neglected.  Relating the Weizs\"acker-Williams field to the gluon TMD gives
\footnote{In making the transition from a quantum-mechanical plane-wave state to a classical state which is localized in position and momentum, care must be taken to account for the normalization of the states.  This can be done rigorously by using quantum-mechanical states constructed along the lines of the Wigner distributions of Chapter~\ref{chap-MVspin} and fixing the normalization so that the spatial integral recovers the momentum-space expectation value as in \eqref{WW2}.  One then sees that the volume factor $1/2 p^+ \mathcal{V}^-$ is replaced by an average over the spatial density as in \eqref{WW1}. }
\begin{align}
 \label{WW3}
 \phi^{WW}(x,k_T) &= \pi x \: f_1^g (x,k_T)
 \\ \nonumber &=
 \frac{(k^+)^2}{4(2\pi)^2} \int d^{2-} x \, d^{2-} y \, e^{i k \cdot (x-y)} \left\langle A_{LC}^{i a}(y) \, A_{LC}^{i a}(x) \right\rangle_A
 \\ \nonumber &=
 \frac{(k^+)^2}{8 \pi^2} \int d^{2-} x \, d^{2-} y \, e^{i k \cdot (x-y)} \left\langle \Tr \left[ 
 \hat{A}_{LC}^{i}(y) \, \hat{A}_{LC}^{i}(x) \right] \right\rangle_A
\end{align}
where we have used the relation \eqref{gluonTMD3} between the gluon creation / annihilation operators and the light-cone gauge fields.  For the nucleus composed of point charges in \eqref{eq-MV current}, the classical fields are given by \eqref{ALC}.

The fields $A_{LC}^{i a}(\ul{x}, x^-)$ are proportional to $\theta(x^- - x_n^-)$ for the contribution of the $n^{th}$ charge, and similarly for $A_{LC}^{i a}(\ul{y}, y^-) \propto \theta(y^- - x_n^-)$.  All of the charges are contained within the highly Lorentz-contracted nucleus and therefore have positions $x_n^-$ that are all very close together, of order $(M_A R_A) / p^+ \sim 1/p^+$.  For the small-$x$ part of the Weizs\"acker-Williams field which is relevant for the Regge limit, $k^+ \ll p^+$, so the typical separation between $x^-$ and $y^-$ in the Fourier transform is much larger than the separation between the individual charges $x_n^-$.  Therefore, for the purposes of the Fourier transform, we approximate the light-cone fields as
\begin{align}
 \label{WW4}
 A_{LC}^{i a}(\ul{x},x^-) \approx \theta(x^-) \, A_{LC}^{i a} (\ul{x}, x^- \rightarrow \infty) \equiv
 \theta(x^-) A_{LC}^{i a} (\ul{x})
\end{align}
which effectively takes all of the charges $x_n^-$ at the same point ($0^-$).  Then the $x^-$ and $y^-$ integrals just generate a combined factor of $4 / (k^+)^2$, so that the Weizs\"acker-Williams field of the nucleus is
\begin{align}
 \label{WW5}
 \phi^{WW}(x,k_T) &= \frac{1}{2\pi^2} \int d^2 x \, d^2 y \, e^{-i \ul{k} \cdot (\ul{x}-\ul{y})}
 \left\langle \Tr\left[\hat{A}_{LC}^i (\ul{y}) \, \hat{A}_{LC}^i (\ul{x}) \right] \right\rangle_A
 \\ \nonumber &=
 \frac{\alpha_s}{2 \pi^3} \int d^2 x \, d^2 y \, e^{-i \ul{k} \cdot (\ul{x}-\ul{y})} \sum_{n, n'}
 \: \bigg\langle \:\:
 \left( \frac{ (\ul{y}-\ul{x}_{n'}) \cdot (\ul{x}-\ul{x}_{n}) }
      {|\ul{y}-\ul{x}_{n'}|_T^2 \, |\ul{x}-\ul{x}_{n}|_T^2} \right)
 \\ \nonumber &\times
 T_n^a T_{n'}^b \, \Tr\bigg[ S(\ul{y},x_{n'}^-) T^b S^{-1}(\ul{y},x_{n'}^-)
 \, S(\ul{x},x_{n}^-) T^a S^{-1}(\ul{x},x_{n}^-) \bigg] \:\: \bigg\rangle_A .						
\end{align}

In this classical calculation, the expectation value corresponds to averaging over the positions and colors of the point charges in the nucleus.  The color averaging simplifies the expression because each point charge $n$ has its color averaged separately, corresponding to a factor of $1/N_c$ and a trace over the color matrices acting in the color space of $n$.  Since $\Tr[T_n^a]=0$, the color averaging gives zero unless $n' = n$; then $(1/N_c) \Tr[T_n^a T_n^b] = (1 / 2 N_c) \delta^{a b}$.  This simplification gives
\begin{align}
 \label{WW6}
 \phi^{WW}(x,k_T) &= \frac{\alpha_s}{4 \pi^3 N_c} \int d^2 x \, d^2 y \, e^{-i \ul{k} \cdot 
 (\ul{x}-\ul{y})} \sum_n \: \bigg\langle \:\:
 \left( \frac{ (\ul{y}-\ul{x}_{n}) \cdot (\ul{x}-\ul{x}_{n}) }
      {|\ul{y}-\ul{x}_{n}|_T^2 \, |\ul{x}-\ul{x}_{n}|_T^2} \right) \,
 \\ \nonumber &\times
 \Tr\bigg[ S(\ul{y},x_{n}^-) T^a S^{-1}(\ul{y},x_{n}^-) \,
           S(\ul{x},x_{n}^-) T^a S^{-1}(\ul{x},x_{n}^-) \bigg] \:\: \bigg\rangle_A ,
\end{align}
where now only the spatial averaging over $\ul{x}_n$ and the sum over $n$ remain.  Note from \eqref{SUN8} that the color rotation $S(\ul{x},x_n^-)$ depends on the positions of all the point charges with longitudinal coordinates $x^-$ \textit{earlier} than the endpoint at $x_n^-$: $x^- < x_n^-$.  To explicitly separate out this dependence, it is convenient to assign the labeling $n$ of the point charges in ascending order of $x^-$, so that $x_1^- < x_2^- < \cdots < x_A^-$, with $A$ the number of point charges composing the nucleus.  Then the color factor in the trace depends only on the coordinates of the charges $ 1 \cdots (n-1)$, while the vector product affects only the endpoint at $n$.  Rewriting \eqref{WW6} using this notation gives
\begin{align}
 \label{WW7}
 \phi^{WW}(x,k_T) &= \frac{\alpha_s}{4 \pi^3 N_c} \int d^2 x \, d^2 y \, e^{-i \ul{k} \cdot 
 (\ul{x}-\ul{y})} \sum_{n=1}^A \: 
 \left\langle \frac{ (\ul{y}-\ul{x}_{n}) \cdot (\ul{x}-\ul{x}_{n}) }
      {|\ul{y}-\ul{x}_{n}|_T^2 \, |\ul{x}-\ul{x}_{n}|_T^2} \right\rangle_{n} 
 \\ \nonumber &\times
 \left\langle \Tr\bigg[ S(\ul{y},x_{n}^-) T^a S^{-1}(\ul{y},x_{n}^-) \,
           S(\ul{x},x_{n}^-) T^a S^{-1}(\ul{x},x_{n}^-) \bigg] \right\rangle_{1 \cdots (n-1)}
\end{align}
The spatial averaging over the $n^{th}$ charge only affects the vector product, so we can carry it out independently.  If the nucleus has an average transverse number density $T(\ul{b})$ of these charges at impact parameter $b$, then the averaging over $\ul{x}_n$ is given by
\begin{align}
 \label{WW8}
 \left\langle \frac{ (\ul{y}-\ul{x}_{n}) \cdot (\ul{x}-\ul{x}_{n}) }
      {|\ul{y}-\ul{x}_{n}|_T^2 \, |\ul{x}-\ul{x}_{n}|_T^2} \right\rangle_n &=
 \int d^2 x_n \frac{T(\ul{x}_n)}{A} \frac{ (\ul{y}-\ul{x}_{n}) \cdot (\ul{x}-\ul{x}_{n}) }
      {|\ul{y}-\ul{x}_{n}|_T^2 \, |\ul{x}-\ul{x}_{n}|_T^2}
 \\ \nonumber &\approx
 \frac{1}{A} T\left(\frac{ \ul{x} + \ul{y} }{2}\right)
 \int d^2 x_n \frac{ (\ul{y}-\ul{x}_{n}) \cdot (\ul{x}-\ul{x}_{n}) }
      {|\ul{y}-\ul{x}_{n}|_T^2 \, |\ul{x}-\ul{x}_{n}|_T^2}
 \\ \nonumber &=
 \frac{2\pi}{A} T\left(\frac{ \ul{x} + \ul{y} }{2}\right) \ln\frac{1}{|\ul{x}-\ul{y}|_T \Lambda}
\end{align}
where we have used the fact that the nuclear density $T(\ul{b})$ varies only over macroscopic distances proportional to $A^{1/3}$ to replace $T(\ul{x}_n) \approx T(\tfrac{x+y}{2})$, which is valid whenever $k_T \gg \Lambda$.  Thus we have
\begin{align}
 \label{WW9}
 \phi^{WW}(x,k_T) &= \frac{\alpha_s}{2 \pi^2 N_c} \int d^2 x \, d^2 y \, e^{-i \ul{k} \cdot 
 (\ul{x}-\ul{y})} \, \frac{1}{A} T\left(\frac{ \ul{x} + \ul{y} }{2}\right)  
 \ln\frac{1}{|\ul{x}-\ul{y}|_T \Lambda}
 \\ \nonumber &\times
 \sum_{n=1}^A \: \left\langle \Tr\bigg[ S(\ul{y},x_{n}^-) T^a S^{-1}(\ul{y},x_{n}^-) \,
 S(\ul{x},x_{n}^-) T^a S^{-1}(\ul{x},x_{n}^-) \bigg] \right\rangle_{1 \cdots (n-1)}
\end{align}
where the last step that remains is to compute the net effect of the color rotation due to all of the charges $1 \cdots (n-1)$ preceding the endpoint $n$.  

To evaluate the color factor from \eqref{WW9}, it is helpful to make use of the following identity that relates Wilson lines in the fundamental and adjoint representations \cite{Kovchegov:2012mbw},
\begin{align}
 \label{Wident}
 U^{a b}(\ul{x},x^-) T^b &= S^{-1} (\ul{x},x^-) T^a S(\ul{x},x^-) \\ \nonumber
 (U^\dagger)^{a b}(\ul{x},x^-) T^b &= S (\ul{x},x^-) T^a S^{-1}(\ul{x},x^-)
\end{align}
where
\begin{align}
 \label{Uadj}
 U(\ul{x}, x^-) &= \mathcal{P} \exp\left[ + \frac{i g}{2} \int\limits_{x^-}^{-\infty} d b^- \hat{\mathcal{A}}_{cov}^+ (\ul{x}, b^-) \right]
 \\ \nonumber &=
 \prod\limits_{-\infty \cdots x^-} \!\!\!\!\!\!{}_n \: \exp\left[ \frac{-i g^2}{2\pi} t^a \otimes 
 T_n^a \theta(x^- - x_n^-) \ln(|\ul{x} - \ul{x}_n|_T \Lambda) \right]
\end{align}
is the adjoint-representation analog of $S(\ul{x},x^-)$ from \eqref{SUN8} and $\hat{\mathcal{A}}_{cov}^+ \equiv A_{cov}^{+ a} t^a$ is a color matrix in the adjoint representation of $SU(N_c)$.  Using \eqref{Wident} allows us to simplify the color factor in \eqref{WW9} by expressing it in the adjoint representation,
\begin{align}
 \label{Wident2}
 \Tr\bigg[S(\ul{y},x^-) T^a S^{-1}(\ul{y},x^-) &\, S(\ul{x},x^-) T^a S^{-1}(\ul{x},x^-) \bigg] =
 \\ \nonumber &= 
 \left(U^\dagger (\ul{y},x^-)\right)^{a b} \left(U^\dagger (\ul{x},x^-)\right)^{a c} \, \Tr[T^b T^c]
 \\ \nonumber &= 
 \frac{1}{2} \left(U^\dagger (\ul{y},x^-)\right)^{a b} \, \Big(U (\ul{x},x^-) \Big)^{b a}
 \\ \nonumber &= 
 \frac{1}{2} \Tr_{adj} \left[ U(\ul{x},x^-) \, U^\dagger (\ul{y},x^-) \right],
\end{align}
where we have used the fact that an adjoint Wilson line is purely real, and $\Tr_{adj}$ stands for a trace over adjoint indices.  Inserting this back into \eqref{WW9} gives
\begin{align}
 \label{WW10}
 \phi^{WW}(x,k_T) &= \frac{\alpha_s}{4 \pi^2 N_c} \int d^2 x \, d^2 y \, e^{-i \ul{k} \cdot 
 (\ul{x}-\ul{y})} \, \frac{1}{A} T\left(\frac{ \ul{x} + \ul{y} }{2}\right)  
 \ln\frac{1}{|\ul{x}-\ul{y}|_T \Lambda}
 \\ \nonumber &\times
 \sum_{n=1}^A \: \left\langle \Tr_{adj} \bigg[ U(\ul{x},x_{n}^-) \, U^\dagger (\ul{y},x_{n}^-) \bigg] \right\rangle_{1 \cdots (n-1)} .
\end{align}

The product form \eqref{Uadj} of the Wilson line $U$ is ordered with the factor corresponding to $x_{n-1}^-$ occurring on the right, since it has the largest value of $x^-$; similarly $U^\dagger$, which is ordered the opposite way, has the factor corresponding to $x_{n-1}^-$ occurring on the left.  Therefore we can isolate the dependence on the $(n-1)^{st}$ charge by pulling out the latest factor:
\begin{align}
 \nonumber
 \Tr_{adj} \bigg[ U(\ul{x},x_{n}^-) &\, U^\dagger (\ul{y},x_{n}^-) \bigg] = 
 \Tr_{adj} \bigg[ U(\ul{x},x_{n-1}^-) \, 
 \exp\left(\frac{-i g^2}{2\pi} t^a \otimes T_{n-1}^a \ln(|\ul{x}-\ul{x}_{n-1}|_T \Lambda)\right)
 \\ \label{UU1} &\times
 \exp\left(\frac{+i g^2}{2\pi} t^a \otimes T_{n-1}^a \ln(|\ul{y}-\ul{x}_{n-1}|_T \Lambda)\right)
 \, U^\dagger (\ul{y},x_{n-1}^-) \bigg] .
\end{align}
In principle the two exponentials contain emissions of an arbitrary number of gluons from each point charge.  However, from our analysis in Sec.~\ref{subsec-GGM}, we only expect the classical Yang-Mills description to agree with the quantum calculation in QCD up to the level of 2 gluons per color charge as illustrated in Fig.~\ref{fig-NN5}.  Beyond 2 gluons per ``nucleon,'' we do not expect the classical calculation to agree with the quantum one, because the quantum calculation will include genuine loop diagrams that cannot be reproduced by the classical formula \eqref{UU1}.  Therefore, in seeking to compare with the quantum calculation of Sec.~\ref{subsec-GGM}, we should truncate the interaction with point charge $n-1$ at $\ord{g^4}$, corresponding to a contribution of 2 gluons per charge to the Wilson line trace.  Performing this expansion yields
\begin{align}
\nonumber
 \Tr_{adj} \bigg[ U(\ul{x},x_{n}^-) &\, U^\dagger (\ul{y},x_{n}^-) \bigg] \approx
 \Tr_{adj} \Bigg[ U(\ul{x},x_{n-1}^-) \, 
 \Bigg( 1 - \frac{i g^2}{2\pi} t^a T^a_{n-1} \ln\frac{|x-x_{n-1}|_T}{|y-x_{n-1}|_T}
 \\ \label{UU2} &-
 \frac{g^4}{2(2\pi)^2} t^a t^b T^a_{n-1} T^b_{n-1} \ln^2 \frac{|x-x_{n-1}|_T}{|y - x_{n-1}|_T} \Bigg)
 \, U^\dagger (\ul{y},x_{n-1}^-) \Bigg] .
\end{align}

Now that we have factored out the $(n-1)^{st}$ point charge from the others, we can perform the average over its color and spatial coordinate.  The color average eliminates the $\ord{g^2}$ term, since $\Tr[T^a_{n-1}]=0$, and in the $\ord{g^4}$ term gives 
\begin{align}
 \label{UUcolor}
 \frac{1}{N_c} \Tr[T^a_{n-1} T^b_{n-1}] t^a t^b = \frac{1}{2 N_c} t^a t^a =
 \frac{C_A}{2 N_c} = \frac{1}{2}
\end{align}
where $C_A = N_c$ is the quadratic Casimir in the adjoint representation.  Then the spatial averaging  over $\ul{x}_{n-1}$ can be carried out along the lines of \eqref{WW8}:
\begin{align}
 \label{UU3}
 \left\langle \ln^2 \frac{|x-x_{n-1}|_T}{|y - x_{n-1}|_T} \right\rangle_{n-1} &=
 \frac{1}{A} T\left(\frac{x+y}{2}\right) \int d^2 x_{n-1} \, \ln^2 \frac{|x-x_{n-1}|_T}{|y - x_{n-1}|_T}
 \\ \nonumber &=
 \frac{1}{A} T\left(\frac{x+y}{2}\right) \int d^2 x_{n-1} \bigg[
 \ln^2\frac{1}{|x-x_{n-1}|_T \Lambda} + \ln^2\frac{1}{|y-x_{n-1}|_T \Lambda} 
 \\ \nonumber &-
 2 \ln\frac{1}{|x-x_{n-1}|_T \Lambda}\ln\frac{1}{|y-x_{n-1}|_T \Lambda} \bigg].
\end{align}
These logarithmic integrals should be familiar; they arose when we calculated the dipole-dipole scattering cross-section and are most easily carried out in momentum space using \eqref{2gamma3}.  For brevity we relabel $\ul{x}_{n-1} \rightarrow \ul{b}$:
\begin{align}
 \label{UU4}
 \left\langle \ln^2 \frac{|x-x_{n-1}|_T}{|y - x_{n-1}|_T} \right\rangle_{n-1} &=
 \frac{1}{A} T\left(\tfrac{x+y}{2}\right) \int d^2 b \, \frac{1}{(2\pi)^2} 
 \int \frac{d^2 \ell \, d^2 q}{\ell_T^2 \, q_T^2} 
 \bigg[ e^{i \ul{\ell}\cdot(\ul{x}-\ul{b})} \, e^{-i \ul{q}\cdot(\ul{x}-\ul{b})}
 \\ \nonumber &+
  e^{i \ul{\ell}\cdot(\ul{y}-\ul{b})} \, e^{-i \ul{q}\cdot(\ul{y}-\ul{b})} -
	e^{i \ul{\ell}\cdot(\ul{x}-\ul{b})} \, e^{-i \ul{q}\cdot(\ul{y}-\ul{b})} -
	e^{i \ul{\ell}\cdot(\ul{y}-\ul{b})} \, e^{-i \ul{q}\cdot(\ul{x}-\ul{b})}
 \bigg]
 \\ \nonumber &=
 \frac{1}{A} T\left(\tfrac{x+y}{2}\right)
 \int \frac{d^2 \ell}{\ell_T^4} \left[ 2 - e^{i \ul{\ell} \cdot (\ul{x}-\ul{y})} - 
 e^{-i \ul{\ell} \cdot (\ul{x}-\ul{y})} \right]
\end{align}
Comparing with \eqref{2gamma9}, we see that this averaged quantity is proportional to $\int d^2\ell \, \ell_T^{-2} \phi^{dip}(\ell)$, which corresponds to the eikonal scattering of a color dipole on the $1/\ell_T^2$ field of a point charge \eqref{eik8}.  But here there is no dynamical scattering occurring; the effective ``scattering'' occurs between the point charge which is generating the field and the adjoint Wilson lines $\Tr_{adj} [U_x U_y^\dagger]$ occurring in \eqref{WW10}.  The Wilson lines are effectively forming a ``dipole'' of size $|x-y|_T$ which is ``scattering'' in the field of the point charges making up the nucleus.  (See also Fig.~\ref{fig-WW}.)

We can evaluate the integral \eqref{UU4} straightforwardly by doing the angular integral and employing the small-argument asymptotics of the Bessel function $J_0$:
\begin{align}
 \label{UU5}
 \left\langle \ln^2 \frac{|x-x_{n-1}|_T}{|y - x_{n-1}|_T} \right\rangle_{n-1} &=
 \frac{1}{A} T\left(\tfrac{x+y}{2}\right) \, (4\pi) \int \frac{d\ell_T}{\ell_T^3} 
 \left[ 1 - J_0 (\ell_T |x-y|_T) \right]
 \\ \nonumber &=
 \frac{1}{A} T\left(\tfrac{x+y}{2}\right) \, 4\pi |x-y|_T^2 
 \int \frac{d \zeta}{\zeta^3} \left[ 1 - J_0 (\zeta) \right]
 \\ \nonumber &=
 \frac{1}{A} T\left(\tfrac{x+y}{2}\right) \, \pi |x-y|_T^2 \!\!\!\!\!\!
 \int\limits_{(|x-y|_T \Lambda)}^{1} \!\!\!\!\!\! \frac{d \zeta}{\zeta}
 \\ \nonumber &=
 \frac{1}{A} T\left(\tfrac{x+y}{2}\right) \, \pi |x-y|_T^2 \ln\frac{1}{|x-y|_T \Lambda}
\end{align}
Substituting \eqref{UU5} and \eqref{UUcolor} back into \eqref{UU2} gives a full evaluation of the $(n-1)^{st}$ point charge as
\begin{align}
 \nonumber
 \bigg\langle \Tr_{adj} \bigg[ U(\ul{x},x_{n}^-) &\, U^\dagger 
 (\ul{y},x_{n}^-) \bigg] \bigg\rangle_{1 \cdots (n-1)} =
 \left\langle \Tr_{adj} \Bigg[ U(\ul{x},x_{n-1}^-) \, U^\dagger \
 (\ul{y},x_{n-1}^-) \Bigg] \right\rangle_{1 \cdots (n-2)} 
 \\ \label{UU6} &\times
 \left(1 - \alpha_s^2 \frac{\pi}{A} T\left(\tfrac{x+y}{2}\right) 
 |x-y|_T^2 \ln\frac{1}{|x-y|_T \Lambda} \right) .
\end{align}
This procedure systematically evaluates the averaging over the colors and positions of the last color-charge, generating the multiplicative factor in parentheses.  Iterating this for all of the charges generates a total of $n-1$ such factors, and the adjoint trace of unity yields a factor of $N_c^2 - 1$:
\begin{align}
 \nonumber
 \bigg\langle \Tr_{adj} \bigg[ U(\ul{x},x_{n}^-) &\, U^\dagger 
 (\ul{y},x_{n}^-) \bigg] \bigg\rangle_{1 \cdots (n-1)} \!\!\!\!\!\!\!\! = (N_c^2 - 1) 
 \left(1 - \alpha_s^2 \frac{\pi}{A} T\left(\tfrac{x+y}{2}\right) 
 |x-y|_T^2 \ln\frac{1}{|x-y|_T \Lambda} \right)^{n-1}
 \\ \label{UU6new} &=
 (N_c^2 - 1) \left[\exp\left(- \alpha_s^2 \frac{\pi}{A} T\left(\tfrac{x+y}{2}\right) 
 |x-y|_T^2 \ln\frac{1}{|x-y|_T \Lambda} \right) \right]^{n-1},
\end{align}
where we switch interchangeably between the factor in parentheses in the first line and the full exponential in the second line.  This is consistent because we have already truncated the exponentials in \eqref{UU2} at $\ord{g^4} = \ord{\alpha_s^2}$ and also because the quantity $\alpha_s^2 T(b)/A$ is parametrically small, of order $\alpha_s^2 A^{-2/3}$.

Substituting \eqref{UU6} back into \eqref{WW10} gives
\begin{align}
 \label{UU7}
 \phi^{WW}(x,k_T) &= \frac{\alpha_s}{2\pi^2} \left(\frac{N_c^2 - 1}{2 N_c} \right)
 \int d^2x \, d^2y \, e^{-i \ul{k} \cdot (\ul{x} - \ul{y})} \,
 \frac{1}{A} T \left(\tfrac{x+y}{2}\right) \ln\frac{1}{|x-y|_T \Lambda} 
 \\ \nonumber &\times
 \sum_{n=1}^{A} \left[\exp\left(- \alpha_s^2 \frac{\pi}{A} T\left(\tfrac{x+y}{2}\right) 
 |x-y|_T^2 \ln\frac{1}{|x-y|_T \Lambda} \right) \right]^{n-1}.
\end{align}
We identify the color factor in parentheses as $C_F$, and we can perform the summation directly as a finite geometric series
\begin{align}
 \label{WW11}
 \sum_{n=1}^{A} \bigg[\exp\bigg(- \alpha_s^2 \frac{\pi}{A} T\left(\tfrac{x+y}{2}\right) &
 |x-y|_T^2 \ln\frac{1}{|x-y|_T \Lambda} \bigg) \bigg]^{n-1} =
 \\ \nonumber &=
 \frac{
 1 - \left[\exp\left(- \alpha_s^2 \tfrac{\pi}{A} T\left(\tfrac{x+y}{2}\right) |x-y|_T^2 \ln\frac{1}{|x-y|_T \Lambda} \right)\right]^A
 }{
 1 - \left[1 - \alpha_s^2 \frac{\pi}{A} T\left(\tfrac{x+y}{2}\right) |x-y|_T^2 \ln\frac{1}{|x-y|_T \Lambda} \right] 
 }
 \\ \nonumber &=
 \frac{
 1 - \exp\left(- \alpha_s^2 \pi T\left(\tfrac{x+y}{2}\right) |x-y|_T^2 \ln\frac{1}{|x-y|_T \Lambda} \right)
 }{
 \alpha_s^2 \frac{\pi}{A} T\left(\tfrac{x+y}{2}\right) |x-y|_T^2 \ln\tfrac{1}{|x-y|_T \Lambda} }
\end{align}
where we have exploited the ability to interchange between the exponential and its expansion.  The denominator of \eqref{WW11} cancels the other factors in the integrand of \eqref{UU7}, giving the fully-
evaluated Weizs\"acker-Williams field as
\begin{align}
 \label{WWfull}
 \phi^{WW}(x,k_T) &= \frac{C_F}{2\pi^3 \alpha_s} \int d^2x \, d^2y \, 
 e^{-i \ul{k} \cdot (\ul{x} - \ul{y})} \, \frac{1}{|x-y|_T^2} 
 \\ \nonumber &\times
 \left[ 1 - \exp\left(- \alpha_s^2 \pi T\left(\tfrac{x+y}{2}\right) |x-y|_T^2 \ln\frac{1}{|x-y|_T \Lambda} \right) \right]
 \\ \nonumber \phi^{WW}(x,k_T) &=
 \frac{C_F}{2\pi^3 \alpha_s} \int d^2b \, d^2r \, e^{-i \ul{k} \cdot \ul{r}} \, \frac{1}{r_T^2}
 \left[ 1 - \exp\left(- \alpha_s^2 \pi T(\ul{b}) r_T^2 \ln\frac{1}{r_T \Lambda} \right) \right] ,
\end{align}
where we have changed variables to the impact parameter $\ul{b} \equiv \tfrac{1}{2}(\ul{x} + \ul{y})$ and dipole separation $\ul{r} \equiv \ul{x} - \ul{y}$.

\subsection{The Color-Glass Condensate}
\label{subsec-CGC}

The form of the Weizs\"acker-Williams field \eqref{WWfull} obtained by solving the classical Yang-Mills equations \eqref{eq-EOM} mirrors the form of the dipole-dipole forward scattering amplitude in the Glauber-Gribov-Mueller multiple scattering approximation \eqref{GGMsoln2} that enters eikonal DIS scattering cross-section.  Defining the \textit{gluon saturation scale} from the exponent in \eqref{WWfull} to mirror the quark saturation scale \eqref{Qsat2},
\begin{align}
 \label{QsG}
 Q_{s,G}^2 (\ul{b}) \equiv 4\pi \alpha_s^2 T(\ul{b})
\end{align}
we see that $Q_{s,G}^2$ differs from $Q_s^2$ in two regards: a factor of 2 and a factor of $C_F / N_c$.  The factor of 2 arises because \eqref{Qsat2} describes the interaction with target nucleus composed of $A$ dipole ``nucleons,'' whereas \eqref{QsG} describes the scattering on a target nucleus composed of $A$ point charges.  The factor of $C_F/N_c$ arises because \eqref{Qsat2} describes the interaction of a projectile composed of a quark dipole in the fundamental representation, whereas \eqref{WWfull} describes the interaction with a pair of gluon Wilson lines in the adjoint representation.  In the adjoint representation, $C_A = N_c$ so that $C_A / N_c = 1$.

\begin{figure}
 \centering
 \includegraphics[width=\textwidth]{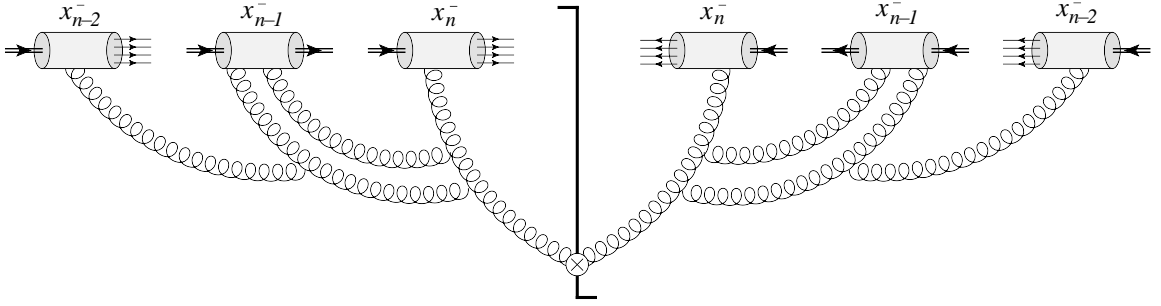}
 \caption{Diagrams contributing to the Weizs\"acker-Williams field \eqref{WWfull}.  The field has been evaluated at the level of 2 gluons per point charge (``nucleon''), and the transformation to light-cone gauge introduces a color factor in \eqref{WW10} that takes the form of the eikonal scattering of a ``gluon dipole'' \eqref{NG}, visualized here.}
 \label{fig-WW}
\end{figure}

This parallel between a scattering cross-section and the classical gluon distribution occurs because the light-cone gauge fields necessary to define the number of gluons have the structure of theta functions \eqref{ALC}.  In the light-cone gauge, gluons emitted at later ``times'' $x^-$ are affected by the pre-existing gluon fields of the earlier sources.  This is the origin of the adjoint Wilson lines 
in \eqref{WW10}: a gluon emitted at $x_n^-$ ``scatters'' in the field generated by the earlier charges, as illustrated in Fig.~\ref{fig-WW}.  We can carry this interpretation further by noting that the Weizs\"acker-Williams field is defined in \eqref{WW3} as the expectation value of light-cone gluon fields in a hadronic state $\bra{A} \vec{A}_{LC}(y) \cdot \vec{A}_{LC}(x) \ket{A}$.  Taking this definition as the classical limit of a quantum operator equation, we can always insert a complete set of final states $\sum_{X} \ket{X} \bra{X}$ to obtain the square of a transition amplitude.  The same is true of the Weizs\"acker-Williams field \eqref{WW10} expressed in terms of adjoint Wilson lines; after the insertion of a complete set of states, this describes the scattering cross-section of a \textit{gluon dipole} on the nuclear target.  In analogy with \eqref{Qsat3}, we define the (imaginary part of the) forward scattering amplitude of a gluon dipole as
\begin{align}
 \label{NG}
 N_G (r_T, \ul{b}) = 1 - \exp\left[-\frac{1}{4} r_T^2 Q_{s,G}^2(\ul{b}) \ln\frac{1}{r_T \Lambda} \right],
\end{align}
in terms of which we can write the Weizs\"acker-Williams field as
\begin{align}
 \label{WWfinal}
 \phi^{WW}(x,k_T) = \frac{C_F}{2\pi^3 \alpha_s} \int d^2b \, d^2r \, e^{-i \ul{k} \cdot r} \, 
 \frac{1}{r_T^2} N_G (r_T , \ul{b}) .
\end{align}
This classical result is independent of the details of the nuclear target and can be obtained even from a continuous charge-density description \cite{Jalilian-Marian:1997xn} of the nucleus.  This is the central result of the McLerran-Venugopalan model of a heavy nucleus in the classical limit.

We can also gain insight from the calculation of the Weizs\"acker-Williams field by interpreting the expectation value $\langle \vec{A}_{LC}(y) \cdot \vec{A}_{LC}(x) \rangle_A$ as a correlation function of the gluon fields.  Comparing \eqref{WW3} and \eqref{WWfull}, we identify the correlator as
\begin{align}
\label{eq-MV correlator}
 \left\langle \underline{A}_{LC}(\ul{b} - \tfrac{1}{2}\ul{r}) \cdot 
 \underline{A}_{LC}(\ul{b} + \tfrac{1}{2}\ul{r}) \right\rangle_A &=
 \frac{2 C_F}{\alpha_s \pi} \, \frac{1}{r_T^2} \: \left[1-e^{-\frac{1}{4} r_T^2 Q_{s,G}^2 (\ul{b})
 \ln\tfrac{1}{r_T \Lambda}} \right],
 \\ \nonumber &=
 \frac{2 C_F}{\alpha_s \pi} \, \frac{1}{r_T^2} \: N_G (r_T, \ul{b}) .
\end{align}
From this expression, we see that $1/Q_{s,G}$ is the \textit{correlation length} of the color fields in the transverse plane.  This strengthens the interpretation given in Fig.~\ref{fig-ColorDomains} of $Q_s^{-1}$ as the typical transverse size of correlated color domains; here we can see this explicitly at the level of a color-color correlation length.  

Furthermore, we observe that the magnitude of the Weizs\"acker-Williams field \eqref{WWfinal} is parametrically large, of order $\ord{1/\alpha_s}$, corresponding to individual gluon fields on the order of $\ord{\tfrac{1}{g}}$.  In comparison, the fields in the Abelian limit given by the covariant-gauge solution \eqref{Acov} are parametrically small: $\ord{g}$.  The uniquely non-Abelian physics, reflected in the nonlinear color rotation matrix $S$ present in the light-cone gauge \eqref{ALC}, has driven the intensity of the fields up by a factor of $1/\alpha_s$.  This is generically expected for any classical solution of the nonlinear Yang-Mills equations, which we can see trivially by equating the linear and nonlinear terms in the equation of motion \eqref{eq-EOM}.  But it is quite \textit{nontrivial} that the same physics generates nuclear shadowing and saturation in the Glauber-Gribov-Mueller formula \eqref{Qsat3}, which was obtained by the resummation of $\alpha_s^2 A^{1/3} \sim \ord{1}$ in a \textit{quantum-mechanical} calculation.  The parametrically strong fields correspond to high gluon occupation numbers consistent with the $\ord{\tfrac{1}{g}}$ classical fields.  This observation confirms the interpretation of Fig.~\ref{fig-NN5} that eikonal scattering on a system of dense color charges corresponds to the interaction with classical gluon fields.

\begin{figure}
 \centering
 \includegraphics[width=0.6\textwidth]{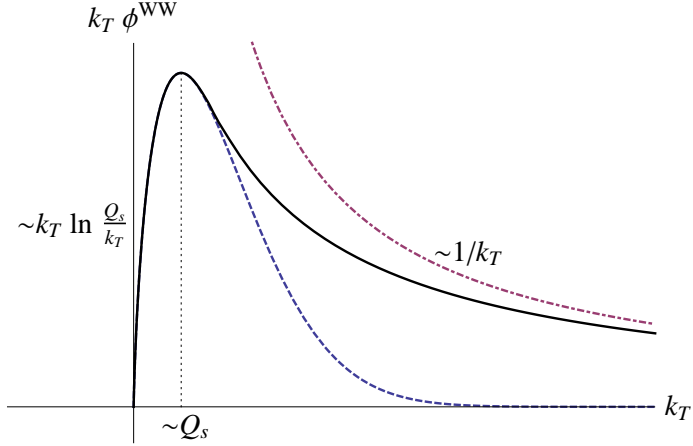}
 \caption{Plot of the phase-space distribution $k_T \phi^{WW}(k_T)$ using \eqref{WWsat}.  The $k_T \sim Q_s$ behavior in the saturation regime is given by \eqref{WWfinitek} (blue dashed curve); the large-$k_T$ asymptotics are given by \eqref{WWlargek} (red dash-dotted curve); and a smooth interpolation between the two limits has been constructed by hand (black solid curve).  The interpolation should only be considered schematic; a calculation of the next subleading term of \eqref{WWlargek}, for example, demonstrates that the full solution should approach the $1/k_T$ asymptotics from above.}
 \label{fig-WWplot}
\end{figure}

Finally, it is instructive to examine the limits of the Weizs\"acker-Williams distribution \eqref{WWfinal} at large and small transverse momentum.  Rewriting this using \eqref{NG} gives
\begin{align}
 \label{WWsat}
 \phi^{WW}(x,k_T) &=
 \frac{C_F}{2\pi^3 \alpha_s} \int d^2b \, d^2r \, e^{-i \ul{k} \cdot \ul{r}} \, \frac{1}{r_T^2}
 \left[ 1 - \exp\left(- \frac{1}{4} r_T^2 Q_{s,G}^2 (\ul{b}) \ln\frac{1}{r_T \Lambda} \right) \right]
 \\ \nonumber &=
 \frac{C_F}{2\pi^2 \alpha_s} R_A^2 \, \int d^2r \, e^{-i \ul{k} \cdot \ul{r}} \, \frac{1}{r_T^2}
 \left[ 1 - \exp\left(- \frac{1}{4} r_T^2 Q_{s,G}^2 \ln\frac{1}{r_T \Lambda} \right) \right] ,
\end{align}
where in the second line we have neglected the impact parameter dependence for simplicity and generated a trivial factor of the transverse area $\pi R_A^2$.  In the limit $k_T^2 \sim 1/r_T^2 \gg Q_{s,G}^2$ the exponent becomes small; we therefore expand the exponential to the lowest nontrivial order:
\begin{align}
 \label{WWlargek}
 \phi^{WW}(k_T \gg Q_{s,G}) &= \frac{C_F}{4\pi \alpha_s} R_A^2 \frac{Q_{s,G}^2}{k_T^2} =
 A \left(\frac{\alpha_s C_F}{\pi} \frac{1}{k_T^2}\right) = \frac{A}{2} \bigg\langle \phi^{dip} \bigg\rangle = A \bigg\langle \phi^{quark} \bigg\rangle
\end{align}
where the angular-averaged dipole field was defined in \eqref{2gamma12}.  Thus the large-$k_T$ limit corresponds to short distances inside a single color domain, where the field strength is just a superposition of the perturbative (linear) fields.  On the other hand, if we consider the small-$k_T$ limit, $k_T^2 \sim 1/r_T^2 \ll Q_{s,G}^2$, the full exponential becomes small; after dropping it, the Fourier transform generates the familiar logarithm \eqref{eik8}
\begin{align}
 \label{WWsmallk}
 \phi^{WW}(k_T \ll Q_{s,G}) &= \frac{C_F}{\pi \alpha_s} R_A^2 \ln\frac{Q_{s,G}}{k_T} .
\end{align}
Deep in the saturation regime, the infrared divergence that scaled perturbatively as $1/k_T^2$ has softened to just a logarithmic one.  This means that the phase-space distribution of gluons $\sim k_T \phi^{WW}(k_T)$ is infrared finite, going to zero as $k_T \rightarrow 0$.  And finally, we can estimate the behavior in the saturation regime by neglecting the logarithm in the exponent and performing the integral over $d^2 r$ analytically.  The logarithm is necessary to obtain the right large-$k_T$ asymptotics, but for $k_T$ in the vicinity of $Q_s$ this approximation should give the right answer:
\begin{align}
 \label{WWfinitek}
 \phi^{WW}(k_T \sim Q_{s,G}) &\approx \frac{C_F}{\pi \alpha_s} R_A^2 \int \frac{d r_T}{r_T} \, J_0(k_T r_T)
 \left[1 - \exp\left(-\frac{1}{4} r_T^2 Q_{s,G}^2\right)\right]
 \\ \nonumber &=
 \frac{C_F}{2\pi \alpha_s} R_A^2 \, \Gamma\left(0,\frac{k_T^2}{Q_{s,G}^2}\right) ,
\end{align}
where $\Gamma(0, k_T^2/Q_{s,G}^2)$ is the incomplete gamma function.  These results are plotted in Fig.~\ref{fig-WWplot}, illustrating the $k_T \sim Q_{s,G}$ behavior, the large-$k_T$ asymptotics, and a smooth interpolation between them.  

From Fig.~\ref{fig-WWplot}, we see that the phase-space distribution of gluons peaks around $k_T \sim Q_{s,G}$.  (In \eqref{WWfinitek}, the peak occurs at $k_T / Q_s \approx 0.32$).  In a (very) loose analogy with phenomena in condensed-matter physics, this parametrically large concentration of gluons in a single momentum state is sometimes referred to as the \textit{color-glass condensate} (CGC) (see, e.g. \cite{Jalilian-Marian:1997xn,Jalilian-Marian:1997jx,Jalilian-Marian:1997gr,Jalilian-Marian:1997dw,Jalilian-Marian:1998cb,Kovner:2000pt,Weigert:2000gi,Iancu:2000hn,Ferreiro:2001qy}).  The nonlinear effects of multiple gluon scattering have shifted the low-$k_T$ gluons up in momentum and depleted the distribution in the far infrared.  The saturation scale $Q_s$ is a measure of the density of the system, scaling with the number of charges as $Q_s^2 \propto T(\ul{b}) \sim A^{1/3}$, so by increasing the density of charges, we further deplete the infrared region.  This leads to a profound conclusion: since $Q_s$ emerges as a dynamical infrared cutoff that increases with the density of the system, for a system of sufficient density that $Q_s^2 \gg \Lambda_{QCD}^2$, the physics of high-energy scattering can be made \textit{perturbative}!  Indeed, calculations of the scale at which the coupling $\alpha_s$ runs in typical high-energy collisions confirm that they are proportional to $Q_s$ (see, e.g. \cite{Horowitz:2010yg}).  Thus, the emergent physics of saturation provides a well-defined resummation of QCD itself in which high-energy scattering becomes perturbative and classical fields dominate:
\begin{align}
 \label{scales}
 A^{1/3} \gg 1 \hspace{1cm}
 Q_s^2 \gg \Lambda_{QCD}^2 \hspace{1cm}
 \alpha_s (Q_s^2) \ll 1 \hspace{1cm}
 \alpha_s^2 A^{1/3} \sim \ord{1} .
\end{align}
Thus the CGC approach, characterized by the Glauber-Gribov-Mueller and McLerran-Venugopalan formalisms, is a powerful tool which can bring high-energy, high-density hadronic processes into the perturbative domain.


\section{Comments on Small-$x$ Quantum Evolution}
\label{sec-Evolution}

In this Chapter we have discussed the role of strong classical gluon fields arising from a system such as a heavy nucleus with a natural parameter $A$, the nucleon number, parameterizing the large transverse density.  This external parameter makes it possible to re-sum the interactions with the large density of partons in a systematic manner.  However, this is not the only manner in which a system with a high density of partons can arise.  A much more physically achievable - if more calculationally difficult - route to the dense limit is through the effects of \textit{quantum evolution}.  A full discussion of the role of quantum corrections is beyond the scope of this document.  However, to put the results of the previous Sections in context, it is necessary to at least describe the role of quantum evolution in generating systems of high-density to which the CGC formalism can be applied.  Thus, in this Section, we will summarize and motivate the physics of high-energy, small-$x$ quantum evolution without explicit derivations.  For discussions of quantum evolution, the reader is referred to the works of \cite{Kovchegov:2012mbw, Collins:2011zzd}.

``Quantum evolution'' refers to the re-ordering of the perturbation series that occurs when certain classes of quantum corrections which are usually suppressed by the coupling become parametrically enhanced.  In deep inelastic scattering in the Bjorken regime, for example, the collinear splitting of partons is suppressed by a power of the coupling $\alpha_s$ but systematically enhanced by a large logarithm of the photon virtuality $Q^2$.  If the virtuality becomes large enough that $\alpha_s \ln Q^2 / \Lambda^2$ becomes $\ord{1}$, then the logarithm can fully offset the suppression by $\alpha_s$, making these quantum corrections equal in importance to the tree-level processes.  When this is true, all such diagrams which are maximally enhanced by any power of $(\alpha_s \ln Q^2/\Lambda^2)^n$ must be re-summed.  This can be done by formulating a differential \textit{evolution equation} describing how observables are affected by one such enhanced quantum correction; for the collinear splitting enhanced by large logarithms of $Q^2$, this evolution equation is the Dokshitzer-Gribov-Lipatov-Altarelli-Parisi (DLGAP) equation \cite{Gribov:1972ri, Altarelli:1977zs, Dokshitzer:1977sg}.  The solution to the DGLAP equation then re-sums all orders of $(\alpha_s \ln Q^2/\Lambda^2)^n$.  

\begin{figure}
 \centering
 \includegraphics[width=0.85\textwidth]{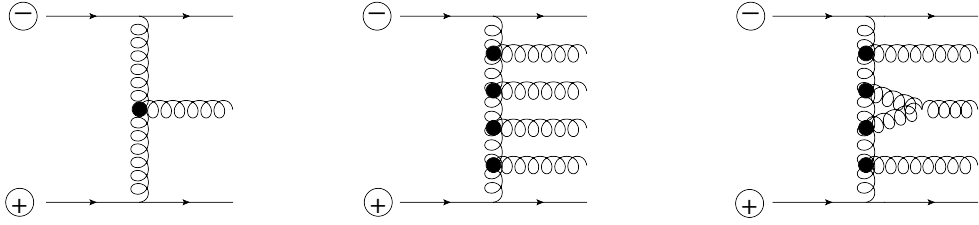}
 \caption{Small-$x$ quantum evolution corrections to the $2 \rightarrow 2$ eikonal quark scattering of Fig.~\ref{fig-quark_scatter}.  Left panel: radiation of a single small-$x$ gluon in addition to the usual eikonal scattering process.  The vertex represents the sum of all possible bremsstrahlung emissions from either of the quarks or the gluon.  Center panel: the development of a small-$x$ gluon cascade through the radiation of many such small-$x$ gluons.  These corrections are re-summed by the BFKL equation.  Right panel: modification of the small-$x$ gluon cascade by nonlinear gluon fusion when the density of gluons becomes large.  These corrections are re-summed systematically by the BK or JIMWLK equations.}
 \label{fig-Evolution}
\end{figure}

A similar situation occurs in high-energy scattering in the Regge limit.  When eikonal scattering processes such as the quark-quark scattering of Fig.~\ref{fig-quark_scatter} take place, they are kinematically insensitive to the emission of additional small-$x$ gluons through bremsstrahlung (Fig.~\ref{fig-Evolution}, left panel).  This radiation of an extra gluon during the eikonal scattering is suppressed by a power of $\alpha_s$ compared to the $2\rightarrow 2$ process of Fig.~\ref{fig-quark_scatter}, but it is enhanced by the phase space available for soft gluons: $\Delta Y \sim \ln 1/x$.  Thus, for the inclusive observables to which these diagrams contribute, the overall power-counting of the quantum correction is $\alpha_s \ln 1/x$.  As the energy $s \sim 1/x$ of the scattering process is increased, the amount of phase space available for the emission of low-$x$ gluons increases; thus, when the energy is sufficiently large that $\alpha_s \ln s/\Lambda^2 \sim \alpha_s\ln 1/x \sim \alpha_s \Delta Y \sim \ord{1}$, all such quantum corrections must be re-summed.  This leads to the quantum evolution equation known as the Balitsky-Fadin-Kuraev-Lipatov (BFKL) equation \cite{Kuraev:1977fs, Balitsky:1978ic}, which describes the development of a small-$x$ gluon cascade (Fig.~\ref{fig-Evolution}, center panel).  

The bremsstrahlung corrections embodied in the BFKL equation lead to a gluon distribution that grows at small-$x$ as \cite{Kovchegov:2012mbw}
\begin{align}
 \label{BFKLpomeron}
 \phi(x, k_T) \propto \left(\frac{1}{x}\right)^{\left( \tfrac{4 \alpha_s N_c}{\pi} \ln 2 \right)}
 \approx \left(\frac{1}{x}\right)^{0.79},
\end{align}
where the numerical value was obtained for $N_c = 3 , \alpha_s = 0.3$.  Since the number of gluons generated through bremsstrahlung increases with energy, the total scattering cross-section \eqref{2gamma10} also grows.  Both the apparent increase in the density of gluons and the rapid growth of the scattering cross-section with energy (and hence, with decreasing $x$) are in tension with the constraints imposed by the unitarity of the $S$-matrix.  The Froissart-Martin bound, for example, derives from unitarity the constraint that the cross-section can grow at most as $\ln^2 1/x$ \cite{Froissart:1961ux, Martin:1969a, Lukaszuk:1967zz}, while \eqref{BFKLpomeron} gives rise to a cross-section that grows as $\approx (1/x)^{0.79}$.  These considerations indicate that the growth of gluon density due to bremsstrahlung cannot continue unabated forever, down to arbitrarily small-$x$ (or arbitrarily high energies).

Eventually, the densities of gluons become sufficiently large that nonlinear effects, like those discussed in this Chapter, become important.  When the density of gluons becomes sufficiently large, newly-radiated gluons are likely to scatter on the dense fields that already exist.  The analogous role to the GGM multiple scattering discussed in Sec.~\ref{subsec-GGM} is played by gluon fusion, which limits the total rate at which the gluon density can grow (Fig.~\ref{fig-Evolution}, right panel).  Since the rate of gluon bremsstrahlung is proportional to the total number of gluons $x G$ in the cascade, but gluon fusion is proportional to its square $[x G]^2$, this can be used to estimate the momentum scale $Q_s$ at which the nonlinear effects become important.  The result is \cite{Gribov:1984tu, Mueller:1986wy, Kovchegov:2012mbw}
\begin{align}
 \label{GLRMQ}
 Q_s^2 = \frac{\alpha_s \pi^2}{2 S_\bot C_F} x G(x, Q_s^2) ,
\end{align}
where $S_\bot$ is the transverse area.  This estimate agrees completely with the calculation from the GGM multiple-scattering formalism \eqref{Qsat1} after the conversion $C_F \rightarrow C_A = N_c$ between the scattering of fundamental-representation quarks and adjoint-representation gluons.

Thus the physics of saturation can also be obtained by the small-$x$ quantum evolution of collisions at very high energies.  When nonlinear effects are systematically included in the small-$x$ evolution equations, the BFKL equation is replaced by the Balitsky-Kovchegov (BK) equation \cite{Balitsky:1996ub, Kovchegov:1999yj}, which explicitly satisfies the unitarity bound.  Equivalently, in the continuous charge-density formulation of the MV model, the evolution equation is known as the Jalilian-Marian--Iancu--McLerran--Weigert--Leonidov--Kovner (JIMWLK) functional differential equation \cite{Jalilian-Marian:1997jx, Jalilian-Marian:1997gr, Jalilian-Marian:1997dw, Iancu:2001ad, Iancu:2000hn}.  The (approximate) solutions to these equations give a saturation scale that grows with energy (or decreasing $x$) as
\begin{align}
 \label{BK}
 Q_s (x) \propto \left(\frac{1}{x} \right)^{2.44 \left(\tfrac{\alpha_s N_c}{\pi}\right)} 
\approx \left(\frac{1}{x} \right)^{0.7} 
\end{align}
where the final numerical value was obtained using $\alpha_s = 0.3$.  As the energy of the process is increased (and $x$ decreased), gluon bremsstrahlung generates more color charges, increasing $Q_s$ and decreasing the radius of the correlated color domains.  Therefore, small-$x$ evolution equations suggest that the perturbative limit described by the CGC formalism is also reached in the limit of high energies through quantum corrections.  The essential physics of saturation at high charge densities applies both to a proton at high energies through quantum evolution equations, and to a heavy nucleus in the initial conditions of those equations \cite{Tribedy:2010ab, Kowalski:2003hm, GolecBiernat:1999qd, GolecBiernat:1998js}.

Therefore we can consider the scattering on a heavy nucleus discussed in this Chapter as a metaphor for the more general situation of scattering on a high-density system.  The physical conclusions reached by using the GGM or MV formulas can be carried over and applied to a proton at very high energies, even though the quantitative details of their application may change \cite{Kovchegov:2012mbw}.  For this reason, we will use the saturation formalism in terms of a heavy nucleus presented here as a window into the application of the spin- and transverse-momentum physics discussed in Chapter~\ref{chap-TMD} to systems of high density.  In Chapter~\ref{chap-odderon} we will discuss saturation effects in the generation of single transverse spin asymmetries by the scattering of a polarized projectile on an unpolarized target.  In Chapter~\ref{chap-MVspin} we will study the reverse situation, examining saturation effects in a polarized target by generalizing the MV model of Sec.~\ref{sec-classical_fields} to include spin and transverse momentum. 
\chapter{Transverse Spin as a Novel Probe of Saturation}
\label{chap-odderon}


Consider the single transverse spin asymmetry (STSA) \eqref{AN1} produced in high-energy proton collisions $p^\uparrow + p \rightarrow h + X$.  The rapidity $y_h$ of the tagged hadron $h$ is defined as
\begin{align}
 \label{rap1}
 y_h \equiv \frac{1}{2} \ln \frac{h^+}{h^-} = \ln \left( \frac{h^+}{\sqrt{h_T^2 + m_h^2}} \right) =
 \ln \left( \frac{\sqrt{h_T^2 + m_h^2}}{h^-} \right),
\end{align}
where we have used the on-shell condition $h^+ h^- - h_T^2 = m_h^2$.  If we choose a frame in which the polarized proton moves along the $x^+$ axis with large momentum $p_1^+$ and the unpolarized proton  moves along the $x^-$ axis with large momentum $p_2^-$, then the total rapidity interval $\Delta Y$ between the colliding protons is
\begin{align}
 \label{rap2}
 \Delta Y \equiv y_1 - y_2 = \ln \left( \frac{p_1^+ p_2^-}{m_N^2} \right) \approx
 \ln \frac{s}{m_N^2},
\end{align}
where $s$ is the center-of-mass energy (squared) of the collision.  

Experimental measurements of the STSA at Fermilab and at RHIC indicate that the asymmetry produced in these collisions is small for most of the kinematic range, but becomes quite large when the rapidity of the tagged hadron is close to the rapidity of the polarized proton \cite{Adams:1991ru, Adams:1991cs, Adams:1991rw, Adams:1995gg, Bravar:1996ki, Adams:1994yu, Abelev:2007ii, Nogach:2006gm, Adler:2005in, Lee:2007zzh, Abelev:2008qb, Wei:2011nt}; that is, when the rapidity interval $\Delta y_h \equiv y_1 - y_h$ is small.  We can translate from the rapidity to the longitudinal momentum fractions $x_1 , x_2$ of the tagged hadron with respect to the polarized and unpolarized protons, respectively, as
\begin{align}
 \label{rap3}
 x_1 &\equiv \frac{h^+}{p_1^+} & 
 x_2 &\equiv \frac{h^-}{p_2^-} \\ \nonumber
 x_1 &= \frac{\sqrt{h_T^2 + m_h^2}}{m_N} \, e^{-\Delta y_h} &
 x_2 &= \frac{\sqrt{h_T^2 + m_h^2}}{m_N} \, e^{-(\Delta Y - \Delta y_h)}
\end{align}
by straightforward application of \eqref{rap1}.  When $\Delta y_h \ll 1$, we can expand \eqref{rap3} to find
\begin{eqnarray}
 \label{rap4}
 x_1 \approx& \frac{\sqrt{h_T^2 + m_h^2}}{m_N} \big( 1 - \Delta y_h \big) &\sim \ord{1} \\ \nonumber
 x_2 \approx& \frac{\sqrt{h_T^2 + m_h^2}}{m_N} \, e^{-\Delta Y} \big( 1 + \Delta y_h \big) &\ll 1.
\end{eqnarray}
Under these highly-asymmetric kinematics, the scattering process is most sensitive to the large-$x$ valence region in the polarized proton, and the small-$x$ bremsstrahlung region in the unpolarized proton.  As we have seen in Chapter~\ref{chap-CGC}, the small-$x$ regime is characterized by a high density of gluons that gives rise to the saturation paradigm of high-energy scattering.

These considerations suggest that the asymmetry generated in these high-energy proton collisions reflects the spin-dependent interaction of the polarized projectile with the dense, classical gluon fields of the unpolarized target.  Since, as discussed in Sec.~\ref{subsubsec-STSA}, the STSA is a $T$-odd observable, this suggests that the polarized proton is acting as a unique probe of the $T$-odd component of the classical gluon fields.  The dominant gluon fields which drive the unpolarized observables described in Chapter~\ref{chap-CGC} are manifestly $T$-even, so the high-density physics which gives rise to STSA must be fundamentally different from the conventional channels of high-energy scattering.  In this Chapter, we will analyze the generation of STSA from these $T$-odd gluon fields, which are known in the literature as the ``odderon.''

Since the asymmetry is generated from the valence region of the polarized proton, a useful proxy for the full proton wave function is a single transversely-polarized valence quark $q^\uparrow$ which proceeds to scatter on the classical gluon fields of a dense unpolarized target.  As we saw in Sec.~\ref{sec-classical_fields}, a useful realization of this dense target is a heavy nucleus $A$ in the McLerran-Venugopalan model.  Thus we can consider the $q^\uparrow A$ scattering of a transversely-polarized quark on a heavy nucleus as a simple implementation of saturation effects in the far forward regime of $p^\uparrow p$ collisions.  The goal is then to find the leading-order channels through which the transverse spin dependence enters in the saturation framework.

\begin{figure}[ht]
\centering
\includegraphics[width=8cm]{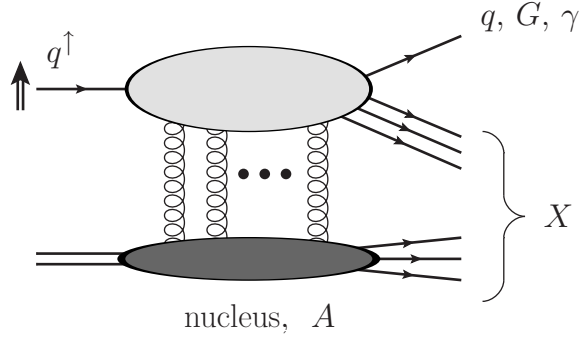}
 \caption{Transversely polarized quark scattering in the field of the 
   nucleus producing either a quark ($q$), a gluon ($G$), or a prompt
   photon ($\gamma$) along with extra hadrons denoted by $X$:
   $q^{\uparrow} +  A \rightarrow (q, \, G, \, \gamma) + X$.} 
\label{fig-QuarkScattering}  
\end{figure} 

\begin{figure}[ht]
\centering
\includegraphics[width=15cm]{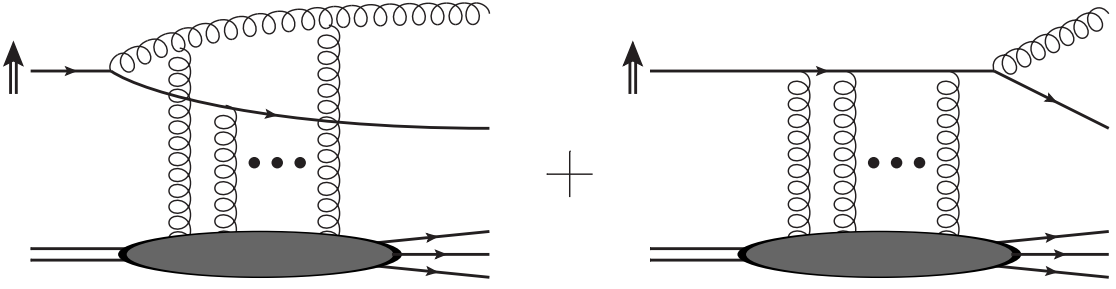}
 \caption{Two contributions to the amplitude for the high energy 
   quark--target scattering in LCPT.}
\label{qtoqGampl}  
\end{figure}

In this Chapter, we will consider the STSA of quarks, gluons, and prompt photons produced in polarized $q^\uparrow A$ collisions: $q^\uparrow + A \rightarrow (q,G,\gamma) + X$.  This process is illustrated in \fig{fig-QuarkScattering}, where the high energy interaction between the projectile quark and the target nucleus is schematically denoted by gluon exchanges.  To introduce the methodology, we will first concentrate on the quark production process, $q^{\uparrow} + A \rightarrow q + X$.  Certainly keeping only the eikonal interaction of the polarized quark with the target would not generate the STSA, since the eikonal scattering \eqref{eik8} is independent of the quark polarization. A non-eikonal correction has to be included somewhere: in the multiple-rescattering Glauber-Gribov-Mueller \cite{Mueller:1989st} approximation discussed in Sec.~\ref{subsec-GGM}, the non-eikonal rescattering corrections are suppressed by powers of energy and are very small. A much larger spin-dependent contribution comes from the non-eikonal splitting of the projectile quark into a quark and a gluon, $q \to q \, G$, which is suppressed only by a power of the strong coupling $\as$. In the language of light-cone perturbation theory (LCPT), the $q \to q \, G$ splitting may take place either before or after the interaction with the target, as shown in \fig{qtoqGampl}. Splitting during the interaction with the target is suppressed by powers of energy due to the instantaneous nature of eikonal scattering \cite{Kovchegov:1998bi}.

 The lowest-order diagrams shown in \fig{qtoqGampl} that contribute to STSA in $q^{\uparrow} + A \rightarrow q + X$ contain the emission of a single gluon from the polarized quark, where both the gluon and quark can scatter in the field of the target.  Multi-gluon non-eikonal emissions and quark pair-production are also possible, but they are higher-order in $\as$ and, hence, outside of the leading-order precision of this work.  The spin dependence of the process illustrated in \fig{qtoqGampl} originates within the light-cone wave function of the quark-gluon system, which couples to the interaction in a way that generates the asymmetry. In Sec.~\ref{sec-General Result}, we will first outline the calculation of the $q \to q \, G$ light-cone wave function using LCPT.  Then we will combine the resulting splitting wave function squared with the quark and gluon interactions in the field of the target and identify the contribution to the asymmetry. In the end we obtain general expressions for quark, gluon, and photon STSA's in our formalism. The incoming light quark has a particular flavor $f$; multiple quark flavors can be incorporated into our formalism by convoluting the obtained cross sections with quark distributions corresponding to different flavors (inserting the appropriate quark masses into our results below).  As discussed above, we choose a frame in which the incoming projectile quark is moving along the light-cone $x^+$-axis, while the target is moving along the $x^-$-axis, and we will work in the light-cone gauge of the projectile, $A^+ = 0$ (which is equivalent to the covariant gauge $\partial_\mu A^\mu = 0$, as discussed in Sec.~\ref{subsec-MV}).  The analysis in this Chapter is original work which follows closely our paper \cite{Kovchegov:2012ga}.


\section{General Result: Coupling Spin to Interaction C-Parity}

\label{sec-General Result} 

Recall that the single transverse spin asymmetry, first defined in \eqref{AN1}, is given by
\begin{equation}
\label{AN1v2}
A_N \equiv \frac{d\sigma^\uparrow (\ul{k}) - d\sigma^\downarrow (\ul{k})} {2 \, d\sigma_{unp}} =
 \frac{d\sigma^\uparrow (\ul{k}) - d\sigma^\uparrow(-\ul{k})}{2 \, d\sigma_{unp}} \equiv
 \frac{d(\Delta\sigma)}{2 d\sigma_{unp}}.
\end{equation}
We will determine which parts of the wave function and the interaction couple to the numerator
$d(\Delta\sigma)$ and the denominator $d\sigma_{unp}$ of the asymmetry.



\subsection{Light-Cone Wave Function and Transverse Polarization}

\label{subsec-Wavefn}

\begin{figure}[ht]
\centering
\includegraphics[width=8cm]{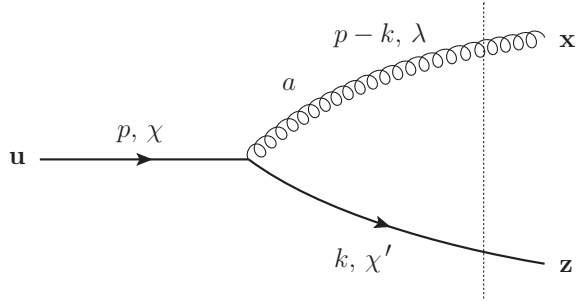}
\caption{The light-cone wave function for the $q \to q \,
  G$ splitting. Vertical dotted line denotes the intermediate state.}
\label{fig-SplittingVertex}  
\end{figure}

Consider the splitting shown in \fig{fig-SplittingVertex} of a
transversely polarized quark with momentum $p$ and polarization $\chi
= \pm 1$ decaying into a gluon (with momentum $p-k$, polarization
$\lambda$, and color $a$) and a recoiling quark (with momentum $k$ and
polarization $\chi'$).  The projectile quark is traveling along the
light-cone $x^+$-direction and the recoiling quark carries a fraction
\begin{equation}
 \label{eq-momentum fraction}
 \alpha \equiv \frac{k^+}{p^+}
\end{equation}
of the incoming quark's longitudinal momentum.  We do not restrict
ourselves to the case of an eikonal quark emitting a soft gluon ($1 -
\alpha \ll 1$), but work in the general case when both the quark and
the gluon can carry comparable longitudinal momenta.  We do, however, assume for simplicity that $\alpha < 1$ to be able to neglect the contribution of virtual corrections proportional to $\delta(1-\alpha)$.  These corrections are discussed in the paper \cite{Kovchegov:2012ga}.

The calculation of the light-cone wave function corresponding to the
diagram in \fig{fig-SplittingVertex} is different from other similar
calculations in the literature (see e.g.
\cite{Itakura:2003jp,Albacete:2006vv}) only in that now the incoming
quark is polarized {\sl transversely}.  For this, we use the transverse spinors with polarization along the $x$-axis first defined in \eqref{spinors2}.  It is convenient
to also use the same spinor basis \eqref{spinors2} for the outgoing quark
in \fig{fig-SplittingVertex} as well. Using the standard rules of LCPT
\cite{Lepage:1980fj,Brodsky:1997de} with this spinor basis, we evaluate the
light-cone wave function shown in \fig{fig-SplittingVertex} as
\begin{equation}
 \label{wf1}
 \psi_{\lambda \chi \chi'}^a ({\ul k}, {\ul p}, \alpha) =
 \frac{g \, T^a}{p^- - k^- - (p-k)^-} \;
 \bigg[ \frac{\bar U_{\chi'} (k)}{\sqrt{k^+}} \, \gamma \cdot \epsilon_{\lambda}^* \,
 \frac{U_\chi (p)}{\sqrt{p^+}} \bigg] \; ,
\end{equation}
where 
\begin{equation}
  \label{eq:pol}
  \epsilon_\lambda^\mu = \left( 0, \frac{2 \, {\ul \epsilon}_\lambda \cdot
 ({\ul p} -{\ul k})}{p^+ - k^+}, {\ul \epsilon}_\lambda \right)
\end{equation}
is the gluon polarization vector with the transverse components ${\ul
  \epsilon}_\lambda = (-1/\sqrt{2}) \, (\lambda, i)$.

In arriving at \eqref{wf1} we have used the fact that the incoming state
in \fig{fig-SplittingVertex} contains only the quark with momentum $p$
while the intermediate state contains the quark and the gluon, as
denoted by the vertical dotted line in \fig{fig-SplittingVertex}. For
diagrams with initial-state interactions where the polarized quark scatters in the nucleus before the
gluon emission, as shown in the right panel of \fig{qtoqGampl}, the
roles are reversed: the quark line $p$ is the intermediate state, and
the quark--gluon system is the final state.  Since $\sum_{init} p_i^-
= \sum_{final} p_i^-$, the energy denominator reverses sign for
final-state splittings.  The wave function \eqref{wf1} is also normalized differently from \cite{Kovchegov:2012mbw} and will require a compensating factor when we calculate the cross-section using the rules of LCPT.

Using the on-shell conditions explicitly gives the terms entering the
energy denominator:
\begin{equation}
 \label{eq-minus components}
 p^- = \frac{p_T^2 + m^2}{p^+}  \;\;,\;\;
 k^- = \frac{k_T^2 + m^2}{k^+} \;\;,\;\;
 (p-k)^- = \frac{(\ul p - \ul k)_T^2}{p^+ - k^+} \;\; .
\end{equation}
The relevant spinor products can be straightforwardly tabulated from \eqref{spinors2} and \eqref{spinors1} as
\footnote{Note again that $U_\chi (p)$ becomes a spinor for a
  transversely polarized particle only for $\ul p =0$:
  Eqs.~\peq{eq-transverse matrix elts} give us the matrix elements for
  spinors related to the Brodsky--Lepage spinors via \eqref{spinors2},
  which do not necessarily correspond to transverse polarizations in
  the general case.}
 \begin{eqnarray}
  \label{eq-transverse matrix elts}
  \frac{\bar U_{\chi'} (k)}{\sqrt{k^+}} \, \gamma^+ \, \frac{U_\chi (p)}{\sqrt{p^+}} &=&
  2 \, \delta_{\chi, \chi'} \\
  \frac{\bar U_{\chi'} (k)}{\sqrt{k^+}} \, \gamma_\bot^i \, \frac{U_\chi (p)}{\sqrt{p^+}} &=&
  \frac{\delta_{\chi,\chi'}}{\alpha \, p^+} \bigg[ 
   (k_\perp^i + \alpha \, p_\perp^i) + (1-\alpha) \, i \, m \, \chi \, \delta^{i2}
  \bigg] \\ \nonumber
  &-& \frac{\delta_{\chi,-\chi'}}{\alpha \, p^+} \bigg[
   i \, \epsilon^{ij} \, (k^j_\perp - \alpha \, p^j_\perp) + (1-\alpha) \, m \, \chi \, \delta^{i1}
  \bigg] \;,
 \end{eqnarray}
and the $\gamma^-$ matrix element does not contribute to $\gamma \cdot
\epsilon^*_\lambda$ since $\epsilon_\lambda^+ = 0$ in the light-cone
gauge. Here $\epsilon^{12} = - \epsilon^{21} =1$, $\epsilon^{11} =
\epsilon^{22} =0$. With the matrix elements \peq{eq-transverse matrix
  elts} it is straightforward to evaluate the light-cone wave function
\eqref{wf1} in momentum space, obtaining
\begin{align}
 \label{eq-momentumwavefn}
 \psi_{\lambda \chi \chi'}^a ( {\ul k} , {\ul p}, \alpha) &= \frac{g \, T^a}{({\ul k} -   
 \alpha \, {\ul p})_T^2 + {\tilde m}^2} \bigg[ {\ul \epsilon}_{\lambda}^* \cdot ({\ul k} - \alpha \, {\ul p}) \, \bigg(
 (1+\alpha) \, \delta_{\chi \chi'} + \lambda \, (1-\alpha) \, \delta_{\chi, -\chi'} \bigg) 
 \\ \nonumber &-
 \frac{\tilde m}{\sqrt 2} \, (1-\alpha) \, \chi \,
 \big (\delta_{\chi \chi'} - \lambda \, \delta_{\chi, -\chi'} \big) \bigg] \;, 
\end{align}
where
\begin{equation}
 \nonumber
 \tilde m \equiv (1-\alpha) m
\end{equation}
is a natural effective mass parameter in the wave function and $T^a$
are the SU($N_c$) generators in the fundamental representation.

Now we can Fourier transform the wave function to coordinate space
\begin{equation}
 \label{eq-Fourier transform}
 \psi_{\lambda \chi \chi'}^a ({\ul x}, {\ul z}, \alpha; {\ul u} ) \equiv 
 \int \frac{d^2 k}{(2\pi)^2} \frac{d^2p }{(2\pi)^2} e^{i \, {\ul k} \cdot ({\ul z} -
 {\ul x})} \, e^{i \, {\ul p} \cdot ({\ul x - \ul u})} \,
 \psi_{\lambda \chi \chi'}^a ({\ul k}, {\ul p}, \alpha) \; 
\end{equation}
with the transverse coordinates defined in \fig{fig-SplittingVertex}.
Since the momentum-space wave function depends only on $\ul k - \alpha
\, \ul p$, one of the two integrals can be performed to yield a delta
function $\delta^2 [(\ul x - \ul u) + \alpha \, (\ul z - \ul x)]$.
Performing the remaining momentum integral in \eqref{eq-Fourier
  transform} yields modified Bessel functions
\begin{align}
 \label{eq-coord wavefn}
 \psi_{\lambda \chi \chi'}^a ({\ul x}, {\ul z}, \alpha ; {\ul u}) &=
\frac{g \, T^a}{2\pi} \, \delta^2 [({\ul x} - {\ul u}) + \alpha ({\ul z - \ul x}) ]  
 \, {\tilde m} \:
 \\ \nonumber &\times
 \bigg\{ i \, {\ul \epsilon}_\lambda^* \cdot \frac{{\ul z - \ul x}}{{|\ul
 z - \ul x|_T}} \,  K_1 ( \tilde m \, |\ul z - \ul x|_T) 
 \bigg[ (1+\alpha) \, \delta_{\chi,\chi'} +  \lambda \, (1-\alpha) \, \delta_{\chi, -\chi'} \bigg] 
 \\ \nonumber &- 
 \frac{\chi \, (1-\alpha)}{\sqrt 2} \, K_0 (\tilde m \, |\ul z - \ul x|_T) \, \bigg[ \delta_{\chi, \chi'} - \lambda \, \delta_{\chi, - \chi'} \bigg] \bigg\}. 
\end{align} 
It is useful to separate out the color factor $T^a$ and the delta
function from the rest of the wave function (denoted by $\Psi_{\lambda
  \chi \chi'}$), such that
\begin{equation}
 \label{eq-Psi vs Phi}
 \psi_{\lambda \chi \chi'}^a (\ul x , \ul z , \alpha ; \ul u) \equiv
 T^a \, \delta^2 [(\ul u - \ul x) - \alpha \, (\ul z - \ul x)] \,
 \Psi_{\lambda \chi \chi'} (\ul z - \ul x, \alpha) \;.
\end{equation}

\begin{figure}[ht]
\centering
\includegraphics[width=12cm]{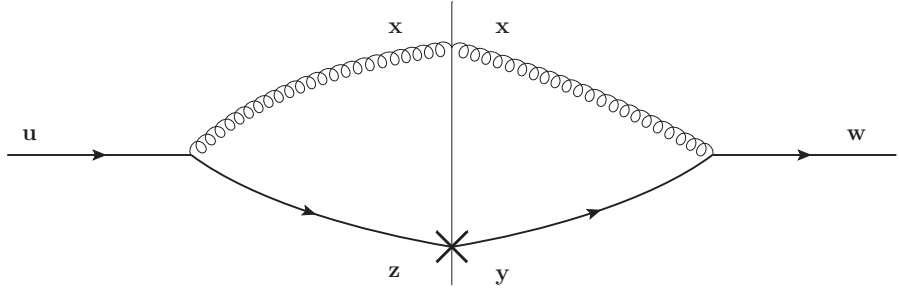}
 \caption{Light-cone wave function from \fig{fig-SplittingVertex} squared. 
   The vertical straight line separates the wave function from its
   conjugate, while the cross denotes the quark that we tag on. The
   untagged gluon's coordinate $x$ is unchanged, but the quark
   coordinates differ ($z$ and $u$ vs. $y$ and $w$, as
   explained in the text.)}
\label{fig-wavefn squared}  
\end{figure}

Finally, we need to square the wave function and sum over the final
particles' polarizations. Here we are interested in producing a quark
with a fixed transverse momentum, while integrating over all
transverse momenta of the produced gluon in \fig{qtoqGampl}. 
As we will show when we calculate the cross-section in Sec.~\ref{subsec-Quark STSA (gen)}, this will set the gluon's transverse
coordinate $\ul x$ to be the same both in the amplitude and in the
complex conjugate amplitude, while the quarks will have different
transverse coordinates between the amplitude and the conjugate
amplitude (since their momentum is tagged).  See
\fig{fig-amplitude squared} below for the illustration of the
full amplitude squared.  The ``square'' of the light cone wave function
\peq{eq-coord wavefn} with the above rule for the quark and gluon
transverse coordinates is illustrated in \fig{fig-wavefn squared}.
     
The wave function \peq{eq-coord wavefn} squared $\Phi_\chi$ as shown
in \fig{fig-wavefn squared} contains one contribution which is
polarization-independent and another which is proportional to the
quark polarization eigenvalue $\chi$
\begin{align}
 \label{eq-wavefn squared}
 \Phi_\chi (\ul z - \ul x , \ul y - \ul x, \alpha) &\equiv \sum_{\lambda \, , \, \chi' = \pm 1} \, 
 \Psi_{\lambda \chi \chi'} (\ul z - \ul x, \alpha) \, \Psi^{*}_{\lambda \chi \chi'} 
 (\ul y - \ul x, \alpha) 
 \\ \nonumber &\equiv
 \Phi_{unp} (\ul z - \ul x , \ul y - \ul x, \alpha) + \chi \, \Phi_{pol} (\ul z - \ul x , 
 \ul y - \ul x, \alpha) \; .
\end{align}
Substituting the wave function \peq{eq-coord wavefn} into \eqref{eq-wavefn
  squared} and performing the sums gives the unpolarized part as
\begin{eqnarray}
 \label{eq-unpolarized wavefn}
 \Phi_{unp} &=& \frac{2 \, \alpha_s}{\pi} \, {\tilde m}^2 \, \bigg[ (1+\alpha^2) \, \frac{(\ul z -   \ul x) \cdot (\ul y - \ul x)}{|\ul z - \ul x|_T \;       
 |\ul  y - \ul x|_T} \, K_1 (\tilde m \, |\ul z - \ul x|_T) \, K_1 (\tilde m \, 
 |\ul y - \ul x|_T) \\ \nonumber
 && + \, (1-\alpha)^2 \, K_0 (\tilde m \, |\ul z - \ul x|_T) \, K_0 (\tilde m \,  
 |\ul y - \ul x|_T) \bigg]
\end{eqnarray}
and the transversely-polarized part as
\begin{align}
 \label{eq-polarized wavefn}
 \Phi_{pol} &= \frac{2 \, \alpha_s}{\pi} \, {\tilde m}^2 \, \alpha \, (1-\alpha) \, \bigg[
 \frac{z_\bot^{2} - x_\bot^{2}}{|\ul z - \ul x|_T} \, K_0 (\tilde m \, |\ul y - 
 \ul x|_T) \, K_1 (\tilde m \, |\ul z - \ul x|_T) 
 \\ \nonumber &+
 \frac{y_\bot^{2} - x_\bot^{2}}{|\ul y - \ul x|_T} \, K_1 (\tilde m \, |\ul y -
 \ul x|_T) \, K_0 (\tilde m \, |\ul z - \ul x|_T) \bigg] .
\end{align}
Note that $\Phi_{unp}$ is a scalar under rotations in the transverse
plane and is parity-even, whereas $\Phi_{pol}$ has an explicitly preferred
azimuthal direction (i.e., the $y$ axis) and is parity-odd since it ``knows'' about the
transverse polarization of the incoming quark. The $x_\bot^{2}$ axis can be
written as the direction of the ${\vec p} \times {\vec S}$ vector,
since the incoming quark with momentum $\vec p$ is moving along the
$z$ axis, while being polarized along the $x = x_\bot^1$ axis, such that
${\vec S} \, \| \, {\hat x}_1$.  We show in Sec.~\ref{subsec-symmetries} that the unpolarized part of
the wave function squared $\Phi_{unp}$ contributes to the unpolarized
quark production cross section $d\sigma_{unp}$, while the
polarization-dependent part of the wave function squared $\Phi_{pol}$
generates the spin-asymmetric cross section $d (\Delta \sigma)$.



\subsection{Target Interactions in Quark Production}
 
\label{subsec-Quark STSA (gen)}

\begin{figure}[ht]
 \centering
 \includegraphics[width=8cm]{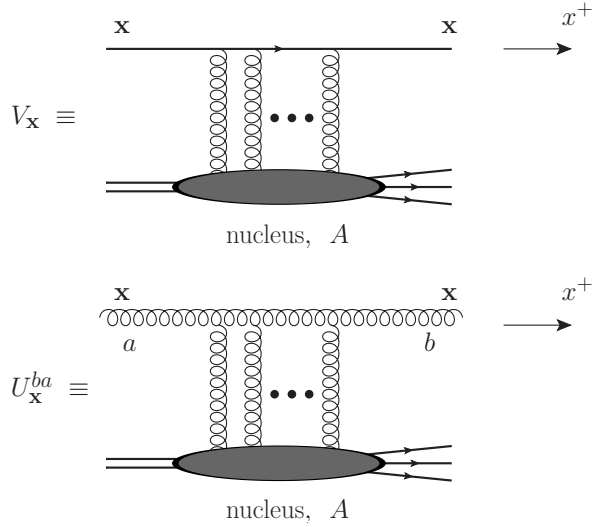}
\caption{Wilson lines resumming scattering in the small-$x$ field of the 
  target.  The quark propagator is in the fundamental representation
  (top), and the gluon propagator is in the adjoint representation
  (bottom).}
\label{fig-Wilson lines} 
\end{figure}

Having computed the $q \to q \, G$ light-cone wave function, we can
now construct the scattering cross section by allowing the wave
function to interact with the small-$x$ field of the target nucleus.
It is well known \cite{Balitsky:1996ub,Jalilian-Marian:1998cb} that
eikonal quark and gluon propagators in the background color field
$A^{\mu \, a}$ can be correspondingly written as fundamental and
adjoint path-ordered Wilson lines as in \eqref{Wilson1}
 \begin{eqnarray}
  \label{eq-Wilson lines}
   V_{\ul x} &\equiv& {\mathcal P} \exp \left[ \frac{i \, g}{2} \, \int\limits_{-\infty}^{+\infty} d x^+ \, T^a \, A^{- \, a} 
     (x^+, x^- =0, \ul x ) \right] \\
   U_{\ul x}^{ba} &\equiv& {\mathcal P} \exp \left[ \frac{i \, g}{2} \, \int\limits_{-\infty}^{+\infty} d x^+ \, t^c \, A^{- \,
       c} (x^+, x^-=0, \ul x) \right]^{ba} \, ,
 \end{eqnarray}
where $t^a$'s are the SU($N_c$) generators in the adjoint
representation and the projectile is moving along the light-cone
$x^+$-axis. In essence, this means that the the projectile's
transverse position is not altered during the scattering, and the
effect of the target field is to perform a net SU($N_c$) color
rotation on the projectile.  The Wilson lines resum these interactions
and give the total phase of that color rotation.  They are illustrated
in \fig{fig-Wilson lines}. Note that the adjoint Wilson line $U_{\ul
  x}^{ba}$ is real-valued.

The Wilson-line approach is quite generic: if the target gluon field
is quasi-classical, as in the case of the McLerran--Venugopalan (MV)
model \cite{McLerran:1993ni,McLerran:1993ka,McLerran:1994vd} discussed in Sec.~\ref{sec-classical_fields}, then
correlators of the Wilson lines resum powers of $\as^2 \, A^{1/3}$
corresponding to the Glauber-Gribov-Mueller (GGM) multiple-rescattering
approximation \cite{Mueller:1989st} of Sec.~\ref{subsec-GGM}. Non-linear small-$x$ evolution
resumming powers of $\as \, Y \sim \as \, \ln s$ can be included into
the correlators of the Wilson lines through the Balitsky--Kovchegov
(BK)
\cite{Balitsky:1996ub,Balitsky:1997mk,Balitsky:1998ya,Kovchegov:1999yj,Kovchegov:1999ua}
and Jalilian-Marian--Iancu--McLerran--Weigert--Leonidov--Kovner
(JIMWLK)
\cite{Jalilian-Marian:1997xn,Jalilian-Marian:1997jx,Jalilian-Marian:1997gr,Jalilian-Marian:1997dw,Jalilian-Marian:1998cb,Kovner:2000pt,Weigert:2000gi,Iancu:2000hn,Ferreiro:2001qy}
evolution equations mentioned in Sec.~\ref{sec-Evolution}. Thus expressing the interaction with the target
in terms of the Wilson lines \peq{eq-Wilson lines} allows for several
different levels of approximation for this interaction.

\begin{figure}[th]
\centering
 \includegraphics[width=\textwidth]{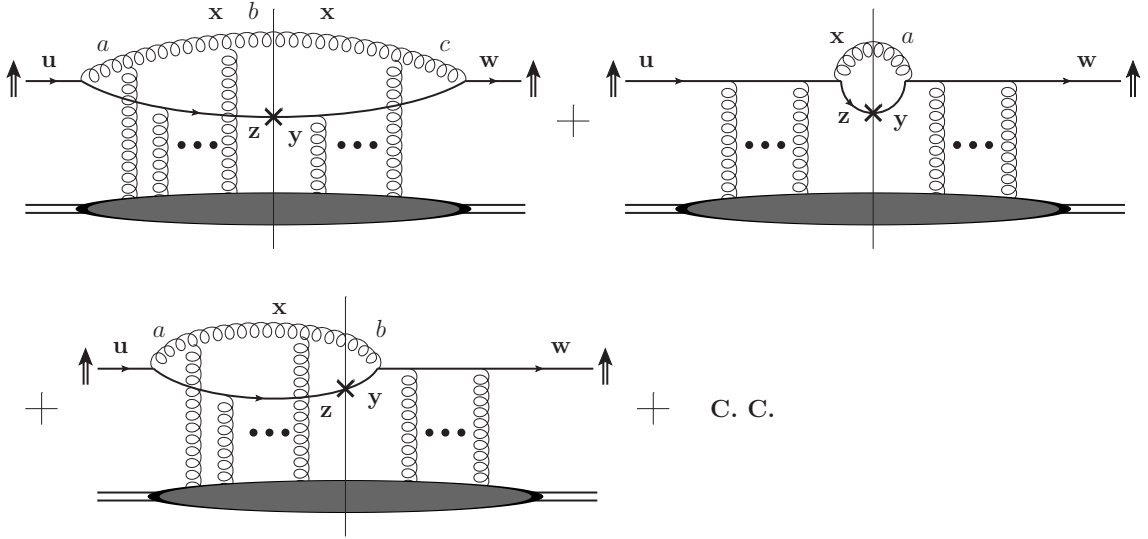}
 \caption{The cross section for quark production in the polarized quark--nucleus scattering.}
\label{fig-amplitude squared} 
\end{figure}

The scattering amplitude for quark production is composed of two
sub-processes: the splitting of \eqref{eq-coord wavefn} calculated in
Sec.~\ref{subsec-Wavefn} and the Wilson line scattering
\eqref{eq-Wilson lines} of the quark and the gluon in the field of the
target.  These elements give two distinct diagrams contributing to the
scattering amplitude shown in \fig{qtoqGampl} above for $\alpha < 1$. To find the quark
production cross section we need to square the diagrams in
\fig{qtoqGampl}, keeping the transverse momentum of the quark fixed,
as depicted in \fig{fig-amplitude squared}. As discussed above, this
implies that the transverse coordinates of the quark are different on
both sides of the cut.  Just like in other similar calculations
\cite{Kovchegov:1998bi,Albacete:2006vv}, the $q \to q \, G$ splitting
may occur with either initial-state or final-state interactions with the
target, both in the amplitude and in the complex conjugate amplitude, resulting in
the four different terms shown in \fig{fig-amplitude squared}.  Diagrams with both
initial- and final-state interactions correspond to $q \to q \, G$ splitting during the brief interaction with the target, which is suppressed by the center-of-mass energy $s$ \cite{Kovchegov:1998bi}.

Using Fig.~\ref{fig-amplitude squared} we can write down the
expression for the energy-rescaled, color-averaged amplitude squared $\left\langle A^2
\right\rangle$ in terms of Wilson lines and the wave function
responsible for the splitting, remembering to reverse the sign in the
wave function for splitting occurring after the interaction:
\begin{eqnarray}
 \label{eq-amplitude squared}
 \left\langle |A|^2 \right\rangle &=& \frac{1}{N_c} \sum_{\lambda, \chi'} 
 \bigg[ 
 \mathrm{Tr} \left[ V_{\ul
     z} \, \psi_{\lambda \chi \chi'}^a \, \psi_{\lambda \chi \chi'}^{c \, \dagger} \, V_{\ul
     y}^\dagger \right] \, U_{\ul x}^{b a} \, U_{\ul x}^{b c} + 
 \mathrm{Tr} \left[ \psi^a_{\lambda
     \chi \chi'} \, V_{\ul u} \, V^\dagger_{\ul w} \, \psi^{a \, \dagger}_{\lambda \chi 
     \chi'} \right]
 \\ \nonumber &-&
 \mathrm{Tr} \left[ V_{\ul z} \, \psi^a_{\lambda \chi \chi'} \, V^\dagger_{\ul w} \,
   \psi^{b \, \dagger}_{\lambda \chi \chi'} \right] \, U^{b a}_{\ul x} - \mathrm{Tr} \left[
   \psi^a_{\lambda \chi \chi'} \, V_{\ul u} \, \psi^{b \, \dagger}_{\lambda \chi \chi'} \, 
   V^\dagger_{\ul y} \right] \,  U^{a b}_{\ul x} 
 \bigg] \; ,
\end{eqnarray}
where $N_c$ is the number of colors, the traces are taken over the
fundamental representation indices, and summation is implied over
repeated adjoint color indices.  Substituting Eq.~\eqref{eq-Psi vs
  Phi} into \eqref{eq-amplitude squared} and using the identities (c.f. \eqref{Wident})
\begin{equation}
 \label{eq-identities}
 U^{b a}_{\ul x} \, T^a = V^\dagger_{\ul x} \, T^b \, V_{\ul x}
 \;\;\;\; , \;\;\;\;
 \mathrm{Tr} \left[ A \, T^a \, B \, T^a \right] = \frac{1}{2} \, 
\mathrm{Tr} A \ \mathrm{Tr} B - \frac{1}{2 N_c} \, \mathrm{Tr} \left[ A \, B \right]
\end{equation}
for arbitrary $N_c \times N_c$ matrices $A, \, B$, we find
\begin{equation}\label{eq-amplitude squared 1}
\left\langle |A|^2 \right\rangle = C_F \, \delta^2 \big[ \ul u - \ul x -
 \alpha \, (\ul z - \ul x) \big] \, \delta^2 \big[ \ul w - \ul x -
 \alpha \, (\ul y - \ul x) \big] \, \Phi_\chi (\ul z - \ul x , \ul y - \ul x) 
 \  \mathcal{I}^{(q)}
\end{equation}
where the factor responsible for the quark's interaction with the
target, denoted by $\mathcal{I}^{(q)}$, is given by
\begin{align}
\mathcal{I}^{(q)} &=  \bigg\langle \frac{1}{N_c} \, \mathrm{Tr} \, 
\left[ V_{\ul z} \, V^\dagger_{\ul y} \right] 
+ \frac{1}{N_c} \, \mathrm{Tr} \, \left[ V_{\ul u} \, V^\dagger_{\ul w} \right]  
- \frac{1}{2 \, N_c \, C_F} \,  \mathrm{Tr} \, \left[ V_{\ul z} \, V^\dagger_{\ul x} \right]  
\, \mathrm{Tr} \, \left[ V_{\ul x} \, V^\dagger_{\ul w} \right]  
\\ \nonumber &+
\frac{1}{2 \, N_c^2 \, C_F} \, \mathrm{Tr} \, \left[ V_{\ul z} \, V^\dagger_{\ul w} \right]
- \frac{1}{2 \, N_c \, C_F} \, \mathrm{Tr} \, \left[ V_{\ul u} \, V^\dagger_{\ul x} \right]  \, 
\mathrm{Tr} \, \left[ V_{\ul x} \, V^\dagger_{\ul y} \right] + 
\frac{1}{2 \, N_c^2 \, C_F} \, \mathrm{Tr} \, \left[ V_{\ul u} \, V^\dagger_{\ul y} \right] 
\bigg\rangle .
\end{align}
Here $C_F = (N_c^2 - 1)/2 N_c$ is the fundamental Casimir operator of
SU($N_c$), and the angle brackets on the right denote averaging over
the field configurations of the target.

Defining the $S$-matrix operator for a fundamental-representation
color dipole by
\begin{equation}
  \label{Ddef}
  {\hat D}_{\ul x \, \ul y} \equiv \frac{1}{N_c} \, \mathrm{Tr} \, \left[ V_{\ul x}
 \, V^\dagger_{\ul y} \right]
\end{equation}
we can rewrite $\mathcal{I}^{(q)}$ more compactly as
\begin{align}
 \label{Iq}
 \mathcal{I}^{(q)} &= \bigg\langle {\hat D}_{\ul z \, \ul y} + {\hat D}_{\ul u \, \ul w} 
 - \frac{N_c}{2 \, C_F} \, {\hat D}_{\ul z \, \ul x} \, {\hat D}_{\ul x \, \ul w} + 
 \frac{1}{2 \, N_c \, C_F} \, {\hat D}_{\ul z \, \ul w} - 
 \frac{N_c}{2 \, C_F} \, {\hat D}_{\ul u \, \ul x} \, {\hat D}_{\ul x \, \ul y} 
 \\ \nonumber &+
 \frac{1}{2 \, N_c \, C_F} \, {\hat D}_{\ul u \, \ul y} \bigg\rangle . 
\end{align}
As we have already mentioned, this interaction with the target can be
evaluated either in the Glauber-Gribov-Mueller multiple-rescattering
approximation of Sec.~\ref{subsec-GGM} or using the JIMWLK evolution equation.

The expression \peq{Iq} simplifies in 't Hooft's large-$N_c$ limit \cite{'tHooft:1973jz}, which corresponds to taking the number of colors $N_c$ to be large and the coupling $\alpha_s$ to be small such that the product is constant:
\begin{align}
 \label{largeNc}
 N_c \gg 1 \hspace{1cm} \alpha_s \ll 1 \hspace{1cm} \alpha_s N_c = const .
\end{align}
In this limit, the correlators of several single-trace operators factorize,
such that, for instance, $\langle {\hat D}_{\ul u \, \ul x} \, {\hat
  D}_{\ul x \, \ul y} \rangle = \langle {\hat D}_{\ul u \, \ul x}
\rangle \, \langle {\hat D}_{\ul x \, \ul y} \rangle$
\cite{Balitsky:1996ub,Kovchegov:1999yj,Weigert:2005us}.  Defining
\begin{equation}
 \label{eq-dipole propagator}
 D_{\ul x \, \ul y} \equiv \left\langle {\hat D}_{\ul x \, \ul y} \right\rangle
= \frac{1}{N_c} \left\langle \mathrm{Tr} \, \left[ V_{\ul x}
 \, V^\dagger_{\ul y} \right] \right\rangle
\end{equation}
we rewrite \eqref{Iq} in the large-$N_c$ limit as
\begin{equation}
 \label{eq-interaction}
 \mathcal{I}^{(q)} \bigg|_{\mbox{large}-N_c} = D_{\ul z \, \ul y} + D_{\ul u \, \ul w} -
 D_{\ul z \, \ul x} \, D_{\ul x \, \ul w} - D_{\ul u \,
 \ul x} \, D_{\ul x \, \ul y} \; .
\end{equation}

\begin{figure}
 \centering
 \includegraphics[width=0.4\textwidth]{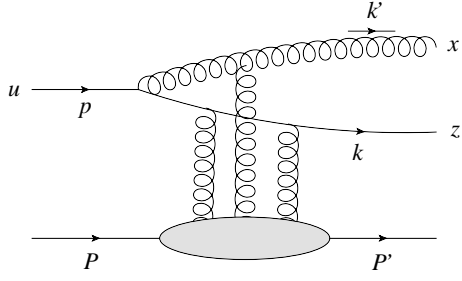}
 \caption{Illustration of the kinematics for the rescaled amplitude $\mathcal{A}$ being related to the cross-section.}
 \label{fig-LCPT_CS}
\end{figure}

To compute the quark production cross sections, we need to Fourier
transform the coordinate space amplitude squared of \eqref{eq-amplitude
  squared 1} back to momentum space and include the appropriate
kinematic factors.  We will now derive the relation between $\langle A^2 \rangle$ and the cross-section; for this it is useful to label the momenta of the incoming and outgoing particles as in Fig.~\ref{fig-LCPT_CS} without imposing momentum conservation a priori, since it will be handled explicitly by the formulas.
The standard expression for the cross-section is \cite{Peskin:1995ev}
\begin{align}
 \label{LCPT_CS1}
 d\sigma &= \frac{1}{8 E_p E_P} \left[\frac{d^{2+}k'}{2(2\pi)^3 k'^+} \right]
 \left[\frac{d^{2+}k}{2(2\pi)^3 k^+} \right] \left[\frac{d^{2-}P'}{2(2\pi)^3 P'^-} \right]
 \left\langle \mathcal{M}(p,k,k') \right\rangle^2 
 \\ \nonumber &\times
 (2\pi)^4 \delta^4 (p + P - k - k' - P'),
\end{align}
where $P'^- \approx 2 E_P$, $p^+ \approx 2 E_p$, and $\mathcal{M}$ is the usual scattering amplitude as calculated from Feynman diagrams.  Light-cone perturbation theory (LCPT) relates the cross-section to an amplitude $A$ which has been rescaled by the center-of-mass energy as in \eqref{eik8}; also, to use the LCPT rules given in \cite{Kovchegov:2012mbw}, we need to account for a different normalization of the wave function \eqref{eq-coord wavefn}.  Thus we define
\begin{align}
 \label{LCPT_CS2}
 A \equiv \frac{1}{2 p^+ P^-} \sqrt{\frac{p^+}{k^+}} \mathcal{M}.
\end{align}
Additionally, we can use the dominant kinematics to simplify the delta function:
\begin{align}
 \label{LCPT_CS3}
 \delta^4(p + P - k - k' - P') = 2 \delta(p^+ - k^+ - k'^+) \, \delta(P^- - P'^-) \,
 \delta^2 (-\ul{k} - \ul{k}' - \ul{P}');
\end{align}
using \eqref{LCPT_CS2} and \eqref{LCPT_CS3} in \eqref{LCPT_CS1} gives
\begin{align}
 \label{LCPT_CS4}
 \frac{d\sigma}{d^2 k dy_q} = \frac{1}{2(2\pi)^3} \frac{k^+}{p^+ - k^+} \int \frac{d^2 k'}{(2\pi)^2}
 \left\langle A(p,k,k')^2 \right\rangle,
\end{align}
where we have used the delta function \eqref{LCPT_CS3} to integrate over $d^{2-}P'$ and $d k'^+$.  Now, Fourier-transforming the amplitude and complex-conjugate amplitude
\begin{align}
 \label{LCPT_CS5}
 A(p,k,k') &= \int d^2 u \, d^2 z \, d^2 x \, e^{i \ul{p} \cdot \ul{u}} \,
 e^{-i \ul{k} \cdot \ul{z}} \, e^{-i \ul{k}' \cdot \ul{x}} \,
 A(x, z, u)
 \\ \nonumber
 A^*(p,k,k') &= \int d^2 w \, d^2 y \, d^2 x' \, e^{-i \ul{p} \cdot \ul{w}} \,
 e^{+i \ul{k} \cdot \ul{y}} \, e^{+i \ul{k}' \cdot \ul{x}'} \, A^*(x', y, w),
\end{align}
we see that the $d^2 k'$ integral in \eqref{LCPT_CS4} will generate a delta function $\delta^2 (\ul{x} - \ul{x}')$ which sets the coordinate of the untagged gluon equal in the amplitude and complex-conjugate amplitude.  Altogether, this gives the expression for the cross-section in terms of \eqref{eq-amplitude squared 1} as
\begin{equation}
 \label{eq-inverse Fourier transform}
 \frac{d\sigma^{(q)}}{d^2 k \, d y_q} 
 = \frac{1}{2 \, (2\pi)^3} \, \frac{\alpha}{1-\alpha} \, 
 \int d^2 x \, d^2 y \, d^2 z \, d^2 u \, d^2 w \, e^{-i \ul k \cdot (\ul z - 
 \ul y)} \, e^{i \ul p \cdot (\ul u - \ul w)} \, 
 \langle A^2 \rangle \; 
\end{equation}
with $\ul k$ and $y_q$ the transverse momentum and rapidity of the
produced quark. Integrating over the delta functions from the wave function in
\eqref{eq-amplitude squared 1} imposes the kinematic constraints
 \begin{eqnarray}
  \label{eq-kinematic constraints}
  \ul u = \ul x + \alpha \, (\ul z - \ul x) \\ 
  \ul w = \ul x + \alpha \, (\ul y - \ul x)
 \end{eqnarray}
which relate the quark coordinates before and after the $q \to q \, G$
splitting and describe the non-eikonal quark recoil.  To make the
incoming quark transversely polarized we need to put the transverse
momentum of the incoming quark to zero: $\ul p = \ul 0$. We thus
obtain the general result for quark production in $q^\uparrow A$ scattering
\begin{equation}
 \label{eq-cross section}
 \frac{d\sigma^{(q)}}{d^2 k \, d y_q} = \frac{C_F}{2 \, (2\pi)^3} \, 
 \frac{\alpha}{1-\alpha} \, \int
 d^2 x \, d^2 y \, d^2 z \, e^{-i \ul k \cdot (\ul z - \ul y)} \,
 \Phi_\chi (\ul z - \ul x \, , \, \ul y - \ul x, \alpha) \
 \mathcal{I}^{(q)} (\ul x \, , \, \ul y \, , \, \ul z)
\end{equation}
with $\Phi_\chi$ from \eqref{eq-wavefn squared} and $\mathcal{I}^{(q)}$
from \eqref{Iq}. The expression \eqref{eq-cross section} contains multiple
rescatterings and non-linear small-$x$ evolution between the
projectile and the target. Note that it does not resum the small-$x$
evolution between the produced quark and the projectile (which can be
included following \cite{Albacete:2006vv}), and hence is not valid for
very small $\alpha$ (i.e., the values of $\alpha$ are restricted by
$\as \, \ln \tfrac{1}{\alpha} \ll 1$). Since, as we will see below, both the
experimental STSA and the STSA resulting from our production mechanism
fall off with decreasing $\alpha$, the region of interest in this work
corresponds to $\alpha$ not being very small, where \eqref{eq-cross
  section} is fully applicable.


\subsection{Spin, Asymmetry, and C-Parity in Quark Production}
\label{subsec-symmetries}

Using \eqref{eq-cross section} we can explicitly determine the parts of the wave function \eqref{eq-wavefn squared} and the interaction \eqref{Iq} which couple to the numerator $d(\Delta\sigma)$ and denominator $d\sigma_{unp}$ of the asymmetry \eqref{AN1v2}.  In Sec.~\ref{subsubsec-STSA} we demonstrated that the reversal of transverse spin $\chi \rightarrow - \chi$ is equivalent to a transverse rotation which reverses the direction of the transverse momentum $\ul{k} \rightarrow -\ul{k}$ (see Fig.~\ref{STSArot}).  This was the reason for the two equivalent definitions of STSA as written in \eqref{AN1v2}.

Therefore we can project out the symmetric part $d\sigma_{unp}$ of the cross-section by explicitly symmetrizing \eqref{eq-cross section} with respect to both spin flip $\chi \rightarrow - \chi$ and momentum reversal $\ul{k} \rightarrow -\ul{k}$.  Similarly, we can project out the antisymmetric part $d(\Delta\sigma)$ by explicitly \textit{anti}symmetrizing \eqref{eq-cross section} under spin flip and momentum reversal.  Since the eikonal interaction \eqref{Iq} is independent of the spin eigenvalue $\chi$, the (anti)symmetrization with respect to spin simply selects the polarized \eqref{eq-polarized wavefn} or unpolarized \eqref{eq-unpolarized wavefn} parts of the wave function squared; thus we write
\begin{align}
 \label{symm/anti1}
 d(\Delta\sigma) &= \frac{C_F}{(2\pi)^3} \frac{\alpha}{1-\alpha} 
 \frac{1}{2} \int d^2 x \, d^2 y \, d^2 z \, e^{- i \ul k \cdot (\ul z - \ul y)} \, 
 \Phi_{pol} (\ul z - \ul x , \ul y - \ul x , \alpha) \, \mathcal{I}^{(q)}(\ul x , \ul y , \ul z) 
 \\ \nonumber & - \: (\ul{k} \rightarrow - \ul{k}) 
 \\ \nonumber \, \\ \nonumber
 d\sigma_{unp} &= \frac{C_F}{2(2\pi)^3} \frac{\alpha}{1-\alpha} 
 \frac{1}{2} \int d^2 x \, d^2 y \, d^2 z \, e^{- i \ul k \cdot (\ul z - \ul y)} \, 
 \Phi_{unp} (\ul z - \ul x , \ul y - \ul x , \alpha) \, \mathcal{I}^{(q)}(\ul x , \ul y , \ul z)
 \\ \nonumber & + \:  (\ul{k} \rightarrow - \ul{k}) ,
\end{align}
where the relative factor of $2$ between $d(\Delta\sigma)$ and $d\sigma_{unp}$ arises because $d(\Delta\sigma)$ is defined as the full difference between the cross-sections $d\sigma^\uparrow - d\sigma^\downarrow$ rather than the half-difference.  

Reversing the transverse momentum $\ul{k} \rightarrow - \ul{k}$ is equivalent to interchanging the coordinates $\ul{z} \leftrightarrow \ul{y}$ in the Fourier factor, and we note that both the polarized \eqref{eq-polarized wavefn} and unpolarized wave functions \eqref{eq-unpolarized wavefn} are invariant under this exchange.  Thus the (anti)symmetrization under spin flip only affects the wave functions, while the (anti)symmetrization under momentum reversal only affects the interactions:
\begin{align}
 \label{symm/anti2}
 d(\Delta\sigma) &= \frac{C_F}{(2\pi)^3} \frac{\alpha}{1-\alpha} \int d^2 x \, d^2 y \, d^2 z \,
 e^{- i \ul k \cdot (\ul z - \ul y)} \, \Phi_{pol} (\ul z - \ul x , \ul y - \ul x , \alpha)
 \\ \nonumber &\times \, \frac{1}{2} \left[
 \mathcal{I}^{(q)}(\ul x , \ul y , \ul z) - (\ul z \leftrightarrow \ul y) \right]
 \\ \nonumber
 d\sigma_{unp} &= \frac{C_F}{2(2\pi)^3} \frac{\alpha}{1-\alpha} \int d^2 x \, d^2 y \, d^2 z \,
 e^{- i \ul k \cdot (\ul z - \ul y)} \, \Phi_{unp} (\ul z - \ul x , \ul y - \ul x , \alpha)
 \\ \nonumber &\times \, \frac{1}{2} \left[
 \mathcal{I}^{(q)}(\ul x , \ul y , \ul z) + (\ul z \leftrightarrow \ul y) \right] .
\end{align}
These (anti)symmetric combinations of spin-dependence in the wave function and $\ul{z}\leftrightarrow\ul{y}$ dependence in the interaction are the only ones which are consistent with rotational invariance as illustrated in Fig.~\ref{STSArot}.  An explicit cross-check verifies that the other combinations are identically zero for precisely this reason \cite{Kovchegov:2012ga}.

Motivated by this observation, we define the symmetric and antisymmetric parts of the interaction as
\begin{equation}
 \label{eq-defn interaction parity}
 \mathcal{I}_{symm \, / \, anti} \equiv \frac{1}{2} \bigg( \mathcal{I} \pm (\ul z 
 \leftrightarrow \ul y) \bigg).
\end{equation}
To understand the meaning of this symmetrization, let us note that in the dipole trace \eqref{eq-dipole propagator}, the Wilson line of a quark in the complex-conjugate amplitude is equivalent to the Wilson line of an antiquark in the amplitude \cite{Kovchegov:2003dm,Hatta:2005as}.  Thus the interaction \eqref{Iq} for this quark production process can be expressed in terms of quark-antiquark color dipoles \eqref{eq-dipole propagator}.  The exchange of coordinates $\ul z \leftrightarrow \ul y$ then is equivalent to swapping the roles of the quark and antiquark.  This is accomplished by \textit{charge conjugation} $C$, which indeed transforms the Wilson lines and dipole operators as
\begin{align}
 \label{WPT7}
 C V_{\ul x} C^\dagger &= V_{\ul x}^* = \left( V_{\ul x}^\dagger \right)^T
 \\ \nonumber
 C \hat{D}_{\ul z \ul y} C^\dagger &= \hat{D}_{\ul y \ul z}.
\end{align}
Thus it is natural to decompose each dipole $S$-matrix into the
even and odd pieces under charge conjugation:
 \begin{eqnarray}
  {\hat D}_{\ul x \, \ul y} &\equiv& {\hat S}_{\ul x \, \ul y} + i \,
  {\hat O}_{\ul x \, \ul y} \\
  {\hat S}_{\ul x \, \ul y} &\equiv& \frac{1}{2} \, ( {\hat D}_{\ul x \, \ul y} +
  {\hat D}_{\ul y \, \ul x}) \label{Sdef} \\
  {\hat O}_{\ul x \, \ul y} &\equiv& \frac{1}{2i} \, ( {\hat D}_{\ul x \, \ul y} -
  {\hat D}_{\ul y \, \ul x}) \; . \label{Odef}
 \end{eqnarray}
The $C$-even real part of the target-field-averaged $S$-matrix $S_{\ul
  x \, \ul y} \equiv \langle {\hat S}_{\ul x \, \ul y} \rangle$ is
responsible for the total unpolarized cross section of the
dipole--target interactions. Its small-$x$ evolution is given by the
BK/JIMWLK equations.  The $C$-odd imaginary part of the
target-averaged $S$-matrix $O_{\ul x \, \ul y} \equiv \langle {\hat
  O}_{\ul x \, \ul y} \rangle$ is known as the odderon interaction
\cite{Lukaszuk:1973nt,Nicolescu:1990ii,Ewerz:2003xi}. The small-$x$
evolution equation for $O_{\ul x \, \ul y}$ was constructed in
\cite{Kovchegov:2003dm,Hatta:2005as,Kovner:2005qj}, and, in the linear
approximation, was found to be identical to the dipole BFKL equation
\cite{Mueller:1994rr} with $C$-odd initial conditions. For the current status of the experimental
searches for the QCD odderon and for an overview of the theory see
\cite{Ewerz:2003xi}.

With these explicitly symmetrized elements, it is straightforward to
construct the symmetric and antisymmetric parts of the interaction
with the target \peq{Iq} for quark production:
 \begin{eqnarray}
  \label{eq-symmetrized interaction}
  \mathcal{I}_{symm}^{(q)} &=& \left\langle {\hat S}_{\ul z \, \ul y} +
  {\hat S}_{\ul u \, \ul w} - \frac{N_c}{2 \, C_F} \, \left( {\hat S}_{\ul z \, \ul x} \, 
  {\hat S}_{\ul x \, \ul w} - {\hat O}_{\ul z \,  \ul x} \, {\hat O}_{\ul x \, \ul w} \right) 
  + \frac{1}{2 \, N_c \, C_F} \, {\hat S}_{\ul z \, \ul w} \right. \notag \\ 
  && \left. - \frac{N_c}{2 \, C_F} \, \left(
  {\hat S}_{\ul u \, \ul x} \, {\hat S}_{\ul x \, \ul y} -  {\hat O}_{\ul u \, \ul x} \, 
  {\hat O}_{\ul x \, \ul y} \right) 
  + \frac{1}{2 \, N_c \, C_F} \, {\hat S}_{\ul u \, \ul y} \right\rangle , \\
  \mathcal{I}_{anti}^{(q)} &=& i \, \left\langle {\hat O}_{\ul z \, \ul y} +
  {\hat O}_{\ul u \, \ul w} - \frac{N_c}{2 \, C_F} \, 
  \left( {\hat O}_{\ul z \, \ul x} \, {\hat S}_{\ul x \, \ul w} 
   + {\hat S}_{\ul z \, \ul x} \, {\hat O}_{\ul x \, \ul w} \right) 
  + \frac{1}{2 \, N_c \, C_F} \, {\hat O}_{\ul z \, \ul w} \right. \notag \\ 
  && \left. - \frac{N_c}{2 \, C_F} \, 
  \left(
  {\hat O}_{\ul u \, \ul x} \, {\hat S}_{\ul x \, \ul y} +  {\hat S}_{\ul u \, \ul x} \,
  {\hat O}_{\ul x \, \ul y} \right) 
  + \frac{1}{2 \, N_c \, C_F} \, {\hat O}_{\ul u \, \ul y} \right\rangle \; .
 \end{eqnarray}
In the large-$N_c$ limit these expressions simplify to
 \begin{align}
  \mathcal{I}_{symm}^{(q)}\bigg|_{\mbox{large}-N_c} &= S_{\ul z \, \ul y} +
  S_{\ul u \, \ul w} - S_{\ul z \, \ul x} \, S_{\ul x \, \ul w} -
  S_{\ul u \, \ul x} \, S_{\ul x \, \ul y} + O_{\ul z \,
  \ul x} \, O_{\ul x \, \ul w} + O_{\ul u \, \ul x} \, 
  O_{\ul x \, \ul y} , \label{qsymm} \\
  \mathcal{I}_{anti}^{(q)} \bigg|_{\mbox{large}-N_c} &= i \left[ O_{\ul z \, \ul y} +
  O_{\ul
  u \, \ul w} - O_{\ul z \, \ul x} \, S_{\ul x \, \ul w} -
  O_{\ul u \, \ul x} \, S_{\ul x \, \ul y} - S_{\ul z \,
  \ul x} \, O_{\ul x \, \ul w} - S_{\ul u \, \ul x} \,
  O_{\ul x \, \ul y} \right] \; . \label{qanti}
 \end{align}
Knowing these symmetry properties, we can summarize our results for the
spin-dependent and spin-averaged cross sections $d(\Delta \sigma)$ and
$d\sigma_{unp}$ as
 \begin{align}
  &d(\Delta\sigma^{(q)}) = \frac{C_F}{(2\pi)^3} \, \frac{\alpha}{1-\alpha} \, 
  \int d^2 (x y z)
  \, e^{-i \ul k \cdot (\ul z - \ul y)} \, \Phi_{pol} (\ul z - \ul x \, , \, \ul y - \ul x, \alpha) \
  \mathcal{I}^{(q)}_{anti} (\ul x \, , \, \ul y \, , \, \ul z) \label{dsigmaq} 
	\\
  &d\sigma^{(q)}_{unp} = \frac{C_F}{2 \, (2\pi)^3} \, \frac{\alpha}{1-\alpha} \, 
  \int d^2 (x y z)
  \, e^{-i \ul k \cdot (\ul z - \ul y)} \, \Phi_{unp} (\ul z - \ul x \, , \, \ul y - \ul x, \alpha) \
  \mathcal{I}^{(q)}_{symm} (\ul x \, , \, \ul y \, , \, \ul z) \label{dsigmaq_unp}
 \end{align}
where the wave functions squared are given by Eqs.
\eqref{eq-unpolarized wavefn}, \eqref{eq-polarized wavefn}, and the
interactions are given by Eqs.~\eqref{eq-symmetrized interaction} (and
by Eqs.~\peq{qsymm} in the large-$N_c$ limit).
Eqs.~\eqref{dsigmaq} and \eqref{dsigmaq_unp} are one of the main results of this work:
together with \eqref{AN1v2} they give the single-transverse spin
asymmetry $A_N$ generated in quark production by the $C$-odd CGC
interactions with the target.  

As we have discussed in \ref{subsubsec-STSA} and has been emphasized in the literature, the time reversal transformation $T$ plays an integral role in the origin of STSA \cite{Collins:1992kk}.  Consider the transformation of a single Wilson line under time reversal $T$:
\begin{align}
 \label{WPT1}
 T V_{\ul x} T^\dagger = \left( V_{\ul x}^\dagger \right)^* = V_{\ul x}^T ;
\end{align}
then the transformation of a dipole operator $\hat{D}_{\ul z \ul y}$ is
\begin{align}
 \label{WPT2}
 T \hat{D}_{\ul z \ul y} T^\dagger = \frac{1}{N_c} \Tr \left[ V_{\ul z}^T \left( V_{\ul y}^\dagger \right)^T \right] = \frac{1}{N_c} \Tr \left[ \left( V_{\ul y}^\dagger V_{\ul z} \right)^T \right] = \hat{D}_{\ul z  \ul y},
\end{align}
where the last step follows from the cyclicity of the trace and its invariance under transposition.  Thus a single dipole operator is invariant under time reversal.  But because of the antilinearity of $T$ (which complex conjugates any c-numbers), the odderon component is explicitly $T$-odd:
\begin{align}
 \label{WPTnew1}
 T {\hat O}_{\ul x \, \ul y} T^\dagger = 
 \left(\frac{1}{2i}\right)^* \, T ( {\hat D}_{\ul x \, \ul y} - {\hat D}_{\ul y \, \ul x}) T^\dagger 
 = - \frac{1}{2i} \, ( {\hat D}_{\ul x \, \ul y} - {\hat D}_{\ul y \, \ul x}) 
 = - {\hat O}_{\ul x \ul y} .
\end{align}
It is interesting to note that in the high-energy approximation
considered here the application of time reversal \eqref{WPT2} to dipole correlators
is equivalent to the application of charge-conjugation \eqref{WPT7}, such that the STSA
arises from the odderon exchange, which is both $T$- and $C$-odd.  Thus we see that the $C,T$-odd odderon exchange leads to the $T$-odd STSA observable $A_N$.

The mechanism for the generation of the STSA in
Eqs.~\eqref{dsigmaq} is different from both the well-known
Sivers \cite{Sivers:1989cc,Sivers:1990fh} and Collins
\cite{Collins:1992kk} effects. It appears difficult (if not
impossible) to absorb the interactions of \fig{fig-amplitude squared}
into the projectile wave function (distribution function): hence our
result is different from the Sivers effect. In the above calculation
the asymmetry is generated before fragmentation; hence the STSA
resulting from Eqs.~\eqref{dsigmaq} cannot be due to the Collins
effect either.  As we will see below, the non-zero part of
\eqref{dsigmaq} stems from the multiple interactions with the target
(higher-twist effects), and its contribution is in fact zero in the
linearized (leading-twist) approximation. In this sense the above
mechanism for generating STSA is similar in spirit to the higher-twist
mechanisms of
\cite{Efremov:1981sh,Efremov:1984ip,Qiu:1991pp,Ji:1992eu,Qiu:1998ia,Brodsky:2002cx,Collins:2002kn,Koike:2011mb,Kanazawa:2000hz,Kanazawa:2000kp},
though a detailed comparison of the diagrams appears to indicate that
the two approaches are, in fact, different.

We have shown explicitly that the single-transverse spin asymmetry
$A_N$ occurs in the CGC framework as a coupling between the transverse
spin of the projectile and a $C$-odd interaction with the target,
driven by the odderon.\footnote{In the past, the relation between the
  odderon and the single and double transverse spin asymmetries was
  investigated in
  \cite{Ahmedov:1999ne,Jarvinen:2006dm,Leader:1999ua,Buttimore:1998rj,Trueman:2007fr}
  in the pomeron and reggeon formalism.} Note that to date there is no
unambiguous experimental evidence for the QCD odderon. If our
mechanism for generating STSA can be isolated experimentally from
other contributions, it may constitute the first direct observation of
the QCD odderon! To make such a distinction possible, one needs to
determine phenomenological characteristics of our mechanism, such as
its rapidity, energy, and centrality dependence; some of this work
will be carried out below, while the rest, along with a proper
phenomenological implementation of our results, is left for future
work.

Finally, the reader may wonder whether the cross section in
\eqref{dsigmaq} is demonstrably non-zero. While it is very difficult to carry
out the integration in \eqref{dsigmaq} exactly, we instead will evaluate
\eqref{dsigmaq} approximately in Sec.~\ref{sec-Estimates}, showing that
the cross section and the corresponding STSA $A_N$ are in fact
non-zero. However, first we would like to derive the analogues of
Eqs.~\eqref{dsigmaq} for gluon and prompt photon production.



\subsection{STSA in Gluon and Photon Production}
 
\label{subsec-Gluon Photon STSA (gen)}

Having laid out the methodology in Section~\ref{subsec-Quark STSA
  (gen)}, we can now perform similar calculations
of STSA for the cases of gluon and photon production.

\begin{figure}[ht]
\centering
  \includegraphics[width=\textwidth]{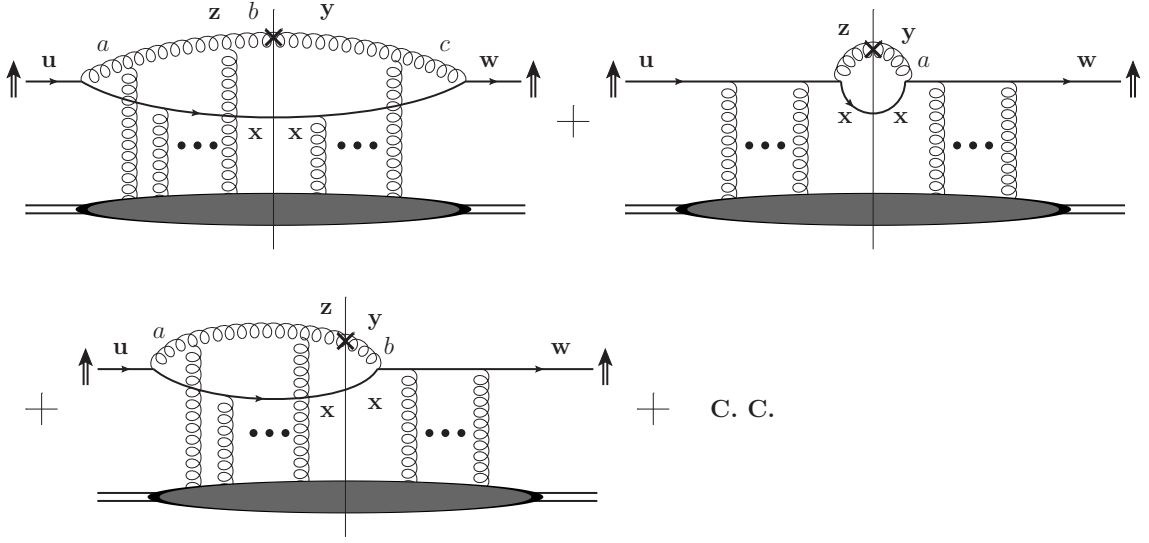}
 \caption{Diagrams contributing to the gluon / photon production cross section.}
\label{fig-gluon diagrams} 
\end{figure}

We begin with the gluon production. The gluon production diagrams are
shown in \fig{fig-gluon diagrams}. Since now we tag on the gluon, its
transverse-space positions are different on both sides of the cut, now
denoted $\ul z$ and $\ul y$, while the untagged quark has the same
transverse positions $\ul x$ in the amplitude and in the complex
conjugate amplitude. We see that to obtain the gluon production cross
section from the quark production expression found in the previous
Section, we need to interchange
\begin{equation}
  \label{interchange}
  {\ul z} \leftrightarrow {\ul x} \ \ \ \mbox{and} \ \ \  {\ul y} \leftrightarrow {\ul x}
\end{equation}
in the wave function and its complex conjugate correspondingly. In
addition, since we are interested in the differential cross section
per unit gluon rapidity $y_G$, we use
\begin{equation}
  \label{Gtoq_rap}
  d y_G = \frac{\alpha}{1 - \alpha} \, d y_q
\end{equation}
(with $\alpha$ still the fraction of the incoming quark's longitudinal
momentum carried by the final-state quark).

These modifications lead to the following expressions for the
polarization-dependent and unpolarized cross sections for gluon
production:
 \begin{align}
  & d(\Delta\sigma^{(G)}) = \frac{C_F}{(2\pi)^3} \,  
  \int d^2 x \, d^2 y \, d^2 z
  \, e^{-i \ul k \cdot (\ul z - \ul y)} \, \Phi_{pol} (\ul x - \ul z \, , \, \ul x - \ul y, \alpha) \
  \mathcal{I}^{(G)}_{anti} (\ul x \, , \, \ul y \, , \, \ul z) \label{dsigmaG} \\
  & d\sigma^{(G)}_{unp} = \frac{C_F}{2 \, (2\pi)^3} \, 
  \int d^2 x \, d^2 y \, d^2 z
  \, e^{-i \ul k \cdot (\ul z - \ul y)} \, \Phi_{unp} (\ul x - \ul z \, , \, \ul x - \ul y, \alpha) \
  \mathcal{I}^{(G)}_{symm} (\ul x \, , \, \ul y \, , \, \ul z) \; , \label{Gunp}
 \end{align}
with $\Phi_{pol}$ and $\Phi_{unp}$ still given by
Eqs.~\peq{eq-polarized wavefn} and \peq{eq-unpolarized wavefn}.

The interaction with the target for the gluon production case can be
calculated along the similar lines to the above calculation of quark
production by using \fig{fig-gluon diagrams}, yielding
\begin{align}
  \label{IG}
  \mathcal{I}^{(G)} &= \bigg\langle {\hat D}_{\ul u \, \ul w} + 
\frac{N_c}{2 \, C_F} \, {\hat D}_{\ul z \, \ul y} \, {\hat D}_{\ul y \, \ul z} 
- \frac{1}{2 \, N_c \, C_F} 
- \frac{N_c}{2 \, C_F} \, {\hat D}_{\ul x \, \ul z} \, {\hat D}_{\ul z \, \ul w}  
+ \frac{1}{2 \, N_c \, C_F} \, {\hat D}_{\ul x \, \ul w} 
\\ \nonumber &-
\frac{N_c}{2 \, C_F} \, {\hat D}_{\ul u \, \ul y} \, {\hat D}_{\ul y \, \ul x}  
+ \frac{1}{2 \, N_c \, C_F} \, {\hat D}_{\ul u \, \ul x} 
\bigg\rangle .
\end{align}
Note that now
 \begin{eqnarray}
  \label{uvG}
  \ul u = \ul x + (1-\alpha) \, (\ul z - \ul x) \\ 
  \ul w = \ul x + (1-\alpha) \, (\ul y - \ul x)
 \end{eqnarray}
due to the interchanges of \eqref{interchange} carried out in
Eqs.~\peq{eq-kinematic constraints}.

Separating the interaction into the symmetric and anti-symmetric
components under ${\ul z} \leftrightarrow {\ul y}$ interchange one
obtains
\begin{align}
\label{IGsymmanti}
\mathcal{I}^{(G)}_{symm} &= \bigg\langle {\hat S}_{\ul u \, \ul w} + 
\frac{N_c}{2 \, C_F} \, \left( {\hat S}_{\ul z \, \ul y} \, {\hat S}_{\ul y \, \ul z} 
- {\hat O}_{\ul z \, \ul y} \, {\hat O}_{\ul y \, \ul z}\right)
- \frac{1}{2 \, N_c \, C_F} 
- \frac{N_c}{2 \, C_F} \, \left( {\hat S}_{\ul x \, \ul z} \, {\hat S}_{\ul z \, \ul w} - {\hat O}_{\ul x \, \ul z} \, {\hat O}_{\ul z \, \ul w} \right)
 \nonumber \\  &+
 \frac{1}{2 \, N_c \, C_F} \, {\hat S}_{\ul x \, \ul w}
 - \frac{N_c}{2 \, C_F} \, \left( {\hat S}_{\ul u \, \ul y} \, {\hat S}_{\ul y \, \ul x} - {\hat O}_{\ul u \, \ul y} \, {\hat O}_{\ul y \, \ul x} \right) 
+ \frac{1}{2 \, N_c \, C_F} \, {\hat S}_{\ul u \, \ul x} 
\bigg\rangle , \\
\mathcal{I}^{(G)}_{anti} &= i \, \bigg\langle {\hat O}_{\ul u \, \ul w} 
- \frac{N_c}{2 \, C_F} \, \left( {\hat S}_{\ul x \, \ul z} \, {\hat O}_{\ul z \, \ul w} + {\hat O}_{\ul x \, \ul z} \, {\hat S}_{\ul z \, \ul w} \right) 
+ \frac{1}{2 \, N_c \, C_F} \, {\hat O}_{\ul x \, \ul w} 
\nonumber \\  &-
\frac{N_c}{2 \, C_F} \, \left( {\hat S}_{\ul u \, \ul y} \, {\hat O}_{\ul y \, \ul x} + {\hat O}_{\ul u \, \ul y} \, {\hat S}_{\ul y \, \ul x} \right)
+ \frac{1}{2 \, N_c \, C_F} \, {\hat O}_{\ul u \, \ul x} 
\bigg\rangle ,
\end{align}
where we have used the fact that ${\hat O}_{\ul y \, \ul z} = - {\hat
  O}_{\ul z \, \ul y}$ which follows from the definition in \eqref{Odef}.

Finally, in the large-$N_c$ limit Eqs.~\peq{IGsymmanti} simplify to
 \begin{align}
 \label{eq-gluon production interaction}
  \mathcal{I}^{(G)}_{symm}\bigg|_{\mbox{large}-N_c} &= S_{\ul u \, \ul w} +
  \left(S_{\ul z \, \ul y}\right)^2 - S_{\ul x \, \ul z} \, S_{\ul z \, \ul w} 
- S_{\ul u \, \ul y} \, S_{\ul y \, \ul x}  + 
  \left( O_{\ul z \, \ul y}\right)^2  + O_{\ul x \, \ul z} \, O_{\ul z \, \ul w} 
	\\ \nonumber &+
  O_{\ul u \, \ul y} \, O_{\ul y \, \ul x}  \\
  \mathcal{I}^{(G)}_{anti}\bigg|_{\mbox{large}-N_c} &= i \, 
  \left[ O_{\ul u \, \ul w} - S_{\ul x \, \ul z} \, 
  O_{\ul z \, \ul w} - O_{\ul x \, \ul z} \, S_{\ul z \, \ul w} -
  S_{\ul u \, \ul y} \, O_{\ul y \, \ul x} - O_{\ul u \,
  \ul y} \, S_{\ul y \, \ul x}  \right] \; .
 \end{align}
Eqs.~\peq{IGsymmanti} and \peq{eq-gluon production interaction}, when used in \eqref{AN1v2},
give an expression for the gluon STSA in the CGC formalism. This is
another main result of this work.

Constructing the cross sections for prompt photon production out of
the gluon production cross sections we have just derived is
straightforward. One has to drop all color factors in the light-cone
wave functions, replace $\as \to \alpha_{EM} \, Z_f^2$ with the
electric charge $Z_f$ of a quark with flavor $f$ in units of the electron
charge, and recalculate the interaction with the target remembering
that the photon, in this lowest order in $\alpha_{EM}$ approximation
does not interact. One obtains the polarization-dependent and
unpolarized cross sections for photon production:
 \begin{align}
  &d(\Delta\sigma^{(\gamma)}) = \frac{1}{(2\pi)^3} \,  
  \int d^2 x \, d^2 y \, d^2 z
  \, e^{-i \ul k \cdot (\ul z - \ul y)} \, \Phi_{pol} (\ul x - \ul z \, , \, \ul x - \ul y, \alpha) \
  \mathcal{I}^{(\gamma)}_{anti} (\ul x \, , \, \ul y \, , \, \ul z) \label{dsigmagamma} \\
  &d\sigma^{(\gamma)}_{unp} = \frac{1}{2 \, (2\pi)^3} \, 
  \int d^2 x \, d^2 y \, d^2 z
  \, e^{-i \ul k \cdot (\ul z - \ul y)} \, \Phi_{unp} (\ul x - \ul z \, , \, \ul x - \ul y, \alpha) \
  \mathcal{I}^{(\gamma)}_{symm} (\ul x \, , \, \ul y \, , \, \ul z) \; ,
 \end{align}
where $\Phi_{pol}$ and $\Phi_{unp}$ are given by
Eqs.~\peq{eq-polarized wavefn} and \peq{eq-unpolarized wavefn} with
the $\as \to \alpha_{EM} \, Z_f^2$ replacement.

The interaction with the target is calculated to be
\begin{equation}
  \label{Igamma}
  \mathcal{I}^{(\gamma)} = 1 + D_{\ul u \, \ul w} - D_{\ul x \, \ul w} -
  D_{\ul u \, \ul x}
\end{equation}
with the symmetric and anti-symmetric under ${\ul z} \leftrightarrow
{\ul y}$ parts
\begin{eqnarray}
 \mathcal{I}^{(\gamma)}_{symm} &=& 1 + S_{\ul u \, \ul w}  - S_{\ul x \, \ul w} - S_{\ul u \,
 \ul x} 
 \\ \label{phanti}
 \mathcal{I}^{(\gamma)}_{anti} &=& i \left[ O_{\ul u \, \ul w} - O_{\ul x \, \ul w} 
 - O_{\ul u \, \ul x}  \right] \; . 
 \end{eqnarray}
Eqs.~\peq{dsigmagamma} and \peq{phanti} along
with \eqref{AN1v2} give us the prompt photon STSA. This is the
third and final main formal result of this work. Note that below we
will show that \eqref{dsigmagamma} leads to $d(\Delta\sigma^{(\gamma)})
=0$ for any target, which implies zero STSA for photons in our
mechanism.

We have constructed general expressions for STSA generated by quark,
gluon, and photon production in $q^\uparrow A$ collisions. Knowing
the light-cone wave functions squared \eqref{eq-unpolarized wavefn},
\eqref{eq-polarized wavefn} and the interactions for the 3 channels
\eqref{eq-symmetrized interaction},
\eqref{IGsymmanti}, \eqref{phanti}, one can
make explicit predictions for the corresponding asymmetries. In
general terms, we have shown that in this formalism the asymmetry is
generated by the coupling of the spin-dependent part of the wave function
to the odderon interaction with the target.


\section{Evaluations and Estimates of the Asymmetry}

\label{sec-Estimates}

Unfortunately, Eqs.~\peq{dsigmaq}, \peq{dsigmaG}, and
\peq{dsigmagamma} are too complicated to be integrated out analytically in
the general case. In this Section, in order to understand the
qualitative behavior of our results, we evaluate the integrals
analytically, taking the interaction with the target in the
quasi-classical Glauber-Gribov-Mueller approximation of Sec.~\ref{subsec-GGM}. In such a
quasi-classical limit, the real part of the $S$ matrix \eqref{Sdef} is
\cite{Mueller:1989st} (c.f. \eqref{GGMsoln2})
\begin{equation}
 \label{eq-GM Pomeron}
 S_{\ul x \, \ul y} = \exp \left[ -\frac{1}{4} \, |\ul x - \ul y|_T^2 \ Q_s^2 \! 
\left( \frac{\ul x + \ul y}{2} \right) \, \ln \frac{1}{|\ul x - \ul y|_T \, \Lambda} \right],
\end{equation}
where the quark saturation scale scale $Q^2 (\ul b)$ is defined in
terms of the nuclear profile function (transverse nuclear density)
$T(\ul b)$ as
\begin{equation}
 \label{eq-saturation scale}
 Q_s^2 (\ul b) \equiv \frac{4 \, \pi \, \alpha_s^2 \, C_F}{N_c} \, T(\ul b) \; 
\end{equation} 
and $\Lambda$ is a non-perturbative IR cutoff (c.f. \eqref{Qsat3}).

In the same quasi-classical approximation the odderon amplitude is
\cite{Kovchegov:2003dm}
\begin{equation}
  \label{Ocl}
  O_{\ul x \, \ul y} = \left\langle c_0 \, \alpha_s^3 \, \ln^3 
 \frac{|\ul x - \ul r|_T}{|\ul y - \ul r|_T} \right\rangle \, 
 \exp \left[ -\frac{1}{4} \, |\ul x - \ul y|_T^2 \ Q_s^2 \!
\left( \frac{\ul x + \ul y}{2} \right) \, \ln \frac{1}{|\ul x - \ul y|_T \, \Lambda} \right]
\end{equation}
with the constant
\cite{Hatta:2005as,Kovner:2005qj,Jeon:2005cf}\footnote{Note that the
  sign is different from that in \cite{Hatta:2005as,Jeon:2005cf}: the
  sign in \eqref{c0} arises when using a consistent convention for the
  sign of the coupling $g$ both in the Wilson lines and in the
  classical gluon field of the target. (Our sign convention is to have
  $+i \, g$ for the quark-gluon vertex, resulting in $+i \, g$ in the
  Wilson lines \eqref{eq-Wilson lines}.) While the physical
  conclusions reached in
  \cite{Kovchegov:2003dm,Hatta:2005as,Kovner:2005qj,Jeon:2005cf} are
  independent of the sign of the odderon amplitude, the direction of
  the asymmetry in question explicitly depends on the sign of $O_{\ul
    x \, \ul y}$.}
\begin{equation}
  \label{c0}
  c_0 = - \frac{(N_c^2 -4) \, (N_c^2 -1)}{12 \, N_c^3}.
\end{equation}
The logarithm cubed in \eqref{Ocl} arises due to the triple gluon
exchange as in \eqref{eik9} between the dipole and some quark in the target nucleus
located at transverse position $\ul r$. Angle brackets in \eqref{Ocl}
denote the averaging over positions of the quark in the nuclear wave
function, along with the summation over all the nucleons in the
nucleus that may contain this quark. This averaging is carried out below.

For simplicity we will also work in the large-$N_c$ limit for the
light-cone wave function. Just like before, we mainly concentrate on
the quark production case in \eqref{dsigmaq}: STSA in the gluon
production channel can be evaluated along similar lines. We will also
consider STSA for the prompt photon production.

\subsection{Averaging the Odderon Amplitude}
\label{A4}

\begin{figure}[ht]
 \centering
 \includegraphics[width=4cm]{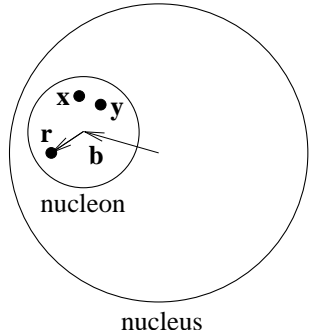}
 \caption{The geometry of the dipole--nucleus scattering as employed in \eqref{Ocl1}.}
\label{odd_ave} 
\end{figure}

Let us construct the dipole odderon amplitude averaged over the target
field.  The triple gluon exchange happens between the dipole and a
nucleon in the target, which, for simplicity we model as a valence
quark in a bag.  The overall factor in front of the averaged odderon
amplitude should depend on the details of the averaging;
however, we believe the coordinate-space dependence would remain the
same for other models of the nuclear wave function. The target
averaging then consists of averaging over the positions of the quark
in the nucleon and over the positions of nucleons in the nucleus,
along with summation over all nucleons. Assuming, again for
simplicity, that the quark has equal probability to be anywhere inside
the nucleon in the transverse plane (a cylindrical ``nucleon''
approximation), we write for the averaged odderon amplitude
\begin{eqnarray}
  \label{Ocl1}
  O_{\ul x \, \ul y} = c_0 \, \alpha_s^3 \, \int d^2 b \ T({\ul b}) \, 
\int \frac{d^2 r}{\pi \, a^2} \, \ln^3 
 \frac{|\ul x - \ul b - \ul r|_T}{|\ul y - \ul b - \ul r|_T}  \, \theta (a-r) \, 
\theta \left( a - \left| \frac{\ul x + \ul y}{2} - \ul b \right| \right)  \notag \\
\times \, \exp \left[ -\frac{1}{4} \, |\ul x - \ul y|_T^2 \ Q_s^2 \! 
\left( \frac{\ul x + \ul y}{2} \right) \, \ln \frac{1}{|\ul x - \ul y|_T \, \Lambda} \right].
\end{eqnarray}
Here $\ul b$ is the position of the center of a nucleon in the
transverse plane with respect to the center of the nucleus, $\ul r$ is
the position of the valence quark in the nucleon, and $a$ is the
radius of the nucleon, as illustrated in \fig{odd_ave}. The two
theta-functions in \eqref{Ocl1} insure that the valence quark and the
center of the ${\ul x}, {\ul y}$-dipole are both located inside the
nucleon in the transverse plane.
In our simple model of the collision the dipole has to hit the nucleon
directly in order to be able to interact with the quarks inside of it.
Since the dipole ${\ul x}, {\ul y}$ is perturbatively small, we
enforce this condition by demanding that only the center of the dipole
is inside the nucleon's transverse extent. 

To integrate over $\ul r$ in \eqref{Ocl1} we will first show that
\begin{equation}
  \label{eq:zero}
  f(\ul{x},\ul{y}) \equiv \int d^2 r \, \ln^3 \frac{|\ul x - \ul b - \ul r|_T}
	{|\ul y - \ul b - \ul r|_T} = 0
\end{equation}
if the integration carries over the whole transverse plane.  To see this, we can first shift the integration variable $\ul{r} \rightarrow \ul{r} - \ul{b} + \ul{y}$, obtaining
\begin{equation}
 \label{ozero1}
 f(\ul{x},\ul{y}) = \int d^2 r \ln^3 \frac{|\ul{x} - \ul{y} - \ul{r}|_T}{r_T} = 
 f(\ul{x} - \ul{y}),
\end{equation}
which shows that the expression \eqref{eq:zero} is a function only of the difference in coordinates $\ul{x} - \ul{y}$.  Equivalently, we could start with \eqref{eq:zero} and instead shift $\ul{r} \rightarrow \ul{r} - \ul{b} + \ul{x}$, followed by inversion $\ul{r} \rightarrow - \ul{r}$, obtaining
\begin{align}
 \label{ozero2}
 f(\ul{x}-\ul{y}) &= \int d^2 r \ln^3 \frac{r_T}{|\ul{y}-\ul{x}-\ul{r}|_T} \\ \nonumber
 &= \int d^2 r \ln^3 \frac{r_T}{|\ul{y}-\ul{x}+\ul{r}|_T} \\ \nonumber
 &= - \int d^2 r \ln^3 \frac{|\ul{x}-\ul{y}-\ul{r}|_T}{r_T} \\ \nonumber
 &= - f(\ul{x}-\ul{y}),
\end{align}
which therefore shows that $f(\ul{x},\ul{y}) = 0$ as in \eqref{eq:zero}.  Using this result we write
\begin{equation}
  \label{inverse}
  \int d^2 r \, \ln^3 \left( \frac{|\ul x - \ul b - \ul r|_T}{|\ul y - \ul b - \ul r|_T} \right)
\, \theta (a-r) = - \int d^2 r \, \ln^3 
\left( \frac{|\ul x - \ul b - \ul r|_T}{|\ul y - \ul b - \ul r|_T} \right) \, \theta (r-a). 
\end{equation}
To approximate the integral on the right-hand-side of \eqref{inverse} we
expand its integrand in powers of $|\ul x - \ul b|_T/r$ and $|\ul y -
\ul b|_T/r$ to the first non-trivial (after integration) order, thus
obtaining
\begin{equation}
  \label{ln3int}
  \int d^2 r \, \ln^3 \left( \frac{|\ul x - \ul b - \ul r|_T}{|\ul y - \ul b - \ul r|_T} \right)
\, \theta (a-r) \approx \frac{3 \, \pi}{8 \, a^2} \, |\ul x - \ul y|_T^2 \ (\ul x - \ul y) 
\cdot (\ul x + \ul y - 2 \, \ul b).
\end{equation}

Substituting \eqref{ln3int} back into \eqref{Ocl1} and defining a new
integration variable
\begin{equation}
  \label{delta_def}
  {\ul {\tilde b}} = {\ul b} - \frac{\ul x + \ul y}{2}
\end{equation}
yields
\begin{eqnarray}
  \label{Ocl2}
  O_{\ul x \, \ul y} \approx - c_0 \, \alpha_s^3 \, \frac{3}{4 \, a^4} \, |\ul x - \ul y|_T^2 \, 
 \exp \left[ -\frac{1}{4} \, |\ul x - \ul y|_T^2 \ Q_s^2 \!
\left( \frac{\ul x + \ul y}{2} \right) \, \ln \frac{1}{|\ul x - \ul y|_T \, \Lambda} \right] \notag \\ \times \, 
(\ul x - \ul y) \cdot \int d^2 {\tilde b} \ {\ul {\tilde b}} \
T \! \left(\frac{\ul x + \ul y}{2} + {\ul {\tilde b}} \right) \, 
   \theta \left( a - {\tilde b}_T \right) .
\end{eqnarray}

In principle this result is as far as one can simplify $O_{\ul x \,
  \ul y}$ without the explicit knowledge of the nuclear profile
function $T ({\ul b})$. To obtain a closed expression for the STSA we expand
\begin{equation}
  \label{expansion}
  T \! \left(\frac{\ul x + \ul y}{2} + {\ul {\tilde b}} \right) = 
T \! \left(\frac{\ul x + \ul y}{2} \right) + {\ul {\tilde b}} \cdot {\ul \nabla} 
T \! \left(\frac{\ul x + \ul y}{2} \right) + \ldots 
\end{equation}
with $\ul \nabla$ the transverse gradient operator.  Such an expansion is
potentially dangerous near the edge of the nucleus profile, where the
derivatives may get large. For instance, for a solid-sphere model of
the nucleus the nuclear profile function is $T ({\ul b}) = \rho \, 2
\, \sqrt{R^2 - b_T^2}$ with $\rho$ the nucleon density and $R$ the
nuclear radius; the derivatives of such $T ({\ul b})$ near $b_T=R$ are
divergent. Using the realistic Woods-Saxon profile would make the
derivatives finite, but they would still be large. Thus we will
proceed by using the expansion \peq{expansion} as a way to simplify
the expression, keeping in mind that in the cases where this expansion
breaks down one has to return back to \eqref{Ocl2}.

Substituting \eqref{expansion} into \eqref{Ocl2} and integrating over $\ul
{\tilde b}$ yields (for the first non-trivial term after integration)
\begin{align}
  \label{Ocl3}
  O_{\ul x \, \ul y} \approx - c_0 \, \alpha_s^3 \, 
 \frac{3 \, \pi}{16} \, |\ul x - \ul y|_T^2 \, 
 & \left[ (\ul x - \ul y) \cdot {\ul \nabla} 
T \! \left(\frac{\ul x + \ul y}{2} \right) \right] \:
\\ \nonumber &\times
\exp \left[ -\frac{1}{4} \, |\ul x - \ul y|_T^2 \ Q_s^2 \!
\left( \frac{\ul x + \ul y}{2} \right) \, 
\ln \frac{1}{|\ul x - \ul y|_T \, \Lambda} \right]
.
\end{align}
This is our final expression for the target-averaged odderon
amplitude. Note an interesting feature of \eqref{Ocl3}: the non-zero
contribution to the odderon amplitude in transverse coordinate
space arises from the gradient of the nuclear profile function. The
odderon interaction with the target is thus only possible if the
target has a non-uniform profile in the transverse space. This is in
stark contrast to the $C$-even exchanges, which are non-zero even for
the Bjorken model of a nucleus of infinite transverse extent with
constant density in the transverse plane.



\subsection{Single Transverse Spin Asymmetry in Quark Production}
\label{quarkSTSAsec}

\subsubsection{Spin-Dependent Quark Production Cross Section}

First let us evaluate the numerator of the STSA in \eqref{AN1v2},
which, in the quark production case, is given by \eqref{dsigmaq}. (For
simplicity we assume that $\alpha < 1$ which allows us to drop virtual corrections.) 
Working in the large-$N_c$ limit for
the light-cone wave function we substitute the interaction from
\eqref{qanti} into \eqref{dsigmaq} to obtain
\begin{align}
 \label{dsigmaq1} 
  d(\Delta\sigma^{(q)}) &= i \, \frac{N_c}{2 \, (2\pi)^3} \, \frac{\alpha}{1-\alpha} \, 
  \int d^2 x \, d^2 y \, d^2 z
  \, e^{-i \ul k \cdot (\ul z - \ul y)} \, \Phi_{pol} (\ul z - \ul x \, , \, \ul y - \ul x, \alpha)
	\\ \nonumber &\times
	\left[ O_{\ul z \, \ul y} + O_{\ul u \, \ul w} - O_{\ul z \, \ul x} \, S_{\ul x \, \ul w} -
  O_{\ul u \, \ul x} \, S_{\ul x \, \ul y} 
	- S_{\ul z \, \ul x} \, O_{\ul x \, \ul w} - S_{\ul u \, \ul x} \, O_{\ul x \, \ul y} \right] . 
\end{align}
Our goal now is to evaluate this expression using the $S$-matrix from
\eqref{eq-GM Pomeron} and the odderon amplitude \peq{Ocl3}. 

The interaction with the target in \eqref{dsigmaq1} is non-linear. It is
tempting to try to simplify the problem by neglecting all the multiple
rescattering saturation effects. In such a linearized approximation
\eqref{dsigmaq1} reduces to
\begin{eqnarray}
  d(\Delta\sigma^{(q)})_{lin} = i \, \frac{N_c}{2 \, (2\pi)^3} \, 
\frac{\alpha}{1-\alpha} \, 
  \int d^2 x \, d^2 y \, d^2 z
  \, e^{-i \ul k \cdot (\ul z - \ul y)} \, \Phi_{pol} (\ul z - \ul x \, , \, \ul y - \ul x, \alpha) 
   \notag \\ \times \, \left[ o_{\ul z \, \ul y} +
  o_{\ul
  u \, \ul w} - o_{\ul z \, \ul x} -
  o_{\ul u \, \ul x} 
- o_{\ul x \, \ul w} - 
  o_{\ul x \, \ul y} \right] \label{dsigmaq_lin} 
\end{eqnarray}
where 
\begin{equation}
  \label{olin}
  o_{\ul x \, \ul y} \approx \alpha_s^3 \, 
 \frac{\pi \, N_c}{64} \, |\ul x - \ul y|_T^2 \, \ (\ul x - \ul y) 
\cdot {\ul \nabla} T \! \left(\frac{\ul x + \ul y}{2} \right)
\end{equation}
is the linear part of the averaged odderon amplitude \peq{Ocl3}.
However, one can easily show that the cross section in
\eqref{dsigmaq_lin} is in fact zero, i.e., that
\begin{equation}
  \label{linzero}
  d(\Delta\sigma^{(q)})_{lin} = 0. 
\end{equation}
We illustrate this by considering the $o_{\ul z \, \ul y}$ term in
\eqref{dsigmaq_lin}. Defining new transverse vectors
\begin{equation}
  \label{tildes}
  {\tilde {\ul z}} = {\ul z} - {\ul x}, \ \ \ {\tilde {\ul y}} = {\ul y} - {\ul x},
\end{equation}
we rewrite the $o_{\ul z \, \ul y}$ contribution to the cross section
in \eqref{dsigmaq_lin} as
\begin{equation}
  \label{zero1}
  i \, \frac{N_c}{2 \, (2\pi)^3} \, \frac{\alpha}{1-\alpha} \, 
  \int d^2 {\tilde y} \, d^2 {\tilde z}
  \, e^{-i \ul k \cdot ({\tilde {\ul z}} - {\tilde {\ul y}})} \, 
\Phi_{pol} ({\tilde {\ul z}}, {\tilde {\ul y}}, \alpha) \, \int d^2 x \, 
   \,  o_{{\tilde {\ul z}} + {\ul x}, \, {\tilde {\ul y}} + {\ul x}}. 
\end{equation}
This expression is zero since 
\begin{equation}
  \label{zero2}
  \int d^2 x \, 
   \,  o_{{\tilde {\ul z}} + {\ul x}, \, {\tilde {\ul y}} + {\ul x}} =0
\end{equation}
due to the fact that the odderon amplitude \peq{Odef} (and, therefore,
the linearized odderon amplitude \peq{olin}) is an anti-symmetric
function of its transverse coordinate arguments,
\begin{equation}
  \label{Oanti}
  O_{\ul x \, \ul y} = - O_{\ul y \, \ul x}.
\end{equation}
The argument goes as follows. Employing \eqref{Oanti} and shifting the
integration variables we write
\begin{align}
  \label{zero_arg}
  f ({\ul \delta}) &\equiv \int d^2 x \ O_{{\ul x},  \, {\ul x} + {\ul \delta}} \\ \nonumber &=
  \int d^2 x \ O_{{\ul x - \ul \delta},  \, {\ul x}} \\ \nonumber &=
	- \int d^2 x \ O_{{\ul x},  \, {\ul x} - {\ul \delta}} \\ \nonumber f ({\ul \delta}) &= 
	- f (-{\ul \delta}). 
\end{align}
Since $f ({\ul \delta})$ depends only on one vector $\ul \delta$ and is a scalar
under the rotations in the transverse plane, it is a function of ${\delta}_T^2$ only, and can satisfy \peq{zero_arg} (i.e., can be an odd
function of $\ul \delta$) only if $f ({\ul \delta}) =0$. This demonstrates that
\begin{equation}
  \label{zero_arg2}
  \int d^2 x \, O_{{\ul x},  \, {\ul x} + {\ul \delta}} = 0.
\end{equation}

Similar arguments can be carried out for other terms in
\eqref{dsigmaq_lin}, leading in the end to \eqref{linzero}. We arrive at an
important conclusion: STSA cannot result from the interaction with the
target mediated by the odderon exchange alone. Neglecting the
interactions contained in the dipole $S$-matrices in \eqref{dsigmaq1}
would lead to zero transverse spin asymmetry. This is an important
observation elucidating the nature of our result \peq{dsigmaq}
and the corresponding STSA: in order to generate a non-zero STSA the
interaction with the target has to contain both the $C$-odd and
$C$-even contributions!

Returning to the general case of \eqref{dsigmaq1} we see that the
argument we have just presented demonstrates that the $O_{\ul z \, \ul
  y}$ and $O_{\ul u \, \ul w}$ terms are zero in the general case as
well, since they are not multiplied by the $S$-matrices. Dropping
these terms yields
\begin{eqnarray}
  d(\Delta\sigma^{(q)}) = - i \, \frac{N_c}{2 \, (2\pi)^3} \, \frac{\alpha}{1-\alpha} \, 
  \int d^2 x \, d^2 y \, d^2 z
  \, e^{-i \ul k \cdot (\ul z - \ul y)} \, \Phi_{pol} (\ul z - \ul x \, , \, \ul y - \ul x, \alpha) \
  \notag \\ \times \,  \left[ O_{\ul z \, \ul x} \, S_{\ul x \, \ul w} +
  O_{\ul u \, \ul x} \, S_{\ul x \, \ul y}  
+ O_{\ul x \, \ul w} \, S_{\ul z \, \ul x} + O_{\ul x \, \ul y} \, S_{\ul u \, \ul x} 
   \right] . \label{dsigmaq2} 
\end{eqnarray}

To evaluate \eqref{dsigmaq2} let us first study its large-$k_T$
asymptotics. Since ${\tilde m} \le m$ and the quark mass $m$ is at
most the constituent quark mass of about $300$~MeV (we assume light
quark flavors), we have $k_T \gg Q_s \gg {\tilde m}$. Changing the
coordinates using \eqref{tildes} reduces it to
\begin{align}
 \label{dsigmaq3}
  d(\Delta\sigma^{(q)}) &= - i \, \frac{N_c}{2 \, (2\pi)^3} \, \frac{\alpha}{1-\alpha} \, 
  \int d^2 x \, d^2 {\tilde y} \, d^2 {\tilde z}
  \, e^{-i \ul k \cdot ({\tilde {\ul z}} - {\tilde {\ul y}})} \, 
  \Phi_{pol} ({\tilde {\ul z}} \, , \, {\tilde {\ul y}}, \alpha) \
 \\ \nonumber &\times 
	\left[ O_{\ul x + {\tilde {\ul z}},  \, \ul x} \ S_{\ul x, \, \ul x + \alpha \, {\tilde {\ul y}}} +
  O_{\ul x + \alpha \, {\tilde {\ul z}}, \ \ul x} \ S_{\ul x, \, \ul x + {\tilde {\ul y}} }  
+ O_{\ul x, \,  \ul x + \alpha \, {\tilde {\ul y}}} \ S_{\ul x, \, \ul x + {\tilde {\ul z}}} + O_{\ul x, \, \ul x + {\tilde {\ul y}} } \ S_{\ul x , \, \ul x + \alpha \, {\tilde {\ul z}}} 
   \right] .  
\end{align}
For each term in the square brackets of \eqref{dsigmaq3} the integrals
over ${\tilde {\ul z}}$ and ${\tilde {\ul y}}$ factorize (c.f. \eqref{eq-polarized wavefn}); taking the large-$k_T$ limit in each of them separately, we see that the
large-$k_T$ asymptotics corresponds to small ${\tilde {z}}_T$ and
${\tilde {y}}_T$. We thus need to expand the interaction with the
target in the square brackets of \eqref{dsigmaq3} to the lowest
non-trivial order in ${\tilde {z}}_T$ and ${\tilde {y}}_T$. Note that
above we have seen that if we keep the dipole $S$-matrices at the
lowest order in the dipole size, $S=1$, then the spin-dependent cross
section would be zero. We thus use Eqs.~\peq{eq-GM Pomeron} and
\peq{Ocl3} to expand the $S$-matrices to the next-to-lowest order,
while keeping the odderon amplitudes at the lowest order given by
\eqref{olin}.  Performing the expansion, substituting the wave function
squared from \eqref{eq-polarized wavefn} (also expanded to the lowest
non-trivial order in ${\tilde {z}}_T$ and ${\tilde {y}}_T$) into
\eqref{dsigmaq3}, and employing \eqref{eq-saturation scale} we obtain
\begin{align}
 \nonumber
  d(\Delta\sigma^{(q)}) \bigg|_{k_T \gg Q_s} &\approx i \, 
	\frac{N_c^2}{1024 \, \pi^2} \, \alpha_s^6 \,  {\tilde m} \, \alpha^4 \,
  \int d^2 x \, d^2 {\tilde y} \, d^2 {\tilde z}
  \, e^{-i \ul k \cdot ({\tilde {\ul z}} - {\tilde {\ul y}})} \, 
  \bigg( \frac{{\tilde z}_\bot^{2} }{{\tilde z}_T^2} \, \ln \frac{1}{\tilde m \, {\tilde y}_T} 
  + \, \frac{{\tilde y}_\bot^{2}}{{\tilde y}_T^2} \, \ln \frac{1}{\tilde m \, {\tilde z}_T} \bigg) 
 \\ \nonumber &\times
 {\tilde z}_T^2 \, {\tilde y}_T^2
 \bigg[ {\tilde {\ul z}} \cdot {\ul \nabla} 
 T \! \left({\ul x} + \frac{{\tilde {\ul z}}}{2} \right) \ 
 T \! \left({\ul x} + \frac{\alpha \, {\tilde {\ul y}}}{2} \right) \, 
 \ln \frac{1}{\alpha \, {\tilde y}_T \, \Lambda}
 \\ \label{dsigmaq4} &+
 \alpha \ {\tilde {\ul z}} \cdot {\ul \nabla} 
 T \! \left({\ul x} + \frac{\alpha \, {\tilde {\ul z}}}{2} \right) \ 
 T \! \left({\ul x} + \frac{{\tilde {\ul y}}}{2} \right) \, 
 \ln \frac{1}{{\tilde y}_T \, \Lambda} - 
 ({\tilde {\ul z}} \leftrightarrow {\tilde {\ul y}}) \bigg].  
\end{align}

Since ${\tilde {z}}_T$ and ${\tilde {y}}_T$ are small, one may think
of neglecting them compared to $\ul x$ in the arguments of $T$'s in
\eqref{dsigmaq4}. However, this would again lead to a zero answer after
integration over $d^2 x$.  The reason for this conclusion is that any
unpolarized target, after averaging over many events, is rotationally
symmetric in the transverse plane. This implies that ${\ul \nabla} T
({\ul x}) = {\ul \nabla} T (x_T) = {\hat x} \, T' (x_T)$ where ${\hat
  x}$ is a unit vector in the direction of $\ul x$ and $T' (x_T) = d
T(x_T)/d x_T$. Integrating ${\hat x}$ over the angles of $\ul x$ would
give zero.

Instead of neglecting ${\tilde {z}}_T$ and ${\tilde {y}}_T$, we shift
${\ul x} \to {\ul x} - {\tilde {\ul z}}/2$ in the first term in the
square brackets of \eqref{dsigmaq4} and expand $T$ along the lines of
\eqref{expansion}, and perform similar operations to the other terms in
the brackets obtaining
\begin{align}
 \nonumber
  d(\Delta\sigma^{(q)}) \bigg|_{k_T \gg Q_s} &\approx i \, 
\frac{N_c^2}{2048 \, \pi^2} \, \alpha_s^6 \,  {\tilde m} \, \alpha^4 \,
  \int d^2 x \, d^2 {y} \, d^2 {z}
  \, e^{-i \ul k \cdot ({{\ul z}} - {{\ul y}})} \, 
  \bigg(
 \frac{{z}_\bot^{2} }{{z}_T^2} \, \ln \frac{1}{\tilde m \, {y}_T} 
  + \, \frac{{y}_\bot^{2}}{{y}_T^2} \, \ln \frac{1}{\tilde m \, {z}_T} \bigg) 
\, 
\\ \nonumber & \times
{z}_T^2 \, {y}_T^2 \bigg[ {{\ul z}} \cdot {\ul \nabla} 
T \! \left({\ul x} \right) \ (\alpha \, {\ul y} - {\ul z}) \cdot {\ul \nabla} 
T \! \left({\ul x} \right) \, 
\ln \frac{1}{\alpha \, y_T \, \Lambda}
\\ \label{dsigmaq5} &+
\alpha \ {{\ul z}} \cdot {\ul \nabla} 
T \! \left({\ul x}\right) \ ({\ul y} - \alpha \, {\ul z}) \cdot {\ul \nabla} 
T \! \left({\ul x}\right) \, 
\ln \frac{1}{y_T \, \Lambda} - ({{\ul z}} \leftrightarrow {{\ul y}}) \bigg],  
\end{align}
where we have dropped the tildes over $\ul y$ and $\ul z$, since now
it would not cause confusion.

Using ${\ul \nabla} T ({\ul x}) = {\ul \nabla} T (x_T) = {\hat x} \,
T' (x_T)$ and integrating over the angles of $\ul x$ reduces
\eqref{dsigmaq5} to
\begin{align}
 \nonumber
 d(\Delta\sigma^{(q)}) \bigg|_{k_T \gg Q_s} &\approx i \, 
 \frac{N_c^2}{4096 \, \pi} \, \alpha_s^6 \,  {\tilde m} \, \alpha^4 \,
 \int\limits_0^\infty  d x_T^2 \, [T' (x_T)]^2 \, \int d^2 {y} \, d^2 {z}
 \, e^{-i \ul k \cdot ({{\ul z}} - {{\ul y}})} \, \bigg(
 \frac{{z}_\bot^{2} }{{z}_T^2} \, \ln \frac{1}{\tilde m \, {y}_T} 
 \\ \nonumber &+
 \frac{{y}_\bot^{2}}{{y}_T^2} \, \ln \frac{1}{\tilde m \, {z}_T} \bigg) 
 \, {z}_T^2 \, {y}_T^2
 %
 %
 \bigg[ {{\ul z}} \cdot  (\alpha \, {\ul y} - {\ul z})  \, 
 \ln \frac{1}{\alpha \, y_T \, \Lambda} + 
 \alpha \ {{\ul z}} \cdot  ({\ul y} - \alpha \, {\ul z}) \, 
 \ln \frac{1}{y_T \, \Lambda} 
 \\ \label{dsigmaq6} &-
 \alpha \ {{\ul y}} \cdot  ({\ul z} - \alpha \, {\ul y})  \, 
 \ln \frac{1}{z_T \, \Lambda} - {{\ul y}} \cdot (\alpha \, {\ul z} -  {\ul y}) \, 
 \ln \frac{1}{\alpha \, z_T \, \Lambda} \bigg]. 
\end{align}
Integrating over $\ul y$ and $\ul z$ in \eqref{dsigmaq6} and discarding
delta-functions of $\ul k$ (since $k_T \neq 0$) yields
\begin{eqnarray}
  d (\Delta\sigma^{(q)}) \bigg|_{k_T \gg Q_s} \approx && \  
\frac{\pi \, N_c^2}{8} \, \alpha_s^6 \,  {\tilde m} \, 
\alpha^4 \, (2 + 3 \, \alpha + 2 \, \alpha^2) \, 
  \int\limits_0^\infty  d x_T^2 \, [T' (x_T)]^2 \, \frac{k_\bot^2}{k_T^{10}}. \label{dsigmaq7} 
\end{eqnarray}
We see that the polarized spectrum falls off rather steeply with
$k_T$, scaling as $1/k_T^9$. This indicates that in the standard
collinear factorization framework our STSA generating mechanism
originates in some higher-twist operator. 

Another important qualitative feature one can see in \eqref{dsigmaq7} is
that the spin-dependent cross section falls off with decreasing
longitudinal momentum fraction $\alpha$, which implies that the
corresponding STSA decreases with decreasing Feynman-$x$ of the
projectile, in qualitative agreement with the experimental data.

To improve on \eqref{dsigmaq7} let us find the spin-dependent
differential cross section $d (\Delta\sigma^{(q)})$ for lower $k_T$,
closer to the saturation scale. To be more specific let us relax the
$k_T \gg Q_s$ restriction and consider a broader region of $k_T
\lesssim Q_s$ and $k_T \gtrsim Q_s$, but still with $k_T \gg {\tilde
  m}$.  For such not very large $k_T$ we can neglect the logarithms in
the exponents of Eqs.~\peq{eq-GM Pomeron} and \peq{Ocl3} as slowly
varying functions compared to the powers they multiply
\cite{Kovchegov:1998bi,Kharzeev:2003wz,Jalilian-Marian:2005jf},
writing
\begin{equation}
 \label{Sapp}
 S_{\ul x \, \ul y} \approx \exp \left[ -\frac{1}{4} \, |\ul x - \ul y|_T^2 \ Q_s^2 \! 
\left( \frac{\ul x + \ul y}{2} \right) \right]
\end{equation}
and
\begin{equation}
  \label{Oapp}
  O_{\ul x \, \ul y} \approx - c_0 \, \alpha_s^3 \, 
 \frac{3 \, \pi}{16} \, |\ul x - \ul y|_T^2 \, 
 \exp \left[ -\frac{1}{4} \, |\ul x - \ul y|_T^2 \ Q_s^2 \!
\left( \frac{\ul x + \ul y}{2} \right) \right] \ (\ul x - \ul y) 
\cdot {\ul \nabla} T \! \left(\frac{\ul x + \ul y}{2} \right).
\end{equation}
Substituting Eqs.~\peq{Sapp} and \peq{Oapp} into \eqref{dsigmaq3},
expanding the polarized wave function squared, and dropping the tildes
yields
\begin{align}
 \nonumber
 d(\Delta\sigma^{(q)}) &\approx - i \, \frac{N_c^2}{512 \, \pi^3} \, \as^4 \, {\tilde m} \, \alpha^2 \, 
 \int d^2 x \, d^2 y \, d^2 z \, e^{-i \ul k \cdot ({{\ul z}} - {{\ul y}})} \, 
 \bigg( \frac{{z}_\bot^{2} }{{z}_T^2} \, \ln \frac{1}{\tilde m \, {y}_T} 
 + \, \frac{{y}_\bot^{2}}{{y}_T^2} \, \ln \frac{1}{\tilde m \, {z}_T} \bigg)
 \\ \nonumber &\times
 \bigg[ z_T^2 \, {\ul z} \cdot {\ul \nabla} T \! \left({\ul x} + \frac{{\ul z}}{2} \right) 
 e^{-\frac{1}{4} \, z_T^2 \, Q_s^2 \left({\ul x} + \frac{{\ul z}}{2} \right)
 -\frac{1}{4} \, \alpha^2 \, y_T^2 \, Q_s^2 \left({\ul x} + \frac{\alpha \, {\ul y}}{2} \right)}
 \\ \label{dsigmaq8} &+ 
 \alpha^3 \, z_T^2 \, {\ul z} \cdot {\ul \nabla} 
 T \! \left({\ul x} + \frac{\alpha \, {\ul z}}{2} \right) \, 
 e^{-\frac{1}{4} \, \alpha^2 \, z_T^2 \, 
 Q_s^2 \left({\ul x} + \frac{\alpha \, {\ul z}}{2} \right)
 - \frac{1}{4} \, y_T^2 \, Q_s^2 \left({\ul x} + \frac{{\ul y}}{2} \right)}
 - ({\ul z} \leftrightarrow {\ul y}) \bigg]. 
\end{align}
Similar to the large-$k_T$ asymptotics, we shift ${\ul x} \to {\ul x}
- {\ul z}/2$ in the first term in the square brackets of \eqref{dsigmaq8}
and expand the resulting exponential with the help of
\eqref{eq-saturation scale} as
\begin{align}
 \label{Expansion}
 e^{-\frac{1}{4} \, z_T^2 \, Q_s^2 ({\ul x})
 -\frac{1}{4} \, \alpha^2 \, y_T^2 \, 
 Q_s^2 \left({\ul x} + \frac{\alpha \, {\ul y} - {\ul z}}{2} \right)} 
 &\approx \left[1 - 
 \frac{\pi}{4} \, \as^2 \, \alpha^2 \, y_T^2 \, 
 (\alpha \, {\ul y} - {\ul z}) \cdot {\ul \nabla} T ({\ul x}) \right] 
 \\ \nonumber &\times
 e^{-\frac{1}{4} \, z_T^2 \, Q_s^2 ({\ul x})
 -\frac{1}{4} \, \alpha^2 \, y_T^2 \, Q_s^2 ({\ul x})}.
\end{align}
The $1$ in the square brackets of \eqref{Expansion} does not contribute
as its contribution vanishes after integration over the angles of $\ul
x$ in \eqref{dsigmaq8}, leaving only the second term to contribute.
Performing similar expansions in the other terms in the square
brackets of \eqref{dsigmaq8} we obtain
\begin{align}
 \nonumber
 d(\Delta\sigma^{(q)}) &\approx i \, \frac{N_c^2}{2048 \, \pi^2} \, 
 \as^6 \, {\tilde m} \, \alpha^4 \, \int d^2 x \, d^2 y \, d^2 z
 \, e^{-i \ul k \cdot ({{\ul z}} - {{\ul y}})} \, \bigg(
 \frac{{z}_\bot^{2} }{{z}_T^2} + \, \frac{{y}_\bot^{2}}{{y}_T^2} \bigg) \, z_T^2 \, y_T^2 
 \\ \nonumber &\times
 \, \bigg[ {\ul z} \cdot {\ul \nabla} T ({\ul x}) (\alpha \, {\ul y} - {\ul z}) 
 \cdot {\ul \nabla} T ({\ul x}) e^{-\frac{1}{4} \, z_T^2 \, Q_s^2 ({\ul x})
 -\frac{1}{4} \, \alpha^2 \, y_T^2 \, Q_s^2 ({\ul x})}
 \\ \label{dsigmaq9} &+
 \alpha \, {\ul z} \cdot {\ul \nabla} T ({\ul x}) \ ({\ul y} - \alpha \, {\ul z}) \cdot 
 {\ul \nabla} T ({\ul x}) \, e^{-\frac{1}{4} \, \alpha^2 \, z_T^2 \, 
 Q_s^2 ({\ul x}) - \frac{1}{4} \, y_T^2 \, Q_s^2 ({\ul x})}
 - ({\ul z} \leftrightarrow {\ul y}) \bigg],  
\end{align}
where we have also dropped $\ln \tfrac{1}{\tilde m y_T}$ and $\ln
\tfrac{1}{\tilde m z_T}$, since, with our precision, similar logarithms
were neglected in Eqs.~\peq{Sapp} and \peq{Oapp} above as slowly
varying functions of their arguments \footnote{We have done the
  calculation without neglecting those logarithms: the resulting
  changes were mainly of quantitative nature, while the obtained
  expression was significantly more complicated than \eqref{dsigmaq11}.
  Since both the expressions with and without the logarithms are
  approximate, we decided to only show the latter in this work due to
  its relative compactness.}. Again, integrating over the angles of
$\ul x$ yields
\begin{align}
 \nonumber
 d(\Delta\sigma^{(q)}) &\approx i \, \frac{N_c^2}{4096 \, \pi } \, \as^6 \, {\tilde m} \, \alpha^4 \, 
 \int\limits_0^\infty  d x_T^2 \, [T' (x_T)]^2 \, \int d^2 y \, d^2 z
 \, e^{-i \ul k \cdot ({{\ul z}} - {{\ul y}})} \, \bigg(\frac{{z}_\bot^{2} }{{z}_T^2} + \, 
 \frac{{y}_\bot^{2}}{{y}_T^2} \bigg) \, z_T^2 \, y_T^2 
 \\ \nonumber \times & 
 \bigg[(\alpha^2 \, y_T^2 - z_T^2)  \, e^{-\frac{1}{4} \, z_T^2 \, Q_s^2 ({x}_T)
 -\frac{1}{4} \, \alpha^2 \, y_T^2 \, Q_s^2 ({x}_T)} 
 \\ \label{dsigmaq10} &+
 (y_T^2 - \alpha^2 \, z_T^2) \,
 e^{-\frac{1}{4} \, \alpha^2 \, z_T^2 \, Q_s^2 ({x}_T) - \frac{1}{4} \, y_T^2 \, Q_s^2 ({x}_T)} \bigg].  
\end{align}
Integrating over $\ul y$ and $\ul z$ we get
\begin{align}
 \nonumber
 d(\Delta\sigma^{(q)}) \approx \frac{\pi \, N_c^2}{4} \, 
 \frac{\as^6 \, {\tilde m}}{\alpha^4} &\int\limits_0^\infty  d x_T^2 \, [T' (x_T)]^2 
 \, \frac{k_\bot^2 \, k_T^2}{Q_s^{14} ({x}_T)} \, \bigg[ (1 - \alpha^2)^2 \, k_T^2 
 - \\ \label{dsigmaq11} &-
 \alpha^2 \, (1 + \alpha^2) \, Q_s^2 ({x}_T) \bigg]  \,
 e^{-\frac{k_T^2}{Q_s^2 ({x}_T)} \, \left( 1 + \frac{1}{\alpha^2} \right)}.
\end{align}

This is the final expression for the STSA-generating cross section for
quark production. Note again that $k_\bot^2$ is the $y$-component of the
quark's transverse momentum ${\ul k} = (k_\bot^1, k_\bot^2)$. Let us point out a
few of the important features of \eqref{dsigmaq11}.  First of all we see
that, similar to \eqref{dsigmaq7}, it decreases with decreasing $\alpha$ for
small $\alpha$, now due to the factor of $1/\alpha^2$ in the exponent.
We also see that for $k_T \to 0$ the spin-difference cross section
$d(\Delta\sigma^{(q)})$ also goes to zero. We also note that the spin-difference cross
section \peq{dsigmaq11} is not a monotonic function of $k_T$. In
particular, for positive $k^2$ it starts out negative at small $k_T$,
becoming positive for $k_T > Q_s \, \alpha \,
\sqrt{1+\alpha^2}/(1-\alpha^2)$, in agreement with the large-$k_T$
asymptotics of \eqref{dsigmaq7}.


\subsubsection{The Unpolarized Cross-Section}
 
\label{subsec-dsigmaunp}

The real hadronic STSA in \eqref{AN1v2} contains contributions
from both quark and gluon production cross sections
\peq{dsigmaq} and \peq{dsigmaG} in the numerator and in the
denominator, convoluted with the fragmentation functions for the
quarks and gluons decaying into a particular hadron species as well as
the transversity distribution of polarized quarks. This is what needs
to be done to have a real comparison of the data with our theoretical
results. While such comparison is beyond the scope of this work, we
would like to assess the main qualitative features of our
STSA-generating mechanism by concentrating on the quark STSA only.

It may be tempting to consider a situation where both the numerator
and the denominator of \eqref{AN1v2} are driven by the quark
contributions. However, the unpolarized valence quark production cross
section \peq{dsigmaq_unp} is known to decrease with decreasing quark
momentum fraction $\alpha$ \cite{Itakura:2003jp,Albacete:2006vv},
while both the unpolarized gluon and sea quark production cross
sections grow with decreasing $\alpha$ in theoretical calculations
\cite{Kovchegov:2001sc,Kovchegov:2006qn}. In the actual experiments
the hadron multiplicity also increases as we move further away from
the projectile in rapidity.

Therefore, in order to get a somewhat realistic evaluation of the
qualitative behavior of the obtained STSA, we will use the unpolarized
gluon production cross section in the denominator of \eqref{AN1v2}. While the evaluation of the unpolarized gluon cross section
\peq{Gunp} along the same lines as were used to obtain \eqref{dsigmaq11}
is somewhat involved, we will approximate the result by assuming that
the produced gluon is soft (i.e., far from the projectile in
rapidity), in which case the corresponding production cross section is
\cite{Kovchegov:1998bi,Kharzeev:2003wz,Jalilian-Marian:2005jf}
\begin{align}
 \label{Gunp2}
 d\sigma_{unp}^{(G)} \approx \frac{\alpha_s \, N_c}{2 \, \pi} \, 
 \, \int\limits_0^\infty dx_T^2 \, &\bigg\{ - \frac{1}{k_T^2} 
 + \frac{2}{k_T^2} \, e^{-\frac{k_T^2}{Q_s^2 (x_T)}} +
 \frac{1}{Q_s^2 (x_T)} \, e^{-\frac{k_T^2}{Q_s^2 (x_T)}} \,
 \bigg[ \mathrm{Ei}\bigg(\frac{k_T^2}{Q_s^2 (x_T)}\bigg) 
 \\ \nonumber &-
 \ln\frac{4 \, k_T^2 \, \Lambda^2}{Q_s^4 (x_T)} \bigg] \bigg\}.
\end{align}

\subsubsection{Single Transverse Spin Asymmetry}


We now have all the essential ingredients to sketch the STSA due to
quark production in the large-$N_c$ limit (for the wave function): we
have Eqs.~\peq{dsigmaq11} and \peq{Gunp2}, giving the numerator and
the denominator of \eqref{AN1v2} correspondingly. We thus write
\begin{align}
 \label{ANq}
  A_N^{(q)} ({\ul k}) &= \frac{\pi^2 \, N_c \, \as^5 \, m}{4} \, 
  \frac{1-\alpha}{\alpha^4} \, 
	\\ \nonumber &\times
  \int\limits_0^\infty  d x_T^2 \, [T' (x_T)]^2 
  \, \frac{k_\bot^2 \, k_T^2}{Q_s^{14} ({x}_T)} \,
  \left[ (1 - \alpha^2)^2 \, k_T^2 - \alpha^2 \, (1 + \alpha^2) \, Q_s^2 ({x}_T) \right]
	%
	%
	e^{-\frac{k_T^2}{Q_s^2 ({x}_T)} \left( 1 + \frac{1}{\alpha^2} \right)} \, 
	\\ \nonumber &\!\!\!\!\!\!\!\!\!\!\!\!  \times
  \Bigg( \int\limits_0^\infty dy_T^2 \, \bigg\{ - \frac{1}{k_T^2} 
  + \frac{2}{k_T^2} \, e^{-\frac{k_T^2}{Q_s^2 (y_T)}} + 
	%
	%
  \frac{1}{Q_s^2 (y_T)} \, e^{-\frac{k_T^2}{Q_s^2 (y_T)}} \,
  \bigg[ \mathrm{Ei}\bigg(\frac{k_T^2}{Q_s^2 (y_T)}\bigg) - 
  \ln\frac{4 \, k_T^2 \, \Lambda^2}{Q_s^4 (y_T)} \bigg] \bigg\} \Bigg)^{-1} . 
\end{align}
The $x_T$- and $y_T$-integrals in \eqref{ANq} appear to be very hard to
evaluate analytically. Instead we evaluate the integrals numerically
assuming a simple Gaussian form of the nuclear profile function,
\begin{equation}
  \label{Tgauss}
  T (\ul b) = \frac{4}{3} \, R \, \rho \ e^{-b_T^2/R^2}
\end{equation}
with $R$ the nuclear radius and $\rho$ the nucleon density. Such
Gaussian profiles are of course not realistic for nuclei, but have
been successfully used to describe protons (see e.g.
\cite{Ayala:1996em}).

In evaluating the STSA in \eqref{ANq} one has to remember that in the
standard convention one has to choose $\ul k$ in the direction left of
the beam, which, in our notation, means along the negative $y$-axis. Hence we need to replace $k_\bot^2 \to - k_T$ in \eqref{ANq} (recall that $k_\bot^2$ is the $y$ component of the transverse vector $\ul{k}$).

\begin{figure}[ht]
  \centering
  \includegraphics[width=10cm]{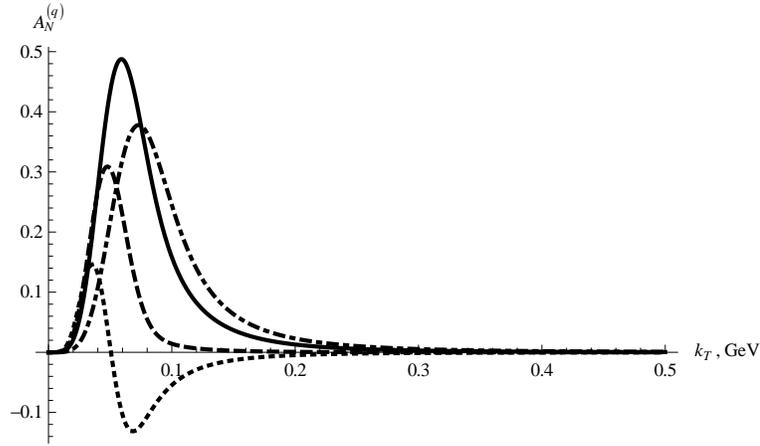}
 \caption{Quark STSA from \eqref{ANq} for the proton target plotted as 
   a function of $k_T$ for different values of the longitudinal
   momentum fraction $\alpha$ carried by the produced quark: $\alpha =
   0.9$ (dash-dotted curve), $\alpha = 0.7$ (solid curve), $\alpha =
   0.6$ (dashed curve), and $\alpha = 0.5$ (dotted curve).  The coordinate-space integrals over
	 $x_T , y_T$ are cut off at 2.1 fm.}
\label{AN_fig} 
\end{figure}

To plot \eqref{ANq} we will attempt to use somewhat realistic numbers,
while realizing that all the theoretically-calculated cross sections
are likely to have non-perturbative normalization corrections, which
may affect the size of the effect. To that end, we will use the
saturation scale (cf. \eqref{eq-saturation scale})
\begin{equation}
  \label{Qsat_real}
  Q_s^2 ({\ul b}) = 2 \, \pi \, \as^2 \, K^2 \, T ({\ul b})
\end{equation}
with the $K$-factor fixed at $K=10$ to make $Q_s \approx 1$~GeV, which
is a realistic value for a proton at; $x \sim 10^{-4}$. (Each $T' (x_T)$ in \eqref{ANq} is
multiplied by the same $K^2$-factor, since it also arises from the
saturation scale.) We put $m=300$~MeV to mimic a constituent quark,
along with $\rho = 0.35$~fm$^{-3}$ for a proton of radius $R =
0.878$~fm, and $\as =0.3$.  We plot the resulting $A_N^{(q)}$ from
\eqref{ANq} in \fig{AN_fig} for different values of $\alpha$ with the IR
cutoff $\Lambda = 100$~MeV and cutting off the $x_T$- and $y_T$
integrals in \eqref{ANq} at $2.1$~fm in the IR. (Note that
strictly-speaking the CGC formalism employed here is valid only for
scattering on a nuclear target, since it resums powers of a large
parameter $\as^2 \, A^{1/3}$. However its applications to a proton
target have been successful phenomenologically in the past
\cite{Albacete:2010sy}, giving one hope that our estimates here could
be relevant for $p^{\uparrow} + p$ collisions.)

From \fig{AN_fig} we see that our STSA is a non-monotonic function of
transverse momentum $k_T$, first rising and then falling off with
$k_T$ in qualitative agreement with the data shown in the right panel
of \fig{fig-Experimental Data}. As one can clearly see from \eqref{ANq}
the maximum of $A_N^{(q)}$ at impact parameter $x_T$ in our formalism
is determined (up to a constant) by the saturation scale, $k_T \sim
Q_s (x_T)$, such that the asymmetry integrated over all impact
parameters peaks at $k_T \sim Q_s$ with $Q_s$ an effective averaged
saturation scale. The conclusion about $A_N$ peaking at $k_T \approx
Q_s$ was previously reached in \cite{Boer:2006rj}. Let us stress again
that the STSA in our case changes sign when plotted as a function of
$k_T$ or $\alpha$ (i.e., it has a ``node'').

Note that, while the magnitude of STSA plotted in \fig{AN_fig} can be as
large as tens of percent, like the data in \fig{fig-Experimental
  Data}, the momentum at which the asymmetry is non-zero appears to be
much smaller in our \fig{AN_fig} than it is in the data of
\fig{fig-Experimental Data}. The discrepancy of the $k_T$-range of the
data and our \fig{AN_fig} signals the following potential problem: the
$x_T$-integral in \eqref{ANq} is dominated by large $x_T$, where $Q_s
(x_T)$ is small, leading to small values of $k_T$ dominating $A_N$,
and potentially making the corresponding physics
non-perturbative. Thus our perturbative calculation appears to be
sensitive to the non-perturbative domain.

\begin{figure}[ht]
  \centering
  \includegraphics[width=10cm]{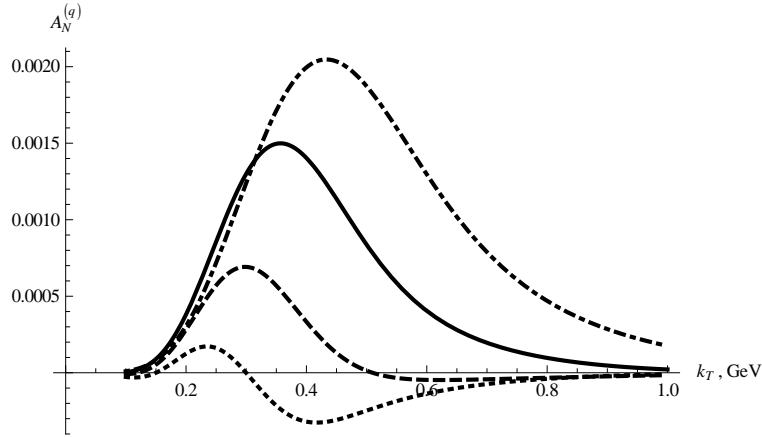}
  \caption{Same as in \fig{AN_fig}, but with a $1.3$~fm upper cutoff on
    the $x_T$- and $y_T$ integrals in \eqref{ANq} and a factor of $\ln
    \tfrac{k_T}{\tilde m}$ inserted.}
\label{AN2_fig} 
\end{figure}

To illustrate the range of spectra that can be obtained by our
estimates, we replot $A_N$ from \fig{AN_fig} in \fig{AN2_fig} cutting off the
$x_T$- and $y_T$-integrals in \eqref{ANq} by $1.3$~fm.  In addition, we
mimic the coordinate-space logarithms, like those that were neglected
after \eqref{dsigmaq8}, by introducing a factor of $\ln\tfrac{k_T}{\tilde
  m}$.  In this plot the $k_T$-range of the asymmetry is broader than
in \fig{AN_fig}, which makes it closer to the experimental data in
\fig{fig-Experimental Data}, but the height of the asymmetry is over
an order-of-magnitude lower than the data.  More work is needed to
assess whether the cutoff dependence is a result of the approximations
made, or whether it actually signals a potential breakdown of the
approach indicating the non-perturbative nature of STSA.

Another important observation one can make from \fig{AN_fig} is that
$A_N^{(q)}$ increases with increasing $\alpha$, except for very large
values of $\alpha$ when it starts to decrease. The increase of
$A_N^{(q)}$ with increasing $\alpha$ is in qualitative agreement with
the data in the left panel of \fig{fig-Experimental Data}, where the
data points increase with increasing Feynman-$x$. 
Inclusion of gluon fragmentation is necessary to perform a quantitative comparison with the
data.

\begin{figure}[ht]
  \centering
  \includegraphics[width=10cm]{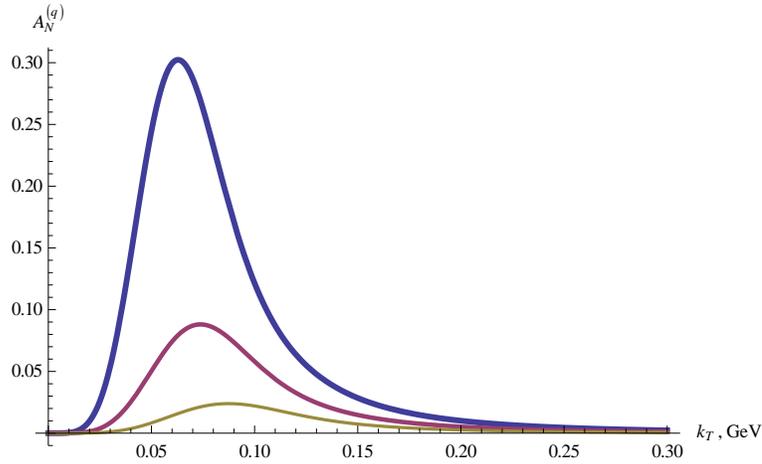}
 \caption{Quark STSA from \eqref{ANq} plotted as a function of $k_T$ for different 
   values of the target radius: $R = 1$~fm (top curve), $R = 1.4$~fm
   (middle curve), and $R = 2$~fm (bottom curve) for $\alpha = 0.7$.  Here the coordinate-space
	 integrals are cut off at 2.4 fm.}
\label{ANrad} 
\end{figure}

Finally, to test the dependence of our STSA in \eqref{ANq} on the size of
the target, we note that for $k_T \approx Q_s$ one gets
\begin{equation}
  \label{Adep}
  A_N^{(q)} (k_T \approx Q_s) \sim \frac{1}{Q_s^7} \sim A^{-7/6},
\end{equation}
if $Q_s^2 \sim A^{1/3}$. This indicates a very steep falloff of STSA
with the atomic number of the nuclear target. Such a conclusion
appears to be supported by the numerical evaluation of \eqref{ANq} for
several different radii of the target shown in \fig{ANrad}. (Now the
$x_T$- and $y_T$ integrals are cut off at $2.4$~fm.) One can see that
$A_N^{(q)}$ drops very rapidly with the size of the target.
If the experimentally observed STSA in $p^{\uparrow} + p$ collisions
are due to our mechanism, our prediction is then that in $p^{\uparrow}
+ A$ collisions STSA should be much smaller than that in $p^{\uparrow}
+ p$. In the case of a heavy ion target like $Au$ the STSA due to our
mechanism is likely to be negligibly small.

While we have demonstrated here the potential for our calculations to
agree with the data, the evaluations presented here have to be
significantly improved to reach a definitive conclusion. For instance
the $k_T$-dependence and the overall normalization in our
Eqs.~\peq{dsigmaq11} and \peq{Gunp2} are overly simple and not
ready to be compared to the data. The equations need to be corrected
for the effects of DGLAP evolution, small-$x$ evolution and for the
running of the coupling for a meaningful quantitative comparison with
the data.  Only such a phenomenological analysis can determine whether
our mechanism for generating STSA is dominant, or whether it is simply
one of the many factors contributing to the asymmetry.



\subsection{STSA in Photon Production}

\label{subsec-Photon STSA (est)} 

Using the methods developed in Sec.~\ref{quarkSTSAsec} we can now
evaluate the photon STSA given by Eqs.~\peq{dsigmaq} and \peq{phanti}. Substituting \eqref{phanti} into
\eqref{dsigmagamma} and performing the variable shift of \eqref{tildes}
while keeping in mind that now $\ul u$ and $\ul w$ are given by
Eqs.~\peq{uvG} yields
\begin{align}
 \nonumber
 d(\Delta\sigma^{(\gamma)}) &= \frac{i}{(2\pi)^3} \,  
 \int d^2 x \, d^2 y \, d^2 z
 \, e^{-i \ul k \cdot (\ul z - \ul y)} \, \Phi_{pol} (- \ul z \, , \, - \ul y, \alpha) \
 \bigg[ O_{{\ul x} + (1-\alpha) \, {\ul z},  \, {\ul x} + (1-\alpha) \, {\ul y}} \:
 \\ \label{dsigmaph1} &-
 O_{{\ul x} , \, {\ul x} + (1-\alpha) \, {\ul y}} 
 - O_{{\ul x} + (1-\alpha) \, {\ul z}, \, {\ul x}} \bigg]
\end{align}
where we have again dropped the tildes for brevity. Using the argument
of Eqs.~\peq{zero_arg} and \peq{zero_arg2} we see that each term in
the square brackets in \eqref{dsigmaph1} is zero after the integration
over $\ul x$. We thus have an exact result that
\begin{equation}
  \label{dsigmaph2}
  d(\Delta\sigma^{(\gamma)}) = 0
\end{equation}
in our mechanism for generating photon STSA. Hence the photon STSA is
zero, $A_N^{(\gamma)} =0$, in the forward production region under
consideration.



\section{Conclusions}

\label{sec-Concl} 

To conclude this Chapter let us summarize the main points discussed herein. Above we
have shown how STSA can be generated in the CGC formalism for quark
and gluon production. The results for the corresponding cross sections
are given in Eqs.~\peq{dsigmaq}, \peq{dsigmaG}.
The same mechanism gives zero STSA for prompt photons.

In our case STSA is generated by both a splitting in the projectile
wave function, and by the combination of the $C$-odd and $C$-even
interactions with the target. Hence our STSA-generating mechanism is
distinctly different from the Collins \cite{Collins:1992kk} and
Sivers \cite{Sivers:1989cc,Sivers:1990fh} effects, and is more akin to
(though still different from) the higher-twist mechanisms of
\cite{Efremov:1981sh,Efremov:1984ip,Qiu:1991pp,Ji:1992eu,Qiu:1998ia,Brodsky:2002cx,Collins:2002kn,Koike:2011mb,Kanazawa:2000hz,Kanazawa:2000kp}.

Evaluating the quark STSA in a simplified quasi-classical model we
found qualitative agreement with the data: quark STSA appears to be a
non-monotonic function of $k_T$, and is an increasing function of
increasing $x_F$ (for most of the $x_F$-range). It is perhaps
encouraging that the obtained asymmetry can be of the
order-of-magnitude of the experimental data.  On the other hand, the plots sketched in
Figs.~\ref{AN_fig} and \ref{AN2_fig} suggest a concerning sensitivity of the perturbative
calculation to nonperturbative cutoffs.  At this level, it is unclear whether this strong 
cutoff dependence is a consequence of the approximations made to the general formulas or
a feature of the approach itself.  Detailed phenomenological
studies of our formulas \peq{dsigmaq} and \peq{dsigmaG} are
needed to resolve these questions and make a meaningful quantitative comparison to the data.

Analyzing the general quark production formula \peq{eq-cross section}
one can see that the contribution to STSA arises from the ${\ul z}
\leftrightarrow {\ul y}$ anti-symmetric part of the integrand, which corresponds
to charge conjugation \eqref{WPT7}.  In arriving at \eqref{dsigmaq} from 
\eqref{eq-cross section} we employed the
lowest-order (order-$\as$) spin-dependent part of the light-cone wave
function squared \peq{eq-polarized wavefn}, which happens to be
$C$-even.  Hence, in our case to obtain a contribution to the STSA, the
interaction with the target had to be $C$-odd, driven by the odderon $O_{\ul x \ul y}$ \eqref{Odef}.  
However, it is possible that higher-order
corrections to the light-cone wave function squared would lead to a $C$-odd
contribution. (By ``wave function corrections'' we understand all the initial- and final-state corrections with rapidities between the projectile and the
particle we tag on.)  In such a case, the interaction with the target
need not be ${\ul z} \leftrightarrow {\ul y}$ anti-symmetric, and could
be mediated by the standard $C$-even exchange $S_{\ul x \ul y}$ \eqref{Sdef}. To test whether such a
scenario is feasible within the CGC/saturation perturbative framework
one has to calculate the higher order corrections to the
polarization-dependent light-cone wave function squared
\peq{eq-polarized wavefn}.  The corrections would need to generate a
relative complex phase between the corrected and uncorrected wave
functions \cite{Brodsky:2002cx,Collins:2002kn}, as discussed in \ref{SIDISF}. 
This can be accomplished in LCPT if the corrections have a contribution from an intermediate state in which the imaginary part of the energy denominator leads to a non-vanishing
polarization-dependent contribution to the scattering amplitude. 

\begin{figure}
 \centering
 \includegraphics[width=0.4\textwidth]{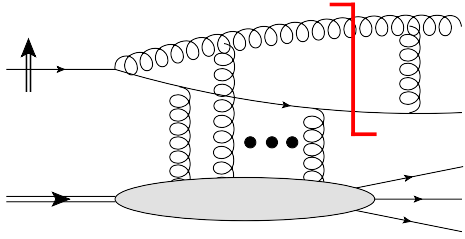}
 \caption{Diagram in which the STSA could be generated by ``lensing'' interaction with spectators
 in the polarized projectile.  Compared to the diagrams shown in Fig.~\ref{qtoqGampl} that
 are driven by the odderon \eqref{Odef}, this process could nominally have the same power-counting
 if mediated by the standard $C$-even exchange \eqref{Sdef} with the target.  As in Fig.~\ref{DIS-DY2},
 the auxiliary cut represents the intermediate state that can contribute its imaginary part by going
 on-shell.}
 \label{fig-pA_Sivers}
\end{figure}

An example of such corrections in our case could be a modification of the
amplitude in \fig{qtoqGampl} resulting from a gluon exchange between
the outgoing quark and gluon formed in the projectile splitting, as illustrated
in Fig.~\ref{fig-pA_Sivers}.  Diagrams of this type are analogs of the ``lensing''
mechanism of \ref{sec:Lensing} in this CGC formalism; they correspond to additional
rescatterings on the remnants of the polarized projectile.
The calculation of such diagrams appears to be rather complicated and is
beyond the scope of this work.  However, the power-counting indicates that 
it potentially may give a contribution comparable to the STSA resulting from \eqref{dsigmaq}: the
latter consists of the order-$\as$ light-cone wave function squared,
convoluted with the target interaction resumming powers of $\as^2 \,
A^{1/3}$ and $\as \, Y$, with one extra power of $\as$ due to the
odderon exchange \peq{Ocl3}.  Our contribution \peq{dsigmaq} is,
therefore, order-$\as^2$, if one assumes that $\as^2 \, A^{1/3} \sim
1$ and $\as \, Y \sim 1$, which is parametrically comparable to the
$C$-odd order-$\as^2$ light-cone wave function squared, interacting
with the target through a $C$-even order-one exchange.  An explicit
calculation is needed to explore this possibility and is left for
future work.


\chapter{Saturation as a Novel Mediator of Transverse Spin}
\label{chap-MVspin}


In Chapter~\ref{chap-TMD}, we discussed the transverse-momentum-dependent parton distribution functions (TMD's) of quarks and gluons in a hadronic state in the Bjorken limit.  The quark TMD's were defined as independent projections of the quark-quark correlator defined in \eqref{DISS-GAUGE3}:
\begin{align}
  \label{eq:q_corr}
  \Phi_{ij}^C (x, \ul{k}; P, S) \equiv \frac{1}{2(2\pi)^3} \int d^{2-} r \, e^{i k \cdot r}
 \bra{P S} \psibar_j (0) \, U_C [0,r] \, \psi_i (r) \ket{P S}_{r^+ = 0},
\end{align}
where $k \cdot r = \tfrac{1}{2} x P^+ r^- - \ul{k} \cdot \ul{r}$, since $r^+ = 0$.  The gauge link $U_C$ makes the correlator gauge-invariant and follows different contours $C$ depending on whether the correlator couples to a process with initial-state interactions \eqref{DISS-GAUGE5} such as the Drell-Yan process (DY) or final-state interactions \eqref{DISS-GAUGE6} such as semi-inclusive deep inelastic scattering (SIDIS).  In this Chapter, we will work in a frame such that the hadronic state $\ket{P S}$ has a large light-cone momentum $P^+$, and we will work in the $A^- =0$ gauge.  As we saw in \eqref{Acov}, in these kinematics at the classical level the gluon field of the hadron / nucleus has zero transverse component, ${\ul A} =0$, such that the only non-zero component is $A^+$.  This gauge choice causes the transverse gauge link at light-cone $\pm \infty$ to become a trivial factor of unity, such that the only contributions to $U_C$ come from the legs of the gauge link directed along the light-cone.  Defining the Wilson line 
\begin{align} 
  \label{Wlines1}
  V_{\ul x} [b^- \, , \, a^-] \equiv \mathcal{P} \, \exp \left[\frac{i g}{2} 
	\int\limits_{a^-}^{b^-} dx^- \hat{A}^+ (x^+ = 0, x^-, \ul{x}) \right]
\end{align}
with Hermitian conjugate
\begin{align}
 \label{Wlines_dagger}
 V_{\ul x}^\dagger [b^- , a^-] \equiv \bigg( V_{\ul x} [b^- , a^-] \bigg)^\dagger 
 = V_{\ul x} [a^- , b^-] ,
\end{align}
we write for the case of final-state interactions in SIDIS \cite{Collins:2002kn,Belitsky:2002sm}
\begin{align}
  \label{eq:U_SIDIS}
  {\cal U}^{SIDIS}[0,r] = V_{\ul 0}^\dagger [+\infty \, , \, 0] \, V_{\ul
    r} [+\infty \, , \, r^-],
\end{align}
while for initial-state interactions in DY we have 
\begin{align}
  \label{eq:U_DY}
  {\cal U}^{DY}[0,r] = V_{\ul 0} [0 \, , \, -\infty ] \, V_{\ul r}^\dagger
  [r^- \, , \, -\infty ].
\end{align}

The correlation function $\Phi_{ij}$ is decomposed into TMD parton distributions, as written in \eqref{DISS-DECOMP1}:
\cite{Boer:1997nt,Boer:2002ju}
\begin{align}
  \label{eq:Phi_dec_5}
  \Phi^C (x, \ul{k} &; P, S) \equiv \left[f_1^q (x, k_T) - \frac{(\ul{k} \times \ul{S})}{M} f_{1T}^{\bot q} 
 (x, k_T) \right] \left[\frac{1}{4} \gamma^- \right] 
 \\ \nonumber &+
 \left[S_L g_1^q (x, k_T) + \frac{(\ul{k} \cdot \ul{S})}
 {M} g_{1T}^q (x, k_T) \right] \left[\frac{1}{4} \gamma^5 \gamma^- \right] 
 \\ \nonumber &+
 \left[S_\bot^i h_{1T}^q (x, k_T) + \left(\frac{k_\bot^i}{M}\right) S_L h_{1L}^{\bot q} (x, k_T) +
 \left(\frac{k_\bot^i}{M} \right) \left(\frac{\ul{k}\cdot\ul{S}}{M}\right) h_{1T}^{\bot q} \right]
 \left[\frac{1}{4} \gamma^5 \gamma_{\bot i} \gamma^- \right] 
 \\ \nonumber &+
 \left[\left(\frac{k_\bot^i}{M}\right) h_1^{\bot q} (x, k_T) \right] \left[\frac{1}{4} i \gamma_{\bot i}
 \gamma^- \right],
\end{align}
where $M$ is the mass of the hadron.  As shown in \eqref{DISS-DECOMP2}, by contracting the Dirac indices of \eqref{eq:q_corr} with $\gamma^+$, we project out two distributions of unpolarized quarks: 
\begin{align}
 \label{eq:Sivers_ext_5}
 \frac{1}{2} \Tr \left[ \Phi^C (x, \ul{k}; P, S) \gamma^+ \right] &= f_1 (x, k_T) - 
 \frac{(\ul{k} \times \ul{S})}{M} f_{1T}^{\bot q} (x , k_T) ,
\end{align}
the unpolarized distribution $f_1 (x, k_T)$ present in an unpolarized hadron and the Sivers function $f_{1T}^\bot (x, k_T)$ reflecting the single transverse spin asymmetry of quarks in a transversely-polarized hadron.  These two functions $f_1 , f_{1T}^\bot$ can be extracted by explicitly symmetrizing or antisymmetrizing \eqref{eq:Sivers_ext_5} with respect to either the spin $\ul{S}$ or transverse momentum $\ul k$.

\begin{figure}
\centering
\includegraphics[width=0.7\textwidth]{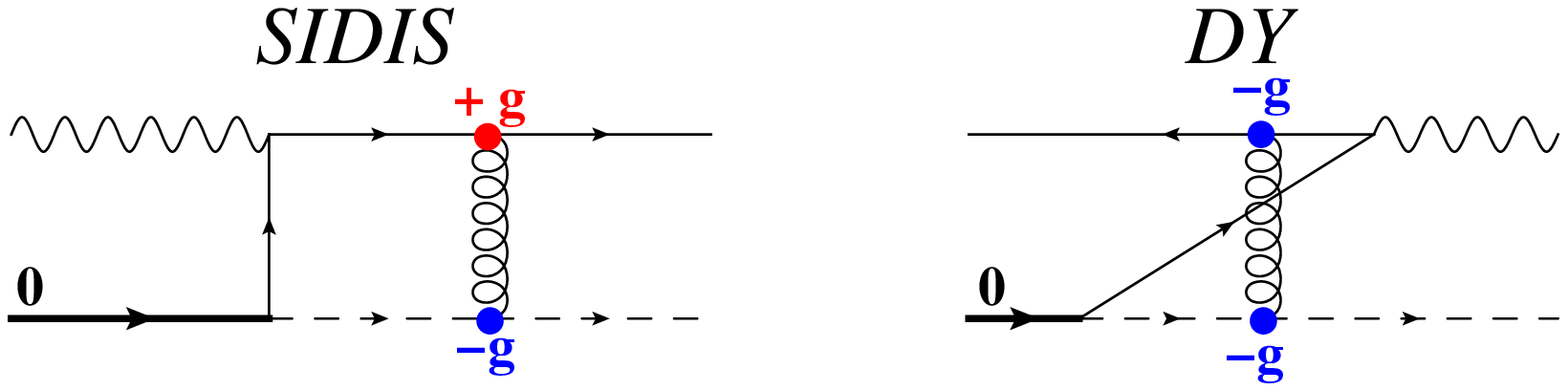}
\caption{The lensing mechanism which generates the asymmetry in the Sivers function
 due to the correlated color-charges of the active quark and the hadronic remnants.  Color
 conservation guarantees that the remnants have the net charge of an antiquark, resulting in
 an attractive final-state interaction in SIDIS and a repulsive initial-state interaction in DY.}
\label{relsign_v2}
\end{figure}

Of particular interest is the Sivers function $f_{1T}^\bot$, since it depends on the presence of the gauge link to be nonzero, as discussed in Sec.~\ref{subsec-sign_flip}.  The Sivers function is therefore sensitive to the different contours $C$ and is predicted to change sign \eqref{DISS-SIVERS7} between SIDIS and DY.  The conventional physical explanation for the Sivers function is through the effects of QCD lensing discussed in Sec.~\ref{sec:Lensing} and illustrated in Fig.~\ref{relsign_v2}.  In this mechanism, the interactions between the projectile (embodied in the gauge link) and the target are color-correlated, giving rise to a net attractive lensing force in the final-state interactions of SIDIS and a net repulsive lensing force in the initial-state interactions of DY.  This physical mechanism is most clearly realized in a simple model such as the scalar diquark model \eqref{DISS-DECOMP4}, in which the hadron fluctuates into a quark and a pointlike scalar particle representing the hadronic remnants.  Essentially, QCD lensing reflects the condition of color conservation on the initial- and final-state interactions; since the hadron is color neutral and a quark is being extracted from its wave function, the remnants must have a net color-charge corresponding to an antiquark.

But in this regard, the scalar diquark model is oversimplified.  QCD lensing depends essentially on the sensitivity of the initial- and final-state interactions to the \textit{total} color-charge of the hadronic remnants.  When the remnants are modeled by a pointlike particle, this feature is trivial, guaranteeing that the rescattering results in a net force in a definite direction.  But in reality, the remnants consist of a large number of partons with a variety of color-charges, with only the total color-charge constrained by color conservation.  As we saw in Chapter~\ref{chap-CGC}, a dense system of color-charges generates a dynamical color correlation length, whose inverse is the saturation scale $Q_s$.  Over transverse distances larger than $1 / Q_s$, the color-charges of the dense system become explicitly \textit{uncorrelated} (see \eqref{eq-MV correlator}); rescattering on these uncorrelated color-charges gives rise to QCD shadowing effects, rather than lensing.

These considerations show that the physics of QCD lensing cannot be a complete explanation of the microscopic origin of the Sivers function, and they suggest that QCD shadowing will play a competing role in the interactions embodied in the gauge link $U_C$.  Moreover, as we saw in Chapter~\ref{chap-CGC}, the effects of QCD shadowing grow as the density increases (c.f. \eqref{Qsat3}), so that QCD lensing should be a negligible effect in the Sivers function of a very dense system.  A natural starting point to analyze the contribution of shadowing to the Sivers function is the quasi-classical McLerran-Venugopalan (MV) model applied to a polarized heavy nucleus.  The analysis we will perform in this Chapter is not necessarily a very realistic description of a polarized heavy nucleus; rather, it is intended to serve as a metaphor for saturation effects generated from quantum evolution in a polarized proton.  This situation is the opposite of the one considered in Chapter~\ref{chap-odderon}; there we considered saturation effects in a dense, unpolarized target being struck by a dilute polarized probe; now we would like to consider saturation effects on the \textit{polarized} side.  
 
Thus, in this Chapter, we will calculate the Sivers function of a polarized heavy nucleus using the framework of Glauber-Gribov-Mueller (GGM) multiple rescattering on a heavy nucleus in the McLerran-Venugopalan model.  This quasi-classical calculation relates the Sivers function of the nucleus to the TMD's of its constituent nucleons; in the process, we will find a nontrivial role played by the orbital motion of the nucleons within the nucleus.  Together with QCD shadowing, this nucleonic orbital angular momentum (OAM) gives rise to a new mechanism, distinct from QCD lensing, which can generate the Sivers function.  In the Sections that follow, we will derive an effective factorization in the quasi-classical limit which allows us to relate the TMD's of the heavy nucleus to the TMD's of its nucleons.  Then we will use this decomposition to explicitly calculate the Sivers function, both for SIDIS and for DY.  In this chapter we present original work and follow closely our paper \cite{Kovchegov:2013cva}.


\section{Quasi-Classical Factorization: SIDIS on a Dense Target}
\label{sec:SIDIS}

We first consider the process of quark production in semi-inclusive
deep inelastic lepton scattering in Bjorken kinematics on a transversely polarized heavy
nucleus: $\ell + A^\uparrow \rightarrow \ell' + q + X$.  As discussed in Sec.~\ref{subsec-SIDIS},
the leptonic tensor can be factorized out in the usual way, so we represent the
process as the scattering of a virtual photon: $\gamma^* + A^\uparrow
\rightarrow q + X$.  This photon carries a large spacelike virtuality
$q_\mu q^\mu = -Q^2$ and knocks out a quark from one of the nucleons,
which may then rescatter on the nuclear remnants. The nucleus is taken
in the classical GGM/MV approximation of Chapter~\ref{chap-CGC}, which we augment by allowing the nucleons to be polarized and the nucleus to rotate around the
transverse polarization axis, which leads to a non-zero OAM.

\begin{figure}[ht]
\centering
\includegraphics[width=0.7\textwidth]{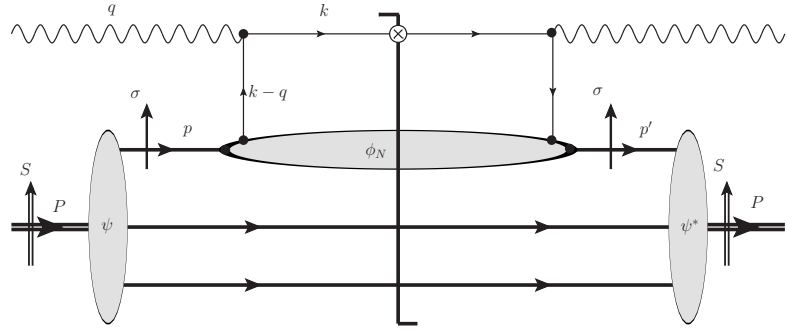}
\caption{The lowest-order SIDIS process in the usual $\alpha_s$
  power-counting. A quark is ejected from a nucleon in the nucleus by
  the high-virtuality photon, which escapes without
  rescattering. Different solid horizontal lines represent valence
  quarks from different nucleons in the nuclear wave function, with
  the latter denoted by the vertical shaded oval.}
\label{fig:DIS1}
\end{figure}

Consider first the lowest-order process shown in Fig.~\ref{fig:DIS1},
in which a quark is ejected without rescattering.\footnote{In
  the Regge limit quark production is dominated by a higher-order in
  $\as$ process, where the virtual photon splits into a $q\bar q$ pair
  before hitting the target (Fig~\ref{fig-Regge_DIS}).   
	Since we now work in the Bjorken limit with $x \sim \ord{1}$, the
  dipole process is not dominant, constituting an order-$\as$
  correction to the channel shown in \fig{fig:DIS1}.} We work in a
frame (such as the photon-nucleus center-of-mass frame) in which the
virtual photon moves along the $x^-$-axis with a large momentum $q^-$
and the nucleus moves along the $x^+$-axis with a large momentum
$P^+$.  In this frame, the kinematics are
\begin{align} \label{DIS1}
 \begin{aligned}
   P^\mu &= \left( P^+ , \frac{M_A^2}{P^+}, \ul{0} \right) \\
   q^\mu &= \left( -\frac{Q^2}{q^-} , q^- , \ul{0} \right) \\
   p^\mu &= \left( \alpha P^+ , \frac{p_T^2 + m_N^2}{\alpha P^+} , \ul{p} \right) \\
   k^\mu &= \left( \frac{k_T^2}{k^-} , k^- , \ul{k} \right),
 \end{aligned}
\end{align}
where $M_A$ is the mass of the nucleus, we have neglected the masses of the quarks, and the on-shell nucleon with momentum $p^\mu$ is a part of the light-cone wave function
of the nucleus.  We will denote the invariant mass of the nucleon remnants (the ``blob'') as  $m_X^2$.

Let us denote the photon-nucleus center-of-mass energy squared by $s_A
\equiv(P+q)^2$ and the photon-nucleon center-of-mass
energy squared by $\hat{s} \equiv (p+q)^2$.  We consider the kinematic
limit $s_A \gg \hat{s} , Q^2 \gg p_T^2 , k_T^2 , M_A^2, m_X^2$ and work to
leading order in the small kinematic quantities
$\tfrac{\bot^2}{\hat{s}} , \tfrac{\bot^2}{Q^2}$, which we denote
collectively as $\mathcal{O}(\tfrac{\bot^2}{Q^2})$.  Since we are 
operating in the limit in which $Q^2 \gg \bot^2 \gg \Lambda^2$, the
formalism of TMD factorization applies \cite{Collins:2011zzd}, justifying the use of the 
correlator \eqref{eq:q_corr} and decomposition \eqref{eq:Phi_dec_5}.
Additionally, to
a good accuracy one can assume that a typical scale for the momentum
fraction $\alpha$ is $\ord{1/A}$, where $A$ is the mass number of the
nucleus.  In this limit,
\begin{align} \label{DIS2}
 \begin{aligned}
   p^+ q^- &= \hat{s} + Q^2 \\
   q^+ &= - \left( \frac{Q^2}{\hat{s}+Q^2} \right) p^+ = - x \, p^+ =
   - \alpha \, x \, P^+
 \end{aligned}
\end{align}
where $x \equiv Q^2 / (2 p \cdot q)$ is the Bjorken scaling variable
per nucleon.  The corresponding scaling variable for the entire
nucleus is $x_A \equiv Q^2 / (2 P \cdot q) = \alpha \, x \approx x/A$.
The kinematic limit at hand, $\hat{s} \sim Q^2 \gg p_T^2 , k_T^2 ,
M_A^2$ corresponds to $x \sim \ord{1}$.  The on-shell condition for
the nucleon remnants is
\begin{align} 
  \label{DIS3} k^- = \frac{k_T^2}{k^+} = q^- + \frac{p_T^2 +
    m_N^2}{\alpha P^+} - \frac{(\ul{p}-{\ul k})_T^2 + m_X^2}{\alpha P^+ -
    \alpha \, x \, P^+ - k^+} \approx q^-
\end{align}
which fixes the struck quark to be ejected along the $x^-$-direction,
so that its light-cone plus momentum
\begin{align} 
  \label{DIS4} k^+ = \frac{k_T^2}{q^-} = \left(\frac{k_T^2}{\hat{s} +
      Q^2}\right) p^+ = \left(\frac{k_T^2}{Q^2} \right) \alpha \, x \,
  P^+
\end{align}
is small since $\sqrt{\hat{s}} \sim Q \sim p^+ \gg k_T$.  This also
fixes the momentum fraction of the active quark just before
interaction with the photon to be $x_F \equiv (k^+ - q^+)/p^+ \approx
- q^+/p^+ = x$ in the usual way. (Note that $q^+ = - Q^2/q^- < 0$.)

In our frame, the $x^-$-extent of the Lorentz-contracted nucleus is
$L^- \sim \tfrac{M_A}{P^+} R$, where $R$ is the radius of the nucleus
in its rest frame.  The incoming virtual photon and outgoing quark
interact with the nucleus based on their corresponding coherence
lengths: $\ell_\gamma^- \sim 1/|q^+|$ and $\ell_k^- \sim 1/k^+$,
respectively.  Comparing these to the size of the nucleus,
\begin{align} \label{DIS5}
 \begin{aligned}
   \frac{\ell_\gamma^-}{L^-} &\sim \frac{1}{x} \frac{1}{\alpha M_A R}
   \sim \ord{A^{-1/3}} \ll 1,
   \\
   \frac{\ell_k^-}{L^-} &\sim \frac{1}{x}
   \left(\frac{Q^2}{k_T^2}\right) \frac{1}{\alpha M_A R} \sim
   \ord{\frac{Q^2 + \hat{s}}{\bot^2} A^{-1/3}} \gg 1,
 \end{aligned}
\end{align}
we see that the photon's coherence length is short, but the coherence
length of the ejected quark is parametrically large for $\hat{s}, Q^2
\gg \perp^2 \, A^{1/3}$.  Thus, for our calculation in which $x \sim
\ord{1}$, the virtual photon interacts incoherently (locally) on a
single nucleon, but the ejected quark interacts coherently with all of
the remaining nucleons it encounters before escaping the nucleus.

This limit thus combines the local ``knockout'' picture of the deep
inelastic scattering process from Chapter~\ref{chap-TMD} with the coherent 
rescattering from Chapter~\ref{chap-CGC} that
usually characterizes the small-$x$ limit.  In the formal limit of a
large nucleus in which $\alpha_s \ll 1$ and $A \gg 1$ such that
$\alpha_s^2 A^{1/3} \sim \ord{1}$, these coherent interactions with
subsequent nucleons must be re-summed according to this
saturation-based power counting.


\subsection{Factorization at Lowest Order}

In general it is rather straightforward to write an answer for the
quasi-classical quark production in SIDIS. As mentioned previously, 
here the problem is a little more subtle than usual
since we are interested in also including the transverse and longitudinal
motion of the nucleons in the nucleus in order to model its OAM. Thus
our quasi-classical description of the nucleus has to provide us both
with the positions and momenta of the nucleons. This can be done using
Wigner distributions, which are overlaps of quantum wave functions that specify the average position and average momentum, consistent with the uncertainty principle \cite{Wigner:1932eb}.  Wigner distributions are thus the quantum-mechanical analog of classical phase-space distributions.

Let us illustrate the method with a simple ``knockout'' process
from \fig{fig:DIS1}.  Such a lowest-order process would dominate for a regime of intermediate density, where the number of nucleons is large enough to justify a mean-field treatment, but not large enough to re-sum scattering corrections; that is, $1/\alpha_s \gg A^{1/3} \gg 1$.
Just like in the parton model, the time scale of
inter-nucleon interactions is Lorentz-dilated in the infinite momentum
frame of the nucleus that we are working in. We can, therefore, write
the scattering amplitude for the process in \fig{fig:DIS1} as a
product of the light-cone wave function $\psi$ of nucleons
in the nucleus (defined according to light-front perturbation theory
rules \cite{Lepage:1980fj,Brodsky:1983gc} in the boost-invariant
convention of \cite{KovchegovLevin}) with the quark--virtual photon
scattering amplitude $M_K$:
\begin{align}
  \label{eq:wf_LO}
  M_{tot} = \psi (p) \, M_K (p, q, k). 
\end{align}
Here $\psi (p) = \psi (p^+/P^+, {\ul p})$ is the boost-invariant
light-cone wave function of a nucleon in the nucleus, while $M_K$ is the scattering amplitude for the
``knock-out'' process $\gamma^* + N \to q + X$. A sum over quantum numbers 
such as spin and color is implied in \eqref{eq:wf_LO}. In calculating
the quark production process we need to square this amplitude,
integrate over the momentum of the final state remnants and sum over
all nucleons in the nucleus. Since the momenta $k$ and $q$ are fixed, this
amounts to integrating over $p$:
\begin{align}
  \label{eq:ampl2_LO}
  \int \frac{d p^+ \, d^2 p}{2 (p^++ q^+) \, (2 \pi)^3} \, |M_{tot}|^2
  = A \, \int \frac{d p^+ \, d^2 p}{2 (p^+ + q^+) \, (2 \pi)^3} \,
  |\psi (p)|^2 \, |M_K (p, q, k)|^2.
\end{align}

First let us introduce a Fourier transform of the nucleon wave
function,
\begin{align}
  \label{eq:Fourier}
  \psi (b) \equiv \psi (b^-, {\ul b}) = \int \frac{d p^+ d^2 p}{2 \, \sqrt{p^+} \,
    (2 \pi)^3} \, e^{- i \, p \cdot b} \, \psi (p),
\end{align}
with $p \cdot b = \tfrac{1}{2} \, p^+ \, b^- - {\ul p} \cdot {\ul b}$.
Next we define the Wigner distribution for the nucleons with the help of the Fourier
transform \eqref{eq:Fourier}:
\begin{align} 
  \label{DIS8} 
  W(p,b) &\equiv W (p^+, {\ul p}; b^- , {\ul b}) \equiv \int d^2 \delta
  b \, d \delta b^- \, e^{i \, p \cdot \delta b} \, \psi(b +
  \tfrac{1}{2} \delta b) \, \psi^*(b - \tfrac{1}{2} \delta b)
	\\ \nonumber & =
	\frac{1}{2(2\pi)^3} \int \frac{d^2 \delta p \, d \delta p^+}
	{\sqrt{(p^+ + \tfrac{1}{2} \delta p^+)(p^+ - \tfrac{1}{2} \delta p^+)}}
	e^{-i \delta p \cdot b} \, \psi(p + \tfrac{1}{2} \delta p) \,
	\psi^* (p - \tfrac{1}{2} \delta p).
\end{align}
Note that the wave function is normalized such that
\begin{align}
\int \frac{d p^+ d^2 p}{2 \, p^+ \,
    (2 \pi)^3} \, |\psi (p)|^2 =1,
\end{align}
which, together with the normalization chosen for the Fourier transform \eqref{eq:Fourier}, gives
\begin{align} \label{rr2}
  \int  \frac{d p^+ \, d^2 p \, d b^- \, d^2 b}{2 (2\pi)^3} \, W(p,b) = 1.
\end{align}
Since
\begin{align}
\int d^2 b \, d b^- \, W (p,b) = |\psi (p)|^2 /p^+
\end{align}
we can recast \eq{eq:ampl2_LO} as
\begin{align}
  \label{eq:ampl2_LO_W}
  \int \frac{d p^+ \, d^2 p}{2 (p^++ q^+) \, (2 \pi)^3} \, |M_{tot}|^2
  = A \, \int \frac{d p^+ \, d^2 p \, d b^- \, d^2 b}{2 \, (2 \pi)^3}
  \, W(p,b) \, \frac{p^+}{p^+ + q^+} \, |M_K (p, q, k)|^2.
\end{align}

Finally, in the following, as usual in the saturation framework, it
will be convenient to calculate the scattering amplitude in (partial)
transverse coordinate space. Writing
\begin{align}
\label{M_Ftr}
M_K (p, q, k) = \int d^2 x \, e^{-i \, {\ul k} \cdot ({\ul x} - {\ul
    b})} \, M_K (p,q, {\ul x} - {\ul b} )
\end{align}
(with $k^-$ and $k^+$ fixed by Eqs.~\eqref{DIS3} and \eqref{DIS4}) we
rewrite \eq{eq:ampl2_LO_W} as
\begin{align}
  \label{eq:ampl2_LO_Wxy}
  \int \frac{d p^+ \, d^2 p}{2 (p^++ q^+) \, (2 \pi)^3} \, |M_{tot}|^2
  = A \, \int \frac{d p^+ \, d^2 p \, d b^- \, d^2 b}{2 \, (2 \pi)^3}
  \, W(p,b) \, \frac{p^+}{p^+ + q^+} \notag \\ \times \, \int d^2x \,
  d^2 y \, e^{-i \, {\ul k} \cdot ({\ul x} - {\ul y})} \, M_K (p, q,
  {\ul x} - {\ul b} ) \, M_K^* (p, q, {\ul y} - {\ul b} ).
\end{align}

Note that the Fourier transform \eqref{M_Ftr} appears to imply that
${\ul b}$ is the transverse position of the outgoing nucleon remnants in
\fig{fig:DIS1}, whereas in the Wigner distribution ${\ul b}$ is the
position of the incoming nucleon $p$. As we will shortly see such an
interpretation is not inconsistent: in the classical limit of a large
nucleus, the Wigner distribution is a slowly varying function of $\ul
b$, with changes in $W$ becoming significant over the variations of
$\ul b$ over distances of the order of the nucleon size 1 fm or
larger. The valence quark and outgoing gluon in \fig{fig:DIS1} are
perturbatively close to each other (being the part of the same Feynman
diagram), and hence the difference in their positions is outside the
precision of $W (p,b)$ and can be taken to be the same in the Wigner
distribution.

In Section~\ref{A5} we will show that the formula \eqref{eq:ampl2_LO_Wxy}
holds not only at the lowest order, but when multiple rescatterings are
included as well.  These rescatterings become important when the density is large enough to offset powers of the coupling and must be re-summed to all orders when $\alpha_s^2 A^{1/3} \sim \ord{1}$.
Let us now preview the generalized form of \eqref{eq:ampl2_LO_Wxy} after these rescatterings have been included.  In the kinematics outlined above, this takes the form
\begin{align}
  \label{eq:ampl2_LO_Wxy_net}
  \int \frac{d p^+ \, d^2 p}{2 (p^++ q^+) \, (2 \pi)^3} \, |A_{tot}|^2
  = A \, \int \frac{d p^+ \, d^2 p \, d b^- \, d^2 b}{2 \, (2 \pi)^3}
  \, W(p,b) \, \frac{p^+}{p^+ + q^+} \notag \\ \times \, \int d^2x \,
  d^2 y \, e^{-i \, {\ul k} \cdot ({\ul x} - {\ul y})} \, A (p, q,
  {\ul x} - {\ul b} ) \, A^* (p, q, {\ul y} - {\ul b} ),
\end{align}
where we define the energy-rescaled $2 \to 2$ scattering amplitudes as in \eqref{LCPT_CS2} by (see also
Eqs.~\eqref{eq:e-ind_ampl} and \eqref{eq:A_Eindep})
\cite{KovchegovLevin}
\begin{align}
  \label{eq:AM}
  A (p, q, k) = \frac{M (p, q, k)}{2 \, p^+ \, q^-}
\end{align}
and $A(p, q, k)$ in \eq{eq:ampl2_LO_Wxy_net} includes the
rescatterings on any number of nucleons in the
nucleus.\footnote{Strictly-speaking we need to include in
  \eq{eq:ampl2_LO_Wxy_net} Wigner function convolutions with the all
  the interacting nucleons in the nucleus: however, since in our
  kinematics only the first ``knockout'' process depends on the
  transverse momentum $p_\perp$ of the nucleon, we only keep one
  convolution with the Wigner function explicitly.} (Note that for a
``nucleus'' made out of a single nucleon, we have $p^+ = P^+$, which allows one
to reduce \eq{eq:ampl2_LO_Wxy} to \eq{eq:ampl2_LO_Wxy_net} by
neglecting the ``spectator'' nucleons.) We therefore conclude that the
quark production cross section for the $\gamma^* + A \to q + X$
process can be written as
\begin{align}\label{xsect_W}
  \frac{d \sigma^{\gamma^* + A \to q + X}}{d^2 k \, dy} = A \, \int
  \frac{d p^+ \, d^2 p \, d b^- \, d^2 b}{2 \, (2 \pi)^3} \, W(p,b) \,
  \frac{d \hat{\sigma}^{\gamma^* + NN\ldots N \to q + X}}{d^2 k \,
    dy},
\end{align}
where the cross section for producing a quark in $\gamma^*$ scattering
on the nucleons is
\begin{align}
 \label{xsectNNN}
 \frac{d \hat{\sigma}^{\gamma^* + NN\ldots N \to q + X}}{d^2 k \, dy} &= 
 {\cal N} \, \int d^2x \, d^2 y \, e^{-i \, {\ul k} \cdot ({\ul x}
   - {\ul y})} \, A_K (p, q, {\ul x} - {\ul b} ) \, A_K^* (p, q, {\ul
    y} - {\ul b} ) 
 \\ \nonumber &\times
 D_{{\ul x} \, {\ul y}} [+\infty, b^-]
\end{align}
with the semi-infinite fundamental dipole scattering amplitude given
by (cf. \eq{eq:U_SIDIS})
\begin{align}
 \label{dipole_def}
  D_{{\ul x} \, {\ul y}} [+\infty, b^-] = \left\langle \frac{1}{N_c}
    \, \mbox{Tr} \left[ V_{\ul x} [+\infty, b^-] \, V^\dagger_{\ul y}
      [+\infty, b^-] \right] \right\rangle
\end{align}
and with some ${\hat s}$ and $Q^2$-dependent prefactor $\cal N$. Here
$y = \ln 1/x$ is the rapidity of the produced quark and a factor of
$A$ in \eq{xsect_W} accounts for the fact that the first scattering
can take place on any of the $A$ nucleons. We fixed the normalization
of \eq{xsect_W} by requiring it to be valid for a nucleus made out of
a single nucleon, which would be described by a trivial Wigner
distribution fixing the momentum and position of the nucleon by simple
delta-functions.

As already mentioned before, with the accuracy of the large-$A$
classical approximation, the argument $\ul b$ in the Wigner
distribution can be replaced by any other transverse coordinate
involved in the scattering process. Hence one can replace $\ul b$ in
$W (b,p)$ from \eq{xsect_W} by either $\ul x$ or $\ul y$ from
\eq{xsectNNN}, or by any linear combination of those
variables. Replacing $\ul b$ in $W (b,p)$ from \eq{xsect_W} by $({\ul
  x} + {\ul y})/2$ and employing \eq{xsectNNN} we write
\begin{align}\label{xsect_W2}
  \frac{d \sigma^{\gamma^* + A \to q + X}}{d^2 k \, dy} = & A \, \int
  \frac{d p^+ \, d^2 p \, d b^- }{2 (2 \pi)^3} \, \int d^2x \, d^2 y \
  W \bigg( p , b^-, \frac{{\ul x} + {\ul y}}{2} \bigg) \notag \\ &
  \times \, e^{- i \, {\ul k} \cdot ({\ul x} - {\ul y})} \, |A_K|^2
  (p, q, {\ul x} - {\ul y} ) \, D_{{\ul x} \, {\ul y}} [+\infty, b^-],
\end{align}
where
\begin{align}
\label{MK2}
|A_K|^2 (p, q, {\ul x} - {\ul y} ) & \equiv \, {\cal N} \, \int d^2 b
\, A_K (p, q, {\ul x} - {\ul b} ) \, A_K^* (p, q, {\ul y} - {\ul b} )
\notag \\ & = \int \frac{d^2 k'}{(2 \pi)^2} \, e^{i \, {\ul k}' \cdot
  ({\ul x} - {\ul y})} \, \frac{d \hat{\sigma}^{\gamma^* + N \to q +
    X}}{d^2 k' \, dy} (p,q).
\end{align}
Substituting \eq{MK2} into \eq{xsect_W2} yields
\begin{align}\label{xsect_W3}
  \frac{d \sigma^{\gamma^* + A \to q + X}}{d^2 k \, dy} = & A \, \int
  \frac{d p^+ \, d^2 p \, d b^- }{2 (2 \pi)^3} \, \int d^2x \, d^2 y \
  W \bigg( p , b^-, \frac{{\ul x} + {\ul y}}{2} \bigg) \notag \\ &
  \times \int \frac{d^2 k'}{(2 \pi)^2} \, e^{- i \,({\ul k} - {\ul
      k}') \cdot ({\ul x} - {\ul y})} \, \frac{d
    \hat{\sigma}^{\gamma^* + N \to q + X}}{d^2 k' \, dy} (p,q) \,
  D_{{\ul x} \, {\ul y}} [+\infty, b^-].
\end{align}
\eq{xsect_W3} is our starting point for exploring the STSA in SIDIS:
it gives the quark production cross section in the quasi-classical
approximation.

\begin{figure}[t]
\centering
\includegraphics[width= .5 \textwidth]{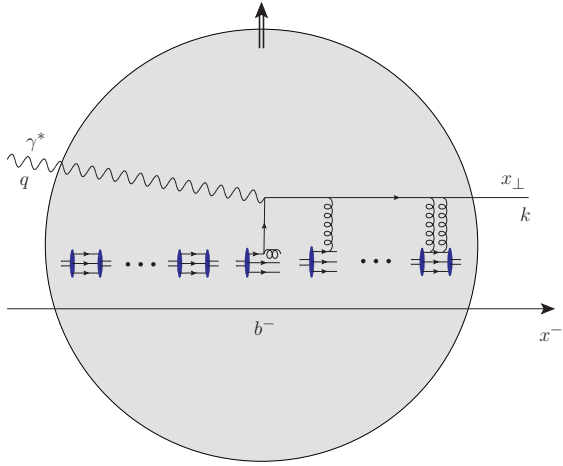}
\caption{Space-time structure of quark production in the
  quasi-classical SIDIS process in the rest frame of the nucleus,
  overlaid with one of the corresponding Feynman diagrams. The shaded
  circle is the transversely polarized nucleus, with the vertical
  double arrow denoting the spin direction.}
\label{SIDIS_qprod}
\end{figure}

The expression \eqref{xsect_W3} is illustrated in \fig{SIDIS_qprod}:
the first interaction between the incident virtual photon and a
nucleon in the transversely polarized nucleus happens at the
longitudinal coordinate $b^-$. A quark is knocked out, which proceeds
to interact with the rest of the nucleons in the nucleus. This latter
interaction is recoilless and is encoded in a Wilson line.

The Wigner distribution in \eq{xsect_W3} allows one to take the
quasi-classical GGM/MV limit of a large nucleus in a controlled
way. For a large ``classical" nucleus we usually can replace $W(p,b)$
by the following classical expression for it (neglecting the longitudinal
orbital motion of the nucleons)
\begin{align}
\label{Wcl}
W_{cl} (p,b) = \frac{4 \, \pi}{A} \, \rho ({\ul b}, b^-) \, \delta
\left( p^+ - \frac{P^+}{A} \right) \, w ( {\ul p} , b),
\end{align}
where $\rho ({\ul b}, b^-)$ is the nucleon number density normalized
such that
\begin{align}
\int d^2 b \, d b^- \rho ({\ul b}, b^-) = A. 
\end{align}
The function $w ( {\ul p} , b)$ in \eq{Wcl} is responsible for the
transverse momentum distribution of the nucleons and, to satisfy
\eq{rr2}, is normalized such that
\begin{align}
\int \frac{d^2 p}{(2 \pi)^2} \, w ( {\ul p} , b) =1. 
\end{align}
As originally formulated
\cite{McLerran:1993ka,McLerran:1994vd,McLerran:1993ni}, the MV model
contained no dependence on the spin or transverse momentum of the
valence quarks (c.f. \eqref{eq-MV current}).  This result is recovered by using $w_{MV} = (2 \pi)^2
\, \delta^2 ({\ul p})$.

Substituting the classical Wigner distribution \eqref{Wcl} into
\eq{xsect_W3} yields
\begin{align}\label{xsect_W4}
  \frac{d \sigma^{\gamma^* + A \to q + X}}{d^2 k \, dy} = & \int
  \frac{d^2 p \, d b^- }{(2 \pi)^2} \, d^2x \, d^2 y \ \rho \left
    (\frac{{\ul x} + {\ul y}}{2}, b^- \right) \, w \bigg( {\ul p} ,
  \frac{{\ul x} + {\ul y}}{2}, b^- \bigg) \notag \\ & \times \, \int
  \frac{d^2 k'}{(2 \pi)^2} \, e^{- i \,({\ul k} - {\ul k}') \cdot
    ({\ul x} - {\ul y})} \, \frac{d \hat{\sigma}^{\gamma^* + N \to q +
      X}}{d^2 k' \, dy} (p,q) \, D_{{\ul x} \, {\ul y}} [+\infty,
  b^-],
\end{align}
which is a simplified version of \eq{xsect_W3}.


\subsection{Factorization with Multiple Rescattering}
\label{A5}

In this Section we will justify the result given in
\eq{eq:ampl2_LO_Wxy_net}.  For simplicity, in this Section we will model nucleons as made out of
single valence quarks; at the end of the calculation, to go back to
the nucleons one simply needs to replace the distribution functions
in a valence quark by the distribution functions in the nucleons.

\begin{figure}[ht]
\centering
\includegraphics[height=4cm]{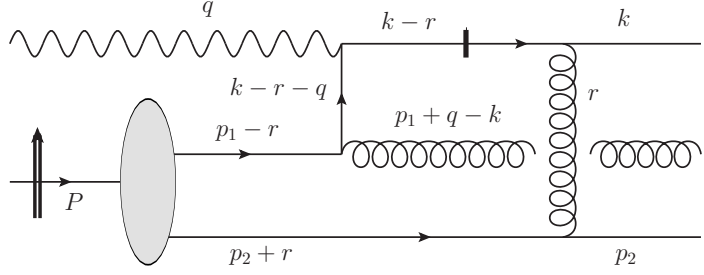}
\caption{
  \label{fig:DIS2} The minimal SIDIS process containing both the
  ``knockout'' of a quark from the nuclear wave function and
  rescattering on a different quark from a second nucleon.  The short
  thick vertical line indicates that the pole of the intermediate
  quark propagator is picked up in the calculation.  }
\end{figure}

To study the interplay between the local
``knockout'' channel of deep inelastic scattering and the coherent
multiple rescattering on the nuclear remnants, it is illustrative to
consider a minimal case with both features.  This process, shown in
Fig.~\ref{fig:DIS2}, consists of the knockout sub-process followed by
a single rescattering on a different quark from a second nucleon in
the nucleus.  Rescattering on a second nucleon receives a combinatoric
enhancement of order $\sim A^{1/3}$ compared to rescattering on the
same nucleon; the former is $\ord{1}$ in the saturation power
counting, while the latter is $\ord{\alpha_s}$.

The total SIDIS amplitude $M_{tot}$ depicted in Fig.~\ref{fig:DIS2}
consists of a loop integral connecting the mean-field single-particle
wave functions $\psi(p)$ of the nucleus to a scattering amplitude
$M_{K+R}$ denoting both the ``knockout'' and rescattering processes:
\begin{align} 
\label{DIS6} 
M_{tot} &= \int \frac{d r^+ \, d^2r }{2 \, (2 \pi)^3} \frac{P^+}{(p_1^+
  - r^+) \, (p_2^+ + r^+)} \psi(p_1 -r) \, \psi(p_2 + r)
\\ \nonumber &\times
M_{K+R}(p_1 - r,p_2 + r,q,k, r),
\end{align}
where we have used the two-particle phase space from light-cone perturbation theory in the conventions of \cite{Kovchegov:2012mbw} and a sum over spins and colors of the participating quarks is
implied. Squaring both sides of \eqref{DIS6} and integrating out the
final-state momenta $p_1$ and $p_2$ gives
\begin{align} \label{DIS7}
  & \langle |M_{tot}|^2 \rangle \equiv A \, (A -1) \int \frac{d p_1^+
    \, d^2 p_1 \, dp_2^+ \, d^2p_2}{[2 \, (2 \pi)^3]^2 \, (p_1^+ +
    q^+) \, p_2^+ } |M_{tot}|^2 \\ = & \int \frac{d p_1^+ \, d^2 p_1 \,
    dp_2^+ \, d^2p_2}{[2 \, (2 \pi)^3]^2 \, (p_1^+ + q^+) \, p_2^+ }
  \frac{d r^+ \, d^2 r}{2 \, (2 \pi)^3} \frac{d r^{\prime +} \, d^2
    r'}{2 \, (2 \pi)^3} \frac{A \, (A -1) \, \left( P^+
    \right)^2}{\sqrt{(p_1^+ - r^+) \, (p_2^+ + r^+) \,
      (p_1^+ - r^{\prime +})  \, (p_2^+ + r^{\prime +})}} \notag \\
  & \times \int d b_1^- \, d^2 b_1 \, d b_2^- \, d^2 b_2 \, e^{- i \,
    (r - r')\cdot(b_1 - b_2)} \, W \left(p_1 - \frac{r + r'}{2},b_1
  \right)
  W \left(p_2 + \frac{r + r'}{2} , b_2 \right) \notag \\
  & \times M_{K+R}(p_1 - r, p_2 + r,q,k, r) \, M^*_{K+R}(p_1 - r',p_2
  + r',q,k, r'), \notag
\end{align}
where we have employed the Wigner distributions defined in \eq{DIS8}
above and summed over all pairs of nucleons.

\eq{DIS7} is still far from \eq{eq:ampl2_LO_Wxy_net} because in
\eqref{DIS7} we do not have the amplitude squared: instead we have the
product of $M_{K+R}$ and $M^*_{K+R}$ with different arguments. It is
easier to further analyze the expression separately for the transverse
and longitudinal degrees of freedom. We proceed by taking the
classical limits, in which case the Wigner distributions give us the
position and momentum distributions of nucleons
simultaneously. Moreover, for the large nucleus at hand the Wigner
distributions depend on ${\ul b}_1$ and ${\ul b}_2$ weakly over the
perturbatively short distances associated with the Feynman
diagrams. We thus define ${\ul b} = ({\ul b}_1 + {\ul b}_2)/2$ and
${\ul \Delta b} = {\ul b}_1 - {\ul b}_2$ and write the Wigner distributions
at the average value $\ul b$, which allows us to simplify the expression as follows:
\begin{align}
  \label{eq:trans}
	\nonumber
  \int d^2 r &\, d^2 r' \, d^2 b_1 \, d^2 b_2 \, e^{i \, ({\ul r} -
  {\ul r}')\cdot({\ul b}_1 - {\ul b}_2)} \, W \left(p_1 - \frac{r +
  r'}{2},b_1 \right) W \left(p_2 + \frac{r + r'}{2} , b_2 \right)
  \\ \nonumber & \times 
	M_{K+R}(p_1 - r, p_2 + r,q,k, r) \, M^*_{K+R}(p_1
  - r',p_2 + r',q,k, r') 
	\\ \nonumber \, \\ \nonumber \approx &
	\int d^2 r \, d^2 r' \, d^2 b \, d^2
  \Delta b \, e^{i \, ({\ul r} - {\ul r}')\cdot {\ul \Delta b}} \,
	W \left(p_1 - \frac{r + r'}{2}, b_1^-, {\ul b} \right) 
	W \left(p_2 + \frac{r + r'}{2} , b_2^-, {\ul b}\right) 
	\\ \nonumber & \times 
	M_{K+R}(p_1 - r, p_2 + r,q,k,r ) \, M^*_{K+R}(p_1 - r',p_2 + r',q,k, r')
	\\ \nonumber \, \\ \nonumber \approx &
	(2 \pi)^2 \int d^2 r \, d^2 b \, W \left(p_1^+ - \frac{r^+ + r^{\prime +}}{2}, 
	{\ul p}_1 - {\ul r}, b_1^-, {\ul b} \right) \, W \left(p_2^+ + \frac{r^+ + r^{\prime
  +}}{2}, {\ul p}_2 + {\ul r} , b_2^-, {\ul b} \right)
	\\ \nonumber & \times 
  M_{K+R}(p_1^+ - r^+, {\ul p}_1 - {\ul r}, p_2^+ + r^+, {\ul p}_2 +
  {\ul r} ,q,k, r^+, {\ul r}) 
	\\ & \times 
	M^*_{K+R}(p_1^+ - r^{\prime +}, {\ul p}_1 - {\ul r}, p_2^+ + r^{\prime +}, {\ul p}_2 +
  {\ul r} , q,k, r^{\prime +}, {\ul r}).
\end{align}

Now the difference in the arguments of $M_{K+R}$ and $M^*_{K+R}$ is
only in the longitudinal momenta $r^+$ and $r^{\prime +}$. To
integrate over these momenta we notice that, as follows from
\fig{fig:DIS2}, in the high energy kinematics at hand the leading
contribution to the amplitude $M_{K+R}$ comes from the region where
$p_1^+, p_2^+ \gg r^+, r^{\prime +}$.  This is because the intermediate quark propagator $(k-r)$
already carries a large light-cone minus momentum $k^- \approx q^-$, and a simultaneously large value of $r^+$ would introduce additional suppression by the virtuality $(k-r)^2 \sim - k^- r^+$.  In this regime we combine
Eqs.~\eqref{DIS7} and \eqref{eq:trans} to write
\begin{align}
  \label{eq:app1} \nonumber
  \langle &|M_{tot}|^2 \rangle = \int \frac{d p_1^+ \, d^2 p_1 \,
  dp_2^+ \, d^2p_2}{[2 \, (2 \pi)^3]^2 \, (p_1^+ + q^+) \, p_2^+ }
  \frac{d r^+ \, d r^{\prime +} \, d^2 r}{4 \, (2 \pi)^4} \frac{\left(
  P^+ \right)^2}{p_1^+ \, p_2^+} d b_1^- \, d b_2^- d^2 b \, e^{-
  i \tfrac{1}{2} \, ( r^+ - r^{\prime +}) \, (b_1^- - b_2^-)} 
	\\ & \times \, 
	A (A -1) \, W \left(p_1^+, {\ul p}_1 - {\ul r}, b_1^-, {\ul b} \right) W \left(p_2^+, 
	{\ul p}_2 + {\ul r} , b_2^-, {\ul b} \right)
	\\ \nonumber & \times \,
  M_{K+R}(p_1^+, {\ul p}_1 - {\ul r}, p_2^+, {\ul p}_2 + {\ul r} ,q,k, r^+, {\ul r}) \,
	M^*_{K+R}(p_1^+, {\ul p}_1 - {\ul r}, p_2^+, {\ul p}_2 + {\ul r} ,q,k, r^{\prime +}, {\ul r}).
\end{align}
In the $p_1^+, p_2^+ \gg r^+, r^{\prime +}$ kinematics, the amplitude
$M_{K+R}$ contains only one pole in $r^+$ resulting from the
denominator of the $k-r$ quark propagator
(cf. \cite{Mueller:1989st,Kovchegov:1998bi,KovchegovLevin}). We can
thus write
\begin{align}
  \label{eq:ampl}
  M_{K+R}(p_1 - r, p_2 + r,q,k) = \frac{i}{(k-r)^2 + i \, \epsilon} \,
  {\tilde M}_{K+R}(p_1 - r, p_2 + r,q,k),
\end{align}
where ${\tilde M}_{K+R}$ denotes the rest of the diagram which does
not contain singularities in $r^+$ in the $p_1^+, p_2^+ \gg r^+,
r^{\prime +}$ approximation. (Note that ${\tilde M}_{K+R}$ also
contains the numerator of the $k-r$ quark propagator.) Since $(k-r)^2
\approx - k^- \, r^+ + {\ul k}^2 - ({\ul k} - {\ul r})^2$ we can use
\eq{eq:ampl} to integrate over $r^+$, as was done in \eqref{dipA6} and illustrated in Fig.~\ref{fig-NN3}:
\begin{align}
  \label{eq:long_int}
  \int\limits_{-\infty}^\infty \frac{d r^+}{2 \pi} \, & e^{- i
    \tfrac{1}{2} \, r^+ \, (b_1^- - b_2^-)} \, M_{K+R}(p_1 - r, p_2 +
  r,q,k) \notag \\ & \approx \frac{1}{k^-} \, \theta (b_2^- - b_1^-)
  \, {\tilde M}_{K+R} (p_1^+ , {\ul p}_1 - {\ul r}, p_2^+ , {\ul p}_2
  + {\ul r} ,q,k) \notag \\ & = \frac{1}{k^-} \, \theta (b_2^- -
  b_1^-) \, M_K(p_1 -r, q, k-r) \, M_R(p_2+r, k-r, k, r).
\end{align}
Here we assumed that $r^+ \sim \bot^2 / Q$ in our kinematics. After putting the $k-r$
quark propagator on mass shell the amplitude ${\tilde M}_{K+R}$
factorizes into a product of separate amplitudes for knockout $M_K(p_1
-r, q, k-r)$ and rescattering $M_R(p_2+r, k-r, k, r)$
\cite{Mueller:1989st,Kovchegov:1998bi,KovchegovLevin}, as employed in
\eq{eq:long_int}, where the sum over quark polarizations and colors is
implicit.

With the help of \eq{eq:long_int} (and a similar one for the
$r^{\prime +}$-integration of $M^*_{K+R}$) we write
\begin{align}
  \label{eq:app2}
  \langle |M_{tot}|^2 \rangle &= \int \frac{d p_1^+ \, d^2 p_1 \,
  dp_2^+ \, d^2p_2}{[2 \, (2 \pi)^3]^2 \, (p_1^+ + q^+) \, p_2^+ }
  \frac{d^2 r}{4 \, (2 \pi)^2} \frac{A \, (A-1) \, \left( P^+
  \right)^2}{p_1^+ \, p_2^+ \, (k^-)^2} \, d b_1^- \, d b_2^- \, d^2
  b \, \theta (b_2^- - b_1^-) 
	\notag \\ &\times \, 
	W \left(p_1^+, {\ul p}_1 - {\ul r}, b_1^-, {\ul b} \right) \, W \left(p_2^+, 
	{\ul p}_2 + {\ul r} , b_2^-, {\ul b} \right) 
	|M_K(p_1 -r, q, k-r)|^2 
	\notag \\ & \times \, 
	|M_R(p_2+r, k-r, k, r)|^2.
\end{align}

Defining the energy-rescaled rescattering amplitude by \cite{Mueller:1989st,KovchegovLevin}
\begin{align}
  \label{eq:e-ind_ampl}
  | A_R (p_2+r, k-r, k, r)|^2 \equiv \frac{1}{4 (p_2^+)^2 \, (k^-)^2}
  \, |M_R(p_2+r, k-r, k, r)|^2
\end{align}
and denoting the average of this object in the Wigner distribution by
the angle brackets
\begin{align}
  \label{eq:Wave}
  \left\langle | A_R (k, r)|^2 \right\rangle (b_1^-, {\ul b}) &= \int
  \frac{dp_2^+ \, d^2p_2 \, d b_2^-}{2 \, (2 \pi)^3} \, \theta (b_2^-
  - b_1^-) \, (A-1) \, W \left(p_2^+, {\ul p}_2 + {\ul r} , b_2^-,
  {\ul b} \right) 
	\notag \\ &\times  
	|A_R(p_2+r, k-r, k, r)|^2
\end{align}
we rewrite \eq{eq:app2} as
\begin{align}
  \label{eq:app3}
  \langle |M_{tot}|^2 \rangle = & A \, \int \frac{d p_1^+ \, d^2 p_1
    \, d b_1^- \, d^2 b}{2 \, (2 \pi)^3} \, \frac{\left( P^+
    \right)^2}{p_1^+ \, (p_1^+ + q^+)} \, W \left(p_1^+, {\ul p}_1,
    b_1^-, {\ul b} \right) \notag \\ & \times \, \int \frac{d^2 r}{(2
    \pi)^2} \, |M_K(p_1, q, k -r)|^2 \, \left\langle | A_R (k, r)|^2
  \right\rangle (b_1^-, {\ul b}).
\end{align}
In arriving at \eq{eq:app3} we have shifted the momentum $p_1 \to p_1
+ r$. 

We now define the energy-rescaled total and ``knockout''
amplitudes \cite{Mueller:1989st,KovchegovLevin}
\begin{align}
  \label{eq:A_Eindep}
  |A_{tot}|^2 \equiv \frac{1}{4 \, \left( P^+ \right)^2 \, (q^-)^2} \,
  |M_{tot}|^2, \ \ \ \ \ |A_k|^2 \equiv \frac{1}{4 \, ( p_1^+ )^2 \,
    (q^-)^2} \, |M_{K}|^2.
\end{align}
Employing the Fourier transform \eqref{M_Ftr} we reduce \eq{eq:app3}
to
\begin{align}
  \label{eq:app4} \nonumber
  \langle |A_{tot}|^2 \rangle = & A \int \frac{d p_1^+ \, d^2 p_1 \, d
    b_1^- \, d^2 b}{2 \, (2 \pi)^3} \, \frac{p_1^+}{p_1^+ + q^+} \, W
  \left(p_1^+, {\ul p}_1, b_1^-, {\ul b} \right)
	\times \\ \nonumber &\times
	\int d^2 x \, d^2 y \, e^{- i \, {\ul k} \cdot ({\ul x} - {\ul y})} 
	A_K(p_1, q, k^-, r^+, {\ul x} - {\ul b}) \, A_K^* (p_1, q, k^-, r^+, {\ul y} - {\ul b})
	\times \\ & \times
	\left\langle | A_R|^2 \right\rangle (k^-, {\ul x} - {\ul y}, b_1^-, {\ul b})
\end{align}
with
\begin{align}
  \label{eq:AR_FT}
  \left\langle | A_R|^2 \right\rangle (k^-, {\ul x} - {\ul y}, b_1^-,
  {\ul b}) = \int \frac{d^2 r}{(2 \pi)^2} \, e^{i \, {\ul r} \cdot ({\ul
      x} - {\ul y})} \,\left\langle | A_R (k, r)|^2 \right\rangle
  (b_1^-, {\ul b}).
\end{align}

Comparing \eq{eq:app3} to \eq{eq:ampl2_LO_Wxy} we see that, just like
in all high energy QCD scattering calculations
\cite{Mueller:1989st,KovchegovLevin,Weigert:2005us,Jalilian-Marian:2005jf,Gelis:2010nm}
the rescattering can be factored out into a multiplicative factor in
transverse coordinate space. Similar to the above one can show that
all further rescatterings would only introduce more multiplicative
factors. Defining a somewhat abbreviated notation
\begin{align}
  \label{eq:M2_def}
  A (p, q, {\ul x} - {\ul b}) \, A^* (p, q, {\ul y} - {\ul b}) &\equiv
  A_K(p, q, k^-, r^+, {\ul x} - {\ul b}) \, A_K^* (p, q, k^-, r^+, {\ul y} - {\ul b}) 
	\notag \\ &\times \, 
	\left\langle | A_R|^2  \right\rangle (k^-, {\ul x} - {\ul y}, b_1^-, {\ul b})
\end{align}
we see that \eq{eq:app4} reduces to \eq{eq:ampl2_LO_Wxy_net}, as
desired. The above discussion also demonstrates how multiple
rescatterings factorize in transverse coordinate space: in the
high energy kinematics they are included through the Wilson lines of
Eqs.~\eqref{xsectNNN} and \eqref{dipole_def}. The Wilson line
correlator $D_{{\ul x} \, {\ul y}} [+\infty, b^-]$ from
\eqref{dipole_def} contains a $b^-$-ordered product of multiple
rescattering factors $\left\langle | A_R|^2 \right\rangle$ from all
the interacting nucleons
\cite{Jalilian-Marian:1997xn,Kovchegov:1998bi}.


\section{SIDIS Sivers Function in the Quasi-Classical Limit}

Imagine a large nucleus with the total angular momentum $\vec J$ such that
\begin{align}
  \label{eq:spin_dec}
  {\vec J} = {\vec L} + {\vec S},
\end{align}
where ${\vec L}$ is the OAM of all the nucleons in the nucleus and
${\vec S}$ is the net spin of all the nucleons. In the quasi-classical
approximation at hand the OAM is generated by the rotation of the nucleons
around a preferred axis. The nucleus is polarized transverse to the
beam; we assume that both ${\vec L}$ and ${\vec S}$ point along the
(positive or negative) $\hat x$-axis.

The result \eqref{xsect_W3} for the quark production cross section in
SIDIS can be utilized to write down an expression for the SIDIS Sivers
function of the large nucleus with the help of \eq{eq:Sivers_ext_5}. We
first note that the quark production cross section in SIDIS is
proportional to the correlator \eqref{eq:q_corr} with the
future-pointing Wilson line given by \eq{eq:U_SIDIS}
(cf. Eqs.~\eqref{xsectNNN} and \eqref{dipole_def}).  The gauge link in
\eqref{dipole_def} begins and ends at the same $b^-$, while the more
general gauge link in \eqref{eq:U_SIDIS} has different endpoints at
$0$ and $x^-$.  The reason is that the nuclear wave function is
composed of color-neutral ``nucleons'' localized in $b^-$; hence there
is only a contribution to the correlator when the gauge link both
begins and ends at the same $b^-$.  The Dirac $\gamma^+$-matrix of
\eq{eq:Sivers_ext_5} is also present in the quark production cross
section since the Dirac structure of the large-$k^-$ outgoing quark
line is given by $\gamma^+ \, k^-$. To obtain the Sivers function one
only needs to eliminate the gamma--matrices stemming from the
quark--photon vertices in the amplitude and in the complex conjugate
amplitude; this can be done by simply contracting the Lorentz indices
of these gamma--matrices, as was done in Sec.~\ref{SIDISF} \cite{Brodsky:2013oya}. 
While such a contraction is not allowed in a calculation of the SIDIS cross section
due to the non-trivial structure of the lepton tensor, it is a legitimate
method of extracting the Sivers function \cite{Brodsky:2013oya}, since
$\gamma_\mu \, \gamma^+ \, \gamma^\mu = -2 \, \gamma^+$. We thus see
that an equation like \eqref{xsect_W3} would still hold for Tr$[\Phi
\, \gamma^+]$ instead of SIDIS cross section, since to obtain the
former one simply needs to repeat all the steps of the cross section
derivation that led to \eq{xsect_W3} without inserting the photon
polarizations (implicit in \eqref{xsect_W3}), and adding a contraction
over Lorentz indices of the gamma--matrices from the quark--photon
vertices in the end.

\begin{figure}[ht]
\centering
\includegraphics[width= .75 \textwidth]{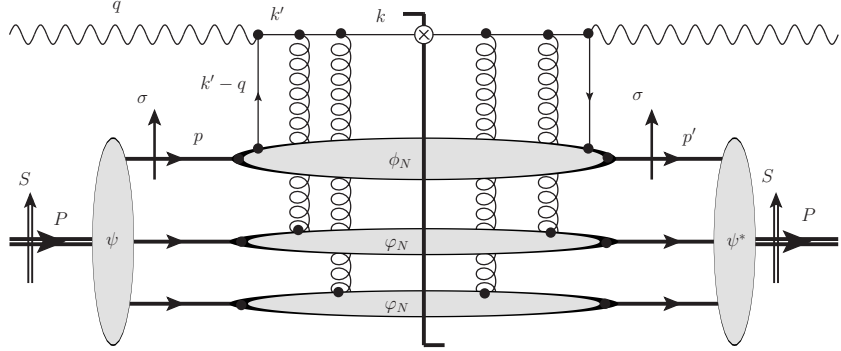}
\caption{
  \label{fig:TMD1} Decomposition of the nuclear quark distribution
  $\Phi_A$ probed by the SIDIS virtual photon into mean-field wave
  functions $\psi, \psi^*$ of nucleons and the quark and gluon
  distributions $\phi_N$ and $\varphi_N$ of the nucleons.  }
\end{figure}

By analogy with \eq{xsect_W3} we can express the quark correlation
function $\Phi_A$ of the nucleus in terms of the quasi-classical
distribution $W_N(p,b)$ of nucleons, the quark correlators $\phi_N$ of
individual nucleons, and the semi-infinite Wilson line trace $D_{{\ul
    x} {\ul y}}[+\infty, b^-]$:
\begin{align} 
 \label{TMD1} \nonumber
 \Tr[\Phi_A({\bar x},\ul{k} &; P, J) \, \gamma^+] = A \int \frac{d p^+
 \, d^2 p \, d b^-}{2(2\pi)^3} \, d^2x \, d^2 y \, \sum_\sigma
 W_N^\sigma \left( p,b^-, \frac{{\ul x} + {\ul y}}{2} \right) 
 \\ &\times
 \int \frac{d^2 k'}{(2\pi)^2} \, e^{ - i \, (\ul{k} - \ul{k'})
 \cdot(\ul{x}-\ul{y})} \,
 %
 %
 \Tr[\phi_N(x ,\ul{k}' -  x \, {\ul p}; p , \sigma) \, \gamma^+] \, 
 D_{{\ul x} {\ul y}}[+\infty, b^-].
\end{align}
\eq{TMD1} is illustrated in \fig{fig:TMD1}. In \eq{TMD1} we explicitly
show the sum over the polarizations $\sigma = \pm 1/2$ of the nucleons 
along the $x$-axis. Note that $x = -
q^+/p+$ and it varies with $p^+$ inside the integral; at the same time
the ``averaged'' value of Bjorken-$x$ per nucleon is ${\bar x} = - A
\, q^+/P^+$.  The quark correlator in the nucleus $\Phi_A$ is defined
by \eq{eq:q_corr},
\begin{align}
  \label{TMD2}
 \Phi_{ij}^A (\bar x, \ul{k}; P, J) \equiv \frac{1}{2(2\pi)^3} \int d^{2-} r \, e^{i k \cdot r}
 \bra{A (P, J)} \psibar_j (0) \, {\cal U}^{SIDIS} [0,r] \, \psi_i (r) \ket{A (P, J)},
\end{align}
and with the corresponding correlator in the nucleon is
\begin{align} 
  \label{TMD3} 
 \phi_{ij}^N ( x, \ul{k}; p, \sigma) \equiv \frac{1}{2(2\pi)^3} \int d^{2-} r \, e^{i k \cdot r}
 \bra{N (p, \sigma)} \psibar_j (0) \, {\cal U}^{SIDIS} [0,r] \, \psi_i (r) \ket{N (p, \sigma)} .
\end{align}
These definitions are made in a frame in which the parent particle's
transverse momentum is zero.  The photon-nucleus center-of-mass frame we are using corresponds to $\ul{q} = \ul{P} = \ul{0}$, but with (possibly) nonzero transverse orbital momentum $\ul p$ of the nucleons.  Thus, to apply the definition \eqref{TMD3}, we must make a transverse boost from the center-of-mass frame to a frame in which the nucleon is at rest and $\ul{p} = \ul{0}$.  This gives rise to the transverse momentum $\ul{k}' - x \, {\ul p}$ in the argument of $\phi_N$ in \eq{TMD1}, which is most directly verified by noting that this transverse momentum preserves the three Lorentz invariants $p^2 , k^2 , (p-k)^2$ between the two frames.  Additionally, the polarization-dependent Wigner functions are normalized as
(cf. \eq{rr2})
\begin{align} 
\begin{aligned}
  \label{Wnorm} \int \frac{d p^+ \, d^2 p \, d b^- \, d^2 b}{2
    (2\pi)^3} \, A \, W^{\uparrow} (p,b) & = \# \ \mbox{of spin-up nucleons} \, ; \\
  \int \frac{d p^+ \, d^2 p \, d b^- \, d^2 b}{2
    (2\pi)^3} \, A \, W^{\downarrow} (p,b) & = \# \ \mbox{of spin-down nucleons} \, .
\end{aligned}
\end{align}

Using the projection \eqref{eq:Sivers_ext_5} onto unpolarized quarks we write
\begin{align} 
  \Tr[\Phi_A ({\bar x}, \ul{k}; P, J) \, \gamma^+] &= 2 \, f_1^A
  ({\bar x}, k_T) + \, \frac{2}{M_A} \, {\hat z} \cdot (\ul{J}
  \times \ul{k})\, f_{1T}^{\perp A} ({\bar x}, k_T) \label{TMD4A} \\
  \Tr[\phi_N (x , \ul{k}'- x \, \ul{p}; p, \sigma) \, \gamma^+] &= 2
  \, f_1^N (x, |\ul{k}'- x \, \ul{p}|)  \notag \\ &+ \, \frac{2}{m_N}
  \, {\hat z} \cdot \left( \ul{\sigma} \times (\ul{k}'- x \, \ul{p})
  \right) \ f_{1T}^{\perp N} (x, |\ul{k}'- x \,
  \ul{p}|), \label{TMD4B}
\end{align}
where we introduced the unpolarized quark TMDs ($f_1^A$ and $f_1^N$)
and Sivers functions ($f_{1T}^{\perp A}$ and $f_{1T}^{\perp N}$) for
the nucleus and nucleons respectively, along with the masses $M_A$ and
$m_N$ of the nucleus and nucleons.

We may extract the Sivers function of the nucleus $f_{1T}^{\perp A}$
by antisymmetrizing \eqref{TMD4A} with respect to either the nuclear
spin or the momentum $\ul{k}$ of the produced quark:
\begin{align} 
  \label{TMD5} {\hat z} \cdot (\ul{J}\times\ul{k}) \, f_{1T}^{\perp A}
  ({\bar x} , k_T) = \frac{1}{4} M_A \, \Tr[\Phi_A ({\bar x},\ul{k}; P
  , J) \, \gamma^+] - (\ul{k} \rightarrow -\ul{k}).
\end{align}
Using \eq{TMD1} in \eq{TMD5} we write
\begin{align} \label{TMD6}
 \begin{aligned}
   {\hat z} \cdot & (\ul{J}\times\ul{k}) \, f_{1T}^{\perp A} ({\bar x}
   , k_T) = \frac{1}{4} M_A \, A \int \frac{d p^+ \, d^2 p \, d
     b^-}{2(2\pi)^3} \, d^2x \, d^2 y \, \sum_\sigma W_N^\sigma \left(
     p,b^-,
     \frac{{\ul x} + {\ul y}}{2} \right) \\
   & \times \int \frac{d^2 k'}{(2\pi)^2} \, e^{ - i \, (\ul{k} -
     \ul{k'}) \cdot(\ul{x}-\ul{y})} \, \Tr[\phi_N(x ,\ul{k}' - x \,
   {\ul p}; p , \sigma) \, \gamma^+] \, D_{{\ul x} {\ul y}}[+\infty,
   b^-] - (\ul{k} \rightarrow - \ul{k}).
 \end{aligned}
\end{align}
The next step will be to analyze the symmetry properties of the factors in \eqref{TMD6} to identify the physical subprocesses that can give rise to the Sivers function within the quasi-classical approximation.

\subsection{Channels Generating the SIDIS Sivers Function}
\label{subsec-channels}

We can decompose the quark correlator in a nucleon $\phi_N$ into the
nucleon's unpolarized quark distribution $f_1^N$ and Sivers function
$f_{1T}^N$ using \eqref{TMD4B}. Substituting this into \eq{TMD6}
yields
\begin{align} 
 \label{TMD7} \nonumber
  {\hat z} \cdot & (\ul{J}\times\ul{k}) \, f_{1T}^{\perp A} ({\bar x}
   , k_T) = \frac{1}{4} M_A \, A \int \frac{d p^+ \, d^2 p \, d
   b^-}{2(2\pi)^3} \, d^2x \, d^2 y \, \sum_\sigma W_N^\sigma \left(p,b^-,
   \frac{{\ul x} + {\ul y}}{2} \right)  \int \frac{d^2 k'}{(2\pi)^2} 
	 \\ \nonumber & \times
	 e^{ - i \, (\ul{k} - \ul{k'}) \cdot(\ul{x}-\ul{y})} \,
   \left[ 2 \, f_1^N (x, |\ul{k}'- x \, \ul{p}|_T) + \frac{2}{m_N} \,
   {\hat z} \cdot \left( \ul{\sigma} \times (\ul{k}'- x \, \ul{p})
   \right) \ f_{1T}^{\perp N} (x, |\ul{k}'- x \, \ul{p}|_T) \right] 
	\\ & \times \, 
	D_{{\ul x} {\ul y}}[+\infty, b^-] - (\ul{k} \rightarrow - \ul{k}).
\end{align}
We can identify the sources of the $T$-odd nuclear Sivers function
$f_{1T}^{\perp A}$ by explicitly (anti)symmetrizing the various terms
on the right of \eq{TMD7}. To start with, perform the nucleon spin sum
$\sum_\sigma$ in a basis parallel or antiparallel to the nuclear spin
$\ul{S}$. This can be done using the definitions
\begin{align} \label{TMD8}
 \begin{aligned}
   \sum_\sigma W_N^\sigma(p,b) & \equiv W_{unp}(p,b) \\
   \sum_\sigma W_N^\sigma(p,b) \, \ul{\sigma}& \equiv \frac{1}{A} \,
   W_{trans}(p,b) \, \ul{S},
 \end{aligned}
\end{align}
where we will refer to $W_{unp}$ as the distribution of unpolarized
nucleons and to $W_{trans}$ as the nucleon transversity
distribution, in analogy with the transversity TMD $h_1 \equiv h_{1T} + \tfrac{1}{2} \tfrac{k_T^2}{M^2} h_{1T}^\bot$.  Note that 
\begin{align} 
  \label{Wnorm2} \int \frac{d p^+ \, d^2 p \, d b^- \, d^2 b}{2
    (2\pi)^3} \, W_{unp} (p,b) = 1, \ \ \ \int \frac{d p^+ \, d^2 p \,
    d b^- \, d^2 b}{2 (2\pi)^3} \, W_{trans} (p,b) = 1,
\end{align}
as follows from the definition \eqref{TMD8} and from \eqref{Wnorm}.

\eq{TMD7} becomes
\begin{align} 
  \label{TMD9} 
	{\hat z} \cdot (\ul{J}\times\ul{k}) \, f_{1T}^{\perp A} 
	({\bar x} , k_T) &= \frac{M_A}{2} \int \frac{d p^+ \, d^2 p \, d
  b^-}{2(2\pi)^3} \, d^2x \, d^2 y \, \frac{d^2 k'}{(2\pi)^2} \, e^{
  - i \, (\ul{k} - \ul{k'}) \cdot(\ul{x}-\ul{y})} 
	\notag \\ & \times \, 
	\bigg[ A \, W_{unp} \left( p,b^-, \frac{{\ul x} + {\ul y}}{2} \right)
	f_1^N (x, |\ul{k}'- x \, \ul{p}|) 
	\notag \\ &+
	W_{trans} \left(p,b^-, \frac{{\ul x} + {\ul y}}{2} \right) \frac{1}{m_N} \, 
	{\hat z} \cdot \left( \ul{S} \times (\ul{k}'- x \, \ul{p})
  \right) \, f_{1T}^{\perp N} (x, |\ul{k}'- x \, \ul{p}|) \bigg] 
	\notag \\ & \times \, 
	D_{{\ul x} {\ul y}}[+\infty, b^-] - (\ul{k} \rightarrow - \ul{k}).
\end{align}
Now, in the terms with $(\ul{k} \rightarrow -\ul{k})$ being
subtracted, we also redefine the dummy integration variables $\ul{x}
\leftrightarrow \ul{y}$, $\ul{k'} \rightarrow - \ul{k'}$, and $\ul{p}
\rightarrow - \ul{p}$.  This leaves the Fourier factors and the
distribution functions $f_{1}^{N}$, $f_{1 T}^{\perp \, N}$ unchanged,
giving
\begin{align} 
  \label{TMD10} 
	&{\hat z} \cdot (\ul{J}\times\ul{k}) \, f_{1T}^{\perp
    A} ({\bar x} , k_T) = \frac{M_A}{2} \int \frac{d p^+ \, d^2 p \, d
    b^-}{2(2\pi)^3} \, d^2x \, d^2 y \, \frac{d^2 k'}{(2\pi)^2} \, e^{
    - i \, (\ul{k} - \ul{k'}) \cdot(\ul{x}-\ul{y})} \, \bigg\{
  f_1^N (x, |\ul{k}'- x \, \ul{p}|) \notag \\
  & \times \, A \, \left[ W_{unp} \left( p^+, \ul{p} ,b^-, \frac{{\ul
          x} + {\ul y}}{2} \right) \, D_{{\ul x} {\ul y}}[+\infty,
    b^-] - W_{unp} \left( p^+ , -\ul{p},b^-, \frac{{\ul x} + {\ul
          y}}{2} \right) \, D_{{\ul y} {\ul x}}[+\infty, b^-] \right]
  \notag \\ & + \frac{1}{m_N} \, {\hat z} \cdot \left( \ul{S} \times
    (\ul{k}'- x \,
    \ul{p}) \right) \, f_{1T}^{\perp N} (x, |\ul{k}'- x \, \ul{p}|) \\
  & \times \, \left[ W_{trans} \left( p^+, \ul{p} ,b^-, \frac{{\ul x}
        + {\ul y}}{2} \right) \, D_{{\ul x} {\ul y}}[+\infty, b^-] +
    W_{trans} \left( p^+, - \ul{p} ,b^-, \frac{{\ul x} + {\ul y}}{2}
    \right) \, D_{{\ul y} {\ul x}}[+\infty, b^-] \right]
  \bigg\}. \notag
\end{align}
At this point it is convenient to explicitly (anti)symmetrize the
distribution functions with respect to $\ul{p} \leftrightarrow
-\ul{p}$ and the Wilson lines with respect to $\ul{x} \leftrightarrow
\ul{y}$. Define the symmetric and antisymmetric parts of the Wilson lines dipole traces as in Sec.~\ref{subsec-symmetries},
\begin{align} \label{Wlines3}
 \begin{matrix}
  S_{{\ul x} {\ul y}} \equiv \tfrac{1}{2}(D_{{\ul x} {\ul y}} + D_{{\ul y} {\ul x}}) \\
  i \, O_{{\ul x} {\ul y}} \equiv \tfrac{1}{2}(D_{{\ul x} {\ul y}} - D_{{\ul y} {\ul x}})
 \end{matrix}
 \hspace{2cm}
 D_{{\ul x} {\ul y}} = S_{{\ul x} {\ul y}} + i \, O_{{\ul x} {\ul y}}
\end{align}
as well as
\begin{align} \label{DIS21}
 \begin{aligned}
   W^{\left( \substack{symm \\ OAM} \right)}(p,b) \equiv \tfrac{1}{2}
   \left[ W(p,b) \pm (\ul{p} \rightarrow - \ul{p}) \right],
 \end{aligned}
\end{align}
where we have used the ``OAM'' label to indicate that the preferred
direction of transverse momentum in the antisymmetric case reflects
the presence of net orbital angular momentum.  We can decompose $W$
into symmetric and OAM parts for both the unpolarized distribution
$W_{unp}$ and the transversity distribution $W_{trans}$.

Using the (anti)symmetrized quantities in \eq{DIS21} we can evaluate
the factors in the square brackets of \eqref{TMD10} as
\begin{align} \label{TMD11}
 \begin{aligned}
   \bigg[W_{unp}(p,b) &\, D_{{\ul x} {\ul y}}[+\infty, b^-] - W_{unp}(-\ul{p},b) \, D_{{\ul y} {\ul x}}[+\infty, b^-] \bigg] = \\
   &= 2 \bigg( W_{unp}^{OAM}(p,b) \, S_{{\ul x} {\ul y}}[+\infty, b^-]
   + W_{unp}^{symm}(p,b) \, i \, O_{{\ul x} {\ul y}}[+\infty, b^-]
   \bigg)
   \\ 
   \bigg[W_{trans}(p,b) &\, D_{{\ul x} {\ul y}}[+\infty, b^-] + W_{trans}(-\ul{p},b) \, D_{{\ul y} {\ul x}}[+\infty, b^-]  \bigg] = \\
   &= 2 \bigg( W_{trans}^{symm}(p,b) \, S_{{\ul x} {\ul y}}[+\infty,
   b^-] + W_{trans}^{OAM}(p,b) \, i \, \, O_{{\ul x} {\ul y}}[+\infty,
   b^-] \bigg)
 \end{aligned}
\end{align}
giving
\begin{align} \label{TMD12} {\hat z} \cdot & (\ul{J}\times\ul{k}) \,
  f_{1T}^{\perp A} ({\bar x}, k_T) = M_A \int \frac{d p^+ \, d^2 p \,
    d b^-}{2(2\pi)^3} \, d^2x \, d^2 y \, \frac{d^2 k'}{(2\pi)^2} \,
  e^{ - i \, (\ul{k} - \ul{k'}) \cdot(\ul{x}-\ul{y})} \, \bigg\{
  f_1^N (x, |\ul{k}'- x \, \ul{p}|) \notag \\
  & \!\!\!\!\!\!\! \times \, A \, \left[ W_{unp}^{OAM} \left( p^+, \ul{p} ,b^-,
      \frac{{\ul x} + {\ul y}}{2} \right) \, S_{{\ul x} {\ul
        y}}[+\infty, b^-] + W_{unp}^{symm} \left( p^+ , \ul{p},b^-,
      \frac{{\ul x} + {\ul y}}{2} \right) \, i \, O_{{\ul x} {\ul
        y}}[+\infty, b^-] \right] \notag \\ & + \frac{1}{m_N} \, {\hat
    z} \cdot \left( \ul{S} \times (\ul{k}'- x \,
    \ul{p}) \right) \, f_{1T}^{\perp N} (x, |\ul{k}'- x \, \ul{p}|) \\
  & \!\!\!\!\!\!\! \times \, \left[ W_{trans}^{symm} \left( p^+, \ul{p} ,b^-,
      \frac{{\ul x} + {\ul y}}{2} \right) \, S_{{\ul x} {\ul
        y}}[+\infty, b^-] + W_{trans}^{OAM} \left( p^+, \ul{p} ,b^-,
      \frac{{\ul x} + {\ul y}}{2} \right) \, i \, O_{{\ul x} {\ul
        y}}[+\infty, b^-] \right] \bigg\}. \notag
\end{align}

Altogether, the symmetry arguments presented above allow us to
decompose the nuclear Sivers function $f_{1T}^{\perp A}$ into four
distinct channels with the right quantum numbers to generate the
$T$-odd asymmetry.  These four channels correspond to the negative
$T$-parity occurring in the nucleon distribution $W^{OAM}$, in the
quark Sivers function of the nucleon $f_{1T}^{\perp N}$, in the
antisymmetric ``odderon'' rescattering $i O_{xy}$, or in all three
simultaneously.  

The odderon was discussed in Chapter~\ref{chap-odderon} as a source of $T$-odd single transverse spin asymmetries in $p^\uparrow A$ collisions.  It naturally appears as a contribution to $T$-odd quantities whenever Wilson lines form the natural degrees of freedom.  However, as we showed in Sec.~\ref{A4}, the preferred direction generated by
odderon-type rescattering couples to transverse gradients of the
nuclear profile function, $\ul{\nabla} T (\ul{b})$.  The length scale
over which these gradients become important is on the order of the
nuclear radius; these gradients are therefore $\ord{A^{-1/3}} \sim
\ord{\alpha_s^2}$ suppressed (in addition to an extra power of $\as$
entering the lowest-order odderon amplitude corresponding to the
triple-gluon exchange
\cite{Lukaszuk:1973nt,Nicolescu:1990ii,Ewerz:2003xi,Bartels:1999yt,Kovchegov:2003dm,Hatta:2005as,Kovner:2005qj,Jeon:2005cf,Kovchegov:2012ga})
and are therefore beyond the precision of the quasi-classical formula
\eqref{TMD12}.

Neglecting the odderon channels in \eqref{TMD12} we arrive at
\begin{align} \label{TMD13}
  \begin{aligned} {\hat z} \cdot & (\ul{J}\times\ul{k}) \,
    f_{1T}^{\perp A} ({\bar x} , k_T) = M_A \, \int \frac{d p^+ \, d^2
      p \, d b^-}{2(2\pi)^3} \, d^2x \, d^2 y \, \frac{d^2
      k'}{(2\pi)^2} \, e^{ - i \, (\ul{k} - \ul{k'})
      \cdot(\ul{x}-\ul{y})}   \\
    & \times \, \bigg\{ A \, W_{unp}^{OAM} \left( p^+, \ul{p} ,b^-,
      \frac{{\ul x} + {\ul y}}{2} \right) \, f_1^N (x, |\ul{k}'- x \,
    \ul{p}|_T) \\ & + \frac{1}{m_N} \, {\hat z} \cdot \left( \ul{S}
      \times (\ul{k}'- x \, \ul{p}) \right) \, W_{trans}^{symm} \left(
      p^+, \ul{p} ,b^-, \frac{{\ul x} + {\ul y}}{2} \right) \,
    f_{1T}^{\perp N} (x, |\ul{k}'- x \, \ul{p}|_T) \bigg\} \, S_{{\ul x}
      {\ul y}}[+\infty, b^-] .
 \end{aligned}
\end{align}
Shifting the integration variable $\ul{k}' \to \ul{k}' + x \, \ul{p}$
we write
\begin{align} \label{TMD14}
  \begin{aligned} {\hat z} \cdot & (\ul{J}\times\ul{k}) \,
    f_{1T}^{\perp A} ({\bar x}, k_T) = M_A \, \int \frac{d p^+ \, d^2
      p \, d b^-}{2(2\pi)^3} \, d^2x \, d^2 y \, \frac{d^2
      k'}{(2\pi)^2} \, e^{ - i \, (\ul{k} - x \, \ul{p} - \ul{k'})
      \cdot(\ul{x}-\ul{y})}   \\
    & \times \, \bigg\{ A \ W_{unp}^{OAM} \left( p^+, \ul{p} ,b^-,
      \frac{{\ul x} + {\ul y}}{2} \right) \, f_1^N (x, k'_T) \\ & +
    \frac{1}{m_N} \, {\hat z} \cdot \left( \ul{S} \times \ul{k}'
    \right) \, W_{trans}^{symm} \left( p^+, \ul{p} ,b^-, \frac{{\ul x}
        + {\ul y}}{2} \right) \, f_{1T}^{\perp N} (x, k'_T) \bigg\} \,
    S_{{\ul x} {\ul y}}[+\infty, b^-] .
 \end{aligned}
\end{align}

\begin{figure}[ht]
\centering
\includegraphics[width= .4 \textwidth]{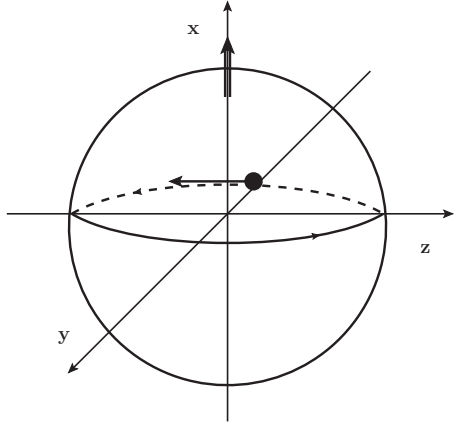}
\caption{This figure demonstrates our axes labeling convention and
  helps illustrate an example discussed in the text.}
\label{nucleus}
\end{figure}

To further simplify the obtained expression \eqref{TMD14} we need to
impose a constraint on the transverse momentum of the
nucleons. Consider the nucleus in its rest frame, as shown in
\fig{nucleus}.
The net OAM $\vec L$ of the transversely-polarized nucleus corresponds
to the rotation of the nucleus around the spin axis (the $x$-axis in
\fig{nucleus}). The rotational invariance about the $x$-axis implies
that the average magnitude of the rotational transverse momentum is
constant for a given distance from the $x$-axis and for fixed
$x$-coordinate. (In Sec.~\ref{B5} we show that in the rest frame of the nucleus, only the rotational motion of the nucleons about the polarization axis is allowed by $PT$ symmetry.)

Consider a nucleon at the point ${\vec x} = (0,-R,0)$ in the $(x,y,z)$
coordinate system, as illustrated by the black circle in
\fig{nucleus}. Its 3-momentum is ${\vec p}_{rest} = (0,0,-p)$, where
$p$ denotes some typical rotational momentum of a nucleon. After a
longitudinal boost along the $z$-axis to the infinite-momentum frame
of \eqref{DIS1} we get the large light-cone component of the momentum
to be
\begin{align}
  p^+ = \frac{P^+}{M_A} \, \left( \sqrt{m_N^2 + p^2} - p \right).
\end{align}
The corresponding Bjorken-$x$ is (see \eq{DIS2})
\begin{align}
  \label{eq:mom2}
  1 \ge x = \frac{- q^+}{p^+} = x_A \, A \, \frac{m_N}{\sqrt{m_N^2 +
      p^2} - p },
\end{align}
where we have used $M_A \approx A \, m_N$. The $x \le 1$ constraint in
\eq{eq:mom2} (cf. \eq{DIS2}) gives
\begin{align}
  \label{eq:mom3}
  p \le m_N \, \frac{1 - x_A^2 \, A^2}{2 \, x_A \, A}.
\end{align}
Since $x_A \, A$ is not a small number, in fact $x_A \, A = \ord{1}$,
we conclude that $p \lesssim m_N$. Therefore, the magnitude of the
rotational momentum in the nuclear rest frame is bounded by $\sim m_N$
from above. The typical transverse momentum $p_T$ in \eq{TMD14}, being
boost-invariant, is also bounded by the nucleon mass from above, $p_T
\lesssim m_N$. Since we assume that $k_T$ is perturbatively large,
$k_T \gg \Lambda_{QCD} \sim m_N$, we do not consistently resum all
powers of $m_N/k_T$. (Recall from Chap.~\ref{chap-CGC} that the saturation approach resums mainly
$A^{1/3}$-enhanced power corrections, that is, powers of
$Q_s^2/k_T^2$, but not powers of $\Lambda_{QCD}^2/k_T^2$.) 

The bound \eqref{eq:mom3} provides us with the condition on when the
SIDIS process on the nucleon highlighted in \fig{nucleus} can take
place. Violation of this bound would imply that SIDIS on that nucleon
is kinematically prohibited, and consequently SIDIS may take place
only on some of the other nucleons in the nucleus. While such a
situation where the nucleus is spinning so fast that SIDIS is only
possible on a subset of its nucleons is highly unlikely in the real
physical experiments, this presents a theoretical example where the
Sivers function \eqref{TMD14} would, in fact, depend on the direction
of $\ul p$ and, hence, of the spin $\ul {J}$, presumably through even
powers of $\ul{J} \cdot \ul{k}$. While such dependence is impossible
for spin-$1/2$ particles such as protons \cite{Bacchetta:1999kz}, it
has been considered for targets with different spin
\cite{Bacchetta:2000jk}; in our case it arises due to the classical
model at hand with the value of spin $J$ not restricted to $1/2$. To
avoid potential formal complications and unrealistic effects
associated with large rotational momentum, below we will assume that
$p_T \lesssim m_N$ such that the bound \eqref{eq:mom3} is
satisfied. Without such an assumption, \eq{TMD14} would be our final
result for the Sivers function in the quasi-classical approximation.

We see that we have to limit the calculation to the lowest non-trivial
power of $p_T / k_T \sim m_N/k_T$ contributing to the Sivers
function. Expanding \eq{TMD14} in the powers of $\ul{p}$ to the lowest
non-trivial order, and remembering that $W^{OAM}$ is an odd function
of $\ul{p}$ we obtain
\begin{align} \label{TMD15}
  \begin{aligned} {\hat z} \cdot & (\ul{J}\times\ul{k}) \,
    f_{1T}^{\perp A} ({\bar x}, k_T) = M_A \,\int \frac{d p^+ \, d^2 p
      \, d b^-}{2(2\pi)^3} \, d^2x \, d^2 y \, \frac{d^2 k'}{(2\pi)^2}
    \, e^{ - i \, (\ul{k} - \ul{k'})
      \cdot(\ul{x}-\ul{y})}   \\
    & \times \, \bigg\{ i\, x \, \ul{p} \cdot(\ul{x}-\ul{y}) \, A \
    W_{unp}^{OAM} \left( p^+, \ul{p} ,b^-, \frac{{\ul x} + {\ul y}}{2}
    \right) \, f_1^N (x, k'_T) \\ & + \frac{1}{m_N} \, {\hat z} \cdot
    \left( \ul{S} \times \ul{k}' \right) \, W_{trans}^{symm} \left(
      p^+, \ul{p} ,b^-, \frac{{\ul x} + {\ul y}}{2} \right) \,
    f_{1T}^{\perp N} (x, k'_T) \bigg\} \, S_{{\ul x} {\ul y}}[+\infty,
    b^-] .
 \end{aligned}
\end{align}

\eq{TMD15} is our main formal result. It relates the Sivers function
of a nucleus to the quark TMD and quark Sivers function in a nucleon.
It shows that within the quasi-classical approximation, there are two
leading channels capable of generating the Sivers function of the
composite nucleus:
\begin{enumerate}
\item \ul{Orbital Angular Momentum (OAM) Channel}: an unpolarized
  nucleon in a transversely polarized nucleus with a preferred
  direction of transverse momentum generated by the OAM of the nucleus
  has a quark knocked out of its symmetric $f_1^N$ transverse momentum
  distribution which rescatters coherently on spectator nucleons. The
  multiple rescatterings bias the initial knockout process to happen
  near the ``back'' of the nucleus, where, due to OAM motion of the
  nucleons, the outgoing quark gets an asymmetric distribution of its
  transverse momentum, generating STSA. (See the left panel of 
	\fig{SIDIS_OAM_vs_Sivers} below.)
\item \ul{Transversity / Sivers Density Channel}: a polarized nucleon
  with its preferred transverse spin direction inherited from the
  nucleus has a quark knocked out of its Sivers $f_{1T}^{\perp N}$
  distribution which rescatters coherently on spectator nucleons. The
  single spin asymmetry is generated at the level of the ``first''
  nucleon, and, unlike the OAM channel, the presence of other nucleons
  is not essential for this channel (see \fig{SIDIS_OAM_vs_Sivers}).
\end{enumerate}

\begin{figure}[ht]
\centering
\includegraphics[width= .5 \textwidth]{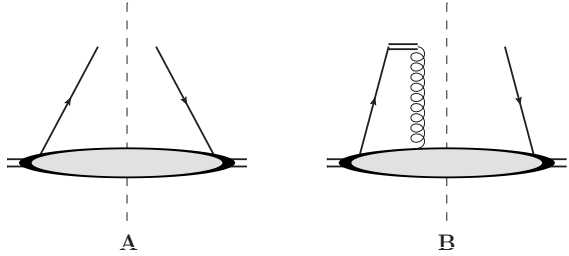}
\caption{Lowest-order diagrams for the quark TMD $f_1$ (panel A) and
  the Sivers function $f_{1T}^\perp$ (panel B). The vertical dashed lines
  denote the final state cut, while the double horizontal line in
  panel B denotes the Wilson line.}
\label{LO_TMDs}
\end{figure}

The OAM and transversity channels are depicted in
\fig{SIDIS_OAM_vs_Sivers} in terms of their space-time structure and
Feynman diagrams. The diagrams resummed in arriving at \eq{TMD15} are
the square of the graph shown in the left panel of
\fig{SIDIS_OAM_vs_Sivers} (OAM channel) and the diagram looking like
the interference between the two panels in \fig{SIDIS_OAM_vs_Sivers}
(transversity channel). The difference between the two channels
outlined above is in the first ``knockout'' interaction: the OAM
channel couples to the quark TMD $f_1^N$, while the transversity channel couples
to the nucleon Sivers function $f_{1T}^{\bot N}$. At the lowest order in perturbation
theory the two functions are illustrated in \fig{LO_TMDs}: indeed the
Sivers function shown in the panel B of \fig{LO_TMDs} requires at
least one more rescattering as compared to the quark TMD in panel A,
as discussed in Chapter~\ref{chap-TMD}.

\begin{figure}[ht]
\centering
\includegraphics[width= 0.9 \textwidth]{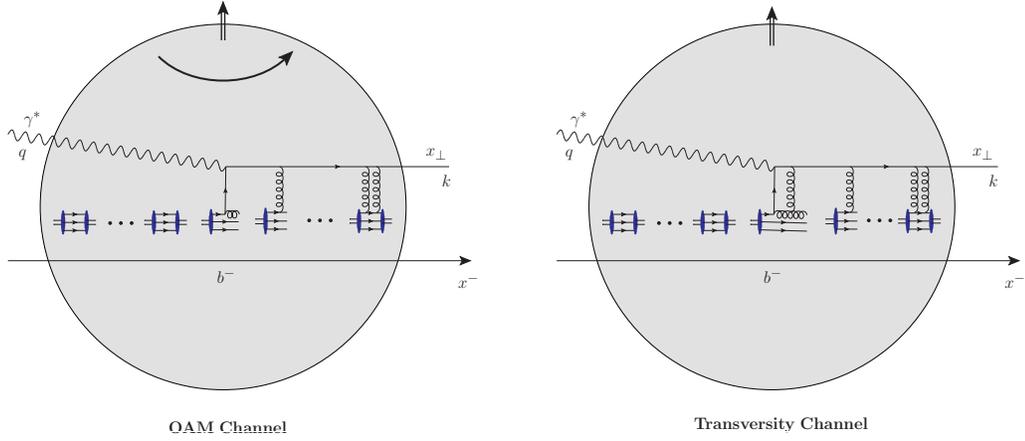}
\caption{Side-by-side comparison of the Feynman diagrams that contribute
  to the OAM and Sivers density channels in the quasi-classical
  approximation (in the rest frame of the nucleus).}
\label{SIDIS_OAM_vs_Sivers}
\end{figure}

Note that, in the OAM channel, the unpolarized quark distribution
$f_1^N$ enters parametrically at $\ord{\alpha_s \, A^{1/3}}$ if
calculated at the lowest-order in the perturbation theory (see panel A
in \fig{LO_TMDs}), which is $\ord{\alpha_s^{-1}}$ in the saturation
power counting (where $\as^2 \, A^{1/3} \sim 1$).  In the transversity
channel, the nucleonic Sivers function $f_{1T}^{\perp N}$ enters at
$\ord{\alpha_s^2 \, A^{1/3}} = \ord{1}$ at the lowest order in
perturbation theory, because it requires an extra $\ord{\alpha_s}$
gluon to be exchanged with the same nucleon to obtain the necessary
lensing effect \cite{Brodsky:2002cx} (see panel B in
\fig{LO_TMDs}). The transversity channel is therefore $\ord{1}$ in the
saturation power counting and is subleading by $\ord{\alpha_s}$ to the
OAM channel in this sense.\footnote{We would like to point out that
  the coupling constant $\as$ in $f_1^N$ runs with some
  non-perturbative momentum scale, and is large, $\as = \as (\sim
  \Lambda_{QCD}^2)$; however, a simple application of the BLM
  \cite{Brodsky:1983gc} prescription to the calculation of
  \cite{Brodsky:2013oya} can show that in $f_{1T}^{\perp N} (x, k_T)$
  the two powers of the coupling run as $\as (k_T^2) \, \as (\sim
  \Lambda_{QCD}^2)$. While one of the couplings is also
  non-perturbatively large, the other one is perturbatively small for
  $k_T \gg \Lambda_{QCD}$, indicating suppression.} Indeed the
non-trivial transverse motion of nucleons due to OAM must be present
for the OAM channel to be non-zero: this channel is leading only if
there is an OAM. In our estimate here we have assumed that the net
spin of our ``nucleons'' scales linearly with the atomic number, $S
\sim A$; perhaps a more realistic (both for protons and nuclei) slower
growth of $S$ with $A$ would introduce extra $A$-suppression for the
transversity channel.

Despite the transversity channel being subleading, it is more dominant
than the $\ord{A^{-1/3}} \sim \ord{\alpha_s^2}$ corrections we
neglected when arriving at the quasi-classical formula \eqref{TMD15}
(again, for $S \sim A$). Order $\alpha_s^1$ quantum corrections to the
OAM channel also enter at the same order as the nucleonic Sivers
function and are also within the precision of the formalism.

To complete \eq{TMD15} we need to construct an expression for the
total nuclear angular momentum ${\vec J} = {\vec L} + {\vec S}$. The OAM of the nucleons
in the nucleus from \fig{nucleus} in the nuclear rest frame is
\begin{align}
  \label{eq:OAM1}
  {\vec L} = A \int \frac{d^3 p \, d^3 b}{2 (2\pi)^3} \, W_{unp}
  \left( {\vec p}, \, {\vec b} \right) \ {\vec b} \times {\vec p} = A
  \int \frac{d^3 p \, d^3 b}{2 (2\pi)^3} \, W_{unp} \left( {\vec p},
    \, {\vec b} \right) \ {\hat x} \, (b_y \, p_z - b_z \, p_y),
\end{align}
where $d^3 p = d p_x \, d p_y \, d p_z$, $d^3 b = d b_x \, d b_y \, d
b_z$, and $W_{unp} \left( {\vec p}, \, {\vec b} \right)$ is the Wigner
distribution in the rest frame of the nucleus expressed in terms of
3-vectors ${\vec p} = (p_x, p_y, p_z)$ and ${\vec b} = (b_x, b_y,
b_z)$.

To boost this into the infinite momentum frame of \eqref{DIS1} we
use the Pauli-Lubanski vector first defined in \eqref{diss-spin4}
\begin{align}
  \label{eq:PL}
  W_\mu = - \frac{1}{2} \, \epsilon_{\mu\nu\rho\sigma} \, J^{\nu\rho}
  \, P^\sigma,
\end{align}
where $J_{\mu\nu} = L_{\mu\nu} + S_{\mu\nu}$ with $L_{\mu\nu}$ and
$S_{\mu\nu}$ the expectation values of the OAM and spin generators of
the Lorentz group in the nuclear state. The OAM generator is
\begin{align}
  \label{eq:generators}
  {\hat L}_{\mu\nu} = {\hat x}_\mu \, {\hat p}_\nu - {\hat x}_\nu \,
  {\hat p}_\mu
\end{align}
as usual, with the hat denoting operators. The nuclear OAM four-vector
is then defined by
\begin{align}
  \label{eq:OAM_4vector}
  L_\mu = - \frac{1}{2} \, \epsilon_{\mu\nu\rho\sigma} \, L^{\nu\rho}
  \, \frac{P^\sigma}{M_A}.
\end{align}
Note that ${\hat p}_\mu$ in \eq{eq:generators} are the momentum
operators of the nucleons, while $P^\sigma$ in Eqs.~\eqref{eq:PL} and
\eqref{eq:OAM_4vector} is the net momentum of the whole nucleus. In
the rest frame of the nucleus \eq{eq:OAM_4vector} gives $L_x = L_{yz}$
as expected (for $\epsilon_{0123} = +1$). The nuclear OAM four-vector
can then be written as
\begin{align}
  \label{eq:L}
  L_\mu = - \frac{1}{2} \, \epsilon_{\mu\nu\rho\sigma} \,
  \frac{P^\sigma}{M_A} \, A \, \int \frac{d p^+
      \, d^2 p \, d b^- \, d^2 b}{2 \, (2 \, \pi)^3} \, W_{unp} (p, b)
    \, (b^\nu \, p^\rho - b^\rho \, p^\nu)
\end{align}
in the infinite momentum frame of the nucleus.

Since boosts preserve transverse components of four-vectors, the boost
along the $\hat z$-axis of the nucleus in \fig{nucleus} would preserve
its OAM three-vector $\vec L$ (which points along the $\hat x$-axis). Hence \eq{eq:OAM1} gives us the
transverse components of OAM in the infinite momentum frame as
well. We thus write
\begin{align}
  \label{eq:J} {\vec J} = {\hat x} \left[ S + A \int \frac{d^3 p \,
      d^3 b}{2 (2\pi)^3} \, W_{unp} \left( {\vec p}, \, {\vec b}
    \right) \ {\hat x} \, (b_y \, p_z - b_z \, p_y) \right],
\end{align}
where the integration over $p$ and $b$ needs to be carried out in the
nucleus rest frame.  

Combining Eqs.~\eqref{TMD15} with \eqref{eq:J} allows one to extract
the Sivers function $f_{1T}^{\perp A}$ of the nucleus.


\subsection{PT-Symmetry and QCD Shadowing}
\label{B5}

The decompositions \eqref{TMD15} and \eqref{DY_TMD8} essentially break
the Wilson line operator $\mathcal{U}$ in the definition
\eqref{eq:q_corr} into two parts: the coherent rescattering $S_{{\ul
    x} {\ul y}}[+\infty, b^-]$ on other spectator nucleons which is a
leading-order contribution in the saturation power counting, and the
subleading lensing interaction with the same nucleon which generates
$f_{1T}^{\perp N}$.  If we neglect the Wilson line operator
$\mathcal{U}$ entirely, then we know that the Sivers function of the
nucleus $f_{1T}^{\perp A}$ must vanish, as discussed in Sec.~\ref{subsec-sign_flip}. 
But if we drop $f_{1T}^{\perp N}$ and $S_{{\ul x} {\ul y}}[+\infty, b^-]$ 
from \eqref{TMD15}, we do not obviously get zero:
\begin{figure}[ht]
\centering
\includegraphics[height=6cm]{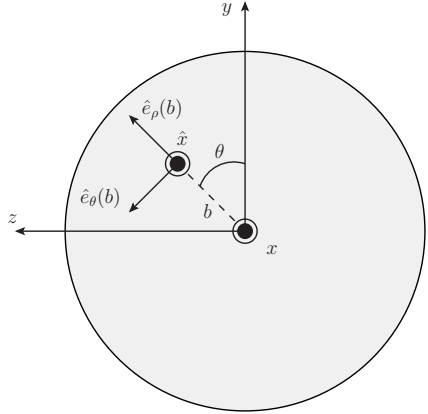}
\caption{
  \label{fig:W_Rot} Definition of the cylindrical coordinate basis
  \eqref{rot1} convenient for formulating the symmetry properties of
  the nucleonic distribution $W_\sigma(p,b)$ in the rest frame of the
  nucleus.  }
\end{figure}
%
\begin{align} \label{TMD16}
  \begin{aligned} {\hat z} \cdot  (\ul{J}\times\ul{k}) \,
    f_{1T}^{\perp A} (x , k_T) &= M_A \, A \int \frac{d p^+ \, d^2 p \,
      d b^-}{2(2\pi)^3} \, d^2x \, d^2 y \, \frac{d^2 k'}{(2\pi)^2} \,
    e^{ - i \, (\ul{k} - \ul{k'})
      \cdot(\ul{x}-\ul{y})}   \\
    & \times \, i \, x \, \ul{p} \cdot (\ul{x}-\ul{y}) \,
    W_{unp}^{OAM} \left( p^+, \ul{p} ,b^-, \frac{{\ul x} + {\ul y}}{2}
    \right) \, f_1^N (x, k'_T) 
		\\ &\stackrel{?}{=} 0.
 \end{aligned}
\end{align}
The right-hand side of this equation must vanish for wave functions
described by $W_{unp}^{OAM}$ that are $PT$ eigenstates
\cite{Collins:1992kk}; we can see this explicitly by considering the
constraints on $W_\sigma(p,b)$ due to rotational invariance and $PT$
symmetry.  It is most convenient to enumerate the rotational symmetry
properties of the nucleon distribution $W_\sigma(p,b)$ in the rest
frame of the nucleus, using a cylindrical vector basis coaxial to the
transverse spin vector $\ul{S}$.  This basis $(\hat{e}_\rho ,
\hat{e}_\theta , \hat{x})$ is shown in Fig.~\ref{fig:W_Rot} and is
defined by
\begin{align} \label{rot1}
 \begin{pmatrix} \hat{e}_\rho \\ \hat{e}_\theta \end{pmatrix} =
 \begin{pmatrix}
  b_y / b_\rho & b_z / b_\rho \\
  -b_z/ b_\rho & b_y / b_\rho 
 \end{pmatrix}
 \begin{pmatrix} \hat{y} \\ \hat{z} \end{pmatrix} =
 \begin{pmatrix}
  \cos\theta & \sin\theta \\
  -\sin\theta & \cos\theta 
 \end{pmatrix}
 \begin{pmatrix} \hat{y} \\ \hat{z} \end{pmatrix}
\end{align}
where $\left( p_\rho(b) , p_\theta(b) \right) = p \cdot
\left(\hat{e}_\rho(b) , \hat{e}_\theta(b)\right)$ and $b_\rho \equiv
\sqrt{b_y^2 + b_z^2}$.

\begin{figure}
\centering
\includegraphics[width=\textwidth]{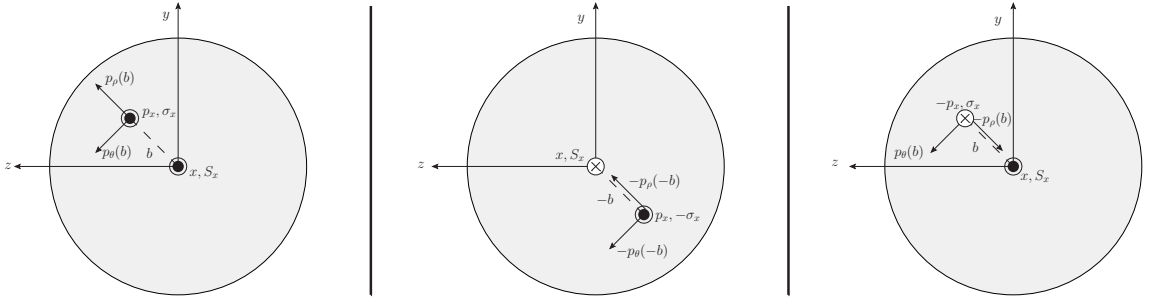}
\caption{
  \label{fig:PT_Rot} Illustration of the $PT$ transformation and
  rotational symmetry in the rest frame used in \eqref{rot3}.  Left
  panel: illustration of the momentum flow represented by
  $W_\sigma(p,b)$.  Center panel: under a $PT$ transformation, the
  spins $S,\sigma$ and coordinate $b$ are reversed, but the momentum
  $p$ is invariant.  Right panel: rotation of the center panel by
  $\pi$ about the vector $\vec{S}\times\vec{b}$ returns the
  distribution to its original position $b$, with $p_\rho$ and $p_x$
  having been reversed.}
\end{figure}

First, the distribution must be symmetric under rotations about the
polarization axis $\hat x$, which are easy to express in this cylindrical
basis:
\begin{align} \label{rot2}
 W_{\sigma} \big(p_x \, ; \, p_\rho (b) \, ; \, p_\theta (b) \, ; \, b \big) = 
 W_{\sigma} \big( p_x \, ; \, p_\rho (b') \, ; \, p_\theta (b') \, ; \, b' \big).
\end{align}
Second, if the nucleus is in a $PT$-symmetric eigenstate of the QCD
Hamiltonian, then $W_\sigma(p,b)$ should be invariant under $PT$
transformations.  These transformations reverse the coordinates $(b
\rightarrow -b)$ and pseudovectors like the spin $(S , \sigma
\rightarrow -S , -\sigma)$, but leave the momentum vector $p$
unchanged.  Using this transformation, together with rotational
invariance as shown in Fig.~\ref{fig:PT_Rot} we obtain
\begin{align} \label{rot3}
 \begin{aligned}
 W_\sigma \left( p_\rho(b) , p_\theta(b) , p_x ; b ; S_x \right) &\overset{PT}{=}
 W_{-\sigma} \left( p_\rho(b) , p_\theta(b) , p_x ; -b ; -S_x \right) \\ &=
 W_{-\sigma} \left( -p_\rho(-b) , -p_\theta(-b) , p_x ; b ; -S_x \right) \\ &\overset{R_b}{=}
 W_\sigma \left( -p_\rho(b) , p_\theta(b) , -p_x ; b ; S_x \right) \\
 \therefore W_\sigma \left( p_\rho(b) , p_\theta(b) , p_x ; b ; S_x \right) &= 
 W_\sigma \left( -p_\rho(b) , p_\theta(b) , -p_x ; b ; S_x \right),
 \end{aligned}
\end{align}
where the rotation $R_b$ is a half-revolution in the $Sb$-plane.  
%
%
%
This means that in a $PT$ eigenstate with transverse spin $S_x$, the
only allowed direction of net momentum flow corresponds to the
azimuthal orbital momentum $p_\theta$ and explains the naming
convention $W^{OAM}$ in \eqref{DIS21}.

The distributions that enter \eqref{TMD15},
however, are the (anti)symmetrized distributions under reversal of the
transverse momenta $(p_x, p_y \rightarrow -p_x , -p_y)$.  For these
purposes, it is more convenient to write the distribution $W_\sigma
(p,b)$ in terms of the Cartesian basis
\begin{align} \label{rot4} W(p_x , p_y, p_z ; b) = W_\sigma \left( p_x
    \,;\, \frac{b_y}{b_\rho} p_\rho(b) - \frac{b_z}{b_\rho}
    p_\theta(b) \,;\, \frac{b_z}{b_\rho} p_\rho(b) +
    \frac{b_y}{b_\rho} p_\theta(b) \,;\, b \right).
\end{align}
Using the symmetry properties \eqref{rot2} and \eqref{rot3}, we can
write the $\ul{p}$-reversed distribution in terms of the distribution
at a point $\overline{b} \equiv (b_x, b_y, -b_z)$ on the opposite side
of the nucleus:
\begin{align} \label{rot5}
 \begin{aligned}
   W_\sigma(-p_x , -p_y, &p_z ; b) = W_\sigma \left( -p_x \,;\,
     -\frac{b_y}{b_\rho} p_\rho(b) + \frac{b_z}{b_\rho} p_\theta(b)
     \,;\, \frac{b_z}{b_\rho} p_\rho(b) + \frac{b_y}{b_\rho}
     p_\theta(b) \,;\, b \right) \\ &\overset{Eq.\eqref{rot2}}{=}
   W_\sigma \left( -p_x \,;\, -\frac{b_y}{b_\rho} p_\rho(\overline{b})
     - \frac{b_z}{b_\rho} p_\theta(\overline{b}) \,;\, -
     \frac{b_z}{b_\rho} p_\rho(\overline{b}) + \frac{b_y}{b_\rho}
     p_\theta(\overline{b}) \,;\, \overline{b} \right) \\
   &\overset{Eq.\eqref{rot3}}{=} W_\sigma \left( p_x \,;\,
     \frac{b_y}{b_\rho} p_\rho(\overline{b}) - \frac{b_z}{b_\rho}
     p_\theta(\overline{b}) \,;\, \frac{b_z}{b_\rho}
     p_\rho(\overline{b}) + \frac{b_y}{b_\rho} p_\theta(\overline{b})
     \,;\, \overline{b} \right) \\ &=
   W_\sigma (p_x , p_y , p_z ; \overline{b}) \\
   \therefore W_\sigma(-p_x , -p_y, &p_z ; b) = W_\sigma (p_x , p_y , p_z ;
   \overline{b}).
\end{aligned}
\end{align}
Thus a nucleon on the back side of the nucleus has an opposite
transverse momentum to a corresponding nucleon in the front of the
nucleus.  Therefore, the (anti)symmetrized distributions have definite
parity under $b_z \rightarrow - b_z$:
\begin{align} \label{rot6}
 \begin{aligned}
   W_\sigma^{symm}(p,b) &\equiv \frac{1}{2} \left[ W_\sigma(p,b) +
     (\ul{p} \rightarrow - \ul{p}) \right] =
   + W_\sigma^{symm}(p,\overline{b}) \\
   W_\sigma^{OAM}(p,b) &\equiv \frac{1}{2} \left[ W_\sigma(p,b) -
     (\ul{p} \rightarrow - \ul{p}) \right] = -
   W_\sigma^{OAM}(p,\overline{b}).
 \end{aligned}
\end{align}

Eq.~\eqref{rot6} tells us that a consequence of $PT$ invariance in the
nucleus is that the transverse momentum due to rotation encountered at any point
in the front of the nucleus is compensated by an equal and opposite
transverse momentum from a corresponding point at the back of the
nucleus.  This is the resolution to the apparent paradox
\eqref{TMD16}: when we neglect all Wilson line contributions (both
$S_{{\ul x} {\ul y}}[+\infty, b^-]$ and $f_{1T}^{\perp N}$), the net
asymmetry in the quark distribution is indeed zero since $\int d b^-
W_{unp}^{OAM}(p,b) = 0$.  Hence neglecting all Wilson line
contributions yields zero for the Sivers function, consistent with
\cite{Collins:1992kk}.

An essential role is played in \eqref{TMD15}, then, by the
rescattering factor $S_{{\ul x} {\ul y}}[+\infty, b^-]$.  In the OAM
channel, even though the rescattering $S_{{\ul x} {\ul y}}[+\infty,
b^-]$ is not the source of a preferred transverse direction, without
it the net contribution to the Sivers function from OAM would vanish
after integration over $b^-$.  This reflects the simple physical interpretation
from the left panel of \fig{SIDIS_OAM_vs_Sivers} that there are as many nucleons 
moving out of the page to the left of the nuclear center as there are nucleons moving
into the page to the right of the nuclear center. 

The rescattering factor $S_{{\ul x} {\ul y}}[+\infty, b^-]$ approaches
$1$ for $b^-$ values near the ``back'' of the nucleus (right end of
the nucleus in \fig{SIDIS_OAM_vs_Sivers}) and is a monotonically
decreasing function of $b^-$. Due to this factor, different $b^-$
regions contribute differently to the integral, allowing it to be non-zero.
This factor is essential because it introduces shadowing
that breaks the front-back symmetry by preferentially screening quarks ejected from
the front of the nucleus more than those ejected from the back.  The Sivers
function relevant for SIDIS is therefore more sensitive to OAM from
the back of the nucleus than from the front, which has the
physical interpretation that it is easier for the quark to escape the
nucleus if it is produced near the edge. 

This analysis is strikingly similar to the arguments that historically
established the existence of the Sivers function and which we have discussed in
Sec.~\ref{subsec-sign_flip}.  As Collins argued
in \cite{Collins:1992kk}, $PT$-invariance of any hadronic eigenstate
prohibits a preferred direction that can generate the Sivers function.
This is directly reflected in the vanishing of \eqref{TMD16} without
the effects of multiple rescattering.  And as Brodsky, Hwang, and
Schmidt demonstrated in \cite{Brodsky:2002cx}, the rescattering
represented by the semi-infinite Wilson lines breaks this symmetry and
permits a preferred direction for the asymmetry.  Unlike that
calculation, however, here the rescattering does not occur as
color-correlated ``lensing'' due to rescattering on the remnants of
the active quark.  Here the interaction is explicitly
color-decorrelated because the rescattering occurs on many nucleons
whose colors are not correlated.  Despite this difference, the
rescattering effects are still sufficient to break the front-back
symmetry and give rise to a net preferred direction for the asymmetry.


\subsection{Evaluation for the Rigid Rotator Toy Model}
\label{toy_model}

We will now illustrate the properties of the Sivers function
\eqref{TMD15} by studying a specific simplified example. Consider the
model of the nucleus as a non-relativistic rigid rotator, with the
rotational momentum in its rest frame being much smaller than the
nucleon mass, $p_T \ll m_N$. The corresponding classical Wigner
distribution is (cf. \eq{Wcl})
\begin{align}
\label{Wcl_oam}
W_{unp} (p,b) \approx \frac{2 \, (2 \, \pi)^3}{A} \, \rho ({\ul b},
b^-) \, \delta^2 \left(\ul{p} - \hat{y} \, p_{max} (b_x) \,
  \frac{b^-}{R^- (b_x)} \right) \, \delta \left( p^+ - \frac{P^+}{A}
\right),
\end{align}
where $2 R^- (b_x)$ is the extent of the nucleus in the
$b^-$-direction at $\ul{b} = (b_x, 0)$ (with $R^-(b_x) = \sqrt{R^2 -
  b_x^2} \ M_A/P^+$), and $p_{max} (b_x) = p_{max} \, \sqrt{R^2 -
  b_x^2}/R$ is the maximum value of the rotational momentum at a given
$b_x$.  In writing down the distribution \eqref{Wcl_oam} we have
neglected possible longitudinal rotational motion of the nucleons,
which is justified in the $p_T \ll m_N$ limit.  We also assume that
a fraction $\beta$ of the nucleons in the nucleus are polarized in the 
$+{\hat x}$-direction, such that their net spin is $S = \beta A/2$ and (see
\eq{TMD8})
\begin{align} 
\label{rr1b} 
W_{trans} (p,b) = \beta W_{unp}(p,b).
\end{align}

Substituting Eqs.~\eqref{Wcl_oam} and \eqref{rr1b} into \eq{TMD15} and
integrating over $p^+$ and $\ul p$ yields
\begin{align} 
  \label{SiversA} 
	J \, k_y \, & f_{1T}^{\perp A} ({\bar x} , k_T) =
  M_A \int d b^- \, d^2 x \, d^2 y \, \rho \left( \frac{{\ul x} + {\ul
  y}}{2} , b^-\right) \, \frac{d^2 k'}{(2\pi)^2} \, e^{ - i \,
  (\ul{k} - \ul{k'}) \cdot ({\ul x} - {\ul y})} 
	\notag \\ & \times 
	\bigg\{ i \, {\bar x} \, p_{max} \left( \frac{({\ul x} +
  {\ul y})_x}{2} \right) \, \frac{b^-}{R^- ((\tfrac{x+y_x}{2})_x)} 
  ({\ul x} - {\ul
  y})_y \, f_1^N ({\bar x} , k'_T) + \frac{\beta}{2 \, m_N} \, k'_y \,
  f_{1T}^{\perp N} ({\bar x}, k'_T ) \bigg\} 
  \notag \\ &\times
	S_{{\ul x} {\ul y}}[+\infty, b^-],
\end{align}
where we also replaced ${\vec J}$ by ${\hat x} \, J$ and ${\vec S}$ by
${\hat x} \, (\beta A/2)$.

To further simplify \eq{SiversA} we need to make specific assumptions
about the form of $f_1^N$ and $f_{1T}^{\perp N}$. Inspired by the
lowest-order expressions for both quantities
\cite{Itakura:2003jp,Boer:2002ju,Brodsky:2013oya} (c.f. \eqref{DISS-DECOMP6} and
\eqref{diquarkSivers}) we write
\begin{align}
  \label{eq:fs}
  f_1^N (x, k_T) = \frac{\as \, C_1}{k_T^2}, \ \ \ f_{1T}^{\perp N}
  (x, k_T) = \frac{\as^2 \, m_N^2 \, C_2}{k_T^4} \, \ln
  \frac{k_T^2}{\Lambda^2},
\end{align}
where $C_1$ and $C_2$ are some $x$-dependent functions and $\Lambda$
is an infrared cutoff. Inserting \eq{eq:fs} into \eq{SiversA} and
integrating over $k'_T$ yields
\begin{align} 
  \label{SiversA2} & J \, k_y \, f_{1T}^{\perp A} ({\bar x} , k_T) =
  \frac{\as \, M_A}{2\pi} \int d b^- \, d^2x \, d^2 y \, \rho
  \left(\frac{{\ul x} + {\ul y}}{2} , b^-\right) \, e^{ - i \,
  \ul{k} \cdot(\ul{x}-\ul{y})} \, (\ul{x}-\ul{y})_y 
  \notag \\ & \times
	\bigg\{ i \, {\bar x} p_{max}\left( \frac{({\ul x} +
  {\ul y})_x}{2} \right) \, \frac{b^- \, C_1}{R^- ((\tfrac{x+y}{2})_x)} \,
  \ln\frac{1}{|\ul{x} -\ul{y}| \Lambda} + \frac{i \, \as \, m_N \, \beta \,
  C_2}{4} \, \ln^2 \frac{1}{|\ul{x} -\ul{y}| \Lambda} \bigg\} 
	\notag \\ & \times
  S_{{\ul x} {\ul y}}[+\infty, b^-].
\end{align}

In the classical MV/GGM approximation \cite{Mueller:1989st} the (symmetric part of the)
dipole scattering matrix is (c.f. \eqref{GGMsoln2})
\begin{align} 
  \label{Wlines4} S_{{\ul x} {\ul y}}[+\infty, b^-] = \exp
  \left[-\frac{1}{4} |\ul{x}- \ul{y}|^2 \, Q_s^2 \left(\frac{{\ul x} +
        {\ul y}}{2} \right) \, \left(\frac{R^- (\ul{b}) - b^-}{2 R^-
        (\ul{b})}\right) \, \ln\frac{1}{|\ul{x} -\ul{y}| \Lambda}
  \right],
\end{align}
where $R^-(\ul{b}) = \sqrt{R^2 - \ul{b}^2} \ M_A/P^+$ and the quark
saturation scale is
\begin{align} \label{Wlines6}
 Q_s^2(\ul{b})~=~4\pi \, \as^2 \, \tfrac{C_F}{N_c} \, T(\ul{b})
\end{align}
with the nuclear profile function
\begin{align}
  \label{eq:nuc_prof}
  T(\ul{b}) = \int d b^- \, \rho \left( \ul{b} , b^-\right).
\end{align}
As usual $N_c$ is the number of colors and $C_F = (N_c^2 -1)/2 N_c$ is
the Casimir operator of SU$(N_c)$ in the fundamental representation.
In arriving at \eq{Wlines4} we assumed that the nuclear density is
constant within the nucleus, such that
\begin{align}
  \label{eq:dens}
  \rho \left( \ul{b} , b^-\right) = \frac{\theta (R^- (\ul{b}) -
    |b^-|)}{2 R^- (\ul{b})} \, T(\ul{b}).
\end{align}

Employing \eq{Wlines4} along with Eqs.~\eqref{eq:dens} and
\eqref{Wlines6}, and neglecting all logarithms $\ln (1/|\ul{x}
-\ul{y}| \Lambda)$ (which is justified as long as $k_T$ is not too
much larger than $Q_s$ \cite{Kovchegov:1998bi}) we can integrate
\eq{SiversA2} over $b^-$ and $\ul{x} -\ul{y}$ obtaining
\begin{align} 
  \label{SiversA3} f_{1T}^{\perp A} ({\bar x} , k_T) = \frac{M_A \,
    N_c}{4\pi \, \as \, J \, C_F} \, \frac{1}{k_T^2} \int d^2b \,
  \bigg\{ & 4 \, {\bar x} \, p_{max} (\ul{b}) \, C_1 \, \left[
    e^{-k_T^2/Q_s^2 ({\ul b})} + 2 \, \frac{k_T^2}{Q_s^2 ({\ul b})} \,
    Ei \left(- \frac{k_T^2}{Q_s^2 ({\ul b})}\right)\right]  \notag \\
  & + \as \, \beta \, m_N \, C_2 \, e^{-k_T^2/Q_s^2 ({\ul b})} \bigg\},
\end{align}
where now $\ul{b} = (\ul{x} + \ul{y})/2$ and $p_{max} (\ul{b}) =
p_{max} \, \sqrt{R^2 - \ul{b}^2}/R$. The $\ul b$-integral appears to
be rather hard to perform for a realistic spherical nucleus: we leave
expression \eqref{SiversA3} in its present form.

To obtain a final expression for the Sivers function we need to
determine the total angular momentum $J$ of the nucleus. For a rigid rotator spinning
around the $\hat x$-axis with the maximum nucleon momentum $p_{max}$
we readily get
\begin{align}
  \label{eq:Lrigid}
  L = \frac{4}{5} \, A \, p_{max} \, R
\end{align}
in the nuclear rest frame. Using this in \eq{eq:J} along with $S =
\beta \, A/2$ we obtain
\begin{align}
  \label{eq:Jmod}
  J = \beta \frac{A}{2} + \frac{4}{5} \, A \, p_{max} \, R.
\end{align}
Inserting \eq{eq:Jmod} into \eq{SiversA3} gives
\begin{align} 
  \label{SiversA4} 
	& f_{1T}^{\perp A} ({\bar x} , k_T) = \frac{m_N \,
  N_c}{2\pi \, \as \, C_F} \, \frac{1}{\beta + \tfrac{8}{5} \, p_{max} \,
  R} \, \frac{1}{k_T^2} 
	\\ \notag & \times 
	\int d^2b \, \bigg\{ 4 \, {\bar x} \, p_{max} (\ul{b}) \,
  C_1 \, \left[ e^{-k_T^2/Q_s^2 ({\ul b})} + 2 \, \frac{k_T^2}{Q_s^2
  ({\ul b})} \, Ei \left(- \frac{k_T^2}{Q_s^2 ({\ul
  b})}\right)\right] + \as \beta \, m_N \, C_2 \, e^{-k_T^2/Q_s^2
  ({\ul b})} \bigg\}.
\end{align}
\eq{SiversA4} is our final expression for the Sivers function of a
nucleus in the quasi-classical approximation with the rigid rotator
model for the nucleus and $k_T$ not too much larger than
$Q_s$. Analyzing this expression we see that the OAM term (the first
term in the curly brackets) does change sign as a function of $k_T$,
while the Sivers density term (the second term in the curly brackets
of \eq{SiversA4}) is positive-definite. Still the first term in the
curly brackets is positive for most of the $k_T$ domain, corresponding to
quarks being produced preferentially into the page in
\fig{SIDIS_fig}. 

To study the $k_T \gg Q_s$ case we have to return to \eq{SiversA2}:
this time we do not neglect the logarithms. The large $k_T$ limit implies
that $|\ul{x} - \ul{y}|$ is small, and we need to expand the
exponential in \eq{Wlines4} to the lowest non-trivial (contributing)
order in each term in \eq{SiversA2}. For the Sivers density term this
corresponds to replacing the exponent by $1$. The remaining evaluation
is easier to carry out in \eq{SiversA}, which yields an intuitively
clear formula
\begin{align}
  \label{eq:transv}
  J \, f_{1T}^{\perp A} ({\bar x}, k_T) \big|_{transversity \ channel, \ k_T
    \gg Q_s} = A \, S \, f_{1T}^{\perp N} ({\bar x}, k_T).
\end{align}
In the first term in the curly brackets of \eq{SiversA2} we need to
expand the exponential in $S_{{\ul x} {\ul y}}[+\infty, b^-]$ one step
further, obtaining after some straightforward algebra for the whole
SIDIS Sivers function
\begin{align}
  \label{eq:Sivers_highk}
  f_{1T}^{\perp A} ({\bar x}, k_T) \big|_{k_T \gg Q_s} &= \frac{S}{J} \left[
  - \frac{4 \, \as \, m_N \, {\bar x} \, C_1}{3 \beta \, k_T^6} \, \ln
  \frac{k_T^2}{\Lambda^2} \, \int d^2 b \, T(\ul{b}) \, p_{max}
  (\ul{b}) \, Q_s^2 (\ul{b}) + A \,
  f_{1T}^{\perp N} ({\bar x}, k_T) \right] 
	\notag \\ &= 
	\frac{\beta}{\beta+ \tfrac{8}{5} \, p_{max} \, R} \bigg[ - \frac{4 \,
  \as \, m_N \, {\bar x} \, C_1}{3 \beta \, k_T^6} \ln \frac{k_T^2}{\Lambda^2}
  \int d^2 b \ T(\ul{b}) \, p_{max} (\ul{b}) \, Q_s^2 (\ul{b}) 
	\notag \\ &+
  \frac{A \, \as^2 \, m_N^2 \, C_2}{k_T^4} \, \ln
  \frac{k_T^2}{\Lambda^2} \bigg].
\end{align}
Since 
\begin{align}
  \label{eq:nucl_prof}
  \int d^2 b \, T(\ul{b}) = A
\end{align}
we see that the OAM channel contribution in \eq{eq:Sivers_highk} (the
first term) is proportional to $A \, \as \, m_N \, p_T \,
Q_s^2/k_T^6$, while the transversity channel contribution (the second
term) scales as $A \, \as^2 \, m_N^2 / k_T^4$. (Note that $x =
\ord{1}$, such that powers of $x$ do not generate suppression.)
Assuming that $p_T \approx m_N$ (see the discussion following
\eq{eq:mom3}) we observe that the ratio of the OAM to transversity
channel contributions is $\sim Q_s^2/(\as \, k_T^2)$. (Note that for
$p_T \approx m_N$ the prefactor of \eq{eq:Sivers_highk} gives a factor
$\sim 1/(m_N \, R) \approx A^{-1/3}$ multiplying both terms, but not
affecting their ratio.) We conclude that the OAM channel dominates for
\begin{align}
  \label{eq:bound}
  k_T < \frac{Q_s}{\sqrt{\as}},
\end{align}
that is both inside the saturation region, and in a sector of phase
space outside that region. For $k_T > Q_s/\sqrt{\as}$ the transversity
channel dominates, mapping onto the expected perturbative QCD result
\eqref{eq:transv}.

While the main aim of this calculation is to model a nucleon at high
energies, a few comments are in order about the application of this
rigid rotator toy model to a realistic nucleus.  Certainly a classical
rigid rotator is a poor model for a real nucleus; a better approach
would be to use our general result \eqref{TMD15} with the Wigner
distribution $W(p,b)$ given by the realistic single-particle wave
functions taken from nuclear structure calculations.  In such
realistic cases, the total angular momentum $J$ of the nucleus is
typically small, and the fraction $\beta$ that comes from the
nucleons' spins is also small due to nucleon spin pairing.  If one
were to approximate a real nucleus with this rigid rotator toy model,
appropriately small $J$ and $\beta$ would need to be used in
\eqref{SiversA4} and \eqref{eq:Sivers_highk}.  The smallness of the
total OAM $J$ does not affect the Sivers function $f_{1T}^{\bot A}$
because the magnitude is contained in the prefactor ${\hat z} \cdot
(\ul{J} \times \ul{k})$ and cancels in the $S/J$ ratio.  The smallness
of the spin contribution $\beta \sim \mathcal{O}(1/A)$, however, would
suppress the transversity channel and ensure the dominance of the OAM
term.  But regardless of its applicability to a real nucleus, the
rigid rotator toy model illustrates the ability of this formalism to
capture the interplay of spin and angular momentum in a dense system
at high energy.


\section{Drell-Yan Sivers Function in the Quasi-Classical Limit}
\label{sec:DY}

We now wish to perform a similar analysis for the Drell-Yan process
$\overline{q} + A^\uparrow \rightarrow \gamma^* + X \rightarrow \ell^+
\ell^- + X$, where the antiquark from the unpolarized hadron scatters
on the transversely polarized hadron/nucleus, producing a space-like
photon which later decays into a di-lepton pair. The annihilation
sub-process $\overline{q} + q^\uparrow \rightarrow \gamma^* + X$ is
related to the SIDIS process by time reversal, which leads to the
famous prediction \cite{Collins:2002kn} as discussed in Sec.~\ref{subsec-sign_flip}
 that the Sivers functions entering observables in the two processes 
should have equal magnitudes and opposite signs.

\begin{figure}[ht]
\centering
\includegraphics[width = 0.6 \textwidth]{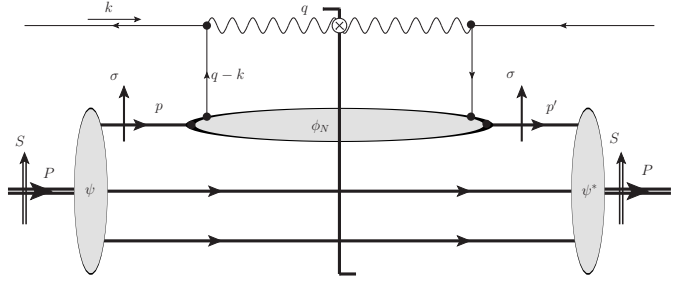}
\caption{
  \label{fig:DY1} Lowest-order DY process in the usual $\alpha_s$
  power-counting.  An antiquark from a projectile hadron annihilates
  with a quark from a nucleon in the target nucleus, producing a
  highly virtual photon which then decays into a di-lepton pair (not
  shown).  }
\end{figure}

The lowest-order Drell-Yan annihilation process is shown in
Fig.~\ref{fig:DY1}, without including initial-state rescattering of
the antiquark on nuclear spectators. Labeling the momenta as in
Fig.~\ref{fig:DY1} and following along the same lines as for SIDIS, we
can write the kinematics in the $\overline{q}+ A^\uparrow$
center-of-mass frame as
\begin{align} \label{DY1}
 \begin{aligned}
   P^\mu &= \big(P^+ , M_A^2 / P^+ , \ul{0}\big) \\
   p^\mu &= \big(p^+ , (p_T^2 + m_N^2) / p^+ , \ul{p}\big) \\
   k^\mu &= \big(\tfrac{m_q^2}{Q^2} \, q^+ , k^- \approx q^-, \ul{0}\big) \\
   q^\mu &= \big(q^+ , q^- \approx Q^2/q^+ , \ul{q}\big),
 \end{aligned}
\end{align}
where 
\begin{align} \label{DY2_5}
 \begin{aligned}
   \hat{s} &\equiv (p+k)^2 \approx p^+ q^- \\
   x &\equiv \frac{Q^2}{2 p \cdot q} \approx \frac{Q^2}{\hat{s}}
   \approx \frac{q^+}{p^+}.
 \end{aligned}
\end{align}
As with SIDIS, we are working in the kinematic limit $s_A = (P+k)^2
\gg \hat{s}, Q^2 \gg \bot^2$, with $\alpha \equiv p^+ / P^+ \approx
s_A / \hat{s} \sim \ord{1/A}$.  Again, we can compare the coherence
lengths $\ell_k^- \sim 1 / k^+$ of the antiquark and $\ell_\gamma^-
\sim 1/q^+$ of
\begin{align} \label{DY3}
 \begin{aligned}
   \frac{\ell_k^-}{L^-} &\sim \frac{1}{x}
   \left(\frac{Q^2}{m_q^2}\right) \frac{1}{\alpha \, M_A \, R}
   \sim \ord{\frac{Q^2}{m_q^2} \, A^{-1/3}} \gg 1 \\
   \frac{\ell_\gamma^-}{L^-} &\sim \frac{1}{x} \frac{1}{\alpha \, M_A
     \, R} \sim \ord{A^{-1/3}} \ll 1.
 \end{aligned}
\end{align}
Analogous to SIDIS, this shows that the coherence length of the
incoming antiquark is large; in fact it would be infinite if we
dropped the quark mass $m_q$ as we have elsewhere in the calculation.
We conclude that the long-lived antiquark is able to rescatter off of
many nucleons before it finally annihilates a quark. The annihilation
occurs locally, as indicated by the short coherence length of the
virtual photon, and thereafter the produced photon / dilepton system
does not rescatter hadronically. This again motivates the resummation
of these initial state rescatterings into a Wilson line dipole trace.


\subsection{Channels Generating the DY Sivers Function}

\begin{figure}[ht]
\centering
\includegraphics[width= .5 \textwidth]{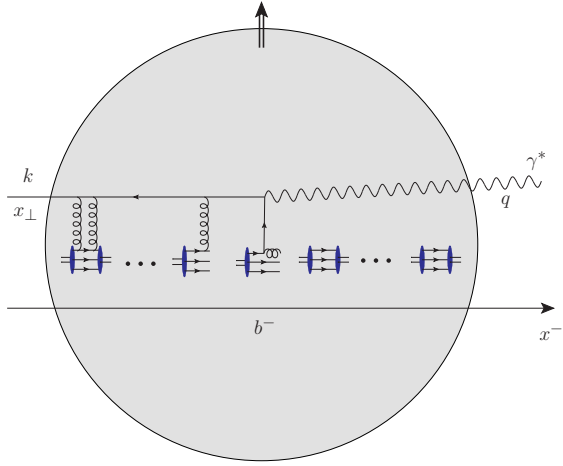}
\caption{Space-time structure of the quasi-classical DY process in the
  rest frame of the nucleus, overlaid with one of the corresponding
  Feynman diagrams. The shaded circle is the transversely polarized
  nucleus, with the vertical double arrow denoting the spin
  direction.}
\label{DY_ampl}
\end{figure}

The entire Drell-Yan (DY) process in the quasi-classical approximation
is shown in \fig{DY_ampl} at the level of the scattering amplitude:
the incoming antiquark coherently scatters on the nucleon in the
transversely polarized nucleons, until the last interaction in which
the virtual photon is produced, which later generates the di-lepton
pair.

\begin{figure}[ht]
\centering
\includegraphics[width = 0.7 \textwidth]{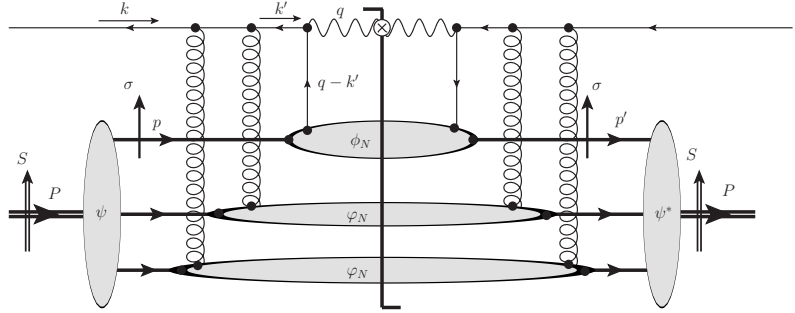}
\caption{
  \label{fig:DY2} Decomposition of the nuclear quark distribution
  $\Phi_A$ probed by the DY process into mean-field wave functions
  $\psi, \psi^*$ of nucleons and the quark and gluon distributions
  $\phi_N$ and $\varphi_N$ of the nucleons.  }
\end{figure}

By analogy with \eq{TMD1} in SIDIS we write the following relation for
the quark correlators in DY,
\begin{align} 
 \label{DYTMD1}
 \begin{aligned}
  \Tr[\Phi_A({\bar x}, \ul{q} &; P, J) \gamma^+ ] = A \int \frac{d
  p^+ \, d^2 p \, d b^-}{2(2\pi)^3} \int \frac{d^2 k' \, d^2 x \, d^2
  y}{(2\pi)^2} \, e^{i \ul{k'} \cdot (\ul{x} - \ul{y})} \,
	\\ &\times 
	\sum_\sigma W_N^\sigma \left(p, b^-, \frac{\ul{x} + \ul{y}}{2} \right)
	\Tr[\phi_N (x , \ul{q} -\ul{k}'-x \, \ul{p}); p, \sigma)
  \gamma^+] \, D_{{\ul{y} \, \ul{x}}} [b^-, -\infty],
 \end{aligned}
\end{align}
where 
\begin{align}
 \label{dipole_def_DY}
 D_{{\ul y} \, {\ul x}} [b^-, -\infty] = \left\langle \frac{1}{N_c} \,
   \mbox{Tr} \left[ V_{\ul y} [b^-, -\infty] \, V^\dagger_{\ul x}
     [b^-, -\infty] \right] \right\rangle
\end{align}
and the quark correlators are defined by equations similar to
\eqref{TMD2} and \eqref{TMD3}, but now using a different gauge link
\eqref{eq:U_DY}:
\begin{align}
  \label{DYcorr2}
	\Phi_{ij}^A (\bar x, \ul{k}; P, J) \equiv \frac{1}{2(2\pi)^3} \int d^{2-} r \, e^{i k \cdot r}
 \bra{A (P, J)} \psibar_j (0) \, {\cal U}^{DY} [0,r] \, \psi_i (r) \ket{A (P, J)},
\end{align}
\begin{align} 
  \label{DYcorr3} 
	\phi_{ij}^N ( x, \ul{k}; p, \sigma) \equiv \frac{1}{2(2\pi)^3} \int d^{2-} r \, e^{i k \cdot r}
  \bra{N (p, \sigma)} \psibar_j (0) \, {\cal U}^{DY} [0,r] \, \psi_i (r) \ket{N (p, \sigma)} .
\end{align}
Here ${\bar x} = A \, q^+/P^+$. \eq{DYTMD1} is illustrated in
\fig{fig:DY2}. The main difference compared to \eq{TMD1} is that now
$\ul{k} =0$ and $\ul{q} \neq 0$.

Projecting out the DY Sivers function of the nucleus $f_{1T}^{\perp
  A}$ using \eqref{TMD5} gives
\begin{align} 
 \label{DY_TMD2}
 {\hat z} \cdot (\ul{J}\times\ul{q}) \, & f_{1T}^{\perp A} ({\bar x} ,
 q_T) = \, \frac{M_A \, A}{4} \int \frac{d p^+ \, d^2 p \, d
 b^-}{2(2\pi)^3} \int \frac{d^2 k' \, d^2 x \, d^2 y}{(2\pi)^2} \,
 e^{i \ul{k'} \cdot (\ul{x} - \ul{y})} 
 \notag \\ &\times 
 \sum_\sigma W_N^\sigma \left(p, b^-, \tfrac{\ul{x} + \ul{y}}{2} \right)
 \Tr[\phi_N (x , \ul{q} -\ul{k}'-x \, \ul{p}); p, \sigma)
 \gamma^+] \, D_{{\ul{y} \, \ul{x}}} [b^-, -\infty] 
 \notag \\ &-
 (\ul{q} \rightarrow -\ul{q}).
\end{align}
With the help of \eq{TMD4B} we write
\begin{align} 
\label{DY_TMD3}
   {\hat z} \cdot & (\ul{J}\times\ul{q}) \, f_{1T}^{\perp A} ( {\bar
     x} , q_T) = \, \frac{M_A \, A}{2} \int \frac{d p^+ \, d^2 p \, d
     b^-}{2(2\pi)^3} \int \frac{d^2 k' \, d^2 x \, d^2 y}{(2\pi)^2} \,
   e^{i \ul{k'} \cdot (\ul{x} - \ul{y})} \, \sum_\sigma W_N^\sigma
   \left(p, b^-, \tfrac{\ul{x} +
       \ul{y}}{2} \right) \notag  \\
   & \times \bigg[ f_1^N (x , |\ul{q} -\ul{k}' - x \, \ul{p}|_T) +
   \frac{1}{m_N} {\hat z} \cdot \left(\ul{\sigma} \times (\ul{q}
     -\ul{k}'-x \, \ul{p})\right) \, f_{1T}^{\perp N} (x , |\ul{q}
   -\ul{k}'- x \ul{p}|_T) \bigg] \notag  \\
   & \times \, D_{{\ul{y} \, \ul{x}}} [b^-, -\infty] - (\ul{q}
   \rightarrow -\ul{q}).
\end{align}
Performing the spin sums \eqref{TMD8} gives
\begin{align} 
  \label{DY_TMD4} {\hat z} \cdot & (\ul{J}\times\ul{q}) \,
  f_{1T}^{\perp A} ( {\bar x} , q_T) = \, \frac{M_A}{2} \int \frac{d
    p^+ \, d^2 p \, d b^-}{2(2\pi)^3} \int \frac{d^2 k' \, d^2 x \,
    d^2 y}{(2\pi)^2} \, e^{i \ul{k'} \cdot (\ul{x} - \ul{y})} \,
  \bigg[ A \, W_{unp} \left(p, b^-, \tfrac{\ul{x} +
      \ul{y}}{2} \right) \notag  \\
  & \times \, f_1^N (x , |\ul{q} -\ul{k}' - x \, \ul{p}|_T) +
  W_{trans} \left(p, b^-, \tfrac{\ul{x} + \ul{y}}{2} \right) \,
  \frac{1}{m_N} {\hat z} \cdot \left(\ul{S} \times (\ul{q}
    -\ul{k}'-x \, \ul{p})\right) \notag  \\
  & \times \, f_{1T}^{\perp N} (x , |\ul{q} -\ul{k}'- x \ul{p}|_T)
  \bigg] \, D_{{\ul{y} \, \ul{x}}} [b^-, -\infty] - (\ul{q}
  \rightarrow -\ul{q}).
\end{align}
In the terms being subtracted in \eq{DY_TMD4} with $(\ul{q}
\rightarrow -\ul{q})$, we can also reverse the dummy integration
variables $\ul{k'} \rightarrow -\ul{k'}$, $\ul{p} \rightarrow -
\ul{p}$, and $\ul{x} \leftrightarrow \ul{y}$.  This leaves the Fourier
factor and the distributions $f_1^N$, $f_{1T}^{\perp N}$ invariant,
giving
\begin{align} 
  \label{DY_TMD5} &{\hat z} \cdot (\ul{J}\times\ul{q}) \,
  f_{1T}^{\perp A} ( {\bar x} , q_T) = \, \frac{M_A}{2} \int \frac{d
    p^+ \, d^2 p \, d b^-}{2(2\pi)^3} \int \frac{d^2 k' \, d^2 x \,
    d^2 y}{(2\pi)^2} \, e^{i \ul{k'} \cdot (\ul{x} - \ul{y})} \,
  \bigg\{ f_1^N (x , |\ul{q} -\ul{k}' - x \, \ul{p}|_T) \notag \\
  & \times \, A \, \bigg[ W_{unp} \left(p^+, \ul{p} , b^-,
    \tfrac{\ul{x} + \ul{y}}{2} \right) \, D_{{\ul{y} \, \ul{x}}} [b^-,
  -\infty] - W_{unp} \left(p^+, - \ul{p} , b^-, \tfrac{\ul{x} +
      \ul{y}}{2} \right) \, D_{{\ul{x} \, \ul{y}}} [b^-, -\infty]
  \bigg] \notag \\ & + \frac{1}{m_N} {\hat z} \cdot \left(\ul{S}
    \times (\ul{q} -\ul{k}'-x \, \ul{p})\right) \, f_{1T}^{\perp N} (x
  , |\ul{q} -\ul{k}'- x \ul{p}|_T) \\
  & \times \bigg[ W_{trans} \left(p^+, \ul{p}, b^-, \tfrac{\ul{x} +
      \ul{y}}{2} \right) \, D_{{\ul{y} \, \ul{x}}} [b^-, -\infty] +
  W_{trans} \left(p^+, - \ul{p}, b^-, \tfrac{\ul{x} + \ul{y}}{2}
  \right) \, D_{{\ul{x} \, \ul{y}}} [b^-, -\infty] \bigg]
  \bigg\}. \notag
\end{align}
We recognize the factors in brackets from the SIDIS case
\eqref{TMD11}, rewriting \eqref{DY_TMD5} as
\begin{align} 
  \label{DY_TMD6} & {\hat z} \cdot (\ul{J}\times\ul{q}) \,
  f_{1T}^{\perp A} ( {\bar x} , q_T) = \, M_A \int \frac{d
    p^+ \, d^2 p \, d b^-}{2(2\pi)^3} \int \frac{d^2 k' \, d^2 x \,
    d^2 y}{(2\pi)^2} \, e^{i \ul{k'} \cdot (\ul{x} - \ul{y})} \,
  \bigg\{ f_1^N (x , |\ul{q} -\ul{k}' - x \, \ul{p}|_T) \notag \\
  & \times \, A \, \bigg[ W_{unp}^{OAM} \left(p^+, \ul{p} , b^-,
    \tfrac{\ul{x} + \ul{y}}{2} \right) \, S_{{\ul{y} \, \ul{x}}} [b^-,
  -\infty] + W_{unp}^{symm} \left(p^+, - \ul{p} , b^-, \tfrac{\ul{x} +
      \ul{y}}{2} \right) \, i \, O_{{\ul{y} \, \ul{x}}} [b^-, -\infty]
  \bigg] \notag \\ & + \frac{1}{m_N} {\hat z} \cdot \left(\ul{S}
    \times (\ul{q} -\ul{k}'-x \, \ul{p})\right) \, f_{1T}^{\perp N} (x
  , |\ul{q} -\ul{k}'- x \ul{p}|_T) \\
  & \times \bigg[ W_{trans}^{symm} \left(p^+, \ul{p}, b^-,
    \tfrac{\ul{x} + \ul{y}}{2} \right) \, S_{{\ul{y} \, \ul{x}}} [b^-,
  -\infty] + W_{trans}^{OAM} \left(p^+, - \ul{p}, b^-, \tfrac{\ul{x} +
      \ul{y}}{2} \right) \, i \, O_{{\ul{y} \, \ul{x}}} [b^-, -\infty]
  \bigg] \bigg\}. \notag
\end{align}
As before, we drop contributions from the odderon $iO_{{\ul{y} \,
    \ul{x}}}$ as being outside the precision of the quasi-classical
formula \eqref{DYTMD1} to get
\begin{align} 
  \label{DY_TMD7} & {\hat z} \cdot (\ul{J}\times\ul{q}) \,
  f_{1T}^{\perp A} ( {\bar x} , q_T) = \, M_A \int \frac{d p^+ \, d^2
  p \, d b^-}{2(2\pi)^3} \int \frac{d^2 k' \, d^2 x \,
  d^2 y}{(2\pi)^2} \, e^{i \ul{k'} \cdot (\ul{x} - \ul{y})} 
	\notag \\ & \times \, 
	\bigg\{ A \, W_{unp}^{OAM} \left(p^+, \ul{p} , b^-,
  \frac{\ul{x} + \ul{y}}{2} \right) \, f_1^N (x , |\ul{q} -\ul{k}' -
  x \, \ul{p}|_T) 
	\\ \notag & + 
	\frac{1}{m_N} {\hat z} \cdot \left(\ul{S}
  \times (\ul{q} -\ul{k}'-x \, \ul{p})\right) \, W_{trans}^{symm}
  \left(p^+, \ul{p}, b^-, \frac{\ul{x} + \ul{y}}{2} \right) \,
  f_{1T}^{\perp N} (x , |\ul{q} -\ul{k}'- x \ul{p}|_T) \bigg\}
	\\ \notag &\times
  S_{{\ul{y} \, \ul{x}}} [b^-, -\infty].
\end{align}

Since the rotational momentum of the nucleons $p_T$ is assumed to be
small, we have to expand in it to the lowest non-trivial
order. Shifting the integration variable $\ul{k}' \to \ul{k}' + \ul{q}
- x \, \ul{p}$ in \eq{DY_TMD7} and expanding the exponential to the
lowest non-trivial order in $p_T$ we obtain (cf. \eq{TMD15})
\begin{align} 
  \label{DY_TMD8} & {\hat z} \cdot (\ul{J}\times\ul{q}) \,
  f_{1T}^{\perp A} ( {\bar x} , q_T) = \, M_A \int \frac{d p^+ \, d^2
    p \, d b^-}{2(2\pi)^3} \int \frac{d^2 k' \, d^2 x \,
    d^2 y}{(2\pi)^2} \, e^{- i \, (\ul{q} - \ul{k'}) \cdot (\ul{x} - \ul{y})} \notag \\
  & \times \, \bigg\{ i \, x \, \ul{p} \cdot (\ul{x} - \ul{y}) \, A \,
  W_{unp}^{OAM} \left(p^+, \ul{p} , b^-, \frac{\ul{x} + \ul{y}}{2}
  \right) \, f_1^N (x , k'_T) \\ & - \frac{1}{m_N} {\hat z} \cdot
  \left(\ul{S} \times \ul{k}' \right) \, W_{trans}^{symm} \left(p^+,
    \ul{p}, b^-, \frac{\ul{x} + \ul{y}}{2} \right) \, f_{1T}^{\perp N}
  (x , k'_T) \bigg\} \, S_{{\ul{x} \, \ul{y}}} [b^-, -\infty], \notag
\end{align}
where we have also interchanged $\ul{x} \leftrightarrow \ul{y}$ and
$\ul{k}' \to - \ul{k}'$.

\eq{DY_TMD8} is our main formal result for the DY Sivers function. We
again see that the Sivers function in DY can arise through two
distinct channels in this quasi-classical approach: the OAM channel
that contains its preferred direction in the distribution
$W_{unp}^{OAM}$ and the transversity/Sivers density channel that
generates is preferred direction through a local lensing mechanism
$f_{1T}^{\perp N}$.

To demonstrate the importance of the Wilson lines for the Sivers
function, for the moment, let us ignore the contribution of the Wilson
lines associated with initial-state rescattering in
\eq{DY_TMD8}. Without any such initial-state interactions, the
nucleonic Sivers function is zero, $f_{1T}^N = 0$
\cite{Brodsky:2002rv,Boer:2002ju,Brodsky:2013oya}, leaving
\begin{align}
  \label{DY10}
  {\hat z} \cdot (\ul{J}\times\ul{q}) \, f_{1T}^{\perp A} ( {\bar x} ,
  q_T) = \, M_A \int \frac{d p^+ \, d^2 p \, d b^-}{2(2\pi)^3} \int
  \frac{d^2 k' \, d^2 x \,
    d^2 y}{(2\pi)^2} \, e^{- i \, (\ul{q} - \ul{k'}) \cdot (\ul{x} - \ul{y})} \notag \\
  \times \, i \, x \, \ul{p} \cdot (\ul{x} - \ul{y}) \, A \,
  W_{unp}^{OAM} \left(p^+, \ul{p} , b^-, \frac{\ul{x} + \ul{y}}{2}
  \right) \, f_1^N (x , k'_T) = 0,
\end{align}
which vanishes after $b^-$ integration because of the rotational and
$PT$-symmetry conditions \eqref{rot6}.


\subsection{QCD Shadowing and the SIDIS/DY Sign-Flip}

Now that the DY Sivers function \eqref{DY_TMD8} is expressed in the
same form as the Sivers function for SIDIS \eqref{TMD15}, we can
compare both expressions to see how the nuclear Sivers functions have
changed between SIDIS and DY and understand the origin of the SIDIS/DY
sign-flip relation \cite{Collins:2002kn}
\begin{align} 
  \label{DYTMD9} f_{1T}^{\perp A} (x , k_T)\bigg|_{SIDIS} = -
  f_{1T}^{\perp A} (x , k_T)\bigg|_{DY}.
\end{align}
First, we notice that the transversity/Sivers density channel (the
second term in the curly brackets) has changed signs as required
between \eqref{TMD15} and \eqref{DY_TMD8}.  Mathematically, this
occurs because of the $\ul{k}' \to - \ul{k}'$ interchange, simply
because the momentum going into the Wilson line in SIDIS corresponds
to the momentum coming from the Wilson line in DY
(cf. Figs.~\ref{fig:TMD1} and \ref{fig:DY2}). The transversity/Sivers
density channel contribution thus automatically satisfies the
sign-flip relation \eqref{DYTMD9}.

The OAM channel contribution to \eq{DY_TMD8} is more subtle; although
the prefactor has not changed as compared to \eq{TMD15}, the
longitudinal coordinate $b^-$ integral entering \eqref{DY_TMD8} for DY
can be modified using $b^- \to - b^-$ substitution along with
\eq{rot6} to give
\begin{align}
  \label{DY_TMD11}
  \int db^- \, W_{unp}^{OAM} (p,b) \, S_{{\ul{x} \,
         \ul{y}}} [b^-, -\infty] = - \int db^- \, W_{unp}^{OAM} (p,b) \, S_{{\ul{x} \,
         \ul{y}}} [- b^-, -\infty]. 
\end{align}
When evaluating the dipole $S$-matrix we neglect the polarization
effects as being energy suppressed. Therefore, for the purpose of this
$S$-matrix, the nucleus has a rotational symmetry around the $z$-axis
(see \fig{nucleus} for axes labels). We thus write
\begin{align}
  \label{DY_TMD12}
  S_{{\ul{x} \, \ul{y}}} [- b^-, -\infty] \overset{PT}{=} S_{-{\ul{x},
      \, -\ul{y}}} [+ \infty, b^-] \overset{R_z}{=} S_{{\ul{x}
      \, \ul{y}}} [+ \infty, b^-],
\end{align}
where we have used the $PT$ transformation \eqref{WPT2} and 
$R_z$ denotes a half-revolution around the
$z$-axis. Using \eq{DY_TMD12} in \eq{DY_TMD11} we arrive at
\begin{align} \label{DYTMD10}
 \begin{aligned}
   \overbrace{\int db^- \, W_{unp}^{OAM}(p,b) \, S_{{\ul{x} \,
         \ul{y}}} [b^-, -\infty]}^{\mathrm{DY}} &= - \overbrace{\int
     db^- \, W_{unp}^{OAM} (p,b) \, S_{{\ul{x} \, \ul{y}}} [+ \infty,
     b^-] }^{\mathrm{SIDIS}}.
 \end{aligned}
\end{align}

One can also simply see that \eq{DYTMD10} is true by using the
quasi-classical GGM/MV dipole $S$-matrix from \eq{Wlines4} on its
right-hand-side, along with
\begin{align} 
  \label{Wlines42} S_{{\ul x} {\ul y}}[b^-, -\infty] = \exp
  \left[-\frac{1}{4} |\ul{x}- \ul{y}|^2 \, Q_s^2 \left(\frac{{\ul x} +
        {\ul y}}{2} \right) \, \left(\frac{b^- + R^-}{2 R^-}\right) \,
    \ln\frac{1}{|\ul{x} -\ul{y}| \Lambda} \right]
\end{align}
on its left-hand-side. We conclude that the OAM channel contributions
to the SIDIS Sivers function \eqref{TMD15} and the DY Sivers function
\eqref{DY_TMD8} also satisfy the sign-flip relation \eqref{DYTMD9}.

Therefore, for any Wigner distribution $W(p,b)$, the Sivers functions
at the quasi-classical level for SIDIS \eqref{TMD15} and for DY
\eqref{DY_TMD8} are equal in magnitude and opposite in sign,
\eqref{DYTMD9}, in agreement with \eqref{DISS-SIVERS7}.  This statement is a direct consequence of the
invariance of $W(p,b)$ under rotations and $PT$-reversal,
\eqref{rot6}, and it mirrors in this context the original derivation
by Collins \cite{Collins:2002kn}.

The rigid-rotator toy model of Sec.~\ref{toy_model} can also be
constructed for the DY Sivers function. However, due to the sign-reversal
relation \eqref{DYTMD9} we can immediately read off the answer for the DY Sivers
function in the rigid-rotator model as being negative of that in
\eq{SiversA4} for moderate $k_T$ and negative of \eq{eq:Sivers_highk}
for $k_T \gg Q_s$. All the conclusions about the relative importance
of the two contributing channels remain the same.

The advantage of our approach here, apart from providing the explicit
formal results \eqref{TMD15} and \eqref{DY_TMD8}, is in the new
physical interpretation of the Sivers function in the OAM
channel.  We will now summarize the physical picture of the SIDIS / DY sign 
flip in the OAM channel.  

Imagine a large spinning nucleus. The nucleus
is so large that it is almost completely opaque to a colored
probe. This strong nuclear shadowing is due to multiple rescatterings
in the nucleus generating a short mean free path for the quark,
anti-quark, or a gluon.

\begin{figure}[ht]
\centering
\includegraphics[height=4cm]{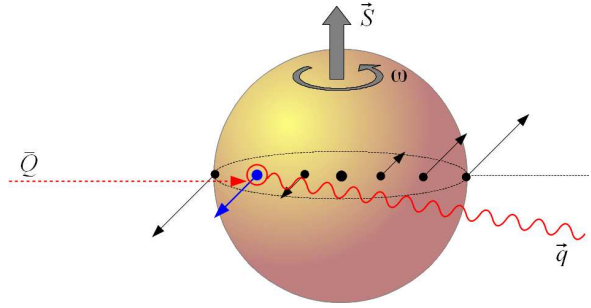}
\caption{The physical mechanism of STSA in DY as envisioned in the
  text.}
\label{DY_fig}
\end{figure}

The Drell-Yan process on such a rotating nucleus with shadowing 
is shown in \fig{DY_fig} in the nuclear rest frame,
with the rotation axis of the nucleus perpendicular to the collision
axis. The incoming anti-quark (generated in the wave function of the
other hadron) scatters on the ``front'' surface of the polarized
nucleus due to the strong shadowing. Since the anti-quark interacts
with the nucleons which, at the ``front'' of the nucleus
preferentially rotate with the nucleus out of the plane of the page in
\fig{DY_fig}, the produced time-like virtual photons are produced
preferentially out of the page, generating a left-of-polarized-beam
single spin asymmetry.

\begin{figure}[ht]
\centering
\includegraphics[height=4cm]{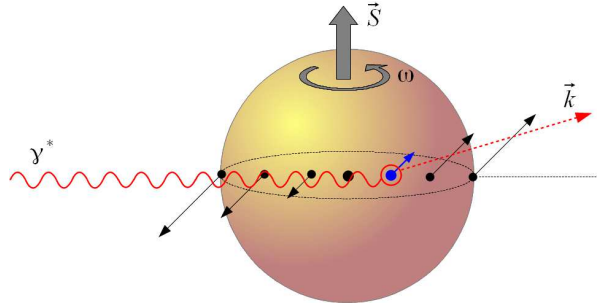}
\caption{The physical mechanism of STSA in SIDIS as envisioned in the
  text.}
\label{SIDIS_fig}
\end{figure}

The same mechanism can be applied to generate the asymmetry in SIDIS, as
illustrated in \fig{SIDIS_fig}, also in the rest frame of the
nucleus. Now the incoming virtual photon interacts with the
transversely polarized nucleus, producing a quark. For the quark to
escape out of the nucleus and be produced the interaction has to take
place at the ``back'' of the nucleus, to minimize the path the quark
needs to travel through the nucleus, maximizing its chances to
escape. The nucleons in the ``back'' of the nucleus rotate
preferentially into the page of \fig{SIDIS_fig}; the scattering of a
virtual photon on such nucleons results in the right-of-beam single
spin asymmetry for the outgoing quarks (quarks produced preferentially
with transverse momentum pointing into the page).

The spin asymmetries in DY and SIDIS shown in Figs.~\ref{DY_fig} and
\ref{SIDIS_fig} are generated through a combination of OAM effects and
nuclear shadowing. The two asymmetries are opposite-sign (left- and
right-of-beam), and, assuming that scattering in the two processes
happens equal distances from the nuclear edge, are likely to be equal
in magnitude, in agreement with the prediction of
\cite{Collins:2002kn,Brodsky:2002rv}.


\section{Discussion}
\label{sec:Discussion}

The main goal of this work was to construct the SIDIS and DY Sivers
functions in the quasi-classical GGM/MV approximation, which models a
proton as a large nucleus, and which we have modified by giving the nucleus
a non-zero OAM. The main formal results are given in
Eqs.~\eqref{TMD15} (SIDIS) and \eqref{DY_TMD8} (DY). We showed that
there are two main mechanisms generating the quasi-classical Sivers
function: the OAM channel and the transversity channel. The former is
leading in the saturation power counting; it also dominates for $k_T <
Q_s/\sqrt{\as}$, that is both inside and, for $Q_s < k_T <
Q_s/\sqrt{\as}$, outside of the saturation region. At higher $k_T$ the
transversity channel dominates. In the future our quasi-classical
calculation can be augmented by including evolution corrections to the
Sivers function, making the whole formalism ready for phenomenological
applications, similar to the successful use of nonlinear small-$x$
evolution equations
\cite{Balitsky:1996ub,Kovchegov:1999yj,Jalilian-Marian:1997dw,Iancu:2000hn,Balitsky:2006wa,Gardi:2006rp,Kovchegov:2006vj}
to the description (and prediction) of high energy scattering data
\cite{Albacete:2010sy,ALbacete:2010ad}.

Perhaps just as important, we constructed a novel physical mechanism
of the STSA generation. This is the OAM channel. The OAM mechanism,
while diagrammatically very similar to the original mechanism proposed by Brodsky, Hwang, and Schmidt
\cite{Brodsky:2002cx}, provides a different interpretation from the
``lensing'' effect \cite{Brodsky:2002cx,Brodsky:2013oya} or the
color-Lorentz force of \cite{Burkardt:2010sy,Burkardt:2008ps}. The OAM
mechanism is based on interpreting the extra rescattering proposed by Brodsky, Hwang, and Schmidt as a shadowing-type
correction. The STSA is then generated by the combination of the OAM
and shadowing. The shadowing makes sure the projectile interacts
differently with the front and the back of the target, generating the
asymmetry of the produced particles.

While shadowing is a high-energy phenomenon, and our calculation was
done in the high-energy approximation ${\hat s} \gg \perp^2$ (though
for $x \sim \ord{1}$), it may be that the OAM mechanism for generating
STSA is still valid for lower-energy scattering, though of course the
formulas derived above would not apply in such a regime. At lower
energies the difference between the interactions of the projectile
with the front/back of the target may result from, say, energy loss of
the projectile as it traverses the target. Again, combined with the
target rotation this would generate STSA, and, hence, the Sivers
function. The formalism needed to describe such a low-energy process
would be quite different from the one presented above; moreover, the
correct degrees of freedom may not be quarks and gluons
anymore. However, the main physics principle of combining OAM with the
difference in interaction probabilities between the projectile and
front/back of the target to generate STSA may be valid at all
energies.

Returning to higher energies and the derived formulas~\eqref{TMD15}
and \eqref{DY_TMD8}, let us point out that these results, when applied
to experimental data, may allow one to determine the distribution of
intrinsic transverse momentum $\ul{p} (\ul{b}, b^-)$ of partons in the
hadronic or nuclear target, along with the transversity/Sivers
function density in the target. This would complement the existing
methods of spatial imaging of quarks and gluons inside the hadrons and
nuclei \cite{Accardi:2012qut}, providing a new independent cross-check
for those methods.


\chapter{Outlook}
\label{chap-Outlook}



\section{Summary}
\label{sec-Summary}

In this Dissertation we have laid out the foundational elements of two paradigms of hadronic structure and explored some of the interconnections between them.  The \textit{spin and transverse momentum paradigm} was presented in Chapter~\ref{chap-TMD}, including the definitions and parameterizations of the transverse-momentum-dependent (TMD) quark (\eqref{DISS-GAUGE3} , \eqref{DISS-DECOMP1}) and gluon (\eqref{gluonTMD1} , \eqref{gluonTMD8}) distribution functions, with a special emphasis on the Sivers function (Sec.~\ref{sec-Sivers}).  The \textit{saturation paradigm} was presented in Chapter~\ref{chap-CGC} in the context of a heavy nucleus, including the role of multiple rescattering \eqref{Qsat3}, the dominance of classical gluon fields \eqref{WWsat}, and how they both give rise to the emergent phenomenon of gluon saturation \eqref{Qsat1}.  

In Chapter~\ref{chap-odderon} we investigated the use of a transversely-polarized projectile to probe the saturated gluon fields of a dense target.  We calculated the leading-order contributions to the single transverse spin asymmetry (STSA) (\eqref{dsigmaq}, \eqref{dsigmaq_unp}) in the saturation formalism, finding that the $T$-odd asymmetry is naturally generated by the $T$-odd, $C$-odd component of the gluon fields known as the odderon \eqref{Ocl3}.  In the quasi-classical approximation, the preferred direction of the odderon couples to the gradients of the nuclear density, so that linear terms involving only the odderon average to zero \eqref{zero_arg2} after integration over impact parameters.  Thus we found that the mechanism responsible for STSA in this formalism requires the nonlinear interference between the odderon and another $C, T$-even rescattering \eqref{dsigmaq3}; consequently, the STSA in the distribution of produced prompt photons is zero \eqref{dsigmaph2}.  To identify the features of this odderon-driven mechanism, we estimated the asymmetry \eqref{ANq} in the quasi-classical approximation, illustrating the range of spectra in Figs.~\ref{AN_fig} and \ref{AN2_fig}.  This mechanism appeared to be dominated by the periphery of the target where the saturation scale is low and perturbation theory begins to break down, leading to the strong cutoff dependence of these plots.

In Chapter~\ref{chap-MVspin} we analyzed the effects of saturation on the TMD quark distributions by calculating the Sivers function of a heavy nucleus in the quasi-classical limit.  We found that, both for semi-inclusive deep inelastic scattering (SIDIS) \eqref{TMD1} and for the Drell-Yan process (DY) \eqref{DYTMD1}, the quasi-classical approximation results in a factorization of the nuclear TMD's into a convolution of three factors: the Wigner distribution \eqref{DIS8} of nucleons in the nucleus, the TMD distributions \eqref{TMD3} of quarks in the nucleons, and nuclear shadowing factors \eqref{dipole_def} for the rescattering of the active quarks on spectator nucleons.  By projecting out the symmetries corresponding to the Sivers function, we identified two contributing channels as visualized in Fig.~\ref{SIDIS_OAM_vs_Sivers}: a conventional ``lensing'' mechanism (transversity channel) which simply aggregates the Sivers functions of the nucleons, and a novel shadowing mechanism (OAM channel) which couples the Sivers function to the rotational motion of the nucleons.  Interestingly, we again found contributions arising from the odderon \eqref{TMD12}, although they are beyond the precision of our quasi-classical factorization formula.  The OAM channel we have derived provides a new interpretation for the Sivers function of a dense system as sensitive to the orbital angular momentum of its constituents, with an intuitive manifestation of the SIDIS / DY sign flip \eqref{DISS-SIVERS7} visualized in Figs.~\ref{DY_fig} and \ref{SIDIS_fig}.


\section{Future Prospects: Spin and Saturation in the Coming EIC Era}
\label{sec-Prospects}

The calculations we have presented here are still only early steps in the work to be done in understanding the interconnection between the spin and saturation paradigms.  Further calculations need to be performed to identify the corrections to the mechanisms presented in Chapters~\ref{chap-odderon} and \ref{chap-MVspin}, and a considerable effort must be made to develop them into robust phenomenological models that are ready to be quantitatively compared with data.

For the odderon mechanism of STSA from Chapter~\ref{chap-odderon}, it is important to emphasize that the plots generated in Figs.~\ref{AN_fig} and \ref{AN2_fig} are the results of considerable approximations made to our general formulas \eqref{dsigmaq} and \eqref{dsigmaq_unp}.  Serious numerical evaluation of the integrals, including the contributions of both quarks and gluons to the polarized and unpolarized cross-sections, are necessary to determine whether the strong cutoff dependence observed is a signal of the breakdown of the applicability of perturbation theory or simply a failure of the approximations.  Equally important, the ``lensing diagram'' of Fig.~\ref{fig-pA_Sivers} (and its permutations) must be evaluated and included in the formulas; its contribution is potentially co-leading with the odderon mechanism of \eqref{dsigmaq3} and could significantly affect the form of the asymmetry.  This calculation would appear to be straightforward but cumbersome, since the ``lensing'' interaction with a spectator from the projectile does not benefit from the simplifications of eikonal high-energy kinematics.  It is interesting to conjecture, however, that for the diagram of Fig.~\ref{fig-pA_Sivers}, the color correlations which drive conventional lensing (see Fig.~\ref{relsign_v2}) may be washed out by the multiple scattering on independent nucleons.  Finally we note that to evaluate the odderon mechanism in the quasi-classical approximation, we also needed to take advantage of the large-$N_c$ limit in \eqref{dsigmaq1} so that the nonlinear interaction terms could be averaged separately in the target field (e.g. $\langle O_{zx} S_{xw} \rangle \rightarrow \langle O_{zx} \rangle \, \langle S_{xw} \rangle$).  This separate averaging eliminated any color correlations between the two terms, so that the only source of a preferred direction came from the gradients of the nuclear profile function.  If these products were evaluated in the quasi-classical approximation at finite $N_c$, the correlation between their colors could provide another source of preferred direction that may be less sensitive to the nonperturbative periphery of the target.  If all of these corrections are considered, then this odderon mechanism could be combined with the known small-$x$ evolution equations discussed in Sec.~\ref{sec-Evolution} and compared with experimental data.  The collision of polarized protons on nuclei $p^\uparrow A$ as considered in Chapter~\ref{chap-odderon} could be performed at the Relativistic Heavy Ion Collider (RHIC) at Brookhaven National Laboratory; presently, this is the only facility in the world capable of performing such a measurement.  However, as discussed in Chapter~\ref{chap-CGC}, the same physics of saturation should also be applicable to high-energy $p^\uparrow p$ collisions.

The quasi-classical factorization formulas \eqref{TMD1} and \eqref{DYTMD1} that were used in Chapter~\ref{chap-MVspin} to calculate the Sivers function of a heavy nucleus open the door to a complete decomposition of the nuclear TMD's into the nucleonic ones.  In principle, this is a straightforward generalization of the procedure implemented in Chapter~\ref{chap-MVspin} to project out the symmetries associated with the Sivers function, and it has the potential to yield an insightful picture of the spin and transverse-momentum structure of a saturated system.  It will be particularly interesting to determine the role of nuclear shadowing and orbital angular momentum in some of the ``exotic'' TMD's such as the Boer-Mulders function, worm-gear $g$ and $h$ functions, and the pretzelosity.  This picture of TMD structure in a dense system would complement other model systems such as a single quark target or the diquark model \eqref{DISS-DECOMP4}.  The analysis we have performed here in the Bjorken limit will also be interesting to extend to the Regge limit of Chapter~\ref{chap-CGC}; there new contributions may arise from processes that become dominant at small-$x$, such as the dipole channel of DIS pictured in Fig.~\ref{fig-Regge_DIS}.  Although the intent of analyzing a polarized nucleus in Chapter~\ref{chap-MVspin} was to simulate the effects of saturation in a polarized nucleon, it would also be possible to search for indications of the OAM and transversity channels in real nuclei as a proof of concept.  Such spin effects may be small for many nuclei, however, due to nucleon spin pairing as discussed in Sec.~\ref{toy_model}.  Generalizing the analysis of Chapter~\ref{chap-MVspin} beyond the resummation of $\alpha_s^2 A^{1/3}$ corresponding to 2 gluons per nucleon is likely to be very difficult, as many subleading effects must be considered such as correlations between nucleons and gradients of the nuclear profile.  Perhaps more interesting and practical would be the inclusion of quantum evolution effects as mentioned in Sec.~\ref{sec-Evolution}.  The semi-infinite Wilson lines that make up the dipole trace \eqref{dipole_def} are different from the infinite Wilson lines such as \eqref{Ddef} which are described by the usual small-$x$ evolution equations; it will be interesting to see whether they result in linear or nonlinear quantum evolution and to identify their diagrammatic origin.

The work presented in this Dissertation has been devoted to studying the interplay of spin, transverse momentum, and saturation in the structure of nucleons and other hadronic systems.  Much of the formalism is only strictly valid in asymptotic limits of high energy or large nuclei, and we have in several places considered the interaction with a heavy nucleus as a proxy for saturation effects in high-energy systems.  Yet these considerations are not solely academic in nature.  A next-generation \textit{electron-ion collider (EIC)} \cite{Accardi:2012qut} has been proposed and endorsed by the U.S. Nuclear Science Advisory Committee as ``embodying the vision for reaching the next QCD frontier.''  The EIC would collide high-energy electron and heavy-ion beams to provide access to an unprecedented depth of small-$x$ kinematics for $e+A$ collisions (on top of the $A^{1/3}$ enhancement from the heavy ions) at unprecedented luminosity, with the added potential of polarizing the colliding beams.  Similar facilities have been proposed in Europe and in China, and any of these discovery machines could be ready to begin taking data as early as 2025.  The EIC represents the future of nuclear physics, and its design specifications are precisely targeted to resolve the proton spin crisis, to unambiguously detect gluon saturation, and to study the three-dimensional structure of the nucleon deep into the intersection of spin and saturation physics.  The work we do now and over the next decade to contribute to the unification of the TMD and saturation paradigms will help lay the theoretical groundwork for the coming EIC era -- and for the new layer of mysteries that we will undoubtedly find in the process.

\backmatter
\bibliographystyle{JHEP} 
\bibliography{references_cumulative}

% ...



\end{document}